\documentclass[a4paper]{JHEP3} 

\JHEPspecialurl{http://jhep.sissa.it/JOURNAL/JHEP3.tar.gz}

\usepackage{epsfig,multicol,bbm}

\newcommand\fverb{\setbox\fverbbox=\hbox\bgroup\verb}
\newcommand\fverbdo{\egroup\medskip\noindent%
            \fbox{\unhbox\fverbbox}\ }
\newcommand\fverbit{\egroup\item[\fbox{\unhbox\fverbbox}]}
\newbox\fverbbox

\title{Diagrammar In Classical Scalar Field Theory}

\author{E. Cattaruzza$^{a}$, E. Gozzi$^{a,b}$ and A. Francisco Neto$^{c}$
  \\
    $^a$Department of Physics (Miramare Campus), University of Trieste, Strada Costiera 11,
    Miramare-Grignano 34014, Trieste\\
    $^b$INFN, Sezione di Trieste, Italy\\
    $^c$DEPRO, Campus Morro do Cruzeiro, UFOP, 35400-000 Ouro Preto MG, Brazil
    E-mail: \email{Enrico.Cattaruzza@gmail.com,gozzi@ts.infn.it,antfrannet@gmail.com}}

\received{\today}       
\accepted{\today}       

\abstract{In this paper we analyze perturbatively a $g \phi^4$
{\it classical} field theory with and without temperature. In
order to do that, we make use of a path-integral approach
developed some time ago for {\it classical} theories. It turns out
that the diagrams appearing at the  classical level are many more
than at the quantum level due to the presence of extra auxiliary
fields in the classical formalism. We shall show that several of
those diagrams cancel against each other due to a universal
supersymmetry present  in the classical path integral mentioned
above. The same supersymmetry allows the introduction of
super-fields and super-diagrams which considerably simplify the
calculations and make the {\it classical} perturbative
calculations almost "identical" formally to  the quantum ones.
Using the super-diagrams technique we develop the classical
perturbation theory up to third order. We conclude the paper with
a perturbative check of the fluctuation-dissipation theorem.}


\usepackage{epsfig}
\usepackage{bbold}
\usepackage{amscd}
\usepackage{amsmath}
\usepackage{graphicx}
\usepackage{psfrag}
\usepackage{amsfonts}
\usepackage{amssymb}
\usepackage{amsfonts}
\usepackage{amsmath}
\usepackage{pst-node}
\usepackage{mathrsfs}
\usepackage{graphics}
\usepackage{mathrsfs}
\usepackage{feynmp}

\newcommand{\ssc}{\scriptscriptstyle}
\newcommand{\be}{\begin{equation}}
\newcommand{\ee}{\end{equation}}
\newcommand{\bea}{\begin{eqnarray}}
\newcommand{\eea}{\end{eqnarray}}

\begin{document}
\unitlength = 1mm
\section{Introduction}\label{Introdu.}

Before starting this paper we would like to apologize with the
authors of the famous paper \cite{Hooftetal} who first made use of the ancient-italian
word ``{\it diagrammar}'' for having copied their idea. We used that same
word because we feel ``diagrammar'' express in the most complete
way the skills one has to develop to draw all possible Feynman
diagrams of a {\it classical} theory which are many more than the
quantum counterpart. We also think that the word express in a nice
way the feeling we had when we realized that the very many
classical diagrams could be encapsulated in few {\it super-diagrams}.

Let us go back to the paper. In the last fifteen years the physics
of heavy ion collisions has become one of the hottest topics in
high energy physics \cite{D'enterria}. The most interesting point
is that this may lead to the formation of state of matter known as
quark-gluon plasma (QGP)(for a review see ref. \cite{Yagietal}).
This topic has attracted the attention of many sectors of theoretical physics, last but not
least, even of strings using the duality AdS/QCD (for example see ref. \cite{Galo} and references therein).

We know that in the first instants after the (QGP) formation, the gluon
occupation numbers is going to be  very high \cite{GriBlai}, so one could
describe the system by a classical field theory \cite{McVenu},
analogously to what happens to QED which, when the photon
occupation number is high, can be described to a good degree of
accuracy by the Maxwell equations. A lot of work started in the
nineties in this direction \cite{AartsSmit,BuckJak,MueSon,Jeon}.
Most of these authors studied classical field theories by just
solving the Hamilton's equations of motion and developing from
there perturbative calculations. Only the author of ref.
\cite{Jeon} used a different route making use of a path-integral
approach for {\it classical } mechanics developed in the eighties
\cite{GoReuTha}. This is basically the functional counterpart of the
operatorial approach to classical mechanics developed in the
thirties by Koopman and von Neumann \cite{KvN}.  We know that the path-integral
is the most natural tool to use in order to develop the
perturbative calculations and the  associated Feynman diagrams. It
was so for the quantum path-integral (QPI) \cite{Feyetal}
and we will show that it will be the same for classical systems via the
classical path-integral (CPI). In this paper, like in ref.
\cite{AartsSmit} and Ref. \cite{Jeon}, we will limit ourselves to
a $g \phi^4$ theory.

The paper is organized as follows. In sec. \ref{Review} we will
briefly review the CPI in order to make this paper self-contained.
We shall show that, besides the phase-space variables, this
formalism makes use of extra auxiliary fields. These have a well
know physical meaning, being associated with the Jacobi fields
\cite{Schul} of the theory plus its symplectic conjugates and the
response fields \cite{DekerHakke}. The presence of all these
fields make the diagrammatic in the CPI rather complicated. In sec.
\ref{PertubnoT} we develop the formalism in the case of a field theory
without temperature and present the associated diagrams which, for
a point particle, have already been worked out in Ref.
\cite{PenMau}. In sec. \ref{PertubwithT} we develop the formalism
in the presence of temperature and make contact with the
calculations developed in Ref. \cite{AartsSmit} and Ref.
\cite{Jeon}. We will notice that there are cancelations among various diagrams.
This cancelation is due to some hidden symmetries present in the
CPI. One of these is a {\it supersymmetry} and this leads to the use of
 super-fields (for a review see Ref. \cite{West}) and
their  associated super-diagrams which we will introduce in sec.
\ref{PertubSuper}. In the same section, we will show how the
super-diagrams will allow us to reproduce the results of sec.
\ref{PertubwithT}, but with fewer diagrams. These
super-diagrams have the same vertices  as in the quantum case. This
is a consequence of the relation between the CPI and the QPI which
was  studied in Ref. \cite{AbriGoMau}. In sec. \ref{FDT} we conclude
the paper with a perturbative  analysis of the
fluctuation-dissipation theorem \cite{DekerHakke} using the
super-diagrams developed in sec. \ref{PertubSuper}. We confine the
detailed derivation of several results to few appendices.

\section{The Path-Integral for Classical Mechanics
(CPI).}\label{Review}

In the thirties Koopman and von Neumann \cite{KvN} developed an
Hilbert space and operatorial approach to classical statistical
mechanics. Their formalism is based on four postulates:

\begin{itemize}
\item[{\bf I.}]A state is given by an element $|\psi\rangle$ of an Hilbert
space.

\item[{\bf II.}]On this Hilbert space the operators $\hat{q}$ and $\hat{p}$
associated to the classical variables $q$ and $p$ commutes, i.e.
$[\hat{q},\hat{p}]=0$. If we indicate with $\varphi^{a}=(q,p)$,
$a=1,2$, then the simultaneous diagonalization of $\hat{q}$ and
$\hat{p}$ can be expressed as
$$
\hat{\varphi}^{a}|\varphi_0^a\rangle = \varphi^{a} |\varphi_0^a\rangle.
$$

\item[{\bf III.}] The evolution of $\psi(q,p)$ is given by the Liouville equation
\begin{equation}
i\frac{\partial\psi(q,p,t)}{\partial t}=\hat{\tilde{\mathcal H}}\psi(q,p,t), \label{9-2}
\end{equation} where $\hat{\tilde{\mathcal H}}\equiv
-i\partial_pH(q,p)\partial_q+i\partial_qH(q,p)\partial_p$ is
the Hamiltonian operator.

\item[{\bf IV.}]The Liouville probability density $\rho(q,p)$ is given by
\begin{equation}
\rho(q,p)=|\psi(q,p)|^2.
\end{equation}

\end{itemize}

As a consequence of the four postulates above also $\rho$ satisfies
the Liouville equation as it should be.

As this is an operatorial formalism there should be an associated
path-integral and, in fact, it is easy to build one
\cite{GoReuTha}. Let us start from the classical transition amplitude $\langle
\varphi^a,t|\varphi^a_0,t_0 \rangle$ which clearly is
\begin{equation}\label{CTransAmp}
\langle
\varphi^a,t|\varphi^a_0,t_0 \rangle=\tilde{\delta}
[\varphi^a-\phi^a_{\textrm{cl}}(t;\varphi_{\ssc 0},t_{\ssc 0})]
\end{equation} where $\phi_{{\rm cl}}$ is the solution at time $t$ of the Hamiltonian
equation of motion
\begin{equation}\label{HamEqMot}
\displaystyle \dot{\varphi}^a=\omega^{ab}\frac{\partial H}{\partial \varphi^b}
\end{equation} with initial configuration $\varphi_0$ at time $t_0$. In Eq. (\ref{HamEqMot}) $\omega^{ab}$
is the symplectic matrix \cite{AbraMar} and $H$ the Hamiltonian of the system.

Slicing the interval of time in Eq. (\ref{CTransAmp}) in $N$ small
intervals and replacing the functional Dirac delta on the r.h.s of
Eq. (\ref{CTransAmp}) with the expression
\begin{equation}\label{FuncDelta}
\displaystyle \widetilde{\delta}[\varphi^a-\phi_{cl}^a(t;\varphi_{\ssc 0},t_{\ssc 0})]=
\widetilde{\delta} [\dot{\varphi}^a-\omega^{ab}\partial_bH]\textrm{det}(
\delta_b^a\partial_t-\omega^{ad}\partial_d\partial_bH),
\end{equation} where $\det[\,\,]$ is a properly regularized functional determinant,
we get that Eq. (\ref{CTransAmp}) can be written as
\begin{equation}\label{CPIComp}
\langle
\varphi^a,t|\varphi^a_0,t_0 \rangle=
\int {\mathscr D}^{\prime\prime}
\varphi^a {\mathscr D}\lambda_a {\mathscr D} c^a {\mathscr D}\bar{c}_a
\; \textrm{exp} \left[i\int_{t_{\ssc 0}}^{t} d\tau \widetilde{\cal L}\right],
\end{equation} where ${\mathscr D}^{\prime\prime}\varphi$ indicates the integration
over all points, except the end points which are fixed, while the symbol
${\mathscr D}$ indicates that the end points are integrated over. The $\widetilde{\cal L}$ is
\begin{equation}\label{TildeL}
\displaystyle \widetilde{\cal L}=\lambda_a\dot{\varphi}^a+i\bar{c}_a\dot{c}^a
-\lambda_a\omega^{ab}\partial_bH-i\bar{c}_a\omega^{ad}\partial_d\partial_bHc^b.
\end{equation} The auxiliar variables $\lambda_a$ have been introduced to do the Fourier
transform of the Dirac delta on the r.h.s of Eq.
(\ref{FuncDelta}), while the Grassmanninan variables $c^a$ and
$\bar{c}_a$ are needed to exponentiate the determinant in Eq.
(\ref{FuncDelta}).

In Ref. \cite{GoReuTha} it was erroneously written that the
determinant above is one. This is not so, but it depends on the
regularization used to calculate it. If we discretize the time and
do the same for the matrix associated to the determinant, then this
turns out to be one, or a constant, if  we use the the
{\it It\^o calculus} \cite{EzaKlau} or pre-point discretization
\cite{Sakita}, while it turns out to be non-constant and
dependent on the ghost fields if we use the
{\it Stratonovich calculus} \cite{Naka}  or mid-point discretization. It is only
with this discretization that the Feynman rules which we derive
from the continuum theory are going to be the same as
those in the discretized case. This was shown in details in
\cite{Sato}. If we had used the It\^o calculus in which the
determinant is constant and could be disregarded, then the Feynman
rules derived from the continuous version of the non-Grassmannian
part could, in principle, not be the same as those derivable from the
discretized version. This is the reason we use the Stratonovich
prescription in which discretized and continuum Feynman rules coincide. In this
prescription, the determinant depends on the fields and so there is a
true interaction between the Grassmannian variables $c^a$,
$\bar{c}_a$ and $\varphi^a$. The vertex coming from this
interaction has to be considered in order  to get the right
results. This is not the only reason to keep the Grassmannian
variables $c^a$ and $\bar{c}_a$. Physically, they are the Jacobi
fields and their correlations give the Lyapunov exponents \cite{Liap}. These
quantities are related to the matrix elements of the determinant
above and to get these elements we need to couple $c^a$,
$\bar{c}_a$ with external currents as explained in ref.
\cite{GoReuTha}.

Let us now go back to Eq. (\ref{CPIComp}) and build the equivalent
operatorial formalism.
Given two variables $O_1$ and $O_2$ from the path-integral
(\ref{CPIComp}) we can derive the commutators defined as follows
\cite{Feyetal,Schul}
\begin{equation}\label{Commu}
\left[\hat{O}_1,\hat{O}_2\right]
=\lim_{\epsilon \rightarrow 0}
\langle O_1(t-\epsilon)O_2(t)-O_2(t-\epsilon)O_1(t)\rangle.
\end{equation} The symbol $\langle \rangle$ means the average under the
path-integral and sandwiched among any state. We have put a hat
$\hat{\,}$ on the variables on the l.h.s. of Eq. (\ref{Commu})
because they become operator via this definition. Applying
(\ref{Commu}) to the basic variables $\varphi^a$, $\lambda_a$,
$c^a$, $\bar{c}_a$ we get:
\begin{equation}\label{Commutators}
\left[\hat{\varphi}^a,\hat{\lambda}_b\right]=i\delta_b^a\,,\;\;\;
\left[\hat{c}^a,\hat{\bar{c}}_b\right]_+=\delta_b^a,
\end{equation} where by $\left[\;\;,\;\;\right]_+$ we indicate the anti-commutators.
All the other commutators and anti-commutators are zero in Eq.
(\ref{Commutators}). In particular,
$[\hat{\varphi}^a,\hat{\varphi}^b]=0$, which implies
$[\hat{q},\hat{p}]=0$, and this confirms that we are doing
classical mechanics. From the first equation in
(\ref{Commutators}) we get that a representation for $\lambda_a$
is
\begin{equation}\label{lambdader}
\hat{\lambda}_a=-i\frac{\partial }{\partial \varphi^a}
\end{equation} and this identifies $\hat{\lambda}_a$ with the response field
(see for example Ref. \cite{DekerHakke}).

Let us first write down the Hamiltonian associated to the
Lagrangian $\mathcal{\tilde{L}}$
\begin{equation}\label{tildeH}
\displaystyle \widetilde{\cal H}=
\lambda_a\omega^{ab}\partial_bH+i\bar{c}_a\omega^{ad}\partial_d\partial_bHc^b.
\end{equation}
Having obtained our operatorial formalism let us see if we can
recover the KvN approach. To do that, let us consider the first
piece of $\tilde{\mathcal{H}}$, which we call Bosonic one $(B)$
$$
\tilde{\mathcal{H}}_B=\lambda_a\omega^{ab}\partial_bH.
$$ Using Eq. (\ref{lambdader}) we get that $\tilde{\mathcal{H}}$ goes into the following operator
\begin{equation}
\begin{array}{lll}
\tilde{\mathcal{H}}_B\rightarrow
\hat{\tilde{\mathcal{H}}}_B&=&-i\omega^{ab}\partial_bH\frac{\partial}{\partial \varphi^a}\\
&=&-i\partial_p H\partial_q+i \partial_q H\partial_p \equiv \hat{L}
\end{array}
\end{equation} and so we obtain the Liouville operator. This confirms that, at
least for the Bosonic part, the path integral behind  the KvN
formalism is exactly the one presented in (\ref{CPIComp}). As we
said after the KvN postulates, both the elements of the KvN
Hilbert space $\psi(\varphi)$ and the probability densities
$\rho(\varphi)$ evolve via the same operator $\tilde{\mathcal
{H}}$. As a consequence also the path-integral for the evolution  of both $\psi$ and $\rho$
 is the same.  In this paper, as was  done  in the first papers
\cite{GoReuTha}, we will consider the averages taken with  the probabilities densities $\rho(\varphi)$ . The
reason for this choice is due to the fact, as it has been proven in Ref.
\cite{GoPa}, the Hilbert space of KvN is actually made of just
Dirac deltas on the phase-space points:
$\delta(\varphi-\varphi_0)=\langle \varphi|\varphi_0 \rangle$, and
the superposition principle does not act on these states. This
is natural in CM. So saying that we use a Hilbert space formalism in CM is
a rather "formal "statement, and for this reason we prefer to use
the probability densities.

Going now back to $\tilde{\mathcal{H}}$ the curious reader may ask what is
the mathematical meaning of
$\tilde{\mathcal{H}}$ if we keep also the Grassmannian
variables as in (\ref{tildeH}). Actually it has been proved in ref.
\cite{GoReuTha} that it becomes a generalization of the Liouville
operator know in the literature \cite{AbraMar} as the Lie derivative
of the Hamiltonian flow. This is easily proved once one realizes
that the $c^a$ behaves under symplectic transformation as basis of
differential forms of $\varphi^a$, while the $\bar{c}_a$ are the
symplectic duals \cite{AbraMar} to $c^a$ or, in other words, the
basis of the totally anti-symmetric tensors. All the geometrical
aspects of the path-integral for classical mechanics have been analyzed in details in
\cite{GoReuTha,Goetal} . In the second reference  it has been proved that the set of
variables $\varphi^a$, $c^a$, $\bar{c}_a$, $\lambda_a$
"coordinatize "a double-bundle over the phase-space \cite{AbraMar}.
The whole Cartan calculus \cite{AbraMar,Goetal} can be reproduced
using these variables and a set of seven charges present in this
formalism \cite{GoReuTha}. Among those charges we will mention
just two which are important for the rest of the paper
\begin{equation}
\begin{array}{ll}
&\hat{Q}_{H}=i \hat{c}^a\hat{\lambda}_a-\hat{c}^a\partial_aH\\
&\hat{\bar Q}_{H}=i\hat{\bar{c}}_a \omega^{ab}\hat{\lambda}_b+\hat{\bar{c}}_a\omega^{ab}\partial_bH.
\end{array}
\end{equation} They make up a universal  $N=2$ {\it supersymmetry} for any Hamiltonian system. In fact, they close on the
Hamiltonian
\begin{equation}
\left[\hat{Q}_{H},\hat{\bar Q}_{H}\right]=2i\hat{\tilde{\mathcal{H}}}.
\end{equation} We will see that, thanks to this supersymmetry, there will be many
cancelations among Feynman diagrams once we embark on the perturbation
theory. The geometrical meaning of these super-charges has also
been investigated: they are basically the exterior derivative on
the constant-energy surfaces of the phase-space. Or, in a more
mathematically technical language, they are associated to the equivariant
cohomology of the Hamiltonian field \cite{Wit}.

In this paper we will make use of another tool typical of
supersymmetry which is the one of  {\it super-fields} (for a review
see ref. \cite{West}). This is an object which allow us to put in
the same multiplet all the variables $\varphi^a$, $\lambda_a$,
$c^a$ and $\bar{c}_a$. To do that we have first to extend time $t$
to two Grassmannian partners of time $\theta$, $\bar{\theta}$.
The super-field  $\Phi^a(t,\theta,\bar{\theta})$
is a function of $t$, $\theta$,
$\bar{\theta}$ defined as
\begin{equation}
\Phi^a(t,\theta,\bar{\theta})
\equiv\varphi^a+\theta c^a+\bar{\theta}\omega^{ab}\bar{c}_b+i\theta \bar{\theta}\omega^{ab}
\lambda_b.
\end{equation} It is then possible to give a much simpler expression for the Lagrangian in
$\tilde{\mathcal{L}}$ (\ref{TildeL}). In fact, it is easy
\cite{AbriGoMau} to prove that
\begin{equation}\label{TildeLandL}
\tilde{\mathcal{L}}=i\int d\theta d\bar{\theta} L\left[\Phi^a\right]+(s.t),
\end{equation} where $L$ is the standard Lagrangian associated to the Hamiltonian $H$
of classical mechanics which we have used in the r.h.s of Eq.
(\ref{TildeL}) and $(s.t.)$ is a surface term.

Using the super-field it is also possible to give a generalization
of expression (\ref{CPIComp}), that is the following (for its
derivation see Ref. \cite{AbriGoMau}),
\begin{equation}\label{GenFuncComp}
\langle \Phi^a,t | \Phi^a_0,t_0\rangle
=\int {\mathscr D}^{\prime\prime}
\Phi^q {\mathscr D}
\Phi^p
\; \textrm{exp} \left\{i\int_{t_{\ssc 0}}^{t} id\tau d\theta d\bar{\theta} L\left[\Phi\right]\right\}.
\end{equation}

The indeces $p$ and $q$ indicate the first and second of the indices of $\varphi^a$ . They are called this way because we know that $\varphi^{1}=q$ and $\varphi^{2}=p$.

An expression similar to (\ref{GenFuncComp}) can also be written for the generating functional, which instead of being
the complicated object:
\begin{equation}\label{GenFuncComponent}
Z[J=0]=\int {\mathscr D}\phi^a {\mathscr D}\lambda_a
{\mathscr D}c^a {\mathscr D}\bar{c}_a
\textrm{exp} \left\{i\int_{t_{\ssc 0}}^{t} d\tau \mathcal{\tilde{L}}\right\}
\end{equation} can be written as:
\begin{equation}\label{GenFuncSuperPhi}
Z[J=0]
=\int {\mathscr D}\Phi^a
\; \textrm{exp} \left\{i\int_{t_{\ssc 0}}^{t} d\tau \int id\theta d\bar{\theta} L\left[\Phi\right]+(s.t.)\right\}.
\end{equation} Note how similar is (\ref{GenFuncSuperPhi}) to the quantum analogue:
\begin{equation}\label{GenFuncQPI}
Z_{QM}[J=0]
=\int {\mathscr D}
\varphi^a
\; \textrm{exp} \left\{\frac{i}{\hbar}\int_{t_{\ssc 0}}^{t} d\tau L\left[\varphi\right]\right\}.
\end{equation}  Basically $\varphi^a$ is replaced by $\Phi^a$ and $1/\hbar$ by $i\int d\theta d\bar{\theta}$ while the
functional weight is given in both cases by the standard
Lagrangian $L$ of classical mechanics.

The similarity between Eq. (\ref{GenFuncSuperPhi}) and Eq.
(\ref{GenFuncQPI}) has been studied in great details in Ref.
\cite{AbriGoMau}. In this paper this similarity will be used to
drastically simplify the perturbation theory which, instead of
making use of the complicated Lagrangian $\tilde{\mathcal{L}}$ of
(\ref{TildeL}) appearing in Eq. (\ref{GenFuncComponent}), will
make use of just $L$ appearing in Eq. (\ref{GenFuncSuperPhi})
borrowing many results from the quantum analog (\ref{GenFuncQPI}).

Before closing this section we would like to bring to the
attention of the reader the physics behind  the variables $c^a$.
We do that because people may consider them useless.
We have already made hints  before to their mathematical meaning, being that of basis for differential
forms (more details can be found in ref.\cite{Goetal}). Their physical
meaning has been explored in Ref. \cite{Liap} and it goes as follows. Let us
look at the equation of motion for $c^a$ which can be derived from the
$\mathcal{\tilde {L}}$ of  Eq. (\ref{TildeL})
\begin{equation}\label{EqMotc}
\dot{c}^a-\omega^{ab}\frac{\partial^2 H}{\partial \varphi^b \partial \varphi^d}c^d=0.
\end{equation} This equation is the same as the one satisfied by the first variations
$\delta \varphi^a$ :
\begin{equation}\label{EqMotdvarphi}
\delta\dot{\varphi}^a-\omega^{ab}\frac{\partial^2 H}{\partial \varphi^b \partial \varphi^d}\delta \varphi^d=0.
\end{equation} These first variations are also known as Jacobi fields.
(See for example Ref. \cite{Schul}.) So we can identify
$c^a\approx \delta \varphi^a$. The correlations of the Jacobi
fields are related to the Lyapunov exponents  of
chaotic systems:
$$
\lim_{t \rightarrow \infty}\langle \delta\varphi(0) \delta \varphi(t)\rangle
\sim e^{-\lambda t}$$
where ($\lambda$) is the highest Lyapunov exponent of the system.
This is equivalent to saying  that the correlations of $c^a, {\bar c}_a$ gives the Lyapunov exponents\cite{Liap}
So we see that the $c^a, {\bar c}_a$  are crucial ingredients to get a full
information on the dynamical system \cite{Liap} we are studying. We talk about correlations between
$c^a$ and $ {\bar c}_a$  because the correlations among only $c^a$ gives zero\cite{GozziFDT}\cite{Liap}\cite{Reuter}. The $ {\bar c}_a$  are just their symplectic dual and,
once they are multiplied by the symplectic matrix, they can also be identified with the Jacobi fields.
So the correlations among $c^a, {\bar c}_a$  are  also equivalent to correlations among jacobi fields.
Of course, the Lyapunov exponents could be obtained in many other
ways \cite{Arno} without using the $c^a, {\bar c}_a$ , but we find this manner
rather elegant. Actually, we know that classical mechanics could
be built by just using the $\varphi^a$, and so the variables
$\lambda_a$, $c^a$ and $\bar{c}_a$ are redundant variables and
this redundancy is signaled by the symmetries present in
$\tilde{\mathcal{L}}$ like the supersymmetry and other invariances.
But, as usual in physics, the redundancies and the symmetries make
things more elegant and allows the use of all the tools that
group theory put at our disposal. In our case,
if we had not used all the variables, we would not be able to
build the super-field and use the expression
(\ref{GenFuncSuperPhi}) for the perturbation theory.

Before starting the next section we should introduce for
completeness a new representation of the commutator algebra
(\ref{Commutators}). More details can be found in Ref.
\cite{AbriGoMau}. This new representation will be very important
in order to do perturbation theory using super-fields.
We had seen that $\mathcal{\tilde L}$ in Eq. (\ref{TildeLandL})
and $L[\Phi]$ differ by a surface term $(s.t)$ and the same will
happen at the level of generating functionals (\ref{GenFuncComp})
and (\ref{GenFuncSuperPhi}). In order to get rid of that surface
term the trick \cite{AbriGoMau} is to change the representation in
(\ref{lambdader}) that we derived from the commutator
(\ref{Commutators}). We could, for example, represent $\hat{q}$
and $\hat{\lambda}_p$ as multiplicative operators and $\hat{p}$
and $\hat{\lambda}_q$ as derivative ones
\begin{equation}
\begin{array}{ll}
&\hat{p}=i\displaystyle\frac{\partial}{\partial \lambda_p}\vspace{2mm}\\
&\hat{\lambda}_q =-i\displaystyle\frac{\partial}{\partial q}.
\end{array}\end{equation} Analogously we can proceed for the Grassmannian
variables: represent  $\hat{\bar{c}}^q$ and $\hat{\bar{c}}_p$ as
multiplicative operators and $\hat{c}^p$ and $\hat{\bar{c}}_q$ as
derivative ones:
\begin{equation}
\begin{array}{ll}
&\hat{c}^p=i\displaystyle\frac{\partial}{\partial \bar{c}_p}\vspace{2mm}\\
&\hat{\bar{c}}_q=\displaystyle\frac{\partial}{\partial c^q}.
\end{array}\end{equation} The generalized states \cite{AbriGoMau} in the extended Hilbert space, including the forms, will then
be $\langle q,\lambda_p,c^q,\bar{c}_p|$. Note that the variables
$q$, $\lambda_p$, $c^q$, $\bar{c}_p$ are exactly those which enter
the super-field $\Phi^q$
\begin{equation}\label{Superfield}
\Phi^q
=q+\theta c^q+\bar{\theta}\bar{c}_p+i\theta \bar{\theta}\lambda_p.
\end{equation} This   is crucial in order to give a
complete super-field representation of our theory without any
surface terms. We invite the reader to master the details contained in ref.\cite{AbriGoMau} before embarking in the rest of this paper.
\section{Perturbation Theory Without Temperature.}\label{PertubnoT}

What we will do in this section for a $g \phi^4$ scalar field
theory, has been partly been done for a point particle in Ref.
\cite{PenMau}. We use the word ``partly'' because those authors
neglected the Grassmannian variables $c^a$ and $\bar{c}_a$. So
their analysis apply only to the zero-form sector of the theory.
We called this sector this way because, as indicated before, the $c^a$ are the basis of the differential forms \cite{AbraMar},\cite{Goetal}.

Let us start with a scalar field theory whose Hamiltonian is
\begin{equation}\label{Hamilton}
H=\int d^4 x\left[\frac{\pi}{2}
+\frac{\left(\nabla \phi\right)^2}{2}
+\frac{m^2 \phi^2}{2}+g\frac{\phi^4}{4!}
\right]
\end{equation}
where $\pi(x)$ is the momentum conjugate to the field $\phi(x)$.
We will indicate with  $\mathscr H$ the Hamiltonian density:
\begin{equation}
\mathscr H(x)=\frac{\pi(x)}{2}
+\frac{\left[\nabla \phi(x)\right]^2}{2}
+\frac{m^2 \phi^2(x)}{2}+g\frac{\phi^4(x)}{4!}
\end{equation}
Let us next build the generating functional associated to the
classical path-integral of the previous section, but we shall
choose only those trajectories that start from a fixed point in
phase-space $\varphi_i=(\phi_i,\pi_i)$ and we will not average
over this initial configuration. The expression is:
\begin{equation}\label{Z}
\begin{array}{lll}
&&\mathcal{Z}_{\varphi_i}
[J_{\phi},J_{\lambda^{\pi}},\bar{J}_{c^{\phi}},J_{\bar{c}_{\pi}}]\vspace{2mm}\equiv\\
&&\equiv\displaystyle\int {\mathscr D}^{\prime}\varphi{\mathscr D}\lambda{\mathscr D}c{\mathscr D}\bar{c}
\exp\left\{i\mathcal{\tilde{S}}+i\displaystyle\int_{t_i}^{t_f}dt \displaystyle\int d^3 x
\left[J_{\phi}\phi+J_{\lambda_{\pi}}\lambda_{\pi}
-i\bar{J}_{c^{\phi}}c^{\phi}-i\bar{c}_{\pi}J_{\bar{c}_{\pi}}\right]\right\}
\end{array}\end{equation} where
\begin{equation}
\mathcal{\tilde{S}}
=\displaystyle\int_{t_i}^{t_f} dt \displaystyle\int d^3x\left(\lambda_a\dot{\varphi}^a+i\bar{c}_a\dot{c}^a
-\lambda_a\omega^{ab}\partial_bH-i\bar{c}_a\omega^{ad}\partial_d\partial_bHc^b\right)
\end{equation} with the $H$ given by Eq. (\ref{Hamilton}) and the symbol ${\mathscr D}^{\prime}$
indicates the ``sum'' over all the trajectories starting from a fixed
$\varphi_i$ and we will not integrate over this initial configuration.
In the $\lambda$, $c$, $\bar{c}$ we have put as
indices $\phi$, $\pi$ to indicate either the first or the second
set of indices ``$a$'' as in postulate ${\bf II} $ of section \ref{Review}.
To the currents we have  coupled only the ``configuration
variables'' in the representation (\ref{Superfield}) where $\phi$,
$\lambda_{\pi}$, $c^{\phi}$, $\bar{c}_{\pi}$ are multiplicative
operators. We did that because, as it is clear from
(\ref{Superfield}), it will be straightforward to pass to the
super-field formulation. The peculiar combination of ``$\pm$'' and
``$i$'' in the coupling of current and fields (\ref{Z}) is also
related to the fact that eventually we also want to build a
supercurrent to couple it to the superfield.

As it is usually done, let us
divide $\mathcal{\tilde S}$ into a free part:
\begin{equation}
\begin{array}{lll}
\mathcal{\tilde S}_0
&\equiv&\displaystyle\int_{t_i}^{t_f}dt\displaystyle\int d^3x
\left[\lambda_{a}\dot{\phi}^a-\lambda_{\phi}\pi
-\lambda_{\pi}\left(\nabla^2-m^2\right)\phi\right.\vspace{2mm}\\
&&\left.+i\bar{c}_a\dot{c}^a-i\bar{c}_{\phi}c^{\pi}
-i\bar{c}_{\pi}\left(\nabla^2-m^2\right)c^{\phi}\right]
\end{array}
\end{equation} and an interaction one:
\begin{equation}
\mathcal{\tilde S}_{V}
\equiv\int d^4x\left(\frac{g}{3!}\lambda_{\pi}\phi^3+\frac{iq}{2!}\bar{c}_{\pi}\phi^2c^{\phi}\right),
\end{equation} where we indicate with $\int d^4 x$
the expression $\int_{t_i}^{t_f}dt \int d^3 x$. To develop the
perturbation theory, as it is usually done in quantum field
theory, in the $\mathcal{\tilde S}_{V}$ of $\mathcal{Z}$ we shall
replace the fields with the derivative operators with respect to
the associated currents as follows:
\begin{equation}
\phi \rightarrow \frac{1}{i}\frac{\delta}{\delta J_{\phi}},\,
\lambda_{\pi} \rightarrow \frac{1}{i}\frac{\delta}{\delta J_{\lambda_{\pi}}},\,
c^{\phi} \rightarrow \frac{\delta}{\delta \bar{J}_{c^{\phi}}},\,
\bar{c}_{\pi} \rightarrow -\frac{\delta}{\delta J_{\bar{c}_{\pi}}}
\end{equation} so we get:
\begin{equation}
\mathcal{\tilde S}_{V}\left[\frac{1}{i}\frac{\delta}{\delta J_{\phi}},
\frac{1}{i}\frac{\delta}{\delta J_{\phi}},\frac{\delta}{\delta \bar{J}_{c^{\phi}}},
-\frac{\delta}{\delta J_{\bar{c}_{\pi}}}\right]
\equiv\int d^4x\left(\frac{g}{3!}\frac{\delta}{\delta J_{\lambda_{\pi}}}
\frac{\delta^3}{\delta J_{\phi}^3}+\frac{ig}{2!}\frac{\delta}{\delta J_{\bar{c}_{\pi}}}
\frac{\delta^2}{\delta J_{\phi}^2}\frac{\delta}{\delta \bar{J}_{c^{\phi}}}\right)
\end{equation} It is then easy to write $\mathcal{Z}_{\varphi_i}
[J_{\phi},J_{\lambda^{\pi}},\bar{J}_{c^{\phi}},J_{\bar{c}_{\pi}}]$
in Eq. (\ref{Z}) as follows:
\begin{equation}\label{Z=expSVZ0}
\mathcal{Z}_{\varphi_i}
[J_{\phi},J_{\lambda_{\pi}},\bar{J}_{c^{\phi}},J_{\bar{c}_{\pi}}]
=\exp\left\{i\mathcal{\tilde{S}}_V\left[\frac{1}{i}\frac{\delta}{\delta J_{\phi}},
\frac{1}{i}\frac{\delta}{\delta J_{\phi}},\frac{\delta}{\delta \bar{J}_{c^{\phi}}},
-\frac{\delta}{\delta J_{\bar{c}_{\pi}}}\right]\right\}\mathcal{Z}_{\varphi_i}^{(0)}
[J_{\phi},J_{\lambda_{\pi}},\bar{J}_{c^{\phi}},J_{\bar{c}_{\pi}}]
\end{equation} where $\mathcal{Z}_{\varphi_i}^{(0)}$ is the free generating functional
built out of the $\mathcal{\tilde{S}}_0$. It is easy to prove that
this part $\mathcal{Z}_{\varphi_i}^{(0)}$ can be factorized into a
"Bosonic" (B) part for the variables $\varphi$ and $\lambda$ and a
"Fermionic "one (F) for the Grassmannian variables $c$, $\bar{c}$:
\begin{equation}\label{Z0}
\mathcal{Z}_{\varphi_i}^{(0)}
[J_{\phi},J_{\lambda_{\pi}},\bar{J}_{c^{\phi}},J_{\bar{c}_{\pi}}]
=\mathcal{Z}_{(B)\;\varphi_i}^{(0)}
[J_{\phi},J_{\lambda_{\pi}}]
\mathcal{Z}_{(F)}^{(0)}
[\bar{J}_{c^{\phi}},J_{\bar{c}_{\pi}}]
\end{equation} where
\begin{equation}\label{ZB}
\begin{array}{lll}
&&\mathcal{Z}_{(B)\;\varphi_i}^{(0)}
[J_{\phi},J_{\lambda_{\pi}}]\vspace{2mm}\\
&&=\displaystyle\int {\mathscr D}^{\prime}\varphi{\mathscr D}\lambda
\exp\left\{i\displaystyle\int_{t_i}^{t_f}dt\displaystyle\int d^3x
\left[\lambda_{a}\dot{\phi}^a-\lambda_{\phi}\pi
-\lambda_{\pi}\left(\nabla^2-m^2\right)\phi+
J_{\phi}\phi+J_{\lambda_{\pi}}\lambda_{\pi}\right]\right\}
\end{array}\end{equation} and
\begin{equation}\label{ZF}
\begin{array}{lll}
&&\mathcal{Z}_{(F)}^{(0)}
[\bar{J}_{c^{\phi}},J_{\bar{c}_{\pi}}]\vspace{2mm}\\
&&=\displaystyle\int {\mathscr D}c{\mathscr D}\bar{c}
\exp\left\{\displaystyle\int_{t_i}^{t_f}dt\displaystyle\int d^3x
\left[i\bar{c}_a\dot{c}^a-i\bar{c}_{\phi}c^{\pi}
-i\bar{c}_{\pi}\left(\nabla^2-m^2\right)c^{\phi}
-i\bar{J}_{c^{\phi}}c^{\phi}-i\bar{c}_{\pi}J_{\bar{c}_{\pi}}\right]\right\}
\end{array}\end{equation} All the fields enter at most in a quadratic form. It is easy to integrate
them out in Eq. (\ref{ZB}) as it has been done in ref.
\cite{PenMau}. The result is:
\begin{equation}
\mathcal{Z}_{(B)\;\varphi_i}^{(0)}
[J_{\phi},J_{\lambda_{\pi}}]
=\exp\left[i\int d^4x J_{\phi}(x)
\phi_0(x)+i\int d^4x d^4x'J_{\phi}(x)G_{R}(x-x')J_{\lambda_{\pi}}(x')\right].
\end{equation} where $\phi_0(x)$ is a solution to the equation
\begin{equation}
\left(\square+m^2\right)\phi=0
\end{equation} and $G_{R}$ is the retarded propagator satisfying
\begin{equation}
\left(\square+m^2\right)G_{R}=-\delta(x).
\end{equation} Its expression is:
\begin{equation}
G_{R}(x)=\displaystyle\int \frac{d^4 p}{(2\pi)^4}\frac{e^{-ip.x}}{(p^0+i\epsilon)^2-\vec{p}^2-m^2}.
\end{equation} Also for the $\mathcal{Z}_{(F)}^{(0)}$ we can easily integrate out the fields
(for details see Ref. \cite{GoPe})and obtain:
\begin{equation}
\mathcal{Z}_{(F)}^{(0)}
[\bar{J}_{c^{\phi}},J_{\bar{c}_{\pi}}]
=\exp\left[\int d^4x d^4x'\bar{J}_{c^{\phi}}(x)G_{R}(x-x')J_{\bar{c}_{\pi}}(x')\right].
\end{equation} We have now all the tools to obtain all the propagators.

Let us start at the zero order in perturbation theory and we will
draw next to them the diagrams:
\begin{equation}\label{<phi>(0)}
\begin{array}{lll}
\langle \phi(x)\rangle^0
&=&\displaystyle\frac{1}{i}\displaystyle\frac{\delta}{\delta J_{\phi}}\mathcal{Z}^{(0)}_{\varphi_i}
[J_{\phi},J_{\lambda_{\pi}},\bar{J}_{c^{\phi}},J_{\bar{c}_{\pi}}]
|_{J_{\phi},J_{\lambda_{\pi}},\bar{J}_{c^{\phi}},J_{\bar{c}_{\pi}}=0}\vspace{2mm}\\
&=&\frac{1}{i}\frac{\delta}{\delta J_{\phi}}\mathcal{Z}^{(0)}_{B\;\varphi_i}
[J_{\phi},J_{\lambda_{\pi}}]|_{J_{\phi},J_{\lambda_{\pi}}=0}\vspace{2mm}\\
&=&\phi_0(x)
\equiv \,\,\,
\begin{minipage}[c]{7mm}
\vspace{0.6cm}\begin{fmffile}{DiagPag25.1}
\begin{fmfgraph*}(7,3)
\fmfkeep{DiagPag25.1}
\fmfleft{i1}
\fmfright{o1}
\fmfdot{i1}
\fmf{plain}{i1,o1}
\fmfv{label=$x$,label.angle=-90}{i1}
\end{fmfgraph*}
\end{fmffile}
\end{minipage}
\end{array}
\end{equation}
\vspace{-0.5cm}
\begin{equation}\label{<phiphi>(0)}
\langle \phi(x_2)\phi(x_1)\rangle^0=\phi_0(x_2)\phi_0(x_1)\equiv \,\,\,\begin{minipage}[c]{16mm}
\vspace{0.6cm}\begin{fmffile}{DiagPag25.2}
\begin{fmfgraph*}(16,1)
\fmfkeep{DiagPag25.2}
\fmfleft{i1}
\fmfright{o1}
\fmfdot{i1}
\fmfdot{o1}
\fmf{phantom,tag=1}{i1,o1}
\fmfposition
\fmfipath{p[]}
\fmfiset{p1}{vpath1(__i1,__o1)}
\fmfi{plain}{subpath (0,length(p1)*0.43) of p1}
\fmfi{plain}{subpath (length(p1)*0.57,length(p1)) of p1}
\fmfv{label=$x_1$,label.angle=-90}{i1}
\fmfv{label=$x_2$,label.angle=-90}{o1}
\end{fmfgraph*}
\end{fmffile}
\end{minipage}
\end{equation}
\begin{equation}\label{lambdaphi}
\begin{array}{lll}
\langle \lambda_{\pi}(x_2)\phi(x_1)\rangle^0
&=&\left(\displaystyle\frac{1}{i}\right)^2
\displaystyle\frac{\delta}{\delta J_{\lambda_{\pi}}(x_2)}
\displaystyle\frac{\delta}{\delta J_{\phi}(x_1)}\mathcal{Z}^{(0)}_{\varphi_i}
[J_{\phi},J_{\lambda_{\pi}},\bar{J}_{c^{\phi}},J_{\bar{c}_{\pi}}]
|_{J_{\phi},J_{\lambda_{\pi}},\bar{J}_{c^{\phi}},J_{\bar{c}_{\pi}}=0}\vspace{2mm}\\
&=&-iG_R(x_1-x_2)=G_{\lambda_{\pi}\phi}\equiv \,\,\,\begin{minipage}[c]{16mm}
\vspace{0.6cm}\begin{fmffile}{DiagPag25.4}
\begin{fmfgraph*}(16,1)
\fmfkeep{DiagPag25.4}
\fmfleft{i1}
\fmfright{o1}
\fmfdot{i1}
\fmfdot{o1}
\fmf{phantom,tag=1}{i1,o1}
\fmfposition
\fmfipath{p[]}
\fmfiset{p1}{vpath1(__i1,__o1)}
\fmfi{dashes}{subpath (0,length(p1)*0.5) of p1}
\fmfi{plain}{subpath (length(p1)*0.5,length(p1)) of p1}
\fmfv{label=$x_1$,label.angle=-90}{i1}
\fmfv{label=$x_2$,label.angle=-90}{o1}
\end{fmfgraph*}
\end{fmffile}
\end{minipage}
\end{array}
\end{equation}
In Eq. (\ref{lambdaphi}) we borrow from ref. \cite{AartsSmit}.
the notation of full-dashed propagator .
In Eq. (\ref{<phiphi>(0)}) we have drawn  a diagram made of two pieces
not linked to each other, because they are actually the product of
two separated fields and not a propagator. We will have them
soldered to each other  once we use the temperature to average over the initial
conditions.

Analogously to Eq. (\ref{lambdaphi}) we can calculate
\begin{equation}\label{philambda}
\begin{array}{lll}
\langle \phi(x_2)\lambda_{\pi}(x_1)\rangle^0
&=&-iG_R(x_2-x_1)=G_{\phi\lambda_{\pi}}\vspace{2mm}\vspace{2mm}\equiv \,\,\,\begin{minipage}[c]{16mm}
\vspace{0.6cm}\begin{fmffile}{DiagPag25.3}
\begin{fmfgraph*}(16,1)
\fmfkeep{DiagPag25.3}
\fmfleft{i1}
\fmfright{o1}
\fmfdot{i1}
\fmfdot{o1}
\fmf{phantom,tag=1}{i1,o1}
\fmfposition
\fmfipath{p[]}
\fmfiset{p1}{vpath1(__i1,__o1)}
\fmfi{plain}{subpath (0,length(p1)*0.40) of p1}
\fmfi{dashes}{subpath (length(p1)*0.40,length(p1)) of p1}
\fmfv{label=$x_1$,label.angle=-90}{i1}
\fmfv{label=$x_2$,label.angle=-90}{o1}
\end{fmfgraph*}
\end{fmffile}
\end{minipage}
\end{array}
\vspace{-0.8cm}
\end{equation}

The correlations among Grassmannian variables give:
\begin{equation}\label{barcc}
\begin{array}{lll}
\langle \bar{c}_{\pi}(x_2)c^{\phi}(x_1)\rangle^0
&=&-
\displaystyle\frac{\delta}{\delta J_{\bar{c}_{\pi}}(x_2)}
\displaystyle\frac{\delta}{\delta \bar{J}_{c^{\phi}}(x_1)}\mathcal{Z}^{(0)}_{F}
[\bar{J}_{c^{\phi}},J_{\bar{c}_{\pi}}]
|_{\bar{J}_{c^{\phi}},J_{\bar{c}_{\pi}}=0}\vspace{2mm}\\
&=&G_R(x_1-x_2)=G_{\bar{c}_{\pi}c^{\phi}}\equiv \,\,\,\begin{minipage}[c]{16mm}
\vspace{0.6cm}\begin{fmffile}{DiagPag25.5}
\fmfset{arrow_len}{3mm}
\begin{fmfgraph*}(16,1)
\fmfkeep{DiagPag25.5}
\fmfleft{i1}
\fmfright{o1}
\fmfdot{i1}
\fmfdot{o1}
\fmfv{label=$x_1$,label.angle=-90}{i1}
\fmfv{label=$x_2$,label.angle=-90}{o1}
\fmf{dots_arrow}{o1,i1}
\end{fmfgraph*}
\end{fmffile}
\end{minipage}
\end{array}\end{equation}
 and
\begin{equation}\label{cbarc}
\begin{array}{lll}
\langle c^{\phi}(x_2)\bar{c}_{\pi}(x_1)\rangle^0
&=&
\displaystyle\frac{\delta}{\delta \bar{J}_{c^{\phi}}(x_2)}
\left[-\displaystyle\frac{\delta}{\delta J_{\bar{c}_{\pi}}(x_1)}\right]\mathcal{Z}^{(0)}_{F}
[\bar{J}_{c^{\phi}},J_{\bar{c}_{\pi}}]
|_{\bar{J}_{c^{\phi}},J_{\bar{c}_{\pi}}=0}\vspace{2mm}\\
&=&-G_R(x_2-x_1)=G_{c^{\phi}\bar{c}_{\pi}}\equiv \,\,\,\begin{minipage}[c]{16mm}
\vspace{0.6cm}\begin{fmffile}{DiagPag25.6}
\fmfset{arrow_len}{3mm}
\begin{fmfgraph*}(16,1)
\fmfkeep{DiagPag25.6}
\fmfleft{i1}
\fmfright{o1}
\fmfdot{i1}
\fmfdot{o1}
\fmfv{label=$x_1$,label.angle=-90}{i1}
\fmfv{label=$x_2$,label.angle=-90}{o1}
\fmf{dots_arrow}{i1,o1}
\end{fmfgraph*}
\end{fmffile}
\end{minipage}.
\end{array}\end{equation}
 Above we have adopted the convention
of putting an arrow which points from $c$ to $\bar{c}$ (see Ref.
\cite{GoPe}).

As both the propagators, of the $\phi \phi$ and $c\bar{c}$, are
related to the $G_{R}$ we will sometimes put an index ``(F)''
or ``(B)'' to indicate if it comes from the Bosonic fields or
``Fermionic'' (Grassmannian) ones. The fact that the two
propagators are equal (modulo $i$) to each other is due to the
supersymmetry present in this formalism \cite{GoReuTha}. We will
see other manifestation of it later on in several cancelations
among diagrams.

Let us now derive the rule for the vertices.
Expanding $e^{i\mathcal{\tilde{S}}_V}$ in Eq. (\ref{Z=expSVZ0}) to
the first order in $g$, we get :
\begin{equation}
i\mathcal{\tilde S}_{V}\left[\frac{1}{i}\frac{\delta}{\delta J_{\phi}},
\frac{1}{i}\frac{\delta}{\delta J_{\phi}},\frac{\delta}{\delta \bar{J}_{c^{\phi}}},
-\frac{\delta}{\delta J_{\bar{c}_{\pi}}}\right]
\equiv\int d^4x\left[ig\left(\frac{1}{3!}\frac{\delta}{\delta J_{\lambda_{\pi}}}
\frac{\delta^3}{\delta J_{\phi}^3}\right)-g\left(\frac{1}{2!}\frac{\delta}{\delta J_{\bar{c}_{\pi}}}
\frac{\delta^2}{\delta J_{\phi}^2}\frac{\delta}{\delta \bar{J}_{c^{\phi}}}\right)\right].
\end{equation} Keeping account of the symmetry factors we get the following rules
for the vertices
:\begin{equation}\label{vertexrules}
\begin{array}{ll}
\begin{minipage}[c]{16mm}
\vspace{0.1cm}\begin{fmffile}{DiagPag28.1}
\begin{fmfgraph*}(15,15)
\fmfkeep{DiagPag28.1}
\fmfleft{i1,i2}
\fmfright{o1,o2}
\fmf{dashes}{i1,v}
\fmf{plain}{v,o2}
\fmf{plain}{i2,v,o1}
\fmfdot{v}
\fmfv{label=$y$,label.angle=180}{v}
\end{fmfgraph*}
\end{fmffile}
\end{minipage}\,\, = i\,g \int d^4y,\quad&
\begin{minipage}[c]{16mm}
\vspace{0.1cm}\begin{fmffile}{DiagPag28.2}
\fmfset{arrow_len}{3mm}
\begin{fmfgraph*}(15,15)
\fmfkeep{DiagPag28.2}
\fmfleft{i1,i2}
\fmfright{o1,o2}
\fmf{dots_arrow}{i1,v}
\fmf{dots_arrow}{v,o1}
\fmf{plain}{v,o2}
\fmf{plain}{i2,v}
\fmfdot{v}
\fmfv{label=$y$,label.angle=180}{v}
\end{fmfgraph*}
\end{fmffile}
\end{minipage}\,\, = -g \int d^4y\\
\end{array}\end{equation}
where the continuos line refer to the field $\phi$, the dashed one to the field $\lambda$ and the dotted one to $c$ or ${\bar c}$.

Let us calculate the first order correction to the expectation
value of the field $\langle \phi_0(x)\rangle$, i.e. $\langle
\phi_0(x)\rangle$ (the super-index $(1)$ on $\langle \phi\rangle$
indicates the first order correction). Using the vertices and the
diagrams we get (where $y$ is integrated over)
\begin{equation}\label{<phi>(1)}
\begin{array}{lll}
\langle \phi(x)\rangle^{(1)}_{\varphi_i}=\Bigg[\,\cfrac{1}{2!}\,\,\,
\begin{minipage}[c]{32mm}
\vspace{0.5cm}
\begin{fmffile}{DiagPag29.1}
\begin{fmfgraph*}(32,10)
\fmfkeep{DiagPag29.1}
\fmfleft{i1}
\fmfright{o1}
\fmfdot{i1}
\fmfv{label=$x$,label.angle=-90}{i1}
\fmf{phantom}{i1,v,o1}
\fmf{phantom,tag=3}{v,o1}
\fmfdot{v}
\fmf{phantom,tag=1}{i1,v}
\fmfv{label=$y$,label.angle=-90}{v}
\fmfposition
\fmfipath{p[]}
\fmfiset{p1}{vpath1(__i1,__v)}
\fmfi{plain}{subpath (0,length(p1)*0.40.) of p1}
\fmfi{plain}{subpath (length(p1)*0.6,length(p1)) of p1}
\fmf{phantom,tension=0.9,tag=2}{v,v}
\fmfposition
\fmfiset{p2}{vpath2(__v,__v)}
\fmfi{plain}{subpath (0,length(p2)/2.) of p2}
\fmfi{dashes}{subpath (length(p2)/2,length(p2)) of p2}
\fmfposition
\fmfiset{p3}{vpath3(__v,__o1)}
\fmfi{plain}{subpath (0,length(p3)*0.50) of p3}
\end{fmfgraph*}
\end{fmffile}
\end{minipage}
\hspace{-0.7cm}+\cfrac{1}{3!}\,&
\begin{minipage}[c]{32mm}
\vspace{0.5cm}
\begin{fmffile}{DiagPag29.2}
\begin{fmfgraph*}(32,20)
\fmfkeep{DiagPag29.2}
\fmfleft{i1,i2,i3}
\fmfright{o1,o2,o3}
\fmf{phantom}{i1,v1,o1}
\fmf{phantom}{i2,v2,o2}
\fmf{phantom}{i3,v3,o3}
\fmfdot{i2}
\fmfdot{v2}
\fmfv{label=$x$,label.angle=-90}{i2}
\fmfv{label=$y$,label.angle=-90}{v2}
\fmffreeze
\fmf{phantom,tag=1}{i2,v2}
\fmf{phantom,tag=2}{v2,o1}
\fmf{phantom,tag=3}{v2,o2}
\fmf{phantom,tag=4}{v2,o3}
\fmfposition
\fmfipath{p[]}
\fmfiset{p1}{vpath1(__i2,__v2)}
\fmfi{plain}{subpath (0,length(p1)*0.40) of p1}
\fmfi{dashes}{subpath (length(p1)*0.4,length(p1)) of p1}
\fmfiset{p2}{vpath2(__v2,__o1)}
\fmfi{plain}{subpath (0,length(p2)*0.7) of p2}
\fmfiset{p3}{vpath3(__v2,__o2)}
\fmfi{plain}{subpath (0,length(p3)*0.7) of p3}
\fmfiset{p4}{vpath4(__v2,__o3)}
\fmfi{plain}{subpath (0,length(p4)*0.7) of p4}
\end{fmfgraph*}
\end{fmffile}
\end{minipage}
\hspace{-0.2cm}+\cfrac{1}{2!}\,\,\,&
\begin{minipage}[c]{32mm}
\vspace{0.5cm}
\begin{fmffile}{DiagPag29.3}
\fmfset{arrow_len}{2mm}
\begin{fmfgraph*}(32,10)
\fmfkeep{DiagPag29.3}
\fmfleft{i1}
\fmfright{o1}
\fmfdot{i1}
\fmf{phantom}{i1,v,o1}
\fmf{phantom,tag=1}{i1,v}
\fmf{phantom,tag=2}{v,o1}
\fmfdot{v}
\fmfv{label=$x$,label.angle=-90}{i1}
\fmfv{label=$y$,label.angle=-90}{v}
\fmfposition
\fmfipath{p[]}
\fmfiset{p1}{vpath1(__i1,__v)}
\fmfi{plain}{subpath (0,length(p1)*0.40) of p1}
\fmfi{plain}{subpath (length(p1)*0.6,length(p1)) of p1}
\fmf{dots_arrow,tension=0.9,tag=2}{v,v}
\fmfiset{p2}{vpath2(__v,__o1)}
\fmfi{plain}{subpath (0,length(p2)*0.5) of p2}
\end{fmfgraph*}
\end{fmffile}
\end{minipage}
\end{array}\hspace{-0.5cm}\Bigg]
\end{equation}
The analytic expression of the first diagram above is
\vspace{0.6cm}
\begin{equation}\label{<phi>(1)1}
\parbox{33mm}{
\fmfreuse{DiagPag29.1}
}\hspace{-0.8cm}=ig \displaystyle\int d^4y\left[\phi_0^2(y) G_{\lambda_{\pi}\phi}(0)\phi_0(x) \right]
\end{equation} while the expression for the third one is
\vspace{0.6cm}
\begin{equation}\label{<phi>(1)2}
\parbox{33mm}{
\fmfreuse{DiagPag29.3}
}\hspace{-0.5cm}=g \displaystyle\int d^4y\left[\phi_0^2(y) G_{c^{\phi}\bar{c}_{\pi}}(0)\phi_0(x) \right].
\end{equation} Let us remember from (\ref{philambda}) and (\ref{cbarc}) that
\begin{equation}
G_{c^{\phi}\bar{c}_{\pi}}=-G_R=-iG_{\lambda_{\pi}\phi}
\end{equation} so (\ref{<phi>(1)1}) and (\ref{<phi>(1)2}) cancel each other and the only correction left
in (\ref{<phi>(1)}) is
\begin{equation}\label{<phi>(1)final}
\langle \phi\rangle^{(1)}_{\varphi_i}=\,\,\parbox{33mm}{
\fmfreuse{DiagPag29.2}
}
\end{equation} whose analytic expression is \vspace{0.4cm}
\begin{equation}
\parbox{33mm}{
\fmfreuse{DiagPag29.2}
}\hspace{-0.4cm}=i\,g \displaystyle\int
d^4y\left[\phi_0^3(y) G_{\phi \lambda_{\pi}}(x-y) \right].
\end{equation}

The reader could, at this point, claim that the loops of the
retarded propagators appearing in Eq. (\ref{<phi>(1)}) are both
zero, because of the $\theta(t)$ which appears in all retarded
propagators. For a loop to be zero we have anyhow to assume that
$\theta(0)=0$. This regularization of the $\theta$ corresponds to
a prepoint choice in a discretized form of $t$ \cite{Sakita}. The
prepoint choice has some problems once it is used in the
path-integral. It has in fact been proved in Ref. \cite{Sato} that
only if one uses the midpoint choice then the Feynman rules which
can be read off the continuum version are the same as the one of
the discretized version. So, for this reason, we have preferred
the midpoint discretization which gives $\theta(0)=1/2$ and this implies that the
retarded loops are not zero.  Let us remember also that it is  with the midpoint  rule that the
determinant is not a constant and the Grassmannian variables are
needed. So, our formalism has an overall logical coherence.

Going now back to our diagrams the reader may be puzzled by
diagrams like (\ref{<phi>(1)final}) where there are just three
lines at the end going nowhere or like the single line in
(\ref{<phi>(0)}) also going nowhere. This is due to the fact that
our generating functional $\mathcal{Z}_{\varphi_i}$ depends on the
initial configuration which is not averaged over. These kind of
diagrams, but without Grassmannian variables, were first developed
in \cite{PenMau} and are somehow similar to those developed for
fluid dynamics in \cite{Wyld}. Perturbation theory and something similar to
Feynman diagrams where also developed long ago in \cite{PenGarr}.
We have not tried to compare our formalism with this other one.
For sure our formalism is different from these other ones because it contains
extra ingredients like the
Grassmaniann variables. Moreover, because of these extra ingredients,
we will be able to develop the super-diagrams which will simplify the diagrammatics.

We will continue now to the second order for the $\langle \phi
\rangle$. The result that can easily be obtained(where $y_1$
and $y_2$ are integrated over) is the following:
\vspace{-0.5cm}
\begin{equation}\nonumber
\langle \phi(x)\rangle_{\varphi_i}=\Bigg[\frac{1}{2!}\,\frac{1}{3!}\,\,\,\parbox{33mm}{\vspace{0.5cm}
\begin{fmffile}{DiagPag32.1}
\begin{fmfgraph*}(33,20)
\fmfleft{i1,i2,i3}
\fmfright{o1,o2,o3}
\fmfdot{i2}
\fmffreeze
\fmf{phantom}{i1,v11,v12,v13,v14,v15,v16,o1}
\fmffreeze
\fmf{phantom,tag=1}{i2,v21}
\fmf{phantom,tag=2}{v21,v22}
\fmf{phantom,tag=3}{v22,o2}
\fmfdot{v21,v22}
\fmfposition
\fmfv{label=$x$,label.angle=-90}{i2}
\fmfv{label=$y_1$,label.angle=-120}{v21}
\fmfv{label=$y_2$,label.angle=-120}{v22}
\fmfipath{p[]}
\fmfiset{p1}{vpath1(__i2,__v21)}
\fmfi{plain}{subpath (0,length(p1)/2.5) of p1}
\fmfi{dashes}{subpath (length(p1)/2.8,length(p1)) of p1}
\fmfiset{p2}{vpath2(__v21,__v22)}
\fmfi{plain}{subpath (0,length(p2)/2.5) of p2}
\fmfi{dashes}{subpath (length(p2)/2.8,length(p2)) of p2}
\fmfiset{p3}{vpath3(__v22,__o2)}
\fmfi{plain}{subpath (0,length(p3)*0.5) of p3}
\fmffreeze
\fmf{phantom}{i3,v31,v32,v33,v34,v35,v36,o3}
\fmffreeze
\fmf{plain}{v21,v33}
\fmf{plain}{v21,v13}
\fmffreeze
\fmf{plain}{v22,v36}
\fmf{plain}{v22,v16}
\end{fmfgraph*}
\end{fmffile}
}
\hspace{-0.4cm}+\frac{1}{2!}\,\frac{1}{3!}\,\,\,
\parbox{33mm}{\vspace{0.5cm}
\begin{fmffile}{DiagPag32.2}
\begin{fmfgraph*}(33,20)
\fmfleft{i1,i2,i3}
\fmfright{o1,o2,o3}
\fmfdot{i2}
\fmffreeze
\fmf{phantom}{i1,v11,v12,v13,v14,v15,v16,o1}
\fmffreeze
\fmf{phantom,tag=1}{i2,v21}
\fmf{phantom,tag=3}{v21,v22}
\fmf{phantom,tag=4}{v22,o2}
\fmfdot{v21,v22}
\fmfposition
\fmfipath{p[]}
\fmfiset{p1}{vpath1(__i2,__v21)}
\fmfi{plain}{subpath (0,length(p1)/2.5) of p1}
\fmfi{dashes}{subpath (length(p1)/2.8,length(p1)) of p1}
\fmfiset{p3}{vpath3(__v21,__v22)}
\fmfi{plain}{subpath (0,length(p3)*0.4) of p3}
\fmfi{plain}{subpath (length(p3)*0.6,length(p3)) of p3}
\fmfiset{p4}{vpath4(__v22,__o2)}
\fmfi{plain}{subpath (0,length(p4)*0.5) of p4}
\fmffreeze
\fmf{phantom}{i3,v31,v32,v33,v34,v35,v36,o3}
\fmffreeze
\fmf{plain}{v21,v33}
\fmf{plain}{v21,v13}
\fmf{phantom,tension=0.7,tag=2}{v22,v22}
\fmfiset{p2}{vpath2(__v22,__v22)}
\fmfi{plain}{subpath (0,length(p2)/2.) of p2}
\fmfi{dashes}{subpath (length(p2)/2,length(p2)) of p2}
\fmfv{label=$x$,label.angle=-90}{i2}
\fmfv{label=$y_1$,label.angle=-120}{v21}
\fmfv{label=$y_2$,label.angle=-90}{v22}
\fmffreeze
\end{fmfgraph*}
\end{fmffile}
}
\hspace{-0.5cm}+\frac{1}{3!}\,\,\,
\parbox{33mm}{\vspace{0.5cm}
\begin{fmffile}{DiagPag32.3}
\begin{fmfgraph*}(33,20)
\fmfleft{i1,i2,i3}
\fmfright{o1,o2,o3}
\fmfdot{i2}
\fmffreeze
\fmf{phantom}{i1,v11,v12,v13,v14,v15,v16,o1}
\fmffreeze
\fmf{phantom,tag=3}{i2,v21}
\fmf{phantom,tension=0.7,tag=1}{v21,v21}
\fmf{phantom,tag=2}{v21,v22}
\fmf{phantom,tag=4}{v22,o2}
\fmfdot{v21,v22}
\fmfposition
\fmfipath{p[]}
\fmfiset{p1}{vpath1(__v21,__v21)}
\fmfi{plain}{subpath (0,length(p1)/2.) of p1}
\fmfi{dashes}{subpath (length(p1)/2,length(p1)) of p1}
\fmfiset{p2}{vpath2(__v21,__v22)}
\fmfi{plain}{subpath (0,length(p2)/2.5) of p2}
\fmfi{dashes}{subpath (length(p2)/2.8,length(p2)) of p2}
\fmfiset{p3}{vpath3(__i2,__v21)}
\fmfi{plain}{subpath (0,length(p3)*0.4) of p3}
\fmfi{plain}{subpath (length(p3)*0.6,length(p3)) of p3}
\fmfiset{p4}{vpath4(__v22,__o2)}
\fmfi{plain}{subpath (0,length(p4)*0.5) of p4}
\fmffreeze
\fmf{phantom}{i3,v31,v32,v33,v34,v35,v36,o3}
\fmffreeze
\fmf{plain}{v22,v36}
\fmf{plain}{v22,v16}
\fmfv{label=$x$,label.angle=-90}{i2}
\fmfv{label=$y_1$,label.angle=-90}{v21}
\fmfv{label=$y_2$,label.angle=-120}{v22}
\end{fmfgraph*}
\end{fmffile}
}\hspace{-0.4cm}+
\end{equation}
\vspace{-1cm}
\begin{equation}\nonumber
\hspace{1.5cm}+\frac{1}{(2!)^3}\,\,\,
\parbox{33mm}{\vspace{0.5cm}
\begin{fmffile}{DiagPag32.4}
\begin{fmfgraph*}(33,20)
\fmfleft{i1,i2,i3}
\fmfright{o1,o2,o3}
\fmfdot{i2}
\fmffreeze
\fmf{phantom}{i1,v11,v12,v13,v14,v15,v16,o1}
\fmffreeze
\fmf{phantom,tag=3}{i2,v21}
\fmf{phantom,tension=0.7,tag=1}{v21,v21}
\fmf{phantom,tension=0.7,tag=2}{v22,v22}
\fmf{phantom,tag=4}{v21,v22}
\fmf{phantom,tag=5}{v22,o2}
\fmfdot{v21,v22}
\fmfposition
\fmfipath{p[]}
\fmfiset{p1}{vpath1(__v21,__v21)}
\fmfi{plain}{subpath (0,length(p1)/2.) of p1}
\fmfi{dashes}{subpath (length(p1)/2,length(p1)) of p1}
\fmfiset{p2}{vpath2(__v22,__v22)}
\fmfi{plain}{subpath (0,length(p2)/2.) of p2}
\fmfi{dashes}{subpath (length(p2)/2,length(p2)) of p2}
\fmfiset{p3}{vpath3(__i2,__v21)}
\fmfi{plain}{subpath (0,length(p3)*0.4) of p3}
\fmfi{plain}{subpath (length(p3)*0.6,length(p3)) of p3}
\fmfiset{p4}{vpath4(__v21,__v22)}
\fmfi{plain}{subpath (0,length(p4)*0.4) of p4}
\fmfi{plain}{subpath (length(p4)*0.6,length(p4)) of p4}
\fmfiset{p5}{vpath5(__v22,__o2)}
\fmfi{plain}{subpath (0,length(p5)*0.5) of p5}
\fmffreeze
\fmf{phantom}{i3,v31,v32,v33,v34,v35,v36,o3}
\fmffreeze
\fmfv{label=$x$,label.angle=-90}{i2}
\fmfv{label=$y_1$,label.angle=-90}{v21}
\fmfv{label=$y_2$,label.angle=-90}{v22}
\end{fmfgraph*}
\end{fmffile}
}
\hspace{-0.5cm}+\frac{1}{(2!)^3}\,\,\,
\parbox{33mm}{\vspace{0.cm}
\begin{fmffile}{DiagPag32.5}
\begin{fmfgraph*}(33,20)
\fmfleft{i1,i2,i3}
\fmfright{o1,o2,o3}
\fmfdot{i2}
\fmffreeze
\fmf{phantom}{i1,v11,v12,v13,v14,v15,v16,o1}
\fmffreeze
\fmf{phantom}{i3,v31,v32,v33,v34,v35,v36,o3}
\fmffreeze
\fmf{phantom,tag=3}{i2,v21}
\fmf{phantom}{v21,v22}
\fmf{phantom}{v22,o2}
\fmffreeze
\fmf{phantom,left=0.7,tag=1}{v21,v22}
\fmf{phantom,right=0.7,tag=2}{v21,v22}
\fmfposition
\fmfipath{p[]}
\fmfiset{p1}{vpath1(__v21,__v22)}
\fmfi{plain}{subpath (0,length(p1)/2.) of p1}
\fmfi{dashes}{subpath (length(p1)/2,length(p1)) of p1}
\fmfiset{p2}{vpath2(__v21,__v22)}
\fmfi{dashes}{subpath (0,length(p2)/2.) of p2}
\fmfi{plain}{subpath (length(p2)/2,length(p2)) of p2}
\fmfiset{p3}{vpath3(__i2,__v21)}
\fmfi{plain}{subpath (0,length(p3)*0.4) of p3}
\fmfi{plain}{subpath (length(p3)*0.6,length(p3)) of p3}
\fmf{plain}{v21,v33}
\fmf{plain}{v22,v36}
\fmf{plain}{v22,v16}
\fmffreeze
\fmfdot{v21,v22}
\fmfposition
\fmffreeze
\fmfv{label=$x$,label.angle=-90}{i2}
\fmfv{label=$y_1$,label.angle=-120}{v21}
\fmfv{label=$y_2$,label.angle=-40}{v22}
\end{fmfgraph*}
\end{fmffile}}
\hspace{-0.7cm}-\frac{1}{(2!)^3}\,\,\,
\parbox{33mm}{\vspace{0.5cm}
\begin{fmffile}{DiagPag32.6}
\begin{fmfgraph*}(33,15)
\fmfset{arrow_len}{2mm}
\fmfleft{i1,i2,i3}
\fmfright{o1,o2,o3}
\fmfdot{i2}
\fmffreeze
\fmf{phantom}{i1,v11,v12,v13,v14,v15,v16,o1}
\fmffreeze
\fmf{phantom,tag=2}{i2,v21}
\fmf{phantom,tension=0.7,tag=1}{v21,v21}
\fmf{dots_arrow,tension=0.7}{v22,v22}
\fmf{phantom,tag=3}{v21,v22}
\fmf{phantom,tag=4}{v22,o2}
\fmfdot{v21,v22}
\fmfposition
\fmfipath{p[]}
\fmfiset{p1}{vpath1(__v21,__v21)}
\fmfi{plain}{subpath (0,length(p1)/2.) of p1}
\fmfi{dashes}{subpath (length(p1)/2,length(p1)) of p1}
\fmfiset{p2}{vpath2(__i2,__v21)}
\fmfi{plain}{subpath (0,length(p2)*0.4) of p2}
\fmfi{plain}{subpath (length(p2)*0.6,length(p2)) of p2}
\fmfiset{p3}{vpath3(__v21,__v22)}
\fmfi{plain}{subpath (0,length(p3)*0.4) of p3}
\fmfi{plain}{subpath (length(p3)*0.6,length(p3)) of p3}
\fmfiset{p4}{vpath4(__v22,__o2)}
\fmfi{plain}{subpath (0,length(p4)*0.5) of p4}
\fmffreeze
\fmf{phantom}{i3,v31,v32,v33,v34,v35,v36,o3}
\fmffreeze
\fmfv{label=$x$,label.angle=-90}{i2}
\fmfv{label=$y_1$,label.angle=-90}{v21}
\fmfv{label=$y_2$,label.angle=-90}{v22}
\end{fmfgraph*}
\end{fmffile}
}\hspace{-0.4cm}+
\end{equation}
\vspace{-1cm}
\begin{equation}\nonumber
\hspace{1.3cm}-\frac{1}{3!}\,\,\,
\parbox{33mm}{\vspace{0.5cm}
\begin{fmffile}{DiagPag32.7}
\begin{fmfgraph*}(33,20)
\fmfset{arrow_len}{2mm}
\fmfleft{i1,i2,i3}
\fmfright{o1,o2,o3}
\fmfdot{i2}
\fmffreeze
\fmf{phantom}{i1,v11,v12,v13,v14,v15,v16,o1}
\fmffreeze
\fmf{phantom,tag=1}{i2,v21}
\fmf{dots_arrow,tension=0.7}{v21,v21}
\fmf{phantom,tag=2}{v21,v22}
\fmf{phantom,tag=3}{v22,o2}
\fmfdot{v21,v22}
\fmfposition
\fmfipath{p[]}
\fmfiset{p1}{vpath1(__i2,__v21)}
\fmfi{plain}{subpath (0,length(p1)*0.4) of p1}
\fmfi{plain}{subpath (length(p1)*0.6,length(p1)) of p1}
\fmfiset{p2}{vpath2(__v21,__v22)}
\fmfi{plain}{subpath (0,length(p2)*0.3) of p2}
\fmfi{dashes}{subpath (length(p2)*0.3,length(p2)) of p2}
\fmfiset{p3}{vpath3(__v22,__o2)}
\fmfi{plain}{subpath (0,length(p3)*0.5) of p3}
\fmffreeze
\fmf{phantom}{i3,v31,v32,v33,v34,v35,v36,o3}
\fmffreeze
\fmf{plain}{v22,v36}
\fmf{plain}{v22,v16}
\fmfv{label=$x$,label.angle=-90}{i2}
\fmfv{label=$y_1$,label.angle=-90}{v21}
\fmfv{label=$y_2$,label.angle=-120}{v22}
\end{fmfgraph*}
\end{fmffile}
}\hspace{-0.5cm}-\frac{1}{2!}\,\frac{1}{3!}\,\,\,
\parbox{33mm}{\vspace{0.5cm}
\begin{fmffile}{DiagPag33.1}
\begin{fmfgraph*}(33,20)
\fmfset{arrow_len}{2mm}
\fmfleft{i1,i2,i3}
\fmfright{o1,o2,o3}
\fmfdot{i2}
\fmffreeze
\fmf{phantom}{i1,v11,v12,v13,v14,v15,v16,o1}
\fmffreeze
\fmf{phantom,tag=1}{i2,v21}
\fmf{phantom,tag=3}{v21,v22}
\fmf{phantom,tag=4}{v22,o2}
\fmfdot{v21,v22}
\fmfposition
\fmfipath{p[]}
\fmfiset{p1}{vpath1(__i2,__v21)}
\fmfi{plain}{subpath (0,length(p1)/2.5) of p1}
\fmfi{dashes}{subpath (length(p1)/2.8,length(p1)) of p1}
\fmfiset{p3}{vpath3(__v21,__v22)}
\fmfi{plain}{subpath (0,length(p3)*0.4) of p3}
\fmfi{plain}{subpath (length(p3)*0.6,length(p3)) of p3}
\fmfiset{p4}{vpath4(__v22,__o2)}
\fmfi{plain}{subpath (0,length(p4)*0.5) of p4}
\fmffreeze
\fmf{phantom}{i3,v31,v32,v33,v34,v35,v36,o3}
\fmffreeze
\fmf{plain}{v21,v33}
\fmf{plain}{v21,v13}
\fmf{dots_arrow,tension=0.7}{v22,v22}
\fmfv{label=$x$,label.angle=-90}{i2}
\fmfv{label=$y_1$,label.angle=-120}{v21}
\fmfv{label=$y_2$,label.angle=-90}{v22}
\fmffreeze
\end{fmfgraph*}
\end{fmffile}
} \hspace{-0.4cm}-\frac{1}{(2!)^3}\,\,\,
\parbox{33mm}{\vspace{0.5cm}
\begin{fmffile}{DiagPag33.2}
\begin{fmfgraph*}(33,20)
\fmfset{arrow_len}{2mm}
\fmfleft{i1,i2,i3}
\fmfright{o1,o2,o3}
\fmfdot{i2}
\fmffreeze
\fmf{phantom}{i1,v11,v12,v13,v14,v15,v16,o1}
\fmffreeze
\fmf{phantom,tag=3}{i2,v21}
\fmf{dots_arrow,tension=0.7}{v21,v21}
\fmf{dots_arrow,tension=0.7}{v22,v22}
\fmf{phantom,tag=4}{v21,v22}
\fmf{phantom,tag=5}{v22,o2}
\fmfdot{v21,v22}
\fmfposition
\fmfipath{p[]}
\fmfiset{p3}{vpath3(__i2,__v21)}
\fmfi{plain}{subpath (0,length(p3)*0.4) of p3}
\fmfi{plain}{subpath (length(p3)*0.6,length(p3)) of p3}
\fmfiset{p4}{vpath4(__v21,__v22)}
\fmfi{plain}{subpath (0,length(p4)*0.4) of p4}
\fmfi{plain}{subpath (length(p4)*0.6,length(p4)) of p4}
\fmfiset{p5}{vpath5(__v22,__o2)}
\fmfi{plain}{subpath (0,length(p5)*0.5) of p5}
\fmffreeze
\fmf{phantom}{i3,v31,v32,v33,v34,v35,v36,o3}
\fmffreeze
\fmfv{label=$x$,label.angle=-90}{i2}
\fmfv{label=$y_1$,label.angle=-90}{v21}
\fmfv{label=$y_2$,label.angle=-90}{v22}
\end{fmfgraph*}
\end{fmffile}
}\hspace{-0.4cm}\vspace{0.1cm}+
\end{equation}
\begin{equation}
\hspace{-0.4cm}-\frac{1}{(2!)^3}\,\,\,\parbox{33mm}{\vspace{0.cm}
\begin{fmffile}{DiagPag33.3}
\begin{fmfgraph*}(33,20)
\fmfset{arrow_len}{2mm}
\fmfleft{i1,i2,i3}
\fmfright{o1,o2,o3}
\fmfdot{i2}
\fmffreeze
\fmf{phantom}{i1,v11,v12,v13,v14,v15,v16,o1}
\fmffreeze
\fmf{phantom}{i3,v31,v32,v33,v34,v35,v36,o3}
\fmffreeze
\fmf{phantom,tag=3}{i2,v21}
\fmf{phantom}{v21,v22}
\fmf{phantom}{v22,o2}
\fmffreeze
\fmf{dots_arrow,left=0.7}{v21,v22}
\fmf{dots_arrow,left=0.7}{v22,v21}
\fmfposition
\fmfipath{p[]}
\fmfiset{p3}{vpath3(__i2,__v21)}
\fmfi{plain}{subpath (0,length(p3)*0.4) of p3}
\fmfi{plain}{subpath (length(p3)*0.6,length(p3)) of p3}
\fmf{plain}{v21,v33}
\fmf{plain}{v22,v36}
\fmf{plain}{v22,v16}
\fmffreeze
\fmfdot{v21,v22}
\fmfposition
\fmffreeze
\fmfv{label=$x$,label.angle=-90}{i2}
\fmfv{label=$y_1$,label.angle=-120}{v21}
\fmfv{label=$y_2$,label.angle=-30}{v22}
\end{fmfgraph*}
\end{fmffile}}\hspace{-0.3cm}\Bigg]
\hspace{6.5cm}
\end{equation}
As before, it is easy to see that there are cancelations. The
second diagram is cancelled  by the eighth one, the third by the seventh, the
fourth is cancelled by  the sum of the sixth and ninth, the fifth one by  the
tenth. So the final result is
\begin{equation}
\langle \phi(x)\rangle_{\varphi_i}=\Bigg[\,\,\begin{minipage}[c]{7mm}
\vspace{0.6cm}\begin{fmffile}{DiagPag33.4}
\begin{fmfgraph*}(7,5)
\fmfleft{i1}
\fmfright{o1}
\fmfdot{i1}
\fmf{plain}{i1,o1}
\end{fmfgraph*}
\end{fmffile}
\end{minipage}+\cfrac{1}{3!}\,\,\,
\begin{minipage}[c]{32mm}
\begin{fmffile}{DiagPag33.5}
\begin{fmfgraph*}(32,20)
\fmfleft{i1,i2,i3}
\fmfright{o1,o2,o3}
\fmf{phantom}{i1,v1,o1}
\fmf{phantom}{i2,v2,o2}
\fmf{phantom}{i3,v3,o3}
\fmfdot{i2}
\fmfdot{v2}
\fmffreeze
\fmf{phantom,tag=1}{i2,v2}
\fmf{phantom,tag=2}{v2,o1}
\fmf{phantom,tag=3}{v2,o2}
\fmf{phantom,tag=4}{v2,o3}
\fmfposition
\fmfipath{p[]}
\fmfiset{p1}{vpath1(__i2,__v2)}
\fmfi{plain}{subpath (0,length(p1)*0.40) of p1}
\fmfi{dashes}{subpath (length(p1)*0.4,length(p1)) of p1}
\fmfiset{p2}{vpath2(__v2,__o1)}
\fmfi{plain}{subpath (0,length(p2)*0.7) of p2}
\fmfiset{p3}{vpath3(__v2,__o2)}
\fmfi{plain}{subpath (0,length(p3)*0.7) of p3}
\fmfiset{p4}{vpath4(__v2,__o3)}
\fmfi{plain}{subpath (0,length(p4)*0.7) of p4}
\end{fmfgraph*}
\end{fmffile}
\end{minipage}
\hspace{-0.3cm}+\frac{1}{2!}\,\frac{1}{3!}\,\,\,\parbox{33mm}{\vspace{0.5cm}
\begin{fmffile}{DiagPag33.6}
\begin{fmfgraph*}(33,15)
\fmfleft{i1,i2,i3}
\fmfright{o1,o2,o3}
\fmfdot{i2}
\fmffreeze
\fmf{phantom}{i1,v11,v12,v13,v14,v15,v16,o1}
\fmffreeze
\fmf{phantom,tag=1}{i2,v21}
\fmf{phantom,tag=2}{v21,v22}
\fmf{phantom,tag=3}{v22,o2}
\fmfdot{v21,v22}
\fmfposition
\fmfipath{p[]}
\fmfiset{p1}{vpath1(__i2,__v21)}
\fmfi{plain}{subpath (0,length(p1)/2.5) of p1}
\fmfi{dashes}{subpath (length(p1)/2.8,length(p1)) of p1}
\fmfiset{p2}{vpath2(__v21,__v22)}
\fmfi{plain}{subpath (0,length(p2)/2.5) of p2}
\fmfi{dashes}{subpath (length(p2)/2.8,length(p2)) of p2}
\fmfiset{p3}{vpath3(__v22,__o2)}
\fmfi{plain}{subpath (0,length(p3)*0.5) of p3}
\fmffreeze
\fmf{phantom}{i3,v31,v32,v33,v34,v35,v36,o3}
\fmffreeze
\fmf{plain}{v21,v33}
\fmf{plain}{v21,v13}
\fmffreeze
\fmf{plain}{v22,v36}
\fmf{plain}{v22,v16}
\end{fmfgraph*}
\end{fmffile}
}\hspace{-0.3cm}\Bigg]
\end{equation}
notice that no {\it loop} is left.

The various symmetry factors have being obtained by working out the
analytical details from the $\mathcal{Z}_{\varphi_i}$ in
(\ref{Z=expSVZ0}) up to second order.

Let us now go back to the two-point functions which, to zero order,
was given in Eq. (\ref{<phiphi>(0)}). After some tedious
calculations with the second derivative of
$\mathcal{Z}_{\varphi_i}[J]$, we get:
\begin{equation}\nonumber
\langle \phi(x_2) \phi(x_1)\rangle_{\varphi_i}^{(1)}=\Bigg[\frac{1}{2!}\quad
\begin{minipage}[c]{32mm}
\vspace{0.55cm}
\begin{fmffile}{DiagPag34.1}
\begin{fmfgraph*}(32,10)
\fmfleft{i1}
\fmfright{o1}
\fmfdot{i1,o1}
\fmfv{label=$x_1$,label.angle=-90}{i1}
\fmfv{label=$x_2$,label.angle=-90}{o1}
\fmf{phantom}{i1,v,o1}
\fmf{phantom,tag=3}{v,o1}
\fmfdot{v}
\fmf{phantom,tag=1}{i1,v}
\fmfposition
\fmfipath{p[]}
\fmfiset{p1}{vpath1(__i1,__v)}
\fmfi{plain}{subpath (0,length(p1)*0.40.) of p1}
\fmfi{plain}{subpath (length(p1)*0.6,length(p1)) of p1}
\fmf{phantom,tension=0.9,tag=2}{v,v}
\fmfposition
\fmfiset{p2}{vpath2(__v,__v)}
\fmfi{plain}{subpath (0,length(p2)/2.) of p2}
\fmfi{dashes}{subpath (length(p2)/2,length(p2)) of p2}
\fmfposition
\fmfiset{p3}{vpath3(__v,__o1)}
\fmfi{plain}{subpath (0,length(p3)*0.40) of p3}
\fmfi{plain}{subpath (length(p3)*0.60,length(p3)) of p3}
\end{fmfgraph*}
\end{fmffile}
\end{minipage}
\,\,+\cfrac{1}{3!}\quad
\begin{minipage}[c]{32mm}
\vspace{0.55cm}
\begin{fmffile}{DiagPag34.2}
\begin{fmfgraph*}(32,20)
\fmfkeep{DiagPag34.2}
\fmfleft{i1,i2,i3}
\fmfright{o1,o2,o3}
\fmf{phantom}{i1,v1,o1}
\fmf{phantom}{i2,v2,o2}
\fmf{phantom}{i3,v3,o3}
\fmfdot{i2,o2}
\fmfdot{v2}
\fmfv{label=$x_1$,label.angle=-90}{i2}
\fmfv{label=$x_2$,label.angle=-90}{o2}
\fmffreeze
\fmf{phantom,tag=1}{i2,v2}
\fmf{phantom,tag=2}{v2,o1}
\fmf{phantom,tag=3}{v2,o2}
\fmf{phantom,tag=4}{v2,o3}
\fmfposition
\fmfipath{p[]}
\fmfiset{p1}{vpath1(__i2,__v2)}
\fmfi{plain}{subpath (0,length(p1)*0.40) of p1}
\fmfi{dashes}{subpath (length(p1)*0.4,length(p1)) of p1}
\fmfiset{p2}{vpath2(__v2,__o1)}
\fmfi{plain}{subpath (0,length(p2)*0.7) of p2}
\fmfiset{p3}{vpath3(__v2,__o2)}
\fmfi{plain}{subpath (0,length(p3)*0.4) of p3}
\fmfi{plain}{subpath (length(p3)*0.6,length(p3)) of p3}
\fmfiset{p4}{vpath4(__v2,__o3)}
\fmfi{plain}{subpath (0,length(p4)*0.7) of p4}
\end{fmfgraph*}
\end{fmffile}
\end{minipage}\hspace{0.3cm}+\vspace{-1.5cm}
\end{equation}
\begin{equation}
\hspace{-0.2cm}+\quad\cfrac{1}{3!}\quad
\begin{minipage}[c]{32mm}
\vspace{0.55cm}
\begin{fmffile}{DiagPag34.3}
\begin{fmfgraph*}(32,20)
\fmfkeep{DiagPag34.3}
\fmfleft{i1,i2,i3}
\fmfright{o1,o2,o3}
\fmf{phantom}{i1,v1,o1}
\fmf{phantom}{i2,v2,o2}
\fmf{phantom}{i3,v3,o3}
\fmfdot{i2,o2}
\fmfdot{v2}
\fmfv{label=$x_1$,label.angle=-90}{i2}
\fmfv{label=$x_2$,label.angle=-90}{o2}
\fmffreeze
\fmf{phantom,tag=1}{i2,v2}
\fmf{phantom,tag=2}{v2,o1}
\fmf{phantom,tag=3}{v2,o2}
\fmf{phantom,tag=4}{v2,o3}
\fmfposition
\fmfipath{p[]}
\fmfiset{p1}{vpath1(__i2,__v2)}
\fmfi{plain}{subpath (0,length(p1)*0.40) of p1}
\fmfi{plain}{subpath (length(p1)*0.6,length(p1)) of p1}
\fmfiset{p2}{vpath2(__v2,__o1)}
\fmfi{plain}{subpath (0,length(p2)*0.7) of p2}
\fmfiset{p3}{vpath3(__v2,__o2)}
\fmfi{dashes}{subpath (0,length(p3)*0.6) of p3}
\fmfi{plain}{subpath (length(p3)*0.6,length(p3)) of p3}
\fmfiset{p4}{vpath4(__v2,__o3)}
\fmfi{plain}{subpath (0,length(p4)*0.7) of p4}
\end{fmfgraph*}
\end{fmffile}
\end{minipage}
\,\,-\frac{1}{2!}
\quad
\begin{minipage}[c]{32mm}
\vspace{0.55cm}
\begin{fmffile}{DiagPag34.4}
\begin{fmfgraph*}(32,10)
\fmfset{arrow_len}{2mm}
\fmfleft{i1}
\fmfright{o1}
\fmfdot{i1,o1}
\fmfv{label=$x_1$,label.angle=-90}{i1}
\fmfv{label=$x_2$,label.angle=-90}{o1}
\fmf{phantom}{i1,v,o1}
\fmf{phantom,tag=3}{v,o1}
\fmfdot{v}
\fmf{phantom,tag=1}{i1,v}
\fmfposition
\fmfipath{p[]}
\fmfiset{p1}{vpath1(__i1,__v)}
\fmfi{plain}{subpath (0,length(p1)*0.40.) of p1}
\fmfi{plain}{subpath (length(p1)*0.6,length(p1)) of p1}
\fmf{dots_arrow,tension=0.9}{v,v}
\fmfposition
\fmfiset{p3}{vpath3(__v,__o1)}
\fmfi{plain}{subpath (0,length(p3)*0.40) of p3}
\fmfi{plain}{subpath (length(p3)*0.60,length(p3)) of p3}
\end{fmfgraph*}
\end{fmffile}
\end{minipage}\,\,\,\,\,\Bigg]
\hspace{-2.6cm}
\end{equation}
The first and the fourth diagram cancel each other
and so the final result is
\begin{equation}\label{phiphi1final}
\langle \phi(x_2) \phi(x_1)\rangle_{\varphi_i}^{(1)}=\Bigg[\cfrac{1}{3!}\quad
\begin{minipage}[c]{32mm}
\vspace{0.55cm}
\begin{fmffile}{DiagPag34.5}
\begin{fmfgraph*}(32,20)
\fmfkeep{DiagPag34.5}
\fmfleft{i1,i2,i3}
\fmfright{o1,o2,o3}
\fmf{phantom}{i1,v1,o1}
\fmf{phantom}{i2,v2,o2}
\fmf{phantom}{i3,v3,o3}
\fmfdot{i2,o2}
\fmfdot{v2}
\fmfv{label=$x_1$,label.angle=-90}{i2}
\fmfv{label=$x_2$,label.angle=-90}{o2}
\fmffreeze
\fmf{phantom,tag=1}{i2,v2}
\fmf{phantom,tag=2}{v2,o1}
\fmf{phantom,tag=3}{v2,o2}
\fmf{phantom,tag=4}{v2,o3}
\fmfposition
\fmfipath{p[]}
\fmfiset{p1}{vpath1(__i2,__v2)}
\fmfi{plain}{subpath (0,length(p1)*0.40) of p1}
\fmfi{dashes}{subpath (length(p1)*0.4,length(p1)) of p1}
\fmfiset{p2}{vpath2(__v2,__o1)}
\fmfi{plain}{subpath (0,length(p2)*0.7) of p2}
\fmfiset{p3}{vpath3(__v2,__o2)}
\fmfi{plain}{subpath (0,length(p3)*0.4) of p3}
\fmfi{plain}{subpath (length(p3)*0.6,length(p3)) of p3}
\fmfiset{p4}{vpath4(__v2,__o3)}
\fmfi{plain}{subpath (0,length(p4)*0.7) of p4}
\end{fmfgraph*}
\end{fmffile}
\end{minipage}
\quad+\cfrac{1}{3!}\quad
\begin{minipage}[c]{32mm}
\vspace{0.55cm}
\begin{fmffile}{DiagPag34.6}
\begin{fmfgraph*}(32,20)
\fmfkeep{DiagPag34.6}
\fmfleft{i1,i2,i3}
\fmfright{o1,o2,o3}
\fmf{phantom}{i1,v1,o1}
\fmf{phantom}{i2,v2,o2}
\fmf{phantom}{i3,v3,o3}
\fmfdot{i2,o2}
\fmfdot{v2}
\fmfv{label=$x_1$,label.angle=-90}{i2}
\fmfv{label=$x_2$,label.angle=-90}{o2}
\fmffreeze
\fmf{phantom,tag=1}{i2,v2}
\fmf{phantom,tag=2}{v2,o1}
\fmf{phantom,tag=3}{v2,o2}
\fmf{phantom,tag=4}{v2,o3}
\fmfposition
\fmfipath{p[]}
\fmfiset{p1}{vpath1(__i2,__v2)}
\fmfi{plain}{subpath (0,length(p1)*0.40) of p1}
\fmfi{plain}{subpath (length(p1)*0.6,length(p1)) of p1}
\fmfiset{p2}{vpath2(__v2,__o1)}
\fmfi{plain}{subpath (0,length(p2)*0.7) of p2}
\fmfiset{p3}{vpath3(__v2,__o2)}
\fmfi{dashes}{subpath (0,length(p3)*0.6) of p3}
\fmfi{plain}{subpath (length(p3)*0.6,length(p3)) of p3}
\fmfiset{p4}{vpath4(__v2,__o3)}
\fmfi{plain}{subpath (0,length(p4)*0.7) of p4}
\end{fmfgraph*}
\end{fmffile}
\end{minipage}\quad\Bigg]
\end{equation}
Going to the second order, after some long but straightforward
calculations, we obtain:
\begin{equation}\nonumber
\langle \phi(x_2) \phi(x_1)\rangle_{\varphi_i}^{(2)}=\Bigg[\cfrac{1}{2!}\,\cfrac{1}{3!}\quad\parbox{33mm}{\vspace{0.5cm}
\begin{fmffile}{DiagPag34.7}
\begin{fmfgraph*}(33,20)
\fmfkeep{DiagPag34.7}
\fmfleft{i1,i2,i3}
\fmfright{o1,o2,o3}
\fmfdot{i2,o2}
\fmffreeze
\fmf{phantom}{i1,v11,v12,v13,v14,v15,v16,o1}
\fmffreeze
\fmf{phantom,tag=1}{i2,v21}
\fmf{phantom,tag=2}{v21,v22}
\fmf{phantom,tag=3}{v22,o2}
\fmfdot{v21,v22}
\fmfposition
\fmfv{label=$x_1$,label.angle=-90}{i2}
\fmfv{label=$x_2$,label.angle=-90}{o2}
\fmfipath{p[]}
\fmfiset{p1}{vpath1(__i2,__v21)}
\fmfi{plain}{subpath (0,length(p1)/2.5) of p1}
\fmfi{dashes}{subpath (length(p1)/2.8,length(p1)) of p1}
\fmfiset{p2}{vpath2(__v21,__v22)}
\fmfi{plain}{subpath (0,length(p2)/2.5) of p2}
\fmfi{dashes}{subpath (length(p2)/2.8,length(p2)) of p2}
\fmfiset{p3}{vpath3(__v22,__o2)}
\fmfi{plain}{subpath (0,length(p3)*0.4) of p3}
\fmfi{plain}{subpath (length(p3)*0.6,length(p3)) of p3}
\fmffreeze
\fmf{phantom}{i3,v31,v32,v33,v34,v35,v36,o3}
\fmffreeze
\fmf{plain}{v21,v33}
\fmf{plain}{v21,v13}
\fmffreeze
\fmf{plain}{v22,v36}
\fmf{plain}{v22,v16}
\end{fmfgraph*}
\end{fmffile}
}
\,\,+\cfrac{1}{2!}\,\cfrac{1}{3!}\quad
\parbox{33mm}{\vspace{0.5cm}
\begin{fmffile}{DiagPag34.8}
\begin{fmfgraph*}(33,20)
\fmfleft{i1,i2,i3}
\fmfright{o1,o2,o3}
\fmfdot{i2,o2}
\fmffreeze
\fmf{phantom}{i1,v11,v12,v13,v14,v15,v16,o1}
\fmffreeze
\fmf{phantom,tag=1}{i2,v21}
\fmf{phantom,tag=3}{v21,v22}
\fmf{phantom,tag=4}{v22,o2}
\fmfdot{v21,v22}
\fmfv{label=$x_1$,label.angle=-90}{i2}
\fmfv{label=$x_2$,label.angle=-90}{o2}
\fmfposition
\fmfipath{p[]}
\fmfiset{p1}{vpath1(__i2,__v21)}
\fmfi{plain}{subpath (0,length(p1)/2.5) of p1}
\fmfi{dashes}{subpath (length(p1)/2.8,length(p1)) of p1}
\fmfiset{p3}{vpath3(__v21,__v22)}
\fmfi{plain}{subpath (0,length(p3)*0.4) of p3}
\fmfi{plain}{subpath (length(p3)*0.6,length(p3)) of p3}
\fmfiset{p4}{vpath4(__v22,__o2)}
\fmfi{plain}{subpath (0,length(p4)*0.4) of p4}
\fmfi{plain}{subpath (length(p4)*0.6,length(p4)) of p4}
\fmffreeze
\fmf{phantom}{i3,v31,v32,v33,v34,v35,v36,o3}
\fmffreeze
\fmf{plain}{v21,v33}
\fmf{plain}{v21,v13}
\fmf{phantom,tension=0.7,tag=2}{v22,v22}
\fmfiset{p2}{vpath2(__v22,__v22)}
\fmfi{plain}{subpath (0,length(p2)/2.) of p2}
\fmfi{dashes}{subpath (length(p2)/2,length(p2)) of p2}
\fmffreeze
\end{fmfgraph*}
\end{fmffile}
}\hspace{0.3cm}+
\end{equation}
\begin{equation}\nonumber
\hspace{2.3cm}+\cfrac{1}{3!}\quad
\parbox{33mm}{\vspace{0.5cm}
\begin{fmffile}{DiagPag34.9}
\begin{fmfgraph*}(33,20)
\fmfleft{i1,i2,i3}
\fmfright{o1,o2,o3}
\fmfdot{i2,o2}
\fmffreeze
\fmf{phantom}{i1,v11,v12,v13,v14,v15,v16,o1}
\fmffreeze
\fmf{phantom,tag=3}{i2,v21}
\fmf{phantom,tension=0.7,tag=1}{v21,v21}
\fmf{phantom,tag=2}{v21,v22}
\fmf{phantom,tag=4}{v22,o2}
\fmfdot{v21,v22}
\fmfv{label=$x_1$,label.angle=-90}{i2}
\fmfv{label=$x_2$,label.angle=-90}{o2}
\fmfposition
\fmfipath{p[]}
\fmfiset{p1}{vpath1(__v21,__v21)}
\fmfi{plain}{subpath (0,length(p1)/2.) of p1}
\fmfi{dashes}{subpath (length(p1)/2,length(p1)) of p1}
\fmfiset{p2}{vpath2(__v21,__v22)}
\fmfi{plain}{subpath (0,length(p2)/2.5) of p2}
\fmfi{dashes}{subpath (length(p2)/2.8,length(p2)) of p2}
\fmfiset{p3}{vpath3(__i2,__v21)}
\fmfi{plain}{subpath (0,length(p3)*0.4) of p3}
\fmfi{plain}{subpath (length(p3)*0.6,length(p3)) of p3}
\fmfiset{p4}{vpath4(__v22,__o2)}
\fmfi{plain}{subpath (0,length(p4)*0.4) of p4}
\fmfi{plain}{subpath (length(p4)*0.6,length(p4)) of p4}
\fmffreeze
\fmf{phantom}{i3,v31,v32,v33,v34,v35,v36,o3}
\fmffreeze
\fmf{plain}{v22,v36}
\fmf{plain}{v22,v16}
\end{fmfgraph*}
\end{fmffile}
}
\hspace{0.3cm}+\,\,\cfrac{1}{(2!)^3}\quad
\parbox{33mm}{\vspace{0.5cm}
\begin{fmffile}{DiagPag34.10}
\begin{fmfgraph*}(33,20)
\fmfleft{i1,i2,i3}
\fmfright{o1,o2,o3}
\fmfdot{i2,o2}
\fmffreeze
\fmf{phantom}{i1,v11,v12,v13,v14,v15,v16,o1}
\fmffreeze
\fmf{phantom,tag=3}{i2,v21}
\fmf{phantom,tension=0.7,tag=1}{v21,v21}
\fmf{phantom,tension=0.7,tag=2}{v22,v22}
\fmf{phantom,tag=4}{v21,v22}
\fmf{phantom,tag=5}{v22,o2}
\fmfdot{v21,v22}
\fmfv{label=$x_1$,label.angle=-90}{i2}
\fmfv{label=$x_2$,label.angle=-90}{o2}
\fmfposition
\fmfipath{p[]}
\fmfiset{p1}{vpath1(__v21,__v21)}
\fmfi{plain}{subpath (0,length(p1)/2.) of p1}
\fmfi{dashes}{subpath (length(p1)/2,length(p1)) of p1}
\fmfiset{p2}{vpath2(__v22,__v22)}
\fmfi{plain}{subpath (0,length(p2)/2.) of p2}
\fmfi{dashes}{subpath (length(p2)/2,length(p2)) of p2}
\fmfiset{p3}{vpath3(__i2,__v21)}
\fmfi{plain}{subpath (0,length(p3)*0.4) of p3}
\fmfi{plain}{subpath (length(p3)*0.6,length(p3)) of p3}
\fmfiset{p4}{vpath4(__v21,__v22)}
\fmfi{plain}{subpath (0,length(p4)*0.4) of p4}
\fmfi{plain}{subpath (length(p4)*0.6,length(p4)) of p4}
\fmfiset{p5}{vpath5(__v22,__o2)}
\fmfi{plain}{subpath (0,length(p5)*0.4) of p5}
\fmfi{plain}{subpath (length(p5)*0.6,length(p5)) of p5}
\fmffreeze
\fmf{phantom}{i3,v31,v32,v33,v34,v35,v36,o3}
\fmffreeze
\end{fmfgraph*}
\end{fmffile}
}\hspace{0.2cm}+\end{equation}
\begin{equation}\nonumber
\hspace{1.8cm}+\,\,\cfrac{1}{(2!)^3}\quad
\parbox{33mm}{\vspace{0.cm}
\begin{fmffile}{DiagPag35.1}
\begin{fmfgraph*}(33,20)
\fmfleft{i1,i2,i3}
\fmfright{o1,o2,o3}
\fmfdot{i2}
\fmffreeze
\fmf{phantom}{i1,v11,v12,v13,v14,v15,v16,o1}
\fmffreeze
\fmf{phantom}{i3,v31,v32,v33,v34,v35,v36,o3}
\fmffreeze
\fmf{phantom,tag=3}{i2,v21}
\fmf{phantom}{v21,v22}
\fmf{phantom}{v22,o2}
\fmffreeze
\fmf{phantom,left=0.7,tag=1}{v21,v22}
\fmf{phantom,right=0.7,tag=2}{v21,v22}
\fmfposition
\fmfv{label=$x_1$,label.angle=-90}{i2}
\fmfv{label=$x_2$,label.angle=0}{v36}
\fmfipath{p[]}
\fmfiset{p1}{vpath1(__v21,__v22)}
\fmfi{plain}{subpath (0,length(p1)/2.) of p1}
\fmfi{dashes}{subpath (length(p1)/2,length(p1)) of p1}
\fmfiset{p2}{vpath2(__v21,__v22)}
\fmfi{dashes}{subpath (0,length(p2)/2.) of p2}
\fmfi{plain}{subpath (length(p2)/2,length(p2)) of p2}
\fmfiset{p3}{vpath3(__i2,__v21)}
\fmfi{plain}{subpath (0,length(p3)*0.4) of p3}
\fmfi{plain}{subpath (length(p3)*0.6,length(p3)) of p3}
\fmf{plain}{v21,v33}
\fmf{phantom,tag=4}{v22,v36}
\fmfiset{p4}{vpath4(__v22,__v36)}
\fmfi{plain}{subpath (0,length(p4)*0.4) of p4}
\fmfi{plain}{subpath (length(p4)*0.6,length(p4)) of p4}
\fmf{plain}{v22,v16}
\fmfdot{v36}
\fmffreeze
\fmfdot{v21,v22}
\fmfposition
\fmffreeze
\end{fmfgraph*}
\end{fmffile}}
\hspace{-0.4cm}+\cfrac{1}{2!}\,\cfrac{1}{3!}
\quad\parbox{33mm}{\vspace{0.5cm}
\begin{fmffile}{DiagPag35.2}
\begin{fmfgraph*}(33,20)
\fmfkeep{DiagPag35.2}
\fmfleft{i1,i2,i3}
\fmfright{o1,o2,o3}
\fmfdot{i2,o2}
\fmffreeze
\fmf{phantom}{i1,v11,v12,v13,v14,v15,v16,o1}
\fmffreeze
\fmf{phantom,tag=1}{i2,v21}
\fmf{phantom,tag=2}{v21,v22}
\fmf{phantom,tag=3}{v22,o2}
\fmfdot{v21,v22}
\fmfposition
\fmfv{label=$x_1$,label.angle=-90}{i2}
\fmfv{label=$x_2$,label.angle=-90}{o2}
\fmfipath{p[]}
\fmfiset{p1}{vpath1(__i2,__v21)}
\fmfi{plain}{subpath (0,length(p1)*0.4) of p1}
\fmfi{plain}{subpath (length(p1)*0.6,length(p1)) of p1}
\fmfiset{p2}{vpath2(__v21,__v22)}
\fmfi{dashes}{subpath (0,length(p2)*0.6) of p2}
\fmfi{plain}{subpath (length(p2)*0.6,length(p2)) of p2}
\fmfiset{p3}{vpath3(__v22,__o2)}
\fmfi{dashes}{subpath (0,length(p3)*0.6) of p3}
\fmfi{plain}{subpath (length(p3)*0.6,length(p3)) of p3}
\fmffreeze
\fmf{phantom}{i3,v31,v32,v33,v34,v35,v36,o3}
\fmffreeze
\fmf{plain}{v21,v33}
\fmf{plain}{v21,v13}
\fmffreeze
\fmf{plain}{v22,v36}
\fmf{plain}{v22,v16}
\end{fmfgraph*}
\end{fmffile}
}\hspace{0.4cm}+
\end{equation}
\begin{equation}\nonumber
+\cfrac{1}{2!}\,\cfrac{1}{3!}\quad
\parbox{33mm}{\vspace{0.5cm}
\begin{fmffile}{DiagPag35.3}
\begin{fmfgraph*}(33,20)
\fmfleft{i1,i2,i3}
\fmfright{o1,o2,o3}
\fmfdot{i2,o2}
\fmffreeze
\fmf{phantom}{i1,v11,v12,v13,v14,v15,v16,o1}
\fmffreeze
\fmf{phantom,tag=3}{i2,v21}
\fmf{phantom,tension=0.7,tag=1}{v21,v21}
\fmf{phantom,tag=2}{v21,v22}
\fmf{phantom,tag=4}{v22,o2}
\fmfdot{v21,v22}
\fmfv{label=$x_1$,label.angle=-90}{i2}
\fmfv{label=$x_2$,label.angle=-90}{o2}
\fmfposition
\fmfipath{p[]}
\fmfiset{p1}{vpath1(__v21,__v21)}
\fmfi{plain}{subpath (0,length(p1)/2.) of p1}
\fmfi{dashes}{subpath (length(p1)/2,length(p1)) of p1}
\fmfiset{p2}{vpath2(__v21,__v22)}
\fmfi{plain}{subpath (0,length(p2)*0.4) of p2}
\fmfi{plain}{subpath (length(p2)*0.6,length(p2)) of p2}
\fmfiset{p3}{vpath3(__i2,__v21)}
\fmfi{plain}{subpath (0,length(p3)*0.4) of p3}
\fmfi{plain}{subpath (length(p3)*0.6,length(p3)) of p3}
\fmfiset{p4}{vpath4(__v22,__o2)}
\fmfi{dashes}{subpath (0,length(p4)*0.6) of p4}
\fmfi{plain}{subpath (length(p4)*0.6,length(p4)) of p4}
\fmffreeze
\fmf{phantom}{i3,v31,v32,v33,v34,v35,v36,o3}
\fmffreeze
\fmf{plain}{v22,v36}
\fmf{plain}{v22,v16}
\end{fmfgraph*}
\end{fmffile}
}
\quad+\cfrac{1}{(3!)^2}\quad\parbox{33mm}{\vspace{0.5cm}
\begin{fmffile}{DiagPag35.4}
\begin{fmfgraph*}(33,20)
\fmfkeep{DiagPag35.4}
\fmfleft{i1,i2,i3}
\fmfright{o1,o2,o3}
\fmfdot{i2,o2}
\fmffreeze
\fmf{phantom}{i1,v11,v12,v13,v14,v15,v16,o1}
\fmffreeze
\fmf{phantom,tag=1}{i2,v21}
\fmf{phantom,tag=2}{v21,v22}
\fmf{phantom,tag=3}{v22,o2}
\fmfdot{v21,v22}
\fmfposition
\fmfv{label=$x_1$,label.angle=-90}{i2}
\fmfv{label=$x_2$,label.angle=-90}{o2}
\fmfipath{p[]}
\fmfiset{p1}{vpath1(__i2,__v21)}
\fmfi{plain}{subpath (0,length(p1)/2.5) of p1}
\fmfi{dashes}{subpath (length(p1)/2.8,length(p1)) of p1}
\fmfiset{p2}{vpath2(__v21,__v22)}
\fmfi{plain}{subpath (0,length(p2)*0.4) of p2}
\fmfi{plain}{subpath (length(p2)*0.6,length(p2)) of p2}
\fmfiset{p3}{vpath3(__v22,__o2)}
\fmfi{dashes}{subpath (0,length(p3)*0.6) of p3}
\fmfi{plain}{subpath (length(p3)*0.6,length(p3)) of p3}
\fmffreeze
\fmf{phantom}{i3,v31,v32,v33,v34,v35,v36,o3}
\fmffreeze
\fmf{plain}{v21,v33}
\fmf{plain}{v21,v13}
\fmffreeze
\fmf{plain}{v22,v36}
\fmf{plain}{v22,v16}
\end{fmfgraph*}
\end{fmffile}
}\hspace{0.5cm}+
\end{equation}
\begin{equation}\nonumber
\hspace{-1cm}+\cfrac{1}{(2!)^3}\quad
\parbox{33mm}{\vspace{0.5cm}
\begin{fmffile}{DiagPag35.5}
\begin{fmfgraph*}(33,20)
\fmfset{arrow_len}{2mm}
\fmfleft{i1,i2,i3}
\fmfright{o1,o2,o3}
\fmfdot{i2,o2}
\fmffreeze
\fmf{phantom}{i1,v11,v12,v13,v14,v15,v16,o1}
\fmffreeze
\fmf{phantom,tag=3}{i2,v21}
\fmf{dots_arrow,tension=0.7}{v21,v21}
\fmf{dots_arrow,tension=0.7}{v22,v22}
\fmf{phantom,tag=4}{v21,v22}
\fmf{phantom,tag=5}{v22,o2}
\fmfdot{v21,v22}
\fmfv{label=$x_1$,label.angle=-90}{i2}
\fmfv{label=$x_2$,label.angle=-90}{o2}
\fmfposition
\fmfipath{p[]}
\fmfiset{p3}{vpath3(__i2,__v21)}
\fmfi{plain}{subpath (0,length(p3)*0.4) of p3}
\fmfi{plain}{subpath (length(p3)*0.6,length(p3)) of p3}
\fmfiset{p4}{vpath4(__v21,__v22)}
\fmfi{plain}{subpath (0,length(p4)*0.4) of p4}
\fmfi{plain}{subpath (length(p4)*0.6,length(p4)) of p4}
\fmfiset{p5}{vpath5(__v22,__o2)}
\fmfi{plain}{subpath (0,length(p5)*0.4) of p5}
\fmfi{plain}{subpath (length(p5)*0.6,length(p5)) of p5}
\fmffreeze
\fmf{phantom}{i3,v31,v32,v33,v34,v35,v36,o3}
\fmffreeze
\end{fmfgraph*}
\end{fmffile}
}
\,\,-\,\,\cfrac{1}{(2!)^3}\quad
\parbox{33mm}{\vspace{0.cm}
\begin{fmffile}{DiagPag35.6}
\begin{fmfgraph*}(33,20)
\fmfset{arrow_len}{2mm}
\fmfleft{i1,i2,i3}
\fmfright{o1,o2,o3}
\fmfdot{i2}
\fmffreeze
\fmf{phantom}{i1,v11,v12,v13,v14,v15,v16,o1}
\fmffreeze
\fmf{phantom}{i3,v31,v32,v33,v34,v35,v36,o3}
\fmffreeze
\fmf{phantom,tag=3}{i2,v21}
\fmf{phantom}{v21,v22}
\fmf{phantom}{v22,o2}
\fmffreeze
\fmf{dots_arrow,left=0.7}{v21,v22}
\fmf{dots_arrow,left=0.7}{v22,v21}
\fmfposition
\fmfv{label=$x_1$,label.angle=-90}{i2}
\fmfv{label=$x_2$,label.angle=0}{v36}
\fmfipath{p[]}
\fmfiset{p3}{vpath3(__i2,__v21)}
\fmfi{plain}{subpath (0,length(p3)*0.4) of p3}
\fmfi{plain}{subpath (length(p3)*0.6,length(p3)) of p3}
\fmf{plain}{v21,v33}
\fmf{phantom,tag=4}{v22,v36}
\fmfiset{p4}{vpath4(__v22,__v36)}
\fmfi{plain}{subpath (0,length(p4)*0.4) of p4}
\fmfi{plain}{subpath (length(p4)*0.6,length(p4)) of p4}
\fmf{plain}{v22,v16}
\fmfdot{v36}
\fmffreeze
\fmfdot{v21,v22}
\fmfposition
\fmffreeze
\end{fmfgraph*}
\end{fmffile}}\hspace{-0.4cm}+
\end{equation}
\begin{equation}\nonumber
\hspace{-1cm}-\,\,\cfrac{1}{(2!)^2}\quad
\parbox{33mm}{\vspace{0.5cm}
\begin{fmffile}{DiagPag35.7}
\begin{fmfgraph*}(33,20)
\fmfset{arrow_len}{2mm}
\fmfleft{i1,i2,i3}
\fmfright{o1,o2,o3}
\fmfdot{i2,o2}
\fmffreeze
\fmf{phantom}{i1,v11,v12,v13,v14,v15,v16,o1}
\fmffreeze
\fmf{phantom,tag=3}{i2,v21}
\fmf{phantom,tension=0.7,tag=1}{v21,v21}
\fmf{dots_arrow,tension=0.7}{v22,v22}
\fmf{phantom,tag=4}{v21,v22}
\fmf{phantom,tag=5}{v22,o2}
\fmfdot{v21,v22}
\fmfv{label=$x_1$,label.angle=-90}{i2}
\fmfv{label=$x_2$,label.angle=-90}{o2}
\fmfposition
\fmfipath{p[]}
\fmfiset{p1}{vpath1(__v21,__v21)}
\fmfi{plain}{subpath (0,length(p1)/2.) of p1}
\fmfi{dashes}{subpath (length(p1)/2,length(p1)) of p1}
\fmfiset{p3}{vpath3(__i2,__v21)}
\fmfi{plain}{subpath (0,length(p3)*0.4) of p3}
\fmfi{plain}{subpath (length(p3)*0.6,length(p3)) of p3}
\fmfiset{p4}{vpath4(__v21,__v22)}
\fmfi{plain}{subpath (0,length(p4)*0.4) of p4}
\fmfi{plain}{subpath (length(p4)*0.6,length(p4)) of p4}
\fmfiset{p5}{vpath5(__v22,__o2)}
\fmfi{plain}{subpath (0,length(p5)*0.4) of p5}
\fmfi{plain}{subpath (length(p5)*0.6,length(p5)) of p5}
\fmffreeze
\fmf{phantom}{i3,v31,v32,v33,v34,v35,v36,o3}
\fmffreeze
\end{fmfgraph*}
\end{fmffile}
}
\,\,-\cfrac{1}{3!}\quad
\parbox{33mm}{\vspace{0.5cm}
\begin{fmffile}{DiagPag35.8}
\begin{fmfgraph*}(33,20)
\fmfset{arrow_len}{2mm}
\fmfleft{i1,i2,i3}
\fmfright{o1,o2,o3}
\fmfdot{i2,o2}
\fmffreeze
\fmf{phantom}{i1,v11,v12,v13,v14,v15,v16,o1}
\fmffreeze
\fmf{phantom,tag=3}{i2,v21}
\fmf{dots_arrow,tension=0.7}{v21,v21}
\fmf{phantom,tag=2}{v21,v22}
\fmf{phantom,tag=4}{v22,o2}
\fmfdot{v21,v22}
\fmfv{label=$x_1$,label.angle=-90}{i2}
\fmfv{label=$x_2$,label.angle=-90}{o2}
\fmfposition
\fmfipath{p[]}
\fmfiset{p2}{vpath2(__v21,__v22)}
\fmfi{plain}{subpath (0,length(p2)/2.5) of p2}
\fmfi{dashes}{subpath (length(p2)/2.8,length(p2)) of p2}
\fmfiset{p3}{vpath3(__i2,__v21)}
\fmfi{plain}{subpath (0,length(p3)*0.4) of p3}
\fmfi{plain}{subpath (length(p3)*0.6,length(p3)) of p3}
\fmfiset{p4}{vpath4(__v22,__o2)}
\fmfi{plain}{subpath (0,length(p4)*0.4) of p4}
\fmfi{plain}{subpath (length(p4)*0.6,length(p4)) of p4}
\fmffreeze
\fmf{phantom}{i3,v31,v32,v33,v34,v35,v36,o3}
\fmffreeze
\fmf{plain}{v22,v36}
\fmf{plain}{v22,v16}
\end{fmfgraph*}
\end{fmffile}
}\hspace{0.3cm}+
\end{equation}
\begin{equation}
\label{phiphiexpansion}
\hspace{-0.8cm}-\cfrac{1}{2!}\,\cfrac{1}{3!}\quad
\parbox{33mm}{\vspace{0.5cm}
\begin{fmffile}{DiagPag35.9}
\begin{fmfgraph*}(33,20)
\fmfset{arrow_len}{2mm}
\fmfleft{i1,i2,i3}
\fmfright{o1,o2,o3}
\fmfdot{i2,o2}
\fmffreeze
\fmf{phantom}{i1,v11,v12,v13,v14,v15,v16,o1}
\fmffreeze
\fmf{phantom,tag=1}{i2,v21}
\fmf{phantom,tag=3}{v21,v22}
\fmf{phantom,tag=4}{v22,o2}
\fmfdot{v21,v22}
\fmfv{label=$x_1$,label.angle=-90}{i2}
\fmfv{label=$x_2$,label.angle=-90}{o2}
\fmfposition
\fmfipath{p[]}
\fmfiset{p1}{vpath1(__i2,__v21)}
\fmfi{plain}{subpath (0,length(p1)/2.5) of p1}
\fmfi{dashes}{subpath (length(p1)/2.8,length(p1)) of p1}
\fmfiset{p3}{vpath3(__v21,__v22)}
\fmfi{plain}{subpath (0,length(p3)*0.4) of p3}
\fmfi{plain}{subpath (length(p3)*0.6,length(p3)) of p3}
\fmfiset{p4}{vpath4(__v22,__o2)}
\fmfi{plain}{subpath (0,length(p4)*0.4) of p4}
\fmfi{plain}{subpath (length(p4)*0.6,length(p4)) of p4}
\fmffreeze
\fmf{phantom}{i3,v31,v32,v33,v34,v35,v36,o3}
\fmffreeze
\fmf{plain}{v21,v33}
\fmf{plain}{v21,v13}
\fmf{dots_arrow,tension=0.7}{v22,v22}
\fmfiset{p2}{vpath2(__v22,__v22)}
\fmffreeze
\end{fmfgraph*}
\end{fmffile}
}
\,\,\,-\cfrac{1}{2!}\,\cfrac{1}{3!}\quad
\parbox{33mm}{\vspace{0.5cm}
\begin{fmffile}{DiagPag35.10}
\begin{fmfgraph*}(33,20)
\fmfset{arrow_len}{2mm}
\fmfleft{i1,i2,i3}
\fmfright{o1,o2,o3}
\fmfdot{i2,o2}
\fmffreeze
\fmf{phantom}{i1,v11,v12,v13,v14,v15,v16,o1}
\fmffreeze
\fmf{phantom,tag=3}{i2,v21}
\fmf{dots_arrow,tension=0.7}{v21,v21}
\fmf{phantom,tag=2}{v21,v22}
\fmf{phantom,tag=4}{v22,o2}
\fmfdot{v21,v22}
\fmfv{label=$x_1$,label.angle=-90}{i2}
\fmfv{label=$x_2$,label.angle=-90}{o2}
\fmfposition
\fmfipath{p[]}
\fmfiset{p2}{vpath2(__v21,__v22)}
\fmfi{plain}{subpath (0,length(p2)*0.4) of p2}
\fmfi{plain}{subpath (length(p2)*0.6,length(p2)) of p2}
\fmfiset{p3}{vpath3(__i2,__v21)}
\fmfi{plain}{subpath (0,length(p3)*0.4) of p3}
\fmfi{plain}{subpath (length(p3)*0.6,length(p3)) of p3}
\fmfiset{p4}{vpath4(__v22,__o2)}
\fmfi{dashes}{subpath (0,length(p4)*0.6) of p4}
\fmfi{plain}{subpath (length(p4)*0.6,length(p4)) of p4}
\fmffreeze
\fmf{phantom}{i3,v31,v32,v33,v34,v35,v36,o3}
\fmffreeze
\fmf{plain}{v22,v36}
\fmf{plain}{v22,v16}
\end{fmfgraph*}
\end{fmffile}
}\quad\Bigg]
\end{equation}
There are the usual cancelations between various diagrams and
the final result is:
\begin{equation}\nonumber
\langle \phi(x_2) \phi(x_1)\rangle_{\varphi_i}^{(2)}=\Bigg[\frac{1}{2!}\,\frac{1}{3!}\,\,\,\parbox{33mm}{
\fmfreuse{DiagPag34.7}}\quad+\frac{1}{2!}\,\frac{1}{3!}\,\,\,\parbox{33mm}{
\fmfreuse{DiagPag35.2}
}
\end{equation}
\begin{equation}
\label{phiphi2final}\hspace{-2cm}+\frac{1}{(3!)^2}\,\,\,\parbox{33mm}{
\fmfreuse{DiagPag35.4}
}\quad\Bigg]
\end{equation} Note that also here no loop is left. This is the real sign
that we are doing a {\it classical } perturbation theory. When we will
introduce the temperature later on loops will appear, but they are due to
temperature and not quantum effects. Actually, loops would appear any time we do
an average over the initial conditions so we can not strictly say
that no loops is a sign of classicality, we can say that no loop
is the sign of a generating functional which has no average over
the initial configurations.

Collecting now all diagrams developed before for the two-point functions ,up to second
order, we get:
\begin{equation}\nonumber\label{<phi phi>phi i}
\langle \phi(x_2) \phi(x_1)\rangle_{\varphi_i}=\,\,\Bigg[\quad
\parbox{33mm}{
\fmfreuse{DiagPag25.2}
}
\hspace{-1.5cm}+\frac{1}{3!}\quad\parbox{33mm}{
\fmfreuse{DiagPag34.2}
}
\,\,\,+\frac{1}{3!}\quad\parbox{33mm}{
\fmfreuse{DiagPag34.3}
}+\end{equation}
\begin{equation}\nonumber
\hspace{1.cm}\,\,+\frac{1}{2!}\,\frac{1}{3!}\quad\parbox{33mm}{
\fmfreuse{DiagPag34.7}}\quad+\frac{1}{2!}\,\frac{1}{3!}\quad\parbox{33mm}{
\fmfreuse{DiagPag35.2}
}\hspace{0.3cm}+
\end{equation}
\begin{equation}
\hspace{-4.4cm}+\frac{1}{(3!)^2}\quad\parbox{33mm}{
\fmfreuse{DiagPag35.4}
}\quad\Bigg]
\end{equation}

The rules for the symmetry factors are the following:
\begin{itemize}

\item[{\bf 1)}]put a factor of $1/n!$ where $n$ are the number of ``free'' legs,
i.e. those not starting or ending at $x_1$ or $x_2$;

\item[{\bf 2)}] add a factor of $1/2!$ for any exchange of propagators which
would leave the diagram invariant.
\end{itemize}

Before concluding this section we would like to answer to a
question that for sure many readers may have. The question is why
all the diagrams in this section are disconnected or have legs
going nowhere. The reason is not related to the fact that we may
have used the wrong generating functional, but to the presence of
the diagram (\ref{<phi>(0)}). To convince the reader, let us give
the analytical expression for some of the most ``strange''
diagrams. For example, the 10th diagram in  (\ref{phiphiexpansion}) where there are both disconnected pieces and lines
going nowhere:
\begin{eqnarray}
\parbox{33mm}{\vspace{0.cm}
\begin{fmffile}{DiagPag38.1}
\begin{fmfgraph*}(33,20)
\fmfset{arrow_len}{2mm}
\fmfleft{i1,i2,i3}
\fmfright{o1,o2,o3}
\fmfdot{i2}
\fmffreeze
\fmf{phantom}{i1,v11,v12,v13,v14,v15,v16,o1}
\fmffreeze
\fmf{phantom}{i3,v31,v32,v33,v34,v35,v36,o3}
\fmffreeze
\fmf{phantom,tag=3}{i2,v21}
\fmf{phantom}{v21,v22}
\fmf{phantom}{v22,o2}
\fmffreeze
\fmf{dots_arrow,left=0.7}{v21,v22}
\fmf{dots_arrow,left=0.7}{v22,v21}
\fmfposition
\fmfv{label=$x_1$,label.angle=-90}{i2}
\fmfv{label=$x_2$,label.angle=0}{v36}
\fmfv{label=$y_1$,label.angle=-120}{v21}
\fmfv{label=$y_2$,label.angle=0}{v22}
\fmfipath{p[]}
\fmfiset{p3}{vpath3(__i2,__v21)}
\fmfi{plain}{subpath (0,length(p3)*0.4) of p3}
\fmfi{plain}{subpath (length(p3)*0.6,length(p3)) of p3}
\fmf{plain}{v21,v33}
\fmf{phantom,tag=4}{v22,v36}
\fmfiset{p4}{vpath4(__v22,__v36)}
\fmfi{plain}{subpath (0,length(p4)*0.4) of p4}
\fmfi{plain}{subpath (length(p4)*0.6,length(p4)) of p4}
\fmf{plain}{v22,v16}
\fmfdot{v36}
\fmffreeze
\fmfdot{v21,v22}
\fmfposition
\fmffreeze
\end{fmfgraph*}
\end{fmffile}}&\sim& \left[\int d^4y_1\,d^4y_2\,\phi_0(y_1)^2\,\phi_0(z_2)^2\,G^{(F)}_R(x_1-x_2)\right.\nonumber\\&&\left.G^{(F)}_R(x_2-x_1)\right]\,\phi_0(x_1)\,\phi_0(x_2)
\nonumber
\end{eqnarray}
We see that we have at $x_1$ and $x_2$ the fields $\phi_0(x_1)$
and $\phi_0(x_2)$ and their diagram is the one of
(\ref{<phi>(0)}). In $y_1$ and $y_2$ we also have two lines going
nowhere and, according to (\ref{<phi>(0)}) there must be two
fields $\phi_0(y)$. The rest, between $y_1$ and $y_2$, are two
``Fermionic'' retarded propagators.

The reason we can not connect $x_1$ with $y_1$ or  $y_2$ with
$x_2$ with a continuous line is because we do not have this kind
of propagator in the formalism with fixed initial condition. The
only propagators we have are those in (\ref{lambdaphi}),
(\ref{philambda}), (\ref{cbarc}) and (\ref{barcc}), which are the
dash-full propagator or the Fermionic dot propagators. For the
$\phi \phi$ we have only the disconnected propagators
(\ref{<phiphi>(0)}). The reader may think that by coupling  in some way  various
dash-full propagators one could get a $\phi \phi$ propagator. If
one tries, it is easy to convince himself that this is not possible.

In this chapter we have studied the "average"  of a single field
$\phi$ or correlations of two fields $\phi \phi$. It is possible to
do the same for the $\lambda_{\pi}$ fields and the $c^{\phi}$,
$\bar{c}_{\pi}$ ones. Instead of doing them explicitly we will
work out the super-field perturbation theory for this
$\mathcal{Z}_{\varphi_i}[J]$ in appendix \ref{Deltabetaborn}. We
can then project out the components of the super-fields and get
all the correlations mentioned above. We advise the reader not to
jump immediately to appendix \ref{Deltabetaborn}, but to wait till
he has read sec. \ref{PertubwithT}.
\section{Perturbation Theory With Temperature.}\label{PertubwithT}

Let us now suppose that, instead of working with some fixed
initial configuration, as in section \ref{PertubnoT}, we do a
thermal average over the initial configuration. Let us use the
following notation: $\vec{x}$ for the 3-dim vector and $x$ for the 4-dimensional one.
Let us remember that in section \ref{PertubnoT} we choose a
solution $\phi_0(\vec{x},t)$ of  the Klein-Gordon equation:
\begin{equation}
\left(\partial_t^2-\nabla^2+m^2\right)\phi_0(\vec{x},t)=0.
\end{equation}  Its  Fourier transform $\tilde{\phi}_0(\vec{p},t)$ is :
\begin{equation}
\phi_0(\vec{x},t)=\displaystyle\int\frac{d^3\vec{p}}{(2\pi)^3}
\tilde{\phi}_0(\vec{p},t)e^{i\vec{p}.\vec{x}}
\end{equation} and  satisfies the equation:
\begin{equation}
\left(\partial_t^2-E_{\vec{p}}^2\right)\tilde{\phi}_0(\vec{p},t)=0,
\end{equation} whose solution has the form
\begin{equation}\label{tildephi0momentum}
\tilde{\phi}_0(\vec{p},t)
=\phi_i(\vec{p})\cos\left[E_{\vec{p}}(t-t_i)\right]
+\frac{\pi_i(\vec{p})}{E_{\vec{p}}}\sin\left[E_{\vec{p}}(t-t_i)\right]
\end{equation} with $t_i$, $\phi_i$, $\pi_i$ the initial time and field configurations. So
the final expression for the field is:
\begin{equation}\label{phi0aphiapi}
\phi_0(\vec{x},t)
=\displaystyle\int\frac{d^3\vec{p}}{(2\pi)^3}\left\{\phi_i(\vec{p})\cos\left[E_{\vec{p}}(t-t_i)\right]
+\frac{\pi_i(\vec{p})}{E_{\vec{p}}}\sin\left[E_{\vec{p}}(t-t_i)\right]e^{i\vec{p}.\vec{x}}
\right\}.\end{equation} The thermal-averaged correlation-
functions that we want to evaluate are defined, for example for a two-point
function , as follows:
\begin{equation}
\langle \phi \phi\rangle_{\beta}
=\frac{\int {\mathscr D}\phi_i(\vec{x}){\mathscr D}\pi_i(\vec{x})
\langle\phi\phi\rangle_{\varphi_i\equiv(\phi_i,\pi_i)} e^{-\beta H(\phi_i,\pi_i)}}
{\int {\mathscr D}\phi_i(\vec{x}){\mathscr D}\pi_i(\vec{x}) e^{-\beta H(\phi_i,\pi_i)}}
\end{equation} where $\phi_i(\vec{x})$, $\pi_i(\vec{x})$ are the Fourier transform of the $\phi_i(\vec{p})$
and $\pi(\vec{p})$ appearing in Eq. (\ref{tildephi0momentum}).

On the correlation $\langle \phi \phi\rangle_{\beta}$ we have not
indicated the argument of the fields. We did that  in order to simplify the
notation for the moment.

The analog of the generating functional (\ref{Z=expSVZ0}) will now
be:
\begin{equation}\label{Z_beta[Jphi]}
\mathcal{Z}_{\beta}[J_{\phi}]
=\frac{\int {\mathscr D}\phi_i(\vec{x}){\mathscr D}\pi_i(\vec{x})
e^{-\beta H(\phi_i,\pi_i)}\mathcal{Z}_{\phi_i}[J]}
{\int {\mathscr D}\phi_i(\vec{x}){\mathscr D}\pi_i(\vec{x}) e^{-\beta H(\phi_i,\pi_i)}}
\end{equation} In order to do the integration over the initial configuration in
(\ref{Z_beta[Jphi]}), let us first do the anti-Fourier transform
of the $\phi(\vec{p})$ and $\pi(\vec{p})$:
\begin{equation}
\phi_i(\vec{p})=\displaystyle\int d^3\vec{x}e^{-i\vec{p}.\vec{x}}\phi_i(\vec{x})
\end{equation}
\begin{equation}
\pi_i(\vec{p})=\displaystyle\int d^3\vec{x}e^{-i\vec{p}.\vec{x}}\pi_i(\vec{x}).
\end{equation} So we can write $\phi_0(\vec{x},t)$ as:
\begin{equation}\label{phi0}
\phi_0(\vec{x},t)
=\displaystyle\int d^3\vec{x}'\phi_i(\vec{x}')a_{\phi}(\vec{x}-\vec{x}',t)
+\displaystyle\int d^3\vec{x}'\pi_i(\vec{x}')a_{\pi}(\vec{x}-\vec{x}',t)
\end{equation} where
\begin{equation}\label{a_phi}
a_{\phi}(\vec{x}-\vec{x}',t)
=\displaystyle\int d^3\vec{x}e^{-i\vec{p}.(\vec{x}-\vec{x}')}
\cos\left[E_{\vec{p}}(t-t_i)\right]
\end{equation}
\begin{equation}\label{a_pi}
a_{\pi}(\vec{x}-\vec{x}',t)
=\displaystyle\int d^3\vec{x}e^{-i\vec{p}.(\vec{x}-\vec{x}')}
\displaystyle\frac{\sin\left[E_{\vec{p}}(t-t_i)\right]}{E_{\vec{p}}}.
\end{equation}

Next, let us extract from (\ref{Z_beta[Jphi]}) the part which
depends only on the initial field configurations which is:
\begin{equation}\label{TildeZ_beta[Jphi]}
\mathcal{\tilde{Z}}_{\beta}[J_{\phi}]
\equiv\frac{\displaystyle\int {\mathscr D}\phi_i(\vec{x}){\mathscr D}\pi_i(\vec{x})
e^{-\beta H_0(\phi_i,\pi_i)+\int d^4x J_{\phi}(x)\phi_0(x)}}
{\int {\mathscr D}\phi_i(\vec{x}){\mathscr D}\pi_i(\vec{x}) e^{-\beta H_0(\phi_i,\pi_i)}}
\end{equation} where we have switched off \cite{Jeon} the interaction in $H$
so that $H_0$ is just:
\begin{equation}
H_0(\phi_i,\pi_i)=\displaystyle\int d^4 x\left[\frac{\pi_i}{2}
+\frac{\left(\nabla \phi_i\right)^2}{2}
+\frac{m^2 \phi^2_i}{2}
\right]
\end{equation} Inserting (\ref{phi0}),
(\ref{a_phi}) and (\ref{a_pi}) in (\ref{TildeZ_beta[Jphi]}) it is
easy to perform the integration over the initial condition (see
Ref. \cite{Jeon} for details or appendix \ref{Deltabetaborn} of
our paper) and the result is:
\begin{equation}\label{TildeZ_beta[Jphi]2}
\mathcal{\tilde{Z}}_{\beta}[J_{\phi}]
=\exp\left[-\frac{1}{2}\int d^4x d^4x'J_{\phi}(x)\Delta_{\beta}(x-x')J_{\phi}(x')
\right]
\end{equation} where:
\begin{equation}
\Delta_{\beta}(x-x')=\int \frac{d^4 p}{(2\pi)^4}\frac{2\pi}{\beta|p^0|}
\delta(p^2-m^2)e^{-ip.(x-x')}.
\end{equation} It is easy to prove that
\begin{equation}
\Delta_{\beta}(x-x')=\Delta_{\beta}(x'-x).
\end{equation} Going now back to the full expression (\ref{Z_beta[Jphi]}) of the generating
functional, the full Bosonic part is:
\begin{equation}\label{ZbetaBJphi}
\begin{array}{lll}
\mathcal{Z}_{(B)\;\beta}[J_{\phi}]
&=&\left[-\frac{1}{2}\displaystyle\int d^4x d^4x'J_{\phi}(x)\Delta_{\beta}(x-x')J_{\phi}(x')\right.
\vspace{2mm}\\
&&\left.+i\displaystyle\int d^4x d^4x'J_{\phi}(x)G_{R}^{(B)}(x-x')J_{\lambda_{\pi}}(x')\right].
\end{array}
\end{equation} Let us calculate the thermal two-point function
\begin{equation}
\langle \phi(x_2) \phi(x_1)\rangle_{\beta}
=\left(\frac{1}{i}\right)^2\frac{\delta }{\delta J_{\phi}(x_2)}
\frac{\delta }{\delta J_{\phi}(x_1)}Z_{\beta}[J_{\phi}]|_{J_{\phi}=0}
=\Delta(x_2-x_1).
\end{equation} We can see that this correlation is not anymore the product of the two fields
at $x_2$ and $x_1$ like in (\ref{<phiphi>(0)}), but it is a
function $\Delta_{\beta}(x_2-x_1)$ which links $x_2$ with $x_1$.
We will use the full line to indicate the Feynman diagram
associated to $\Delta_{\beta}(x_2-x_1)$
\begin{equation}
\Delta_{\beta}(x_2-x_1)=\,\,\begin{minipage}[c]{16mm}
\vspace{0.55cm}\begin{fmffile}{DiagPag44.1}
\begin{fmfgraph*}(16,1)
\fmfkeep{DiagPag44.1}
\fmfleft{i1}
\fmfright{o1}
\fmfdot{i1}
\fmfdot{o1}
\fmf{plain}{i1,o1}
\fmfv{label=$x_1$,label.angle=-90}{i1}
\fmfv{label=$x_2$,label.angle=-90}{o1}
\end{fmfgraph*}
\end{fmffile}
\end{minipage}..
\end{equation}
 Let us now see how to get the first order correction to the
two-point function. Let us, for example, look at the first diagram
in (\ref{phiphi1final}), and let us take its thermal average
indicated by $\langle \;\;\rangle_{\beta}$:
\begin{equation}\nonumber
\hspace{-2cm}\frac{1}{3!}\, \Bigg\langle \quad
\parbox{32mm}{
\fmfreuse{DiagPag34.2}
}
\,\,\quad \Bigg\rangle_{\beta} =\frac{g}{3!}\,\int\,dy\,G_R(x_1-y)\,
\big\langle\phi_0(y)^3\,\phi_0(x_2)\big\rangle_{\beta}
\end{equation}
\vspace{-0.4cm}
\begin{equation}\nonumber
\hspace{5.8cm}=\frac{g}{3!}\,\int\,dy\,G_R(x_1-y)\,\frac{1}{i^4}\,
\frac{\delta^3}{\delta J_{\phi}(y)^3}\frac{\delta}{\delta J_{\phi}(x_2)}Z_{\beta}[J_{\phi}]\Big|_{J_{\phi}=0}
\end{equation}
\vspace{0.2cm}
\begin{equation}\nonumber
\hspace{4.5cm}=\frac{g}{3!}\,\int\,dy\,G_R(x_1-y)\,\Big[3\,\Delta_{\beta}(y,y)\,\Delta_{\beta}(y-x_2)\Big]
\end{equation}
\begin{equation}
\label{thermal1storder1}
\hspace{1.cm}=\frac{1}{2!}\quad
\begin{minipage}[c]{32mm}
\vspace{0.55cm}
\begin{fmffile}{DiagPag44.3}
\begin{fmfgraph*}(32,10)
\fmfkeep{DiagPag44.3}
\fmfleft{i1}
\fmfright{o1}
\fmfdot{i1,o1}
\fmfv{label=$x_1$,label.angle=-90}{i1}
\fmfv{label=$x_2$,label.angle=-90}{o1}
\fmf{phantom}{i1,v,o1}
\fmf{plain}{v,o1}
\fmfdot{v}
\fmfv{label=$y$,label.angle=-90}{v}
\fmf{phantom,tag=1}{i1,v}
\fmfposition
\fmfipath{p[]}
\fmfiset{p1}{vpath1(__i1,__v)}
\fmfi{plain}{subpath (0,length(p1)*0.40.) of p1}
\fmfi{dashes}{subpath (length(p1)*0.4,length(p1)) of p1}
\fmf{plain,tension=0.9}{v,v}
\end{fmfgraph*}
\end{fmffile}
\end{minipage}
\end{equation}
Let us notice  that we do not have anymore disconnected diagrams. The
tree legs in $y$ going nowhere and the one in $x_2$ get soldered
to each other in all possible combination producing the loops in
(\ref{thermal1storder1}) in $y$ and the  propagator between $y$ and
$x_2$.

The same can be done for the second diagram in (\ref{phiphi1final})
\begin{equation}
\label{thermal<>1storder2}
\frac{1}{3!}\, \Bigg\langle \quad
\parbox{32mm}{
\fmfreuse{DiagPag34.3}
}
\quad \Bigg\rangle_{\beta}=\frac{1}{2!}\quad
\begin{minipage}[c]{32mm}
\vspace{0.55cm}
\begin{fmffile}{DiagPag45.1}
\begin{fmfgraph*}(32,10)
\fmfkeep{DiagPag45.1}
\fmfleft{i1}
\fmfright{o1}
\fmfdot{i1,o1}
\fmfv{label=$x_1$,label.angle=-90}{i1}
\fmfv{label=$x_2$,label.angle=-90}{o1}
\fmf{phantom}{i1,v,o1}
\fmf{phantom,tag=2}{v,o1}
\fmfdot{v}
\fmf{phantom,tag=1}{i1,v}
\fmfposition
\fmfipath{p[]}
\fmfiset{p1}{vpath1(__i1,__v)}
\fmfi{plain}{subpath (0,length(p1)*0.40.) of p1}
\fmfi{plain}{subpath (length(p1)*0.4,length(p1)) of p1}
\fmf{plain,tension=0.9}{v,v}
\fmfiset{p2}{vpath2(__v,__o1)}
\fmfi{dashes}{subpath (0,length(p2)*0.60.) of p2}
\fmfi{plain}{subpath (length(p2)*0.6,length(p2)) of p2}
\end{fmfgraph*}
\end{fmffile}
\end{minipage}
\end{equation}
The same for the second order contribution in
(\ref{phiphi2final}). After some long but straightforward
calculations, we get from the first diagram in
(\ref{phiphi2final}):
\newpage
\vspace{8cm}
 \begin{equation}\label{thermal<>2ndtorder1}
\hspace{-4cm}\frac{1}{2!}\,\frac{1}{3!}\,\Bigg\langle \quad \parbox{33mm}{
 \fmfreuse{DiagPag34.7}
 } \quad \Bigg \rangle_{\beta}=\,\,\Bigg[\frac{1}{2!}\quad
\begin{minipage}[c]{30mm}
\vspace{-0.63cm}
 \begin{fmffile}{DiagPag45.2}
\begin{fmfgraph*}(30,15)
\fmfkeep{DiagPag45.2}
\fmftop{t0,t1}
\fmfbottom{b0,b1}
\fmfv{label=$x_1$,label.angle=-90}{b0}
\fmfv{label=$x_2$,label.angle=-90}{b1}
\fmf{phantom}{t0,v1,t1}
\fmf{phantom}{b0,v2,b1}
\fmffreeze
\fmf{plain}{v2,b1}
\fmf{phantom,tag=2}{b0,v2}
\fmfdot{b1}
\fmfdot{b0}
\fmf{plain,tension=0.8}{v1,v1}
\fmf{dashes,tension=0.8,right=0.6,tag=1}{v1,v2}
\fmf{plain,tension=0.8,left=0.6}{v1,v2}
\fmfdot{v1,v2}
\fmfposition
\fmfipath{p[]}
\fmfiset{p1}{vpath1(__v1,__v2)}
 \fmfi{plain}{subpath (length(p1)*0.6,length(p1)) of p1}
\fmfiset{p2}{vpath2(__b0,__v2)}
 \fmfi{plain}{subpath (0,length(p2)*0.4) of p2}
 \fmfi{dashes}{subpath (length(p2)*0.4,length(p2)) of p2}
\end{fmfgraph*}
 \end{fmffile}
 \end{minipage}\hspace{0.5cm}+
\end{equation}
\begin{equation}\nonumber
\hspace{-1.6cm}+\frac{1}{(2!)^2}\quad
\parbox{33mm}{\vspace{0.5cm}
\begin{fmffile}{DiagPag45.3}
\begin{fmfgraph*}(50,20)
\fmfkeep{DiagPag45.3}
\fmfleft{i1,i2,i3}
\fmfright{o1,o2,o3}
\fmfdot{i2,o2}
\fmffreeze
\fmf{phantom}{i1,v11,v12,v13,v14,v15,v16,o1}
\fmffreeze
\fmf{phantom,tag=3}{i2,v21}
\fmf{plain,tension=0.9}{v21,v21}
\fmf{plain,tension=0.9}{v22,v22}
\fmf{phantom,tag=4}{v21,v22}
\fmf{plain}{v22,o2}
\fmfdot{v21,v22}
\fmfv{label=$x_1$,label.angle=-90}{i2}
\fmfv{label=$x_2$,label.angle=-90}{o2}
\fmfposition
\fmfipath{p[]}
\fmfiset{p3}{vpath3(__i2,__v21)}
\fmfi{plain}{subpath (0,length(p3)*0.4) of p3}
\fmfi{dashes}{subpath (length(p3)*0.4,length(p3)) of p3}
\fmfiset{p4}{vpath4(__v21,__v22)}
\fmfi{plain}{subpath (0,length(p4)*0.4) of p4}
\fmfi{dashes}{subpath (length(p4)*0.4,length(p4)) of p4}
\fmffreeze
\fmf{phantom}{i3,v31,v32,v33,v34,v35,v36,o3}
\fmffreeze
\end{fmfgraph*}
\end{fmffile}
}
\hspace{2cm}+\frac{1}{2!}\quad
\parbox{33mm}{\vspace{0.cm}
\begin{fmffile}{DiagPag45.4}
\begin{fmfgraph*}(50,20)
\fmfkeep{DiagPag45.4}
\fmfleft{i1,i2,i3}
\fmfright{o1,o2,o3}
\fmfdot{i2,o2}
\fmffreeze
\fmf{phantom}{i1,v11,v12,v13,v14,v15,v16,o1}
\fmffreeze
\fmf{phantom}{i3,v31,v32,v33,v34,v35,v36,o3}
\fmffreeze
\fmf{phantom,tag=3}{i2,v21}
\fmf{phantom}{v21,v22}
\fmf{plain}{v22,o2}
\fmffreeze
\fmf{plain,left=0.9}{v21,v22}
\fmf{plain,right=0.9}{v21,v22}
\fmf{phantom,tag=1}{v21,v22}
\fmfposition
\fmfv{label=$x_1$,label.angle=-90}{i2}
\fmfv{label=$x_2$,label.angle=-90}{o2}
\fmfipath{p[]}
\fmfiset{p3}{vpath3(__i2,__v21)}
\fmfi{plain}{subpath (0,length(p3)*0.4) of p3}
\fmfi{dashes}{subpath (length(p3)*0.4,length(p3)) of p3}
\fmfiset{p1}{vpath1(__v21,__v22)}
\fmfi{plain}{subpath (0,length(p1)*0.4) of p1}
\fmfi{dashes}{subpath (length(p1)*0.4,length(p1)) of p1}
\fmffreeze
\fmfdot{v21,v22}
\fmfposition
\fmffreeze
\end{fmfgraph*}
\end{fmffile}}\hspace{2cm}\Bigg]
\end{equation}
For the second diagram in (\ref{phiphi2final}) we obtain:

\vspace{0.5cm}\begin{equation}
\label{thermal<>2ndtorder2}
\hspace{-4cm}\frac{1}{2!}\,\frac{1}{3!}\,\Bigg\langle \quad \parbox{33mm}{
\fmfreuse{DiagPag35.2}
} \quad \Bigg \rangle_{\beta} =\,\,\Bigg[\frac{1}{2!}\quad
\begin{minipage}[c]{30mm}
\vspace{-0.63cm}
 \begin{fmffile}{DiagPag46.1}
\begin{fmfgraph*}(30,15)
\fmfkeep{DiagPag46.1}
\fmftop{t0,t1}
\fmfbottom{b0,b1}
\fmfv{label=$x_1$,label.angle=-90}{b0}
\fmfv{label=$x_2$,label.angle=-90}{b1}
\fmf{phantom}{t0,v1,t1}
\fmf{phantom}{b0,v2,b1}
\fmffreeze
\fmf{plain}{b0,v2}
\fmfdot{b1}
\fmfdot{b0}
\fmf{phantom,tag=2}{v2,b1}
\fmf{plain,tension=0.8}{v1,v1}
\fmf{dashes,tension=0.8,right=0.6,tag=1}{v1,v2}
\fmf{plain,tension=0.8,left=0.6}{v1,v2}
\fmfdot{v1,v2}
\fmfposition
\fmfipath{p[]}
\fmfiset{p1}{vpath1(__v1,__v2)}
 \fmfi{plain}{subpath (length(p1)*0.6,length(p1)) of p1}
\fmfiset{p2}{vpath2(__v2,__b1)}
\fmfi{dashes}{subpath (0,length(p2)*0.6) of p2}
\fmfi{plain}{subpath (length(p2)*0.6,length(p2)) of p2}
\end{fmfgraph*}
 \end{fmffile}
 \end{minipage}\hspace{0.4cm}+
\end{equation}
\begin{equation}\nonumber
\hspace{-1.4cm}+\frac{1}{(2!)^2}\quad
\parbox{33mm}{\vspace{0.5cm}
\begin{fmffile}{DiagPag46.2}
\begin{fmfgraph*}(50,20)
\fmfkeep{DiagPag46.2}
\fmfleft{i1,i2,i3}
\fmfright{o1,o2,o3}
\fmfdot{i2,o2}
\fmffreeze
\fmf{phantom}{i1,v11,v12,v13,v14,v15,v16,o1}
\fmffreeze
\fmf{plain}{i2,v21}
\fmf{plain,tension=0.9}{v21,v21}
\fmf{plain,tension=0.9}{v22,v22}
\fmf{phantom,tag=4}{v21,v22}
\fmf{phantom,tag=3}{v22,o2}
\fmfdot{v21,v22}
\fmfv{label=$x_1$,label.angle=-90}{i2}
\fmfv{label=$x_2$,label.angle=-90}{o2}
\fmfposition
\fmfipath{p[]}
\fmfiset{p3}{vpath3(__v22,__o2)}
\fmfi{dashes}{subpath (0,length(p3)*0.6) of p3}
\fmfi{plain}{subpath (length(p3)*0.6,length(p3)) of p3}
\fmfiset{p4}{vpath4(__v21,__v22)}
\fmfi{dashes}{subpath (0,length(p4)*0.6) of p4}
\fmfi{plain}{subpath (length(p4)*0.6,length(p4)) of p4}
\fmffreeze
\fmf{phantom}{i3,v31,v32,v33,v34,v35,v36,o3}
\fmffreeze
\end{fmfgraph*}
\end{fmffile}
}
\hspace{2cm}+\frac{1}{2!}\quad
\parbox{33mm}{\vspace{0.cm}
\begin{fmffile}{DiagPag46.3}
\begin{fmfgraph*}(50,20)
\fmfkeep{DiagPag46.3}
\fmfleft{i1,i2,i3}
\fmfright{o1,o2,o3}
\fmfdot{i2,o2}
\fmffreeze
\fmf{phantom}{i1,v11,v12,v13,v14,v15,v16,o1}
\fmffreeze
\fmf{phantom}{i3,v31,v32,v33,v34,v35,v36,o3}
\fmffreeze
\fmf{phantom,tag=3}{i2,v21}
\fmf{phantom}{v21,v22}
\fmf{phantom,tag=2}{v22,o2}
\fmffreeze
\fmf{plain,left=0.9}{v21,v22}
\fmf{plain,right=0.9}{v21,v22}
\fmf{phantom,tag=1}{v21,v22}
\fmfposition
\fmfv{label=$x_1$,label.angle=-90}{i2}
\fmfv{label=$x_2$,label.angle=-90}{o2}
\fmfipath{p[]}
\fmfiset{p3}{vpath3(__i2,__v21)}
\fmfi{plain}{subpath (0,length(p3)*0.4) of p3}
\fmfi{plain}{subpath (length(p3)*0.4,length(p3)) of p3}
\fmfiset{p1}{vpath1(__v21,__v22)}
\fmfi{dashes}{subpath (0,length(p1)*0.6) of p1}
\fmfi{plain}{subpath (length(p1)*0.6,length(p1)) of p1}
\fmfiset{p2}{vpath2(__v22,__o2)}
\fmfi{dashes}{subpath (0,length(p2)*0.6) of p2}
\fmfi{plain}{subpath (length(p2)*0.6,length(p2)) of p2}
\fmffreeze
\fmfdot{v21,v22}
\fmfposition
\fmffreeze
\end{fmfgraph*}
\end{fmffile}}\hspace{2cm}\Bigg]
\end{equation}

and the third term in (\ref{phiphi2final}) gives:
\begin{equation}\nonumber\label{thermal<>2ndtorder3}
\frac{1}{(3!)^2}\,\Bigg\langle \quad \parbox{33mm}{
\fmfreuse{DiagPag35.4}
}\quad \Bigg \rangle_{\beta} =\Bigg[\,\,\frac{1}{(2!)^3}\quad
\parbox{33mm}{\vspace{0.5cm}
\begin{fmffile}{DiagPag46.4}
\begin{fmfgraph*}(50,20)
\fmfkeep{DiagPag46.4}
\fmfleft{i1,i2,i3}
\fmfright{o1,o2,o3}
\fmfdot{i2,o2}
\fmffreeze
\fmf{phantom}{i1,v11,v12,v13,v14,v15,v16,o1}
\fmffreeze
\fmf{phantom,tag=1}{i2,v21}
\fmf{plain,tension=0.9}{v21,v21}
\fmf{plain,tension=0.9}{v22,v22}
\fmf{phantom,tag=4}{v21,v22}
\fmf{phantom,tag=3}{v22,o2}
\fmfdot{v21,v22}
\fmfv{label=$x_1$,label.angle=-90}{i2}
\fmfv{label=$x_2$,label.angle=-90}{o2}
\fmfposition
\fmfipath{p[]}
\fmfiset{p1}{vpath1(__i2,__v21)}
\fmfi{plain}{subpath (0,length(p1)*0.4) of p1}
\fmfi{dashes}{subpath (length(p1)*0.4,length(p1)) of p1}
\fmfiset{p3}{vpath3(__v22,__o2)}
\fmfi{dashes}{subpath (0,length(p3)*0.6) of p3}
\fmfi{plain}{subpath (length(p3)*0.6,length(p3)) of p3}
\fmfiset{p4}{vpath4(__v21,__v22)}
\fmfi{plain}{subpath (0,length(p4)*0.6) of p4}
\fmfi{plain}{subpath (length(p4)*0.6,length(p4)) of p4}
\fmffreeze
\fmf{phantom}{i3,v31,v32,v33,v34,v35,v36,o3}
\fmffreeze
\end{fmfgraph*}
\end{fmffile}
}\hspace{2cm}+\nonumber\\
\end{equation}
\begin{equation}
\hspace{6.5cm}+\frac{1}{3!}\quad
\parbox{33mm}{\vspace{0.cm}
\begin{fmffile}{DiagPag46.5}
\begin{fmfgraph*}(50,20)
\fmfkeep{DiagPag46.5}
\fmfleft{i1,i2,i3}
\fmfright{o1,o2,o3}
\fmfdot{i2,o2}
\fmffreeze
\fmf{phantom}{i1,v11,v12,v13,v14,v15,v16,o1}
\fmffreeze
\fmf{phantom}{i3,v31,v32,v33,v34,v35,v36,o3}
\fmffreeze
\fmf{phantom,tag=3}{i2,v21}
\fmf{phantom}{v21,v22}
\fmf{phantom,tag=2}{v22,o2}
\fmffreeze
\fmf{plain,left=0.9}{v21,v22}
\fmf{plain,right=0.9}{v21,v22}
\fmf{phantom,tag=1}{v21,v22}
\fmfposition
\fmfv{label=$x_1$,label.angle=-90}{i2}
\fmfv{label=$x_2$,label.angle=-90}{o2}
\fmfipath{p[]}
\fmfiset{p3}{vpath3(__i2,__v21)}
\fmfi{plain}{subpath (0,length(p3)*0.4) of p3}
\fmfi{dashes}{subpath (length(p3)*0.4,length(p3)) of p3}
\fmfiset{p1}{vpath1(__v21,__v22)}
\fmfi{plain}{subpath (0,length(p1)*0.6) of p1}
\fmfi{plain}{subpath (length(p1)*0.6,length(p1)) of p1}
\fmfiset{p2}{vpath2(__v22,__o2)}
\fmfi{dashes}{subpath (0,length(p2)*0.6) of p2}
\fmfi{plain}{subpath (length(p2)*0.6,length(p2)) of p2}
\fmffreeze
\fmfdot{v21,v22}
\fmfposition
\fmffreeze
\end{fmfgraph*}
\end{fmffile}}\hspace{2cm}\Bigg]
\end{equation}
Summing up all the terms up to 2nd order we get:
\vspace{0.5cm}
\begin{equation}\nonumber
\langle \phi(x_1) \phi(x_2)\rangle_{\beta}
=\Bigg[\quad\parbox{33mm}{\fmfreuse{DiagPag44.1}}\hspace{-1.4cm}+\frac{1}{2!}\quad
\begin{minipage}[c]{32mm}
\vspace{0.55cm}
\begin{fmffile}{DiagPag44.3YWithout}
\begin{fmfgraph*}(32,10)
\fmfkeep{DiagPag44.3YWithout}
\fmfleft{i1}
\fmfright{o1}
\fmfdot{i1,o1}
\fmfv{label=$x_1$,label.angle=-90}{i1}
\fmfv{label=$x_2$,label.angle=-90}{o1}
\fmf{phantom}{i1,v,o1}
\fmf{plain}{v,o1}
\fmfdot{v}
\fmf{phantom,tag=1}{i1,v}
\fmfposition
\fmfipath{p[]}
\fmfiset{p1}{vpath1(__i1,__v)}
\fmfi{plain}{subpath (0,length(p1)*0.40.) of p1}
\fmfi{dashes}{subpath (length(p1)*0.4,length(p1)) of p1}
\fmf{plain,tension=0.9}{v,v}
\end{fmfgraph*}
\end{fmffile}
\end{minipage}
\,\,+\frac{1}{2!}\quad\parbox{33mm}{\fmfreuse{DiagPag45.1}}\hspace{0.3cm}+
\end{equation}
\newpage
\parbox{33mm}{
\vspace{2cm}
\begin{equation}\nonumber
\hspace{0.4cm}+\frac{1}{2!}\quad\parbox{33mm}{\vspace{-1.2cm}\fmfreuse{DiagPag45.2}}\,\,+\frac{1}{2!}\quad\parbox{33mm}{\vspace{-1.2cm}\fmfreuse{DiagPag46.1}}+
\end{equation}
}
\vspace{0.4cm}
\begin{equation}\nonumber
\hspace{0.9cm}+\frac{1}{(2!)^2}\quad\parbox{33mm}{\fmfreuse{DiagPag45.3}}\,\,\hspace{1.8cm}+\frac{1}{(2!)^2}\,\,\,\,\parbox{33mm}{\fmfreuse{DiagPag46.2}}\hspace{2cm}+
\end{equation}
\vspace{0.5cm}
\begin{equation}\nonumber
\hspace{0.4cm}+\frac{1}{(2!)^2}\quad\parbox{33mm}{\fmfreuse{DiagPag46.4}}\,\,\hspace{1.8cm}+\frac{1}{3!}\,\,\,\,\parbox{33mm}{\fmfreuse{DiagPag46.5}}\hspace{2cm}+
\end{equation}
\begin{equation}
\label{<phi phi>betafinal}
\hspace{01.2cm}+\frac{1}{3!}\quad\parbox{33mm}{\fmfreuse{DiagPag45.4}}\,\,\hspace{1.8cm}+\frac{1}{3!}\,\,\,\,\parbox{33mm}{\fmfreuse{DiagPag46.3}}\hspace{2cm}\Bigg]
\end{equation}
 The reader may wonder that there could be some extra diagram coming from the
Fermionic part of the generating functional that we have not
considered. This is not so because in this calculation we started
from the result we had got for the analog diagrams without thermal
averages (\ref{<phi phi>phi i}) and for those diagrams we had
already taken account of the Fermionic diagrams cancelation.
For example, just using topological considerations we could have
for the two-point functions also the following diagrams besides the
(\ref{<phi phi>betafinal})
\vspace{0.5cm}
\begin{equation}\nonumber
\begin{minipage}[c]{32mm}
\vspace{0.55cm}
\begin{fmffile}{DiagPag47.1}
\begin{fmfgraph*}(32,10)
\fmfleft{i1}
\fmfright{o1}
\fmfdot{i1,o1}
\fmfv{label=$x_1$,label.angle=-90}{i1}
\fmfv{label=$x_2$,label.angle=-90}{o1}
\fmf{phantom}{i1,v,o1}
\fmf{phantom,tag=2}{v,o1}
\fmfdot{v}
\fmf{phantom,tag=1}{i1,v}
\fmfposition
\fmfipath{p[]}
\fmfiset{p1}{vpath1(__i1,__v)}
\fmfi{plain}{subpath (0,length(p1)*0.40.) of p1}
\fmfi{plain}{subpath (length(p1)*0.4,length(p1)) of p1}
\fmf{phantom,tension=0.9,tag=3}{v,v}
\fmfiset{p2}{vpath2(__v,__o1)}
\fmfi{plain}{subpath (0,length(p2)*0.60.) of p2}
\fmfi{plain}{subpath (length(p2)*0.6,length(p2)) of p2}
\fmfiset{p3}{vpath3(__v,__v)}
\fmfi{plain}{subpath (0,length(p3)*0.5) of p3}
\fmfi{dashes}{subpath (length(p3)*0.5,length(p3)) of p3}
\end{fmfgraph*}
\end{fmffile}
\end{minipage}\quad,
\quad \parbox{33mm}{\vspace{0.5cm}
\begin{fmffile}{DiagPag47.2}
\begin{fmfgraph*}(50,20)
\fmfleft{i1,i2,i3}
\fmfright{o1,o2,o3}
\fmfdot{i2,o2}
\fmffreeze
\fmf{phantom}{i1,v11,v12,v13,v14,v15,v16,o1}
\fmffreeze
\fmf{phantom,tag=1}{i2,v21}
\fmf{phantom,tension=0.9,tag=2}{v21,v21}
\fmf{phantom,tension=0.9,tag=5}{v22,v22}
\fmf{phantom,tag=4}{v21,v22}
\fmf{phantom,tag=3}{v22,o2}
\fmfdot{v21,v22}
\fmfv{label=$x_1$,label.angle=-90}{i2}
\fmfv{label=$x_2$,label.angle=-90}{o2}
\fmfposition
\fmfipath{p[]}
\fmfiset{p1}{vpath1(__i2,__v21)}
\fmfi{plain}{subpath (0,length(p1)*0.4) of p1}
\fmfi{plain}{subpath (length(p1)*0.4,length(p1)) of p1}
\fmfiset{p2}{vpath2(__v21,__v21)}
\fmfi{plain}{subpath (0,length(p2)*0.5) of p2}
\fmfi{dashes}{subpath (length(p2)*0.5,length(p2)) of p2}
\fmfiset{p3}{vpath3(__v22,__o2)}
\fmfi{plain}{subpath (0,length(p3)*0.6) of p3}
\fmfi{plain}{subpath (length(p3)*0.6,length(p3)) of p3}
\fmfiset{p4}{vpath4(__v21,__v22)}
\fmfi{plain}{subpath (0,length(p4)*0.6) of p4}
\fmfi{plain}{subpath (length(p4)*0.6,length(p4)) of p4}
\fmfiset{p5}{vpath5(__v22,__v22)}
\fmfi{plain}{subpath (0,length(p5)*0.5) of p5}
\fmfi{dashes}{subpath (length(p5)*0.5,length(p5)) of p5}
\fmffreeze
\fmf{phantom}{i3,v31,v32,v33,v34,v35,v36,o3}
\fmffreeze
\end{fmfgraph*}
\end{fmffile}
}\hspace{2cm},\hspace{0.5cm}
\begin{minipage}[c]{30mm}
\vspace{-0.63cm}
 \begin{fmffile}{DiagPag47.3}
\begin{fmfgraph*}(30,15)
\fmftop{t0,t1}
\fmfbottom{b0,b1}
\fmfv{label=$x_1$,label.angle=-90}{b0}
\fmfv{label=$x_2$,label.angle=-90}{b1}
\fmf{phantom}{t0,v1,t1}
\fmf{phantom}{b0,v2,b1}
\fmffreeze
\fmf{plain}{v2,b1}
\fmf{phantom,tag=2}{b0,v2}
\fmfdot{b1}
\fmfdot{b0}
\fmf{plain,tension=0.8}{v1,v1}
\fmf{phantom,tension=0.8,right=0.6,tag=1}{v1,v2}
\fmf{phantom,tension=0.8,left=0.6,tag=3}{v1,v2}
\fmfdot{v1,v2}
\fmfposition
\fmfipath{p[]}
\fmfiset{p1}{vpath1(__v1,__v2)}
 \fmfi{plain}{subpath (0,length(p1)*0.5) of p1}
 \fmfi{dashes}{subpath (length(p1)*0.5,length(p1)) of p1}
\fmfiset{p2}{vpath2(__b0,__v2)}
 \fmfi{plain}{subpath (0,length(p2)*0.4) of p2}
 \fmfi{plain}{subpath (length(p2)*0.4,length(p2)) of p2}
\fmfiset{p3}{vpath3(__v1,__v2)}
\fmfi{dashes}{subpath (0,length(p3)*0.5) of p3}
\fmfi{plain}{subpath (length(p3)*0.5,length(p3)) of p3}
\end{fmfgraph*}
 \end{fmffile}
 \end{minipage}
\end{equation}
but all these would be canceled by the diagrams with Fermionic loops like:
\vspace{0.7cm}
\begin{equation}
\begin{minipage}[c]{32mm}
\vspace{0.55cm}
\begin{fmffile}{DiagPag47.4}
\begin{fmfgraph*}(32,10)
\fmfset{arrow_len}{2mm}
\fmfleft{i1}
\fmfright{o1}
\fmfdot{i1,o1}
\fmfv{label=$x_1$,label.angle=-90}{i1}
\fmfv{label=$x_2$,label.angle=-90}{o1}
\fmf{phantom}{i1,v,o1}
\fmf{phantom,tag=2}{v,o1}
\fmfdot{v}
\fmf{phantom,tag=1}{i1,v}
\fmfposition
\fmfipath{p[]}
\fmfiset{p1}{vpath1(__i1,__v)}
\fmfi{plain}{subpath (0,length(p1)*0.40.) of p1}
\fmfi{plain}{subpath (length(p1)*0.4,length(p1)) of p1}
\fmf{dots_arrow,tension=0.9,tag=3}{v,v}
\fmfiset{p2}{vpath2(__v,__o1)}
\fmfi{plain}{subpath (0,length(p2)*0.60.) of p2}
\fmfi{plain}{subpath (length(p2)*0.6,length(p2)) of p2}
\end{fmfgraph*}
\end{fmffile}
\end{minipage}\quad,
\quad \parbox{33mm}{\vspace{0.5cm}
\begin{fmffile}{DiagPag47.5}
\begin{fmfgraph*}(50,20)
\fmfset{arrow_len}{2mm}
\fmfleft{i1,i2,i3}
\fmfright{o1,o2,o3}
\fmfdot{i2,o2}
\fmffreeze
\fmf{phantom}{i1,v11,v12,v13,v14,v15,v16,o1}
\fmffreeze
\fmf{phantom,tag=1}{i2,v21}
\fmf{dots_arrow,tension=0.9,tag=2}{v21,v21}
\fmf{dots_arrow,tension=0.9,tag=5}{v22,v22}
\fmf{phantom,tag=4}{v21,v22}
\fmf{phantom,tag=3}{v22,o2}
\fmfdot{v21,v22}
\fmfv{label=$x_1$,label.angle=-90}{i2}
\fmfv{label=$x_2$,label.angle=-90}{o2}
\fmfposition
\fmfipath{p[]}
\fmfiset{p1}{vpath1(__i2,__v21)}
\fmfi{plain}{subpath (0,length(p1)*0.4) of p1}
\fmfi{plain}{subpath (length(p1)*0.4,length(p1)) of p1}
\fmfiset{p3}{vpath3(__v22,__o2)}
\fmfi{plain}{subpath (0,length(p3)*0.6) of p3}
\fmfi{plain}{subpath (length(p3)*0.6,length(p3)) of p3}
\fmfiset{p4}{vpath4(__v21,__v22)}
\fmfi{plain}{subpath (0,length(p4)*0.6) of p4}
\fmfi{plain}{subpath (length(p4)*0.6,length(p4)) of p4}
\fmffreeze
\fmf{phantom}{i3,v31,v32,v33,v34,v35,v36,o3}
\fmffreeze
\end{fmfgraph*}
\end{fmffile}
}\hspace{2cm},\hspace{0.5cm}
\begin{minipage}[c]{30mm}
\vspace{-0.63cm}
 \begin{fmffile}{DiagPag47.6}
\begin{fmfgraph*}(30,15)
\fmfset{arrow_len}{2mm}
\fmftop{t0,t1}
\fmfbottom{b0,b1}
\fmfv{label=$x_1$,label.angle=-90}{b0}
\fmfv{label=$x_2$,label.angle=-90}{b1}
\fmf{phantom}{t0,v1,t1}
\fmf{phantom}{b0,v2,b1}
\fmffreeze
\fmf{plain}{v2,b1}
\fmf{phantom,tag=2}{b0,v2}
\fmfdot{b1}
\fmfdot{b0}
\fmf{plain,tension=0.8}{v1,v1}
\fmf{dots_arrow,tension=0.8,right=0.6,tag=1}{v1,v2}
\fmf{dots_arrow,tension=0.8,right=0.6,tag=3}{v2,v1}
\fmfdot{v1,v2}
\fmfposition
\fmfipath{p[]}
\fmfiset{p2}{vpath2(__b0,__v2)}
 \fmfi{plain}{subpath (0,length(p2)*0.4) of p2}
 \fmfi{plain}{subpath (length(p2)*0.4,length(p2)) of p2}
\end{fmfgraph*}
 \end{fmffile}
 \end{minipage}
\end{equation}
while no one of those in (\ref{<phi phi>betafinal}) could be
canceled by diagrams with Fermionic loops.

The diagrams we get by
averaging over the initial conditions with the Boltzmann weight
are basically based on the following propagators and vertices
\begin{eqnarray}
\label{rule1}
\Delta_{\beta}(p)&=&\frac{2\pi}{\beta|p^0|}
\delta(p^2-m^2)=\,\,\parbox{20mm}{\fmfreuse{DiagPag44.1}}\vspace{0.4cm}\\&&\nonumber\\
\label{rule2}-iG_{R}^{(B)}(p)
&=&-\frac{i}{p^2-m^2+i\epsilon p^0}=\,\,\parbox{20mm}{\fmfreuse{DiagPag25.3}}\vspace{0.4cm}\\&&\nonumber\\
\label{rule3}
-i\,G_{A}^{(F)}(p)&=&-i\,G_{R}^{(B)}(-p)=\,\,\parbox{20mm}{\fmfreuse{DiagPag25.4}}\\&&\nonumber\\
\label{rule4}
-G_{R}^{(F)}(p)&=&-G_{R}^{(B)}(p)
=\,\,\parbox{20mm}{\fmfreuse{DiagPag25.5}}\vspace{0.4cm}\\&&\nonumber\\
\label{rule5}
-G_{A}^{(F)}(p)&=&-G_{R}^{(B)}(-p)=\,\,\parbox{20mm}{\fmfreuse{DiagPag25.6}}\vspace{0.4cm}\\&&\nonumber\\
\label{rule6} i\,g&=&\,\,\parbox{20mm}{\fmfreuse{DiagPag28.1}}\vspace{2cm}\\&&\nonumber\\
\label{rule7}-g&=&\,\,\parbox{20mm}{\fmfreuse{DiagPag28.2}}
\end{eqnarray}

\section{Perturbation Theory With Superfields.}\label{PertubSuper}

If we add to the $\mathcal{Z}_{\beta}^{(B)}$ of (\ref{ZbetaBJphi})
the Grassmannian piece (\ref{ZF}) (which is not affected by the
thermal average) we get the full generating functional which, for the free theory, is :

\begin{equation} \label{Z0betaComponents}
\begin{array}{lll}
\mathcal{Z}_{\beta}^0[J]
&=&\left[-\frac{1}{2}\displaystyle\int d^4x d^4x'J_{\phi}(x)\Delta_{\beta}(x-x')J_{\phi}(x')\right.+
\vspace{2mm}\\
&& +i\displaystyle\int d^4x d^4x'J_{\phi}(x)G_{R}^{(B)}(x-x')J_{\lambda_{\pi}}(x')+
\vspace{2mm}\\
&& -i\left.\displaystyle\int d^4x d^4x'\bar{J}_{c^{\phi}}(x)G_{R}^{(F)}(x-x')J_{\bar{c}_{\pi}}(x')\right].
\end{array}
\end{equation}
Let us now move quickly to the superfield formalism. The super-field analog of (\ref{Superfield}) is the following one:
\begin{equation}\label{Superfield2}
\Phi=\phi+\theta c^{\phi}+\bar{\theta}\bar{c}_{\pi}+i\bar{\theta}\theta\lambda_{\pi}
\end{equation} and in order to have the current-field coupling that we have in (\ref{ZB}) and (\ref{ZF}),
we need to introduce a super-current of the form:
\begin{equation}\label{Supercurrent}
\mathbb{J}_{\phi}=-J_{\lambda_{\pi}}+\theta J_{{\bar c}_{\pi}}+\bar{\theta}{\bar J}_{c^{\phi}}-i\bar{\theta}\theta
J_{\phi}.
\end{equation} The term:
\begin{equation}\label{manca1}
\exp\left[i \int d^4x d \theta d \bar{\theta}\mathbb{J}_{\phi}(x,\theta,\bar{\theta})
\Phi(x,\theta,\bar{\theta})\right]
\end{equation} will then give  the four couplings of currents and fields present
in (\ref{ZB}) and (\ref{ZF}).
If we want to get not the free generating functional but the full
one, its formal expression will be
\begin{equation}\label{Zsupercurrent}
\mathcal{Z}[\mathbb{J}_{\phi}]
=\exp\left[\displaystyle\int dz V\left(-\frac{\delta}{\delta \mathbb{J}_{\phi}(z)}\right)\right]
\mathcal{Z}^0_{\beta}[\mathbb{J}_{\phi}]
\end{equation} where $V$ is the potential of the $g\phi^4/4!$ theory, so
\begin{equation}
V\left(-\frac{\delta}{\delta \mathbb{J}_{\phi}(z)}\right)
=\frac{g}{4!}
\frac{\delta^4}{\delta \mathbb{J}_{\phi}^4(z)}\Longrightarrow
\vspace{0.1cm}\begin{minipage}[c]{16mm}
\begin{fmffile}{DiagPag50.1}
\begin{fmfgraph*}(15,15)
\fmfkeep{DiagPag50.1}
\fmfleft{i1,i2}
\fmfright{o1,o2}
\fmf{dbl_plain}{i1,v}
\fmf{dbl_plain}{v,o2}
\fmf{dbl_plain}{i2,v,o1}
\fmfdot{v}
\fmfv{label=$z$,label.angle=180}{v}
\end{fmfgraph*}
\end{fmffile}
\end{minipage}=g\int \,d^4y\,d\theta\,d\bar{\theta}=g\int dz,
\end{equation} where the collective variable $z=(x,\theta,\bar{\theta})$ has been introduced. The reason we can use this potential and not the $\mathcal{\tilde{S}}_V$ of Eq. (\ref{Z0})
is because, as we proved in Ref. \cite{AbriGoMau}, the
$\mathcal{\tilde{S}}_{V}$ or any $\mathcal{\tilde{S}}$ can be
obtained via the  usual $\mathcal{S}$ with $\phi$ replaced by $\Phi$
as in formula (\ref{GenFuncComponent}) and
(\ref{GenFuncSuperPhi}). A super-field formalism and the associated Feynman diagrams had been developed for the Langevin stochastic equation in ref.\cite{Marculescu}. This has inspired us to do an analogous formalism for deterministic systems which evolve via  Hamilton's equation.

Going now back to the coupling between supercurrents and superfields in (\ref{manca1}),
it is straightforward to prove that:
\begin{equation}\label{superJgotoJphi}
J_{\phi}=
i\int d\theta d\bar{\theta}
\mathbb{J}_{\phi}(x,\theta, \bar{\theta})
\end{equation}
\begin{equation}\label{superJgotoJpi}
J_{\lambda_{\pi}}=
-\int d\theta d\bar{\theta} \bar{\theta}\theta
\mathbb{J}_{\phi}(x,\theta, \bar{\theta})
\end{equation}
\begin{equation}\label{superJgotoJbarc}
{\bar J}_{c^{\phi}}=
-\int d\theta d\bar{\theta}
\mathbb{J}_{\phi}(x,\theta, \bar{\theta})\theta
\end{equation}
\begin{equation}\label{superJgotoJc}
J_{{\bar c}_{\pi}}=
\int d\theta d\bar{\theta}\bar{\theta}
\mathbb{J}_{\phi}(x,\theta, \bar{\theta}).
\end{equation} Using the rules (\ref{superJgotoJphi}), (\ref{superJgotoJpi}),
(\ref{superJgotoJbarc}), (\ref{superJgotoJc}) and the super-field
we can write the free generating functional
(\ref{Z0betaComponents}) as
\begin{equation}\label{Z0beta}
\mathcal{Z}^0_{\beta}[\mathbb{J}_{\phi}]
=\exp\left[\displaystyle\frac{1}{2}\displaystyle
\int dz dz' \mathbb{J}_{\phi}(z)\mathbb{G}(z,z')\mathbb{J}_{\phi}(z')\right]
\end{equation} where $z=(x,\theta,\bar{\theta})$ and $\mathbb{G}(z,z')$ is defined
as:
\begin{equation}
\mathbb{G}(z,z')=\Delta_{\beta}(x-x')+\mathcal{G}(z,z')=\,\,\begin{minipage}[c]{16mm}
\vspace{0.55cm}\begin{fmffile}{DiagPag50bis.1}
\begin{fmfgraph*}(16,1)
\fmfkeep{DiagPag50bis.1}
\fmfleft{i1}
\fmfright{o1}
\fmfdot{i1}
\fmfdot{o1}
\fmf{dbl_plain}{i1,o1}
\fmfv{label=$z$,label.angle=-90,label.dist=9}{i1}
\fmfv{label=$z'$,label.angle=-90}{o1}
\end{fmfgraph*}
\end{fmffile}
\end{minipage}
\end{equation} with
\begin{equation}\label{vitto}
\mathcal{G}(z,z')
\equiv G_{R}^{(B)}(x-x')\bar{\theta}'\theta'
+G_{R}^{(B)}(x'-x)\bar{\theta}\theta
+G_{R}^{(F)}(x-x')\theta \bar{\theta}'
+G_{R}^{(F)}(x'-x)\theta'\bar{\theta}.
\end{equation} The $\mathbb{G}(z,z')$ and $\mathcal{G}(z,z')$ have several nice
properties that are the following (with $m\geq 1$) :

{\bf 1.}
\begin{equation}\label{Prop1}\mathbb{G}(z,z')=\mathbb{G}(z',z).\end{equation}

{\bf 2.}
\begin{equation}\label{Prop2}\mathbb{G}(z,z)=\Delta_{\beta}(x-x).\end{equation}

{\bf 3.}
\begin{equation}\label{Prop3}
\mathcal{G}^n(z,z')=\delta_{n1}\mathcal{G}^n(z,z')[1-\delta(z-z')]\,, n\geq 1
\end{equation}

{\bf 4.}
\begin{equation}\label{Prop4}
\mathbb{G}^m(z,z')=\Delta_{\beta}^m(x-x')+m\,\Delta_{\beta}^{m-1}(x-x')\mathcal{G}(z,z'),
\end{equation}

While ${\bf 1)}$ and  ${\bf 2)}$  are trivial to prove from the
symmetry properties of $\Delta_{\beta}$ and $\mathcal{G}$, ${\bf 4)}$
can be proved as follows:
\begin{equation}
\begin{array}{lll}
\mathbb{G}^m(z,z')&=&\displaystyle\sum_{0\leq n\leq m}
\left(\begin{array}{c}m\\
n\end{array}\right)
\Delta_{\beta}^{m-n}(x-x')\mathbb{G}^n(z,z')\vspace{2mm}\\
&=&\left(\begin{array}{c}m\\
0\end{array}\right)
\Delta_{\beta}^m(x-x')+
\left(\begin{array}{l}m\\
1\end{array}\right)\Delta_{\beta}^{m-1}(x-x')\mathcal{G}(z,z')\vspace{2mm}\\
&=&\Delta_{\beta}^m(x-x')+m\,\Delta_{\beta}^{m-1}(x-x')\mathcal{G}(z,z').
\end{array}
\end{equation} In the last step we have used
$\mathcal{G}^{n}(z,z')=0 (n\geq 2)$ which is a consequence of the property number ${\bf 3)}$.
Let us now prove property number ${\bf 3)}$. We have, for $n\geq 1$:
\begin{equation}\begin{array}{lll}
\mathcal{G}^n(z,z')&=&
\mathcal{G}^n(z,z')\left\{\delta(z-z')+\left[1-\delta(z-z')\right]\right\}\vspace{2mm}\\
&=&\mathcal{G}^n(z,z)\delta(z-z')+
\mathcal{G}^n(z,z')[1-\delta(z-z')]\vspace{2mm}\\
&=&\left[\delta_{n1}+\left(1-\delta_{n1}\right)\right]\mathcal{G}^n(z,z')[1-\delta(z-z')]\vspace{2mm}\\
&=&\delta_{n1}\mathcal{G}^n(z,z')[1-\delta(z-z')]\vspace{2mm}\\
&&+(1-\delta_{n1})\mathcal{G}^n(z,z')[1-\delta(z-z')]\vspace{2mm}\\
&=&\delta_{n1}\mathcal{G}^n(z,z')[1-\delta(z-z')].
\end{array}\end{equation}
The property no. ${\bf 3)}$ is  at the root of the fact that loops made of
dash-full line cancel against Grassmannian loops. So it is at the
root of ``{\it classicality}'', i.e. that without temperature we would
not have any loops.

Going back to the expressions (\ref{Zsupercurrent}) and
(\ref{Z0beta}) we have seen that the Feynman diagrams are the same
as those of a $g\phi^4/4!$ theory, but with the field replaced by
{\it super-field}, and the propagator replaced by the super
$\mathbb{G}(z,z')$. We call the reader's attention that, in contrast to what happens in standard quantum field theory, where, due to translational invariance we have $G_{\scriptstyle \tilde F}(z,z')\equiv G_{\scriptstyle \tilde F}(z-z')$, where $G_{\scriptstyle \tilde F}$ stays for the Feynman propagator, here we have $\mathbb{G}(z,z')\neq \mathbb{G}(z-z')$. That is why we use the notation $\mathbb{G}(z,z')$ in the perturbative analytic expressions.

Let us start deriving the two-point function:
\begin{equation}\label{superGbeta}
\mathbb{G}_{\beta}(z_1,z_2)
=\langle \Phi(z_1)\Phi(z_2)\rangle_{\beta}
=(-i)^2\displaystyle\frac{\delta_{\mathbb{J}_{\phi(z_1)}}
\delta_{\mathbb{J}_{\phi(z_2)}} \mathcal {Z}[\mathbb{J}_{\phi}]}{\mathcal {Z}[\mathbb{J}_{\phi}]}|_{\mathbb{J}_{\phi}=0}.
\end{equation} Doing the analog of standard $g\phi^4/4!$ QFT perturbation theory,
we have that the first order correction to the two-point function
is
\begin{equation}\label{<superphisuperphi>(1)}
\frac{1}{2}\int dz
\mathbb{G}(z_1,z)\mathbb{G}(z,z)\mathbb{G}(z,z_2)=\,\,\frac{1}{2}\quad
\begin{minipage}[c]{32mm}
\vspace{0.55cm}
\begin{fmffile}{DiagPag53.1}
\begin{fmfgraph*}(32,10)
\fmfkeep{DiagPag53.1}
\fmfleft{i1}
\fmfright{o1}
\fmfdot{i1,o1}
\fmfv{label=$z_1$,label.angle=-90}{i1}
\fmfv{label=$z_2$,label.angle=-90}{o1}
\fmf{dbl_plain}{i1,v,o1}
\fmfdot{v}
\fmfv{label=$z$,label.angle=-90}{v}
\fmfposition
\fmfipath{p[]}
\fmf{dbl_plain,tension=0.9}{v,v}
\end{fmfgraph*}
\end{fmffile}
\end{minipage}
\end{equation} which is the super-analog of the QFT diagram with the same symmetry factor.
The second order correction, following the analogy with QFT, which
has the three diagrams:
\vspace{0.5cm}
\begin{equation}\nonumber
\Bigg[\frac{1}{3!}\,\,\parbox{33mm}{\vspace{0.cm}
\begin{fmffile}{DiagPag53.2}
\begin{fmfgraph*}(40,20)
\fmfleft{i1,i2,i3}
\fmfright{o1,o2,o3}
\fmfdot{i2,o2}
\fmffreeze
\fmf{phantom}{i1,v11,v12,v13,v14,v15,v16,o1}
\fmffreeze
\fmf{phantom}{i3,v31,v32,v33,v34,v35,v36,o3}
\fmffreeze
\fmf{plain}{i2,v21}
\fmf{plain}{v21,v22}
\fmf{plain}{v22,o2}
\fmffreeze
\fmf{plain,left=0.9}{v21,v22}
\fmf{plain,right=0.9}{v21,v22}
\fmfposition
\fmffreeze
\fmfdot{v21,v22}
\fmfposition
\fmffreeze
\end{fmfgraph*}
\end{fmffile}}
\hspace{1.cm}+\frac{1}{(2!)^2}\,\,
\parbox{33mm}{\vspace{0.5cm}
\begin{fmffile}{DiagPag53.3}
\begin{fmfgraph*}(40,20)
\fmfleft{i1,i2,i3}
\fmfright{o1,o2,o3}
\fmfdot{i2,o2}
\fmffreeze
\fmf{phantom}{i1,v11,v12,v13,v14,v15,v16,o1}
\fmffreeze
\fmf{phantom,tag=3}{i2,v21}
\fmf{plain,tension=0.8}{v21,v21}
\fmf{plain,tension=0.8}{v22,v22}
\fmf{phantom,tag=4}{v21,v22}
\fmf{plain}{v22,o2}
\fmfdot{v21,v22}
\fmfposition
\fmfipath{p[]}
\fmfiset{p3}{vpath3(__i2,__v21)}
\fmfi{plain}{subpath (0,length(p3)*0.4) of p3}
\fmfi{plain}{subpath (length(p3)*0.4,length(p3)) of p3}
\fmfiset{p4}{vpath4(__v21,__v22)}
\fmfi{plain}{subpath (0,length(p4)*0.4) of p4}
\fmfi{plain}{subpath (length(p4)*0.4,length(p4)) of p4}
\fmffreeze
\fmf{phantom}{i3,v31,v32,v33,v34,v35,v36,o3}
\fmffreeze
\end{fmfgraph*}
\end{fmffile}
}
\hspace{1cm}+\frac{1}{(2!)^2}\hspace{0.2cm}
\begin{minipage}[c]{30mm}
\vspace{-1.2cm}
 \begin{fmffile}{DiagPag53.4}
\begin{fmfgraph*}(25,15)
\fmftop{t0,t1}
\fmfbottom{b0,b1}
\fmf{phantom}{t0,v1,t1}
\fmf{phantom}{b0,v2,b1}
\fmffreeze
\fmf{plain}{v2,b1}
\fmf{phantom,tag=2}{b0,v2}
\fmfdot{b1}
\fmfdot{b0}
\fmf{plain,tension=0.8}{v1,v1}
\fmf{phantom,tension=0.8,right=0.6,tag=1}{v1,v2}
\fmf{phantom,tension=0.8,left=0.6,tag=3}{v1,v2}
\fmfdot{v1,v2}
\fmfposition
\fmfipath{p[]}
\fmfiset{p1}{vpath1(__v1,__v2)}
 \fmfi{plain}{subpath (0,length(p1)*0.5) of p1}
 \fmfi{plain}{subpath (length(p1)*0.5,length(p1)) of p1}
\fmfiset{p2}{vpath2(__b0,__v2)}
 \fmfi{plain}{subpath (0,length(p2)*0.4) of p2}
 \fmfi{plain}{subpath (length(p2)*0.4,length(p2)) of p2}
\fmfiset{p3}{vpath3(__v1,__v2)}
\fmfi{plain}{subpath (0,length(p3)*0.5) of p3}
\fmfi{plain}{subpath (length(p3)*0.5,length(p3)) of p3}
\end{fmfgraph*}
 \end{fmffile}
 \end{minipage}\hspace{-0.4cm}\Bigg]
\end{equation}
is made of the following three super-diagrams:
\begin{equation}
\label{<superphisuperphi>(2)1}
\frac{1}{3!}\,\,\,\parbox{33mm}{\vspace{0.cm}
\begin{fmffile}{DiagPag53.5}
\begin{fmfgraph*}(40,20)
\fmfkeep{DiagPag53.5}
\fmfleft{i1,i2,i3}
\fmfright{o1,o2,o3}
\fmfdot{i2,o2}
\fmffreeze
\fmf{phantom}{i1,v11,v12,v13,v14,v15,v16,o1}
\fmffreeze
\fmf{phantom}{i3,v31,v32,v33,v34,v35,v36,o3}
\fmffreeze
\fmf{dbl_plain}{i2,v21}
\fmf{dbl_plain}{v21,v22}
\fmf{dbl_plain}{v22,o2}
\fmffreeze
\fmf{dbl_plain,left=0.9}{v21,v22}
\fmf{dbl_plain,right=0.9}{v21,v22}
\fmfposition
\fmffreeze
\fmfdot{v21,v22}
\fmfposition
\fmffreeze
\fmfv{label=$z_1$,label.angle=-90}{i2}
\fmfv{label=$z_2$,label.angle=-90}{o2}
\fmfv{label=$z$,label.angle=-120}{v21}
\fmfv{label=$z'$,label.angle=-60,label.dist=3}{v22}
\end{fmfgraph*}
\end{fmffile}}\hspace{1cm}=\frac{1}{3!}\int dz \,dz'\,
\mathbb{G}(z_1,z)\,\left[\mathbb{G}(z,z')\right]^3\,\mathbb{G}(z',z_2)
\end{equation}
\begin{equation}
\label{<superphisuperphi>(2)2}
\frac{1}{(2!)^2}\,\,
\parbox{33mm}{\vspace{0.5cm}
\begin{fmffile}{DiagPag53.6}
\begin{fmfgraph*}(40,20)
\fmfkeep{DiagPag53.6}
\fmfleft{i1,i2,i3}
\fmfright{o1,o2,o3}
\fmfdot{i2,o2}
\fmffreeze
\fmf{phantom}{i1,v11,v12,v13,v14,v15,v16,o1}
\fmfv{label=$z_1$,label.angle=-90}{i2}
\fmfv{label=$z_2$,label.angle=-90}{o2}
\fmfv{label=$z$,label.angle=-90}{v21}
\fmfv{label=$z'$,label.angle=-90,label.dist=2}{v22}
\fmffreeze
\fmf{dbl_plain}{i2,v21}
\fmf{dbl_plain,tension=0.7}{v21,v21}
\fmf{dbl_plain,tension=0.7}{v22,v22}
\fmf{dbl_plain}{v21,v22}
\fmf{dbl_plain}{v22,o2}
\fmfdot{v21,v22}
\fmf{phantom}{i3,v31,v32,v33,v34,v35,v36,o3}
\fmffreeze
\end{fmfgraph*}
\end{fmffile}
}
\quad\quad\quad=\frac{1}{4}\int \,dz \,dz'
\mathbb{G}(z_1,z)\,\mathbb{G}(z,z)\,\mathbb{G}(z,z')
\mathbb{G}(z',z')\,\mathbb{G}(z',z_2)
\end{equation}
\vspace{0.3cm}
\begin{equation}
\label{<superphisuperphi>(2)3}\frac{1}{(2!)^2}\hspace{0.2cm}
\begin{minipage}[c]{30mm}
\vspace{-1.2cm}
 \begin{fmffile}{DiagPag53.7}
\begin{fmfgraph*}(25,15)
\fmfkeep{DiagPag53.7}
\fmftop{t0,t1}
\fmfbottom{b0,b1}
\fmf{phantom}{t0,v1,t1}
\fmf{dbl_plain}{b0,v2,b1}
\fmffreeze
\fmf{dbl_plain}{b0,v2}
\fmfdot{b1}
\fmfdot{b0}
\fmf{dbl_plain,tension=0.8}{v1,v1}
\fmf{dbl_plain,tension=0.8,right=0.6}{v1,v2}
\fmf{dbl_plain,tension=0.8,left=0.6}{v1,v2}
\fmfdot{v1,v2}
\fmfv{label=$z'$,label.angle=180}{v1}
\fmfv{label=$z$,label.angle=-90}{v2}
\fmfv{label=$z_1$,label.angle=-90}{b0}
\fmfv{label=$z_2$,label.angle=-90}{b1}
\end{fmfgraph*}
 \end{fmffile}
 \end{minipage}=\frac{1}{4}\int \,dz \,dz'\,
\mathbb{G}(z_1,z)\,\mathbb{G}(z,z')\,\mathbb{G}(z',z')\,\mathbb{G}(z',z)\,
\mathbb{G}(z,z_2).
\end{equation}
Let us now check if from the super-diagram (\ref{<superphisuperphi>(1)}) we can
derive, for example, the $\langle \phi \phi\rangle$ correlation at
the first order which were the sum of (\ref{thermal1storder1})
and (\ref{thermal<>1storder2}). The manner to do that is to
extract from the external super-fields of Eq.
(\ref{<superphisuperphi>(1)}) the components $\varphi^a$.
Remembering the formula of the super-field (\ref{Superfield2}),
this can be done via the following integration
\vspace{0.5cm}
\begin{equation}
\label{Projsuperphisuperphi}
\int d\theta_1 d \bar{\theta}_1 \left(\bar{\theta}_1 \theta_1\right) d\theta_2 d \bar{\theta}_2
\left(\bar{\theta}_2 \theta_2\right)
\Bigg[\,\,\frac{1}{2!}\quad\parbox{30mm}{\vspace{-0cm}\fmfreuse{DiagPag53.1}}\quad \,\,\Bigg].
\end{equation}
The variables inside the brackets  $\left(\bar{\theta}_1 \theta_1\right)$ and  $\left(\bar{\theta}_2 \theta_2\right)$  basically act as
projectors from the super-field $\Phi$ to the field $\phi$.

Let us now work out (\ref{Projsuperphisuperphi}) in details
using  (\ref{<superphisuperphi>(1)}) and the analytic
expression of the super-propagators presented on the l.h.s. of
(\ref{vitto}). Below we will go pedantically through the details of the calculations. We do that in order
to get the reader familiar with the formalism.

The explicit expression of  (\ref{Projsuperphisuperphi}) turns out
to be
\vspace{0.5cm}
\begin{equation}
\begin{array}{lll}
&&\displaystyle\int d\theta_1 d \bar{\theta}_1 \left(\bar{\theta}_1 \theta_1\right) d\theta_2 d \bar{\theta}_2
\left(\bar{\theta}_2 \theta_2\right)\,\Bigg[\,\,\frac{1}{2!}\quad\parbox{30mm}{\vspace{-0cm}\fmfreuse{DiagPag53.1}}\quad \,\,\Bigg]\vspace{2mm}\\
&&=\frac{1}{2} \displaystyle\int d\theta_1 d \bar{\theta}_1 \left(\bar{\theta}_1 \theta_1\right) d\theta_2 d \bar{\theta}_2
\left(\bar{\theta}_2 \theta_2\right)
\displaystyle\int d^4x d\theta d \bar{\theta}\vspace{2mm}\\
&&\times \left[\Delta_{\beta}(x_1-x)+
G_{R}^{(B)}(x-x_1)\bar{\theta}_1\theta_1+G_{R}^{(B)}(x_1-x)\bar{\theta}\theta
+G_{R}^{(F)}(x-x_1)\theta \bar{\theta}_1
+G_{R}^{(F)}(x_1-x)\theta_1\bar{\theta}\right]\vspace{2mm}\\
&&\times\left[\Delta_{\beta}(x-x)+\mathcal{G}(z,z)\right]\vspace{2mm}\\
&&\times\left[\Delta_{\beta}(x-x_2)+
G_{R}^{(B)}(x-x_2)\bar{\theta}_2\theta_2+G_{R}^{(B)}(x_2-x)\bar{\theta}\theta
+G_{R}^{(F)}(x-x_2)\theta \bar{\theta}_2
+G_{R}^{(F)}(x_2-x)\theta_2\bar{\theta}\right]\vspace{2mm}\\
&&=\frac{1}{2} \displaystyle\int d^4 x d \theta d \bar{\theta}
\left[\Delta_{\beta}(x_1-x)+G_{R}^{(B)}(x_1-x)\bar{\theta}\theta\right]\Delta_{\beta}(x-x')
\left[\Delta_{\beta}(x-x_2)+G_{R}^{(B)}(x_2-x)\bar{\theta}\theta\right]\vspace{2mm}\\
&&=\frac{1}{2} \displaystyle\int d^4 x d \theta d \bar{\theta}
\left[\Delta_{\beta}(x_1-x)+G_{R}^{(B)}(x_1-x)\bar{\theta}\theta\right]\Delta_{\beta}(x-x')\vspace{2mm}\\
&&\times\left[\Delta_{\beta}(x-x_2)+G_{R}^{(B)}(x_2-x)\bar{\theta}\theta\right]
\end{array}
\end{equation}
\vspace{0.cm}
\begin{equation}\nonumber
\hspace{-5.3cm}=\Bigg[\frac{1}{2!}\quad\parbox{32mm}{\vspace{-0cm}\fmfreuse{DiagPag44.3YWithout}}\,\,\,+\frac{1}{2!}\quad\parbox{32mm}{\vspace{-0cm}\fmfreuse{DiagPag45.1}}\quad\Bigg]
\end{equation}
 and these last diagrams are exactly the sum of (\ref{thermal1storder1}) and
(\ref{thermal<>1storder2}) which were the first order correction
to the $\langle \phi \phi\rangle$ propagator. So the lesson we
learn from here is the following: it is enough to project the external
super-legs of the super-diagrams on the fields we want, in order to
get also the correct internal part of the super-diagrams.

Let us now check the second order. That means let us check if the
three super-diagrams in Eqs. (\ref{<superphisuperphi>(2)1}),
(\ref{<superphisuperphi>(2)2}) and (\ref{<superphisuperphi>(2)3}),
once the external legs are projected on $\phi$, reproduce the
eight  second order diagrams contained in (\ref{<phi
phi>betafinal}).

Again, in order to get the reader familiar with this formalism,
let us calculate explicitly here  the first of the three
super-diagrams in (\ref{<superphisuperphi>(2)1}), while the
calculation of the other two will be confined in appendix
\ref{Projectorsuperphi2ndorder}
\begin{equation}\label{projzz'zz'zz'}
\hspace{-3.3cm}\displaystyle\int d \theta_1 d \bar{\theta}_1 d \theta_2 d \bar{\theta}_2
\left(\bar{\theta}_1 \theta_1\right)\left(\bar{\theta}_2
\theta_2\right)\Bigg[\,\,\frac{1}{3!}\quad\parbox{30mm}{\vspace{-0cm}\fmfreuse{DiagPag53.5}}\quad\quad\quad \,\,\Bigg]=
\end{equation}
\vspace{-0.8cm}
\begin{eqnarray*}
&&\hspace{0cm}=\displaystyle\int d \theta_1 d \bar{\theta}_1 d \theta_2 d \bar{\theta}_2
\left(\bar{\theta}_1 \theta_1\right)\left(\bar{\theta}_2
\theta_2\right) \displaystyle\int d^4x d\theta d\bar{\theta}
\displaystyle\int d^4x' d\theta'
d\bar{\theta}'\vspace{2mm}\\
&&\hspace{0cm}\left[\Delta_{\beta}(x_1-x)+G_{R}^{(B)}(x_1-x)\bar{\theta}\theta
+G_{R}^{(B)}(x-x_1)\bar{\theta}_1\theta_1
+G_{R}^{(F)}(x_1-x)\theta_1 \bar{\theta}
+G_{R}^{(F)}(x-x_1)\theta\bar{\theta}_1\right]\times\vspace{2mm}\\
&&\hspace{0cm}\left[\Delta_{\beta}(x-x')+\mathcal{G}(z,z')\right]^3\times\vspace{2mm}\\
&&\hspace{-0cm}\left[\Delta_{\beta}(x'-x_2)+G_{R}^{(B)}(x'-x_2)\bar{\theta}_2\theta_2
+G_{R}^{(B)}(x_2-x')\bar{\theta}'\theta'
+G_{R}^{(F)}(x'-x_2)\theta' \bar{\theta}_2
+G_{R}^{(F)}(x_2-x')\theta_2\bar{\theta}'\right]\vspace{2mm}\\
&&\hspace{0cm}=\displaystyle \int d \theta_1 d \bar{\theta}_1 d \theta_2 d \bar{\theta}_2
\left(\bar{\theta}_1 \theta_1\right)\left(\bar{\theta}_2
\theta_2\right) \displaystyle\int d^4x d\theta d\bar{\theta}
\displaystyle\int d^4x' d\theta'
d\bar{\theta}'\times\\
\end{eqnarray*}
\begin{equation}\nonumber
\begin{array}{lll}
&&\left[\Delta_{\beta}(x_1-x)+G_{R}^{(B)}(x_1-x)\bar{\theta}\theta
+G_{R}^{(B)}(x-x_1)\bar{\theta}_1\theta_1
+G_{R}^{(F)}(x_1-x)\theta_1 \bar{\theta}
+G_{R}^{(F)}(x-x_1)\theta\bar{\theta}_1\right]\vspace{2mm}\\
&&\times \left[\Delta_{\beta}^3(x-x')+3\Delta_{\beta}^2(x-x')\mathcal{G}(z,z')\right]^3\vspace{2mm}\\
&&=\cfrac{1}{2}\displaystyle\int d^4xd^4x'
\Delta_{\beta}(x_1-x)\Delta_{\beta}^2(x-x')G_{R}^{(B)}(x'-x)G_{R}^{(B)}(x_2-x')\vspace{2mm}\\
&&+\cfrac{1}{2}\displaystyle\int d^4xd^4x'
G_{R}^{(B)}(x_1-x)\Delta_{\beta}^2(x-x')G_{R}^{(B)}(x-x')\Delta_{\beta}(x'-x_2)\vspace{2mm}\\
&&+\cfrac{1}{6}\displaystyle\int d^4xd^4x' G_{R}^{(B)}(x_1-x)\Delta_{\beta}^3(x-x')G_{R}^{(B)}(x_2-x')=\vspace{2mm}\\
\end{array}
\end{equation}
\vspace{-0.3cm}
\begin{equation}
\label{projzz'zz'zz''}
\hspace{0.3cm}=\Bigg[\frac{1}{2!}\quad\parbox{30mm}{\vspace{-0cm}\fmfreuse{DiagPag46.3}}\hspace{2.2cm}+
\frac{1}{2!}\quad\parbox{30mm}{\vspace{-0cm}\fmfreuse{DiagPag45.4}}\hspace{2.4cm}+
\end{equation}
\vspace{-0.3cm}
\begin{equation}\nonumber
\hspace{-7.8cm}+\frac{1}{3!}\quad\parbox{30mm}{\vspace{-0cm}\fmfreuse{DiagPag46.5}}\hspace{2.3cm}\Bigg].
\end{equation} These are exactly the last three diagrams of (\ref{<phi phi>betafinal}) and exactly with the same coefficients.

Analogously we can prove (see appendix
\ref{Projectorsuperphi2ndorder}) that
\vspace{0.3cm}
\begin{equation}\label{projsuperzzzz'z'z'}
\hspace{-3.5cm}\displaystyle\int d \theta_1 d \bar{\theta}_1 d \theta_2 d \bar{\theta}_2
\left(\bar{\theta}_1 \theta_1\right)\left(\bar{\theta}_2
\theta_2\right)\,\Bigg[\,\,\frac{1}{(2!)^2}\quad\parbox{30mm}{\vspace{-0cm}\fmfreuse{DiagPag53.6}}\quad\quad\quad \,\,\Bigg]=
\end{equation}
\begin{equation}\nonumber
\hspace{-1cm}=\Bigg[\frac{1}{(2!)^2}\quad\parbox{30mm}{\vspace{-0cm}\fmfreuse{DiagPag46.2}}\hspace{2.3cm}+
\frac{1}{(2!)^2}\quad\parbox{30mm}{\vspace{-0cm}\fmfreuse{DiagPag45.3}}\hspace{2.3cm}+
\end{equation}
\begin{equation}\nonumber
\hspace{-6.7cm}+\quad\frac{1}{(2!)^2}\quad\parbox{30mm}{\vspace{-0cm}\fmfreuse{DiagPag46.4}}\hspace{2.5cm}\Bigg]\,,
\end{equation}
which are the 6th, 7th, 8th diagrams presented in (\ref{<phi phi>betafinal}).

Along the same lines we can easily get (see appendix
\ref{Projectorsuperphi2ndorder})
\vspace{1.4cm}
\begin{equation}\label{projsuperzz'z'z'z'z}
\hspace{-4.3cm}\displaystyle \int d \theta_1 d \bar{\theta}_1 d \theta_2 d \bar{\theta}_2
\left(\bar{\theta}_1 \theta_1\right)\left(\bar{\theta}_2
\theta_2\right)\,\Bigg[\,\,\frac{1}{(2!)^2}\quad\parbox{30mm}{\vspace{-0.9cm}\fmfreuse{DiagPag53.7}}\,\,\,\Bigg]=
\end{equation}
\vspace{1.6cm}
\begin{equation}\nonumber
\hspace{-2cm}=\Bigg[\frac{1}{2!}\quad\parbox{30mm}{\vspace{-1.2cm}\fmfreuse{DiagPag46.1}}\hspace{0.5cm}+
\frac{1}{2!}\quad\parbox{30mm}{\vspace{-1.2cm}\fmfreuse{DiagPag45.2}}\,\,\,\,\,\,\Bigg].
\end{equation}
 Summing up (\ref{projsuperzzzz'z'z'}), (\ref{projsuperzz'z'z'z'z})
and (\ref{projzz'zz'zz'}) we get all the second order diagrams
contained in Eq. (\ref{<phi phi>betafinal}). These last ones were eight in
numbers and we got the from just the three super-diagrams in Eqs.
(\ref{<superphisuperphi>(2)1}), (\ref{<superphisuperphi>(2)2}) and
(\ref{<superphisuperphi>(2)3}).
This decrease in the number of diagrams is only one of the
virtue of the super-diagram technique. A second one is that we
only have to perform integrations over the Grassmaniann variables
(which are easy to do) to get all the old  diagrams of $\mathcal{\tilde L}$
without bothering with their complicated symmetry
factors, without bothering with soldering in the correct manner the legs and vertices of $\mathcal{\tilde L}$ and so on. Above all, the main advantage of the
super-diagrams is that they can be derived from the analog {\it quantum}
theory whose Lagrangian $L$ is in general simpler than
$\mathcal{\tilde L}$. So, once we have the {\it quantum} theory and its
associated Feynman diagrams and symmetry factors, we are already
more than halfway also with the perturbation theory of the
associated {\it classical} theory. We have just to substitute the fields
with the super-fields and the propagators with the
super-propagators and perform, as we said above, some very simple
Grassmaniann integrations. Moreover, the super-diagram
automatically perform for us the cancelation between Fermion and
dash-full loops.

Last, but not least, by changing the projectors on the external
legs, we can get not only $\langle \varphi \varphi\rangle$
correlations, but also $\langle \lambda \varphi\rangle$, the
$\langle \bar{c} c\rangle$, the $\langle cc\rangle$, the $\langle
\bar{c} \bar{c}\rangle$ and the $\langle \lambda \lambda\rangle$
correlations. So just three second  order super-diagrams will produce
tens of standard diagrams.

We will start now with the $\langle \phi\lambda_{\pi}\rangle$
correlations. Looking at the super-field expression in
(\ref{Superfield2}), we realize immediately that, differently than
the $\langle \phi \phi\rangle$ correlation, we have to project out
only the $\phi(x_1)$ field because the $\lambda_{\pi}(x_2)$ is
already equipped with its own $\bar{\theta}_2\theta_2$ and so it
naturally makes in appearance once we integrate over the final
points. From the vertices and propagators that we can build from
the formalism in components, it is easy to see that $\langle \phi
\lambda_{\pi}\rangle$ correlation will have the zero-order component of the
form:
\begin{equation}
\langle \phi(x_1)\lambda_{\pi}(x_2)\rangle_0=\,\,\parbox{16mm}{\fmfreuse{DiagPag25.3}}.
\end{equation} While the  first order correction is:
\begin{equation}
\langle \phi(x_1)\lambda_{\pi}(x_2)\rangle_1=
\displaystyle\int d\theta_1 d\bar{\theta}_1 d\theta_2 d\bar{\theta}_2 \left(-i\bar{\theta}_1\theta_1\right)\,\Bigg[\,\,\frac{1}{2!}\quad
\parbox{32mm}{\fmfreuse{DiagPag53.1}}
\quad\Bigg]=
\end{equation}
\begin{equation}\nonumber
\hspace{-3.8cm}=\frac{1}{2!}\quad\begin{minipage}[c]{32mm}
\vspace{0.55cm}
\begin{fmffile}{DiagPag61.1}
\begin{fmfgraph*}(32,10)
\fmfkeep{DiagPag61.1}
\fmfleft{i1}
\fmfright{o1}
\fmfdot{i1,o1}
\fmfv{label=$x_1$,label.angle=-90}{i1}
\fmfv{label=$x_2$,label.angle=-90}{o1}
\fmf{phantom}{i1,v,o1}
\fmf{phantom,tag=2}{v,o1}
\fmfdot{v}
\fmf{phantom,tag=1}{i1,v}
\fmfposition
\fmfipath{p[]}
\fmfiset{p1}{vpath1(__i1,__v)}
\fmfi{plain}{subpath (0,length(p1)*0.40.) of p1}
\fmfi{dashes}{subpath (length(p1)*0.4,length(p1)) of p1}
\fmfiset{p2}{vpath2(__v,__o1)}
\fmfi{plain}{subpath (0,length(p2)*0.40.) of p2}
\fmfi{dashes}{subpath (length(p2)*0.4,length(p2)) of p2}
\fmf{plain,tension=0.9}{v,v}
\end{fmfgraph*}
\end{fmffile}
\end{minipage}\,\,.
\end{equation}
Here we have used the projector $\left(-i\bar{\theta}_1\theta_1\right)$
in order to project  $\phi(x_1)$ out of the super-field.

The second order corrections have the form:
\begin{equation}\label{gianni}
\langle \phi(x_1)\lambda_{\pi}(x_2)\rangle_2=
\displaystyle\int d\theta_1 d\bar{\theta}_1 d\theta_2 d\bar{\theta}_2 \left(-i\bar{\theta}_1\theta_1\right)\,\Bigg[\,\,\,\frac{1}{3!}
\quad\parbox{32mm}{\fmfreuse{DiagPag53.5}}\hspace{1cm}+
\end{equation}
\vspace{0.8cm}
\begin{equation}\nonumber
\hspace{3.2cm}+\frac{1}{(2!)^2}
\quad\parbox{32mm}{\fmfreuse{DiagPag53.6}}
\hspace{1.2cm}+\frac{1}{(2!)^2}\quad\parbox{32mm}{\vspace{-1.2cm}\fmfreuse{DiagPag53.7}}\Bigg]\,\,.
\end{equation}
By long but straightforward calculations similar
to those performed in appendix \ref{Projectorsuperphi2ndorder}, the result that we get for the diagrams in
(\ref{gianni}) is the following one:

\begin{equation}
\label{comparisonsupercomp2ndorder}\hspace{-2cm}\Bigg[\frac{1}{3!}\quad
\parbox{33mm}{\vspace{0.cm}
\begin{fmffile}{DiagPag62.1}
\begin{fmfgraph*}(50,20)
\fmfkeep{DiagPag62.1}
\fmfleft{i1,i2,i3}
\fmfright{o1,o2,o3}
\fmfdot{i2,o2}
\fmffreeze
\fmf{phantom}{i1,v11,v12,v13,v14,v15,v16,o1}
\fmffreeze
\fmf{phantom}{i3,v31,v32,v33,v34,v35,v36,o3}
\fmffreeze
\fmf{phantom,tag=3}{i2,v21}
\fmf{phantom}{v21,v22}
\fmf{phantom,tag=2}{v22,o2}
\fmffreeze
\fmf{plain,left=0.9}{v21,v22}
\fmf{plain,right=0.9}{v21,v22}
\fmf{phantom,tag=1}{v21,v22}
\fmfposition
\fmfv{label=$x_1$,label.angle=-90}{i2}
\fmfv{label=$x_2$,label.angle=-90}{o2}
\fmfipath{p[]}
\fmfiset{p3}{vpath3(__i2,__v21)}
\fmfi{plain}{subpath (0,length(p3)*0.4) of p3}
\fmfi{dashes}{subpath (length(p3)*0.4,length(p3)) of p3}
\fmfiset{p1}{vpath1(__v21,__v22)}
\fmfi{plain}{subpath (0,length(p1)*0.4) of p1}
\fmfi{dashes}{subpath (length(p1)*0.4,length(p1)) of p1}
\fmfiset{p2}{vpath2(__v22,__o2)}
\fmfi{plain}{subpath (0,length(p2)*0.4) of p2}
\fmfi{dashes}{subpath (length(p2)*0.4,length(p2)) of p2}
\fmffreeze
\fmfdot{v21,v22}
\fmfposition
\fmffreeze
\end{fmfgraph*}
\end{fmffile}}
\hspace{2cm}+\frac{1}{(2!)^2}\quad
\parbox{33mm}{\vspace{0.5cm}
\begin{fmffile}{DiagPag62.2}
\begin{fmfgraph*}(45,20)
\fmfkeep{DiagPag62.2}
\fmfleft{i1,i2,i3}
\fmfright{o1,o2,o3}
\fmfdot{i2,o2}
\fmffreeze
\fmf{phantom}{i1,v11,v12,v13,v14,v15,v16,o1}
\fmffreeze
\fmf{phantom,tag=3}{i2,v21}
\fmf{plain,tension=0.9}{v21,v21}
\fmf{plain,tension=0.9}{v22,v22}
\fmf{phantom,tag=4}{v21,v22}
\fmf{phantom,tag=1}{v22,o2}
\fmfdot{v21,v22}
\fmfv{label=$x_1$,label.angle=-90}{i2}
\fmfv{label=$x_2$,label.angle=-90}{o2}
\fmfposition
\fmfipath{p[]}
\fmfiset{p1}{vpath1(__v22,__o2)}
\fmfi{plain}{subpath (0,length(p1)*0.4) of p1}
\fmfi{dashes}{subpath (length(p1)*0.4,length(p1)) of p1}
\fmfiset{p3}{vpath3(__i2,__v21)}
\fmfi{plain}{subpath (0,length(p3)*0.4) of p3}
\fmfi{dashes}{subpath (length(p3)*0.4,length(p3)) of p3}
\fmfiset{p4}{vpath4(__v21,__v22)}
\fmfi{plain}{subpath (0,length(p4)*0.4) of p4}
\fmfi{dashes}{subpath (length(p4)*0.4,length(p4)) of p4}
\fmffreeze
\fmf{phantom}{i3,v31,v32,v33,v34,v35,v36,o3}
\fmffreeze
\end{fmfgraph*}
\end{fmffile}
}\hspace{1.4cm}+
\end{equation}
\vspace{0.3cm}
\vspace{0.8cm}
\begin{equation}\nonumber
\hspace{-10cm}+\frac{1}{2!}\quad
\begin{minipage}[c]{30mm}
\vspace{-0.63cm}
 \begin{fmffile}{DiagPag62.3}
\begin{fmfgraph*}(30,15)
\fmfkeep{DiagPag62.3}
\fmftop{t0,t1}
\fmfbottom{b0,b1}
\fmfv{label=$x_1$,label.angle=-90}{b0}
\fmfv{label=$x_2$,label.angle=-90}{b1}
\fmf{phantom}{t0,v1,t1}
\fmf{phantom}{b0,v2,b1}
\fmffreeze
\fmf{phantom,tag=3}{v2,b1}
\fmf{phantom,tag=2}{b0,v2}
\fmfdot{b1}
\fmfdot{b0}
\fmf{plain,tension=0.8}{v1,v1}
\fmf{dashes,tension=0.8,right=0.6,tag=1}{v1,v2}
\fmf{plain,tension=0.8,left=0.6}{v1,v2}
\fmfdot{v1,v2}
\fmfposition
\fmfipath{p[]}
\fmfiset{p1}{vpath1(__v1,__v2)}
 \fmfi{plain}{subpath (length(p1)*0.6,length(p1)) of p1}
\fmfiset{p2}{vpath2(__b0,__v2)}
 \fmfi{plain}{subpath (0,length(p2)*0.4) of p2}
 \fmfi{dashes}{subpath (length(p2)*0.4,length(p2)) of p2}
\fmfiset{p3}{vpath3(__v2,__b1)}
 \fmfi{plain}{subpath (0,length(p3)*0.4) of p3}
 \fmfi{dashes}{subpath (length(p3)*0.4,length(p3)) of p3}
\end{fmfgraph*}
 \end{fmffile}
 \end{minipage}\,\,\,\Bigg]\,\,.
\end{equation}
We checked and found out that the diagrams on the r.h.s. of (\ref{comparisonsupercomp2ndorder})
are exactly those that we would have obtained using the component
formalism with the rules given in Eqs. (\ref{rule1}),(\ref{rule2}),(\ref{rule3}),(\ref{rule4}),(\ref{rule5}),(\ref{rule6}),(\ref{rule7}) and taking into account of the cancelations between
dash-full propagators and Fermionic ones.

From now on we will trust the results given by super-fields once they are
properly projected on the components fields we are interested in.
We will trust them because they seem to give
the same results as if we have done the calculations in components
using the rules in Eqs. (\ref{rule1}) (\ref{rule2}),(\ref{rule3}),(\ref{rule4}),(\ref{rule5}), (\ref{rule6}),(\ref{rule7}).

Let us now calculate the Grassmaniann variables correlations. That
means we consider them as physical fields and not like ``ghosts'',
as it is done in gauge theories, where they contribute only to
internal loops and are never physical (i.e. external ) fields. The
reason we consider them as physical fields is because, as we said before,  they are
the Jacobi fields and they enter in the calculations
of the Lyapunov exponents\cite{Liap}.

Let us start from the
correlation $\langle c^{\phi}(x_1)\bar{c}_{\pi}(x_2)\rangle$. The
zero order is given in Eq. (\ref{rule3}). The first order, via
super-fields , will be obtained from the super-diagram of  Eq.
(\ref{<superphisuperphi>(1)}) by projecting out the
$c^{\phi}(x_1)$ and $ \bar{c}_{\pi}(x_2)$. This is achieved with
the projector $\theta_2 \bar{\theta}_1$. The first one extracts
from the super-field the field $ \bar{c}_{\pi}(x_2)$, while the second
one extracts the $c^{\phi}(x_1)$. So what we have to do is:
\parbox{33cm}{
\vspace{1cm}
\begin{equation}
\hspace{-21cm}
\displaystyle\int d\theta_1 d\bar{\theta}_1 d\theta_2 d\bar{\theta}_2
\left(\theta_2\bar{\theta}_1\right)\,\Bigg[\,\,\frac{1}{2!}\quad
\parbox{32mm}{\fmfreuse{DiagPag53.1}}
\quad\Bigg].
\end{equation}}
Performing the long but straightforward Grassmannian integration above , we get:
\begin{equation}
\frac{1}{2!}\quad
\begin{minipage}[c]{32mm}
\vspace{0.6cm}
\begin{fmffile}{DiagPag64.1}
\begin{fmfgraph*}(32,10)
\fmfset{arrow_len}{2mm}
\fmfleft{i1}
\fmfright{o1}
\fmfdot{i1,o1}
\fmfv{label=$x_1$,label.angle=-90}{i1}
\fmfv{label=$x_2$,label.angle=-90}{o1}
\fmf{dots_arrow}{i1,v,o1}
\fmfdot{v}
\fmfv{label=$z$,label.angle=-90}{v}
\fmfposition
\fmf{plain,tension=0.9}{v,v}
\end{fmfgraph*}
\end{fmffile}
\end{minipage}
\end{equation}
For the second order we just have to perform the following calculations:
\vspace{0.5cm}
\begin{equation}
\hspace{-5cm}\displaystyle\int d\theta_1 d\bar{\theta}_1 d\theta_2 d\bar{\theta}_2
\left(\theta_2\bar{\theta}_1\right)\,\Bigg[\,\,\,\frac{1}{3!}
\quad\parbox{32mm}{\fmfreuse{DiagPag53.5}}\hspace{1cm}+
\end{equation}
\begin{equation}\nonumber
\hspace{3.2cm}+\frac{1}{(2!)^2}
\quad\parbox{32mm}{\fmfreuse{DiagPag53.6}}
\hspace{1.2cm}+\frac{1}{(2!)^2}\quad\parbox{32mm}{\vspace{-1.2cm}\fmfreuse{DiagPag53.7}}\Bigg]
\end{equation}
and the result is:
\begin{equation}\nonumber
\Bigg[-\frac{1}{2!}\,\,\parbox{33mm}{\vspace{0.cm}
\begin{fmffile}{DiagPag64bis.1}
\begin{fmfgraph*}(40,20)
\fmfset{arrow_len}{2mm}
\fmfleft{i1,i2,i3}
\fmfright{o1,o2,o3}
\fmfdot{i2,o2}
\fmffreeze
\fmf{phantom}{i1,v11,v12,v13,v14,v15,v16,o1}
\fmffreeze
\fmf{phantom}{i3,v31,v32,v33,v34,v35,v36,o3}
\fmffreeze
\fmf{dots_arrow}{i2,v21}
\fmf{dots_arrow}{v21,v22}
\fmf{dots_arrow}{v22,o2}
\fmffreeze
\fmf{plain,left=0.9}{v21,v22}
\fmf{plain,right=0.9}{v21,v22}
\fmfposition
\fmffreeze
\fmfdot{v21,v22}
\fmfposition
\fmffreeze
\end{fmfgraph*}
\end{fmffile}}
\hspace{1.cm}-\frac{1}{2!}\,\,
\parbox{33mm}{\vspace{0.5cm}
\begin{fmffile}{DiagPag64bis.2}
\begin{fmfgraph*}(40,20)
\fmfset{arrow_len}{2mm}
\fmfleft{i1,i2,i3}
\fmfright{o1,o2,o3}
\fmfdot{i2,o2}
\fmffreeze
\fmf{phantom}{i1,v11,v12,v13,v14,v15,v16,o1}
\fmffreeze
\fmf{dots_arrow}{i2,v21}
\fmf{plain,tension=0.9}{v21,v21}
\fmf{plain,tension=0.9,label=}{v22,v22}
\fmf{dots_arrow}{v21,v22}
\fmf{dots_arrow}{v22,o2}
\fmfdot{v21,v22}
\fmf{phantom}{i3,v31,v32,v33,v34,v35,v36,o3}
\fmffreeze
\end{fmfgraph*}
\end{fmffile}
}
\hspace{1cm}-\frac{1}{2!}\hspace{0.2cm}
\begin{minipage}[c]{30mm}
\vspace{-1.2cm}
 \begin{fmffile}{DiagPag64bis.3}
\begin{fmfgraph*}(25,15)
\fmfset{arrow_len}{2mm}
\fmftop{t0,t1}
\fmfbottom{b0,b1}
\fmf{phantom}{t0,v1,t1}
\fmf{dots_arrow}{b0,v2,b1}
\fmffreeze
\fmfdot{b1}
\fmfdot{b0}
\fmf{plain,tension=0.8}{v1,v1}
\fmf{phantom,tension=0.8,right=0.6,tag=1}{v1,v2}
\fmf{phantom,tension=0.8,left=0.6,tag=3}{v1,v2}
\fmfdot{v1,v2}
\fmfposition
\fmfipath{p[]}
\fmfiset{p1}{vpath1(__v1,__v2)}
 \fmfi{dashes}{subpath (0,length(p1)*0.5) of p1}
 \fmfi{plain}{subpath (length(p1)*0.5,length(p1)) of p1}
\fmfiset{p3}{vpath3(__v1,__v2)}
\fmfi{plain}{subpath (0,length(p3)*0.5) of p3}
\fmfi{plain}{subpath (length(p3)*0.5,length(p3)) of p3}
\end{fmfgraph*}
 \end{fmffile}
 \end{minipage}\Bigg]\,.
\end{equation}

We leave this calculation to the reader and we let him compare it
with the calculations in components which we can get from the rules stated in Eqs.
(\ref{rule1}), (\ref{rule2}), (\ref{rule3}),
(\ref{rule5}), (\ref{rule6}).

Next, let us calculate
\begin{equation}
\langle c^{\phi}(x_1)c^{\phi}(x_2)\rangle_{\beta}.
\end{equation}
The projector in this case is $\left(\bar{\theta}_1\bar{\theta}_2\right)$:
\begin{equation}\label{Proj<superphisuperphi>=<cc>}
\displaystyle\int d\theta_1 d\bar{\theta}_1 d\theta_2 d\bar{\theta}_2
\left(\bar{\theta}_1\bar{\theta}_2\right)\langle \Phi(z_1)\Phi(z_2)\rangle_{\beta}
= \langle c^{\phi}(x_1)c^{\phi}(x_2)\rangle_{\beta}.
\end{equation}
By an argument presented in appendix \ref{averagecciszero} we can prove
,to all orders in perturbation theory, that:
\begin{equation}\label{<cc>iszero}
\langle c^{\phi}(x_1)c^{\phi}(x_2)\rangle_{\beta}=0.
\end{equation} With similar arguments we can also prove that
\begin{equation}\label{otherszero<>}
\langle \phi(x_1)c^{\phi}(x_2)\rangle_{\beta}
=\langle \phi(x_1)\bar{c}_{\pi}(x_2)\rangle_{\beta}
=\langle \lambda(x_1)c^{\phi}(x_2)\rangle_{\beta}
=\langle \lambda(x_1)\bar{c}_{\pi}(x_2)\rangle_{\beta}=0
\end{equation}

Let us just work out  the first one of  the relations in (\ref{otherszero<>}). The
projector needed to derive things from the super-fields is $\left(\bar{\theta}_2
\bar{\theta}_1 \theta_1 \right)$, so:
\begin{equation}
\begin{array}{lll}
&&\displaystyle\int d\theta_1 d\bar{\theta}_1 d\theta_2 d\bar{\theta}_2
\left(\bar{\theta}_2\bar{\theta}_1 \theta_1\right)\langle \Phi(z_1)\Phi(z_2)\rangle_{\beta}\vspace{2mm}\\
&&=\displaystyle\int d\theta_1 d\bar{\theta}_1 d\theta_2 d\bar{\theta}_2
\left(\bar{\theta}_2\bar{\theta}_1 \theta_1\right)\displaystyle \int dzdz'\mathbb{G}(z_1,z)\mathbb{F}^{(n)}
(z,z')\mathbb{G}(z',z_2)\vspace{2mm}\\
&&=\displaystyle\int d\theta_1d\bar{\theta}_1 d\theta_2 d\bar{\theta}_2
\left(\bar{\theta}_2\bar{\theta}_1 \theta_1\right)
\displaystyle\int d^4 x d\theta d\bar{\theta}
\displaystyle\int d^4 x' d\theta' d\bar{\theta}'
\left[\Delta_{\beta}(x_1-x)+G_{R}^{(B)}(x_1-x)\bar{\theta}\theta\right.\vspace{2mm}\\
&&\left.+G_{R}^{(B)}(x-x_1)\bar{\theta}_1\theta_1+G_{R}^{(F)}(x_1-x)\theta_1 \bar{\theta}
+G_{R}^{(F)}(x-x_1)\theta\bar{\theta}_1\right]
\mathbb{F}^{(n)}(z,z')\vspace{2mm}\\
&&\times\left[\Delta_{\beta}(x'-x_2)+G_{R}^{(B)}(x'-x_2)\bar{\theta}_2\theta_2
+G_{R}^{(B)}(x_2-x')\bar{\theta}'\theta'\right.\vspace{2mm}\\
&&\left.+G_{R}^{(F)}(x'-x_2)\theta' \bar{\theta}_2
+G_{R}^{(F)}(x_2-x')\theta_2\bar{\theta}'\right]\vspace{2mm}\\
&&\equiv \mathbb{A}+\mathbb{B}.
\end{array}
\end{equation} where $\mathbb{F}^{(n)}$ is defined in appendix \ref{Loopstuff}, while $\mathbb{A}$ and
$\mathbb{B}$ are:
\begin{equation}
\mathbb{A}\equiv \displaystyle\int dzdz' \bar{\theta}'\Delta_{\beta}(x_1-x)\mathbb{F}^{(n)}(z,z')G_{R}^{(F)}(x_2-x')
\end{equation} and
\begin{equation}
\mathbb{B}\equiv \displaystyle\int dzdz'
\bar{\theta}\theta\bar{\theta}'G_{R}^{(F)}(x_1-x)\mathbb{F}^{(n)}(z,z')G_{R}^{(F)}(x_2-x').
\end{equation} In appendix \ref{Loopstuff} we proved that the Grassmannian coefficients of $\mathbb{F}^{(n)}$ can
only be one of the following six forms: 1)
$\theta\bar{\theta}\theta'\bar{\theta}'$, 2) $\theta\bar{\theta}$,
3) $\theta'\bar{\theta}'$, 4) $\theta\bar{\theta}'$, 5)
$\bar{\theta}\theta'$ and 6) $\mathbb{I}$. Inserting any of them
in $\mathbb{A}$ or $\mathbb{B}$ we get  that
$\mathbb{A}=0=\mathbb{B}$ in all cases and this proves the first of the Eq.
(\ref{otherszero<>}). The proof of the other relations in Eq.
(\ref{otherszero<>}) are analogous and we leave them to the
reader.

Let us now prove  that, to all orders in perturbation theory, we have:
\begin{equation}
\langle \lambda_{\pi}(x_1)\lambda_{\pi}(x_2)\rangle_{\beta}=0,
\end{equation} To simplify the calculations  let us
use the notation $d\mu \equiv d\theta d\bar{\theta}$. In order to project out
the $\lambda$ field from the super-field we need, as projector, the operator
$-\mathbb{I}$, so
\begin{equation}\label{projsuperlambdalambda}
\displaystyle\int d\theta_1 d\bar{\theta}_1 d\theta_2 d\bar{\theta}_2
(-\mathbb{I})\langle \Phi(z_1)\Phi(z_2)\rangle_{\beta}
\equiv \langle \lambda_{\pi}(x_1)\lambda_{\pi}(x_2)\rangle_{\beta}.
\end{equation} Let us now consider (\ref{projsuperlambdalambda}) at n-th
order in perturbation theory
\begin{equation}\label{Proj<superphisuperphi>=<lambdalambda>}
\begin{array}{lll}
&&\displaystyle\int d\theta_1 d\bar{\theta}_1 d\theta_2 d\bar{\theta}_2
(-\mathbb{I})\langle\Phi(z_1)\Phi(z_2)\rangle_{\beta}^{(n)}\vspace{2mm}\\
&&=-\displaystyle\int d\theta_1 d\bar{\theta}_1 d\theta_2 d\bar{\theta}_2
\displaystyle \int dzdz'
\mathbb{G}(z_1,z)\mathbb{F}^{(n)}(z,z')\mathbb{G}(z',z_2)\vspace{2mm}\\
&&=-\displaystyle\int d\theta_1 d\bar{\theta}_1 d\theta_2 d\bar{\theta}_2
\displaystyle\int d^4 x d\theta d\bar{\theta}
\displaystyle\int d^4 x' d\theta' d\bar{\theta}'
\left[\Delta_{\beta}(x_1-x)+G_{R}^{(B)}(x_1-x)\bar{\theta}\theta\right.\vspace{2mm}\\
&&\left.+G_{R}^{(B)}(x-x_1)\bar{\theta}_1\theta_1+G_{R}^{(F)}(x_1-x)\theta_1 \bar{\theta}
+G_{R}^{(F)}(x-x_1)\theta\bar{\theta}_1\right]
\mathbb{F}^{(n)}(z,z')\vspace{2mm}\\
&&\times\left[\Delta_{\beta}(x'-x_2)+G_{R}^{(B)}(x'-x_2)\bar{\theta}_2\theta_2
+G_{R}^{(B)}(x_2-x')\bar{\theta}'\theta'\right.\vspace{2mm}\\
&&\left.+G_{R}^{(F)}(x'-x_2)\theta' \bar{\theta}_2
+G_{R}^{(F)}(x_2-x')\theta_2\bar{\theta}'\right]\vspace{2mm}\\
&&=-
\displaystyle\int dz dz'
\displaystyle\int d\theta_1 d\bar{\theta}_1 d\theta_2 d\bar{\theta}_2
\left[G_{R}^{(B)}(x-x_1)\bar{\theta}_1\theta_1\right]\mathbb{F}^{(n)}(z,z')
\left[G_{R}^{(B)}(x'-x_2)\bar{\theta}_2\theta_2\right]\vspace{2mm}\\
&&=-\displaystyle\int d^4x d^4x'
G_{R}^{(B)}(x-x_1)
G_{R}^{(B)}(x'-x_2)\left[\displaystyle\int d\mu d\mu'\mathbb{F}^{(n)}(z,z')\right]
\end{array}
\end{equation}
In appendix \ref{Loopstuff} we will prove several identities which
will lead to
\begin{equation}\label{LoopInternalField}
\displaystyle\int d\mu d\mu'\mathbb{F}^{(n)}(z,z')=0.
\end{equation} Because of this, from  (\ref{Proj<superphisuperphi>=<lambdalambda>})
we obtain, to all orders, that :
\begin{equation}
\langle \lambda_{\pi}(x_1)\lambda_{\pi}(x_2)\rangle_{\beta}=0,
\end{equation}  The correlation above is zero not only because of the identies that we proved but
also because we somehow are using here the closed-time-path (CTP)  formalism for classical thermal field theory\cite{GoPe}. We will not expand on this here but advice the interested reader to read ref.\cite{GoPe}.

For the correlation $ \langle \phi(x_2) \phi(x_1)\rangle_{\beta}$ we have calculated
things up to second order which is what has been done in the
literature up to now. We will show in appendix
\ref{thirdorder} that with our super-field technique it is not
difficult to reach the {\it third} order.

\section{Fluctuation-Dissipation Theorem.}\label{FDT}

This is a well known theorem that holds for many
systems like those obeying the Langevin equation and similar equations. It basically
relates the two-point function (fluctuation) of the system to the
manner the system responds to external perturbation (dissipation).
It was proved to hold \cite{DekerHakke} also for deterministic
systems whose initial distribution  was the Boltzmann one, so it
should hold also for the approach to classical field theory with
temperature that we have explored in this paper.

Several years ago it was showed \cite{GozziFDT} that the
fluctuation-dissipation theorem (FDT) for Langevin equation could
be proved non-perturbatively as a ``Ward identity'' of a hidden supersymmetry present
in the Langevin equation. As that supersymmetry is very similar to
the one we have in the classical formalism presented here, also in
the classical case the FDT can be derived as a Ward identity of
the supersymmetry. In appendix \ref{BRS} we will present that
non-perturbative derivation generously provided to us by Martin Reuter. Of course,
the FDT for deterministic systems can be proved in a much simpler
way \cite{DekerHakke}than going through the Ward identities of susy, nevertheless,
we like to look at it  as a Ward identity because it actually relates
different correlation functions  like Ward identities do.

For the moment , anyhow, let us leave aside its relations to Susy and Ward identities
and  let us start verifying its validity at the perturbative level, so that we can test the tools developed in this paper. Let us follow
the notation of Ref. \cite{DekerHakke}, that from now on we will indicate with the
initial letters ``DH'', while  we will indicate ours by the acronym
CPI. In the table below we write down  the correspondence
between the notation of DH and the one of the CPI.
We urge the reader to read Ref. \cite{DekerHakke} in order to
understand the symbols contained in the  table below:

$$\begin{array}{|c|c|}
\hline [DH] & [CPI]  \\
\hline  D_{ij} & \omega^{ab}  \\
\hline R_{ij}(t-t') & -i\theta(t-t')\langle \varphi^{a}(t)\lambda_b(t')\rangle  \\
\hline C_{ij}(t-t') & \langle \varphi^{a}(t) \varphi^{b}(t')\rangle  \\
\hline \psi_i& \varphi^a \\
\hline \hat{\psi}_i\equiv -\frac{\partial}{\partial \psi_i}&-i\lambda_a\\
\hline
\end{array}.$$ The $\varphi^a$ indicates all the phase-space coordinates. According to
Eq. (2.34) of DH the FDT is
\begin{equation}\label{FDTDH}
R_{ij}(t)=-\beta \theta(t)\displaystyle\frac{\partial}{\partial t}C_{i\bar{k}}D_{\bar{k}j},
\end{equation} where we follow the convention adopted in DH: the bar over
an index, like $\bar{k}$, means they are summed over.

Eq. (\ref{FDTDH}) can be written in an alternative form as
\begin{equation}\label{FDTDHbis}
\displaystyle\frac{\partial}{\partial t}C_{ij}(t)=\displaystyle\frac{1}{\beta}R_{i\bar{k}}(t)D_{\bar{k}j}
-\displaystyle\frac{1}{\beta}D_{\bar{k}i}R_{j\bar{k}}(-t).
\end{equation} We can prove it in the following way. For $t>0$, Eq.
(\ref{FDTDH}) becomes
\begin{equation}
R_{ij}(t)=-\beta\frac{\partial }{\partial t}C_{i\bar{k}}D_{\bar{k}j}.
\end{equation} If we multiply the relation above by $D$ we get:
\begin{equation}
\displaystyle\frac{1}{\beta}R_{i\bar{\ell}}(t)D_{\bar{\ell}j}
=-\displaystyle\frac{\partial}{\partial t}C_{i\bar{k}}D_{\bar{k}\bar{\ell}}D_{\bar{\ell}j}.
\end{equation} Next, using the fact that $D_{k\bar{\ell}}D_{\bar{\ell}j}=-\delta_{kj}$,
we obtain:
\begin{equation}\label{FDTt>}
\displaystyle\frac{\partial}{\partial t}C_{ij}(t)=\frac{1}{\beta}R_{i\bar{k}}(t)D_{\bar{k}j}.
\end{equation}
For $t<0$ Eq. (\ref{FDTDH}) gives
\begin{equation}
\begin{array}{lll}
R_{ji}(-t)&=&-\beta \theta(-t)\displaystyle\frac{\partial }{\partial (-t)}C_{j\bar{k}}(-t)D_{\bar{k}i}\\
&=&\beta \theta(-t)\displaystyle\frac{\partial }{\partial t}C_{j\bar{k}}(t)D_{\bar{k}i}
\end{array}
\end{equation} where we have used the fact that
\begin{equation}\label{FDTt<}
\displaystyle\frac{\partial }{\partial t}C_{jk}(t)
=-\displaystyle\frac{\partial }{\partial t}C_{kj}(-t).
\end{equation} Multiplying (\ref{FDTt<}) by $D$ and
operating as before,  we get for $t<0$
\begin{equation}\label{FDTt<2}
\displaystyle\frac{\partial}{\partial t}C_{ij}(t)=-\displaystyle\frac{1}{\beta}D_{\bar{k}i}(t)R_{j\bar{k}}(-t).
\end{equation} Collecting (\ref{FDTt>}) and (\ref{FDTt<2}) for
$t\gtrless 0$ we get exactly (\ref{FDTDHbis}).

Making use of the table of comparison between DH and CPI, we can turn
(\ref{FDTDHbis}) into the following relation
\begin{equation}\label{FDTCPI}
\displaystyle\frac{\partial}{\partial t_1}\langle \varphi^a(t_1)\varphi^b(t_2)\rangle
=\displaystyle\frac{1}{\beta}\theta(t_1-t_2)\langle \varphi^a(t_1)(-i)\lambda_d(t_2)\rangle\omega^{db}
-\displaystyle\frac{1}{\beta}\theta(t_2-t_1)\omega^{da}\langle \varphi^b(t_2)(-i)\lambda_d(t_1)\rangle.
\end{equation} This is the full fluctuation-dissipation theorem.

Let us choose $a=b=1$ in Eq. (\ref{FDTCPI}). We will then obtain  for
a field theory, where $\varphi^1=\phi$ and $\varphi^2=\pi$, the relation :
\begin{equation}\label{FDTCPI2}
\displaystyle\frac{\partial}{\partial t_1}\langle \varphi^a(t_1)\varphi^b(t_2)\rangle
=i\displaystyle\frac{1}{\beta}\left[\theta(t_1-t_2)\langle \phi(t_1)\lambda_{\pi}(t_2)\rangle
-\theta(t_2-t_1)\langle \phi(t_2)\lambda_{\pi}(t_1)\rangle\right].
\end{equation} At {\it order zero} in perturbation theory, in terms of
Feynman diagrams, Eq. (\ref{FDTCPI2}) can be written as follows:
\begin{equation}\label{FDT(0)}
\frac{\partial}{\partial t_1}\,\,\,\,\begin{minipage}[c]{16mm}
\vspace{0.55cm}\begin{fmffile}{DiagPag75.1}
\begin{fmfgraph*}(20,1)
\fmfleft{i1}
\fmfright{o1}
\fmfdot{i1}
\fmfdot{o1}
\fmf{plain,label=${\small \Delta_{\beta}(x_1-x_2)}$, label.side=left}{i1,o1}
\fmfv{label=$x_1$,label.angle=-90}{i1}
\fmfv{label=$x_2$,label.angle=-90}{o1}
\end{fmfgraph*}
\end{fmffile}
\end{minipage}\,\,\quad=\,\,\frac{i}{\beta}\quad\begin{minipage}[c]{16mm}
\vspace{0.6cm}\begin{fmffile}{DiagPag75.2}
\begin{fmfgraph*}(16,1)
\fmfkeep{DiagPag75.2}
\fmfleft{i1}
\fmfright{o1}
\fmfdot{i1}
\fmfdot{o1}
\fmf{phantom,tag=1,label=$G^{(B)}_R(x_1-x_2)$,label.side=left}{i1,o1}
\fmfposition
\fmfipath{p[]}
\fmfiset{p1}{vpath1(__i1,__o1)}
\fmfi{plain}{subpath (0,length(p1)*0.40) of p1}
\fmfi{dashes}{subpath (length(p1)*0.40,length(p1)) of p1}
\fmfv{label=$x_1$,label.angle=-90}{i1}
\fmfv{label=$x_2$,label.angle=-90}{o1}
\end{fmfgraph*}
\end{fmffile}
\end{minipage}\,\,\,-\frac{i}{\beta}\quad
\begin{minipage}[c]{16mm}
\vspace{0.6cm}\begin{fmffile}{DiagPag75.3}
\begin{fmfgraph*}(16,1)
\fmfkeep{DiagPag75.3}
\fmfleft{i1}
\fmfright{o1}
\fmfdot{i1}
\fmfdot{o1}
\fmf{phantom,tag=1,label=$G^{(B)}_R(x_2-x_1)$,label.side=left}{i1,o1}
\fmfposition
\fmfipath{p[]}
\fmfiset{p1}{vpath1(__i1,__o1)}
\fmfi{dashes}{subpath (0,length(p1)*0.5) of p1}
\fmfi{plain}{subpath (length(p1)*0.5,length(p1)) of p1}
\fmfv{label=$x_1$,label.angle=-90}{i1}
\fmfv{label=$x_2$,label.angle=-90}{o1}
\end{fmfgraph*}
\end{fmffile}
\end{minipage}.
\end{equation}
 We will provide a proof of this in appendix \ref{ProofFDT}.

At the {\it first order} in perturbation theory the diagrammatic form of
the FDT (\ref{FDTCPI2}) is:
\vspace{0.5cm}
\begin{equation}\label{FDT(1)}
\frac{1}{2}\frac{\partial}{\partial t_1}\quad\parbox{32mm}{\fmfreuse{DiagPag45.1}}\quad+\frac{1}{2}\frac{\partial}{\partial t_1}\quad\parbox{32mm}{\fmfreuse{DiagPag44.3YWithout}}\,=
\end{equation}
\begin{equation}\nonumber
\hspace{-1cm}=\frac{1}{2}\,\frac{i}{\beta}\quad\parbox{32mm}{\fmfreuse{DiagPag61.1}}\quad-\frac{1}{2}\,\frac{i}{\beta}\quad
\begin{minipage}[c]{32mm}
\vspace{0.55cm}
\begin{fmffile}{DiagPag75.4}
\begin{fmfgraph*}(32,10)
\fmfkeep{DiagPag75.4}
\fmfleft{i1}
\fmfright{o1}
\fmfdot{i1,o1}
\fmfv{label=$x_1$,label.angle=-90}{i1}
\fmfv{label=$x_2$,label.angle=-90}{o1}
\fmf{phantom}{i1,v,o1}
\fmf{phantom,tag=2}{v,o1}
\fmfdot{v}
\fmf{phantom,tag=1}{i1,v}
\fmfposition
\fmfipath{p[]}
\fmfiset{p1}{vpath1(__i1,__v)}
\fmfi{dashes}{subpath (0,length(p1)*0.60.) of p1}
\fmfi{plain}{subpath (length(p1)*0.6,length(p1)) of p1}
\fmfiset{p2}{vpath2(__v,__o1)}
\fmfi{dashes}{subpath (0,length(p2)*0.60.) of p2}
\fmfi{plain}{subpath (length(p2)*0.6,length(p2)) of p2}
\fmf{plain,tension=0.9}{v,v}
\end{fmfgraph*}
\end{fmffile}
\end{minipage}
\end{equation}
 and also this will be proved in appendix \ref{ProofFDT}.

The {\it second order} result is given by the following three
diagrammatic relations.
\begin{equation}\nonumber
\Bigg[\frac{1}{2}\frac{\partial}{\partial t_1}\quad\parbox{32mm}{\fmfreuse{DiagPag46.3}} \,\,\quad\quad\quad \quad\quad +\frac{1}{2}\frac{\partial}{\partial t_1}\quad\parbox{32mm}{\fmfreuse{DiagPag45.4}}\hspace{2cm}+
\end{equation}
\begin{equation}\label{FDT(2)1}\hspace{-6.5cm}
+\frac{1}{3!}\frac{\partial}{\partial t_1}\quad\parbox{32mm}{\fmfreuse{DiagPag46.5}}\hspace{2cm}\Bigg]=
\end{equation}
\begin{equation}\nonumber
\hspace{-0.2cm}=\Bigg[\frac{1}{2}\,\frac{i}{\beta}\quad\parbox{32mm}{\fmfreuse{DiagPag62.1}}
\hspace{2cm}-\frac{1}{2}\,\frac{i}{\beta}\quad
\parbox{33mm}{\vspace{0.cm}
\begin{fmffile}{DiagPag76.1}
\begin{fmfgraph*}(50,20)
\fmfkeep{DiagPag76.1}
\fmfleft{i1,i2,i3}
\fmfright{o1,o2,o3}
\fmfdot{i2,o2}
\fmffreeze
\fmf{phantom}{i1,v11,v12,v13,v14,v15,v16,o1}
\fmffreeze
\fmf{phantom}{i3,v31,v32,v33,v34,v35,v36,o3}
\fmffreeze
\fmf{phantom,tag=3}{i2,v21}
\fmf{phantom}{v21,v22}
\fmf{phantom,tag=2}{v22,o2}
\fmffreeze
\fmf{plain,left=0.9}{v21,v22}
\fmf{plain,right=0.9}{v21,v22}
\fmf{phantom,tag=1}{v21,v22}
\fmfposition
\fmfv{label=$x_1$,label.angle=-90}{i2}
\fmfv{label=$x_2$,label.angle=-90}{o2}
\fmfipath{p[]}
\fmfiset{p3}{vpath3(__i2,__v21)}
\fmfi{dashes}{subpath (0,length(p3)*0.6) of p3}
\fmfi{plain}{subpath (length(p3)*0.6,length(p3)) of p3}
\fmfiset{p1}{vpath1(__v21,__v22)}
\fmfi{dashes}{subpath (0,length(p1)*0.6) of p1}
\fmfi{plain}{subpath (length(p1)*0.6,length(p1)) of p1}
\fmfiset{p2}{vpath2(__v22,__o2)}
\fmfi{dashes}{subpath (0,length(p2)*0.6) of p2}
\fmfi{plain}{subpath (length(p2)*0.6,length(p2)) of p2}
\fmffreeze
\fmfdot{v21,v22}
\fmfposition
\fmffreeze
\end{fmfgraph*}
\end{fmffile}}\hspace{2cm}\Bigg]
\end{equation}
\begin{center}
\line(1,0){150}\\
\vspace{-0.3cm}
\line(1,0){140}
\end{center}
\vspace{1cm}
\begin{equation}\nonumber
\hspace{0.5cm}\Bigg[\frac{1}{(2!)^2}\frac{\partial}{\partial t_1}\quad\parbox{32mm}{\fmfreuse{DiagPag46.2}}\hspace{2.cm}+\frac{1}{(2!)^2}\frac{\partial}{\partial t_1}\quad\parbox{32mm}{\fmfreuse{DiagPag45.3}}\hspace{2cm}+
\end{equation}
\begin{equation}
\label{FDT(2)2}
\hspace{-6.5cm}+\frac{1}{(2!)^2}\frac{\partial}{\partial t_1}\quad\parbox{32mm}{\fmfreuse{DiagPag46.4}}\hspace{2cm}\Bigg]=
\end{equation}
\begin{equation}\nonumber
\hspace{-1.6cm}=\Bigg[\frac{1}{2}\,\frac{i}{\beta}\quad\parbox{32mm}{\fmfreuse{DiagPag62.2}}
\hspace{1.5cm}-\frac{1}{2}\,\frac{i}{\beta}\quad
\parbox{33mm}{\vspace{0.5cm}
\begin{fmffile}{DiagPag76.2}
\begin{fmfgraph*}(45,20)
\fmfkeep{DiagPag76.2}
\fmfleft{i1,i2,i3}
\fmfright{o1,o2,o3}
\fmfdot{i2,o2}
\fmffreeze
\fmf{phantom}{i1,v11,v12,v13,v14,v15,v16,o1}
\fmffreeze
\fmf{phantom,tag=3}{i2,v21}
\fmf{plain,tension=0.9}{v21,v21}
\fmf{plain,tension=0.9}{v22,v22}
\fmf{phantom,tag=4}{v21,v22}
\fmf{phantom,tag=1}{v22,o2}
\fmfdot{v21,v22}
\fmfv{label=$x_1$,label.angle=-90}{i2}
\fmfv{label=$x_2$,label.angle=-90}{o2}
\fmfposition
\fmfipath{p[]}
\fmfiset{p1}{vpath1(__v22,__o2)}
\fmfi{dashes}{subpath (0,length(p1)*0.6) of p1}
\fmfi{plain}{subpath (length(p1)*0.6,length(p1)) of p1}
\fmfiset{p3}{vpath3(__i2,__v21)}
\fmfi{dashes}{subpath (0,length(p3)*0.6) of p3}
\fmfi{plain}{subpath (length(p3)*0.6,length(p3)) of p3}
\fmfiset{p4}{vpath4(__v21,__v22)}
\fmfi{dashes}{subpath (0,length(p4)*0.6) of p4}
\fmfi{plain}{subpath (length(p4)*0.6,length(p4)) of p4}
\fmffreeze
\fmf{phantom}{i3,v31,v32,v33,v34,v35,v36,o3}
\fmffreeze
\end{fmfgraph*}
\end{fmffile}
}\hspace{1.4cm}\Bigg]
\end{equation}
\begin{center}
\line(1,0){150}\\
\vspace{-0.3cm}
\line(1,0){140}
\end{center}
\vspace{1.7cm}
\begin{equation}\label{FDT(2)3}
\hspace{-2cm}\Bigg[\frac{1}{2!}\,\frac{\partial}{\partial t_1}\quad\parbox{32mm}{\vspace{-1.2cm}\fmfreuse{DiagPag46.1}}\hspace{0.cm}+\frac{1}{2!}\frac{\partial}{\partial t_1}\quad\parbox{32mm}{\vspace{-1.2cm}\fmfreuse{DiagPag45.2}}\Bigg]=\hspace{2cm}
\end{equation}
\vspace{1.5cm}
\begin{equation}\nonumber
\hspace{-5.2cm}=\Bigg[\frac{1}{2}\,\frac{i}{\beta}\quad\parbox{32mm}{\vspace{-1.2cm}\fmfreuse{DiagPag62.3}}
\hspace{0.cm}-\frac{1}{2}\,\frac{i}{\beta}\quad
\begin{minipage}[c]{30mm}
\vspace{-0.63cm}
 \begin{fmffile}{DiagPag76.3}
\begin{fmfgraph*}(30,15)
\fmfkeep{DiagPag76.3}
\fmftop{t0,t1}
\fmfbottom{b0,b1}
\fmfv{label=$x_1$,label.angle=-90}{b0}
\fmfv{label=$x_2$,label.angle=-90}{b1}
\fmf{phantom}{t0,v1,t1}
\fmf{phantom}{b0,v2,b1}
\fmffreeze
\fmf{phantom,tag=3}{v2,b1}
\fmf{phantom,tag=2}{b0,v2}
\fmfdot{b1}
\fmfdot{b0}
\fmf{plain,tension=0.8}{v1,v1}
\fmf{dashes,tension=0.8,right=0.6,tag=1}{v1,v2}
\fmf{plain,tension=0.8,left=0.6}{v1,v2}
\fmfdot{v1,v2}
\fmfposition
\fmfipath{p[]}
\fmfiset{p1}{vpath1(__v1,__v2)}
 \fmfi{plain}{subpath (length(p1)*0.6,length(p1)) of p1}
\fmfiset{p2}{vpath2(__b0,__v2)}
 \fmfi{dashes}{subpath (0,length(p2)*0.6) of p2}
 \fmfi{plain}{subpath (length(p2)*0.6,length(p2)) of p2}
\fmfiset{p3}{vpath3(__v2,__b1)}
 \fmfi{dashes}{subpath (0,length(p3)*0.6) of p3}
 \fmfi{plain}{subpath (length(p3)*0.6,length(p3)) of p3}
\end{fmfgraph*}
 \end{fmffile}
 \end{minipage}\,\,\,\Bigg]
\end{equation}



We feel that all these diagrammatic identities will turn out to be very very usefull in simplyfing long sum of diagrams which may appear in phenomenological applications of our formalism.

\section{Conclusions and Outlooks.}

In this paper we have developed the Feynman diagrams for a
{\it classical} $g\frac{\phi^4(x)}{4!}$
field theory either with and without temperature. This
topic is of high interest in the field of heavy ions scattering
where classical field theory with temperature seems to be
playing a central role. Besides this application, the Feynman diagrams for a
classical theory can be useful in many fields from planetary
motion, to fluid dynamics, to chaotic motion, etc. We have been
able do develop the Feynman diagrams in a very easy way because a
path integral for classical systems had been previously developed
\cite{GoReuTha}. With respect to other approaches to perturbation
theory for classical systems, we have been able to give the
Feynman diagrams not only for the Bosonic variables of the system
and for  the response fields, but also for the Jacobi fields
associated to the dynamics. Actually are these last fields
(the Jacobi ones) which can provide us
with indications on the cahotic behavior of the system.
Our Feynman diagrams could for example be used
to calculate perturbatively the Lyapunov exponents of a dynamical system.

Besides developing this diagrammatics for all these fields and their interactions,
we have showed  that the  many different {\it classical} diagrams of
the various different fields mentioned above can be unified
in few super-diagrams which have the same kind of vertices and kinetic term  as the
{\it quantum} one associated to $\varphi^a$.

We hope that this
super-diagram formalism can  be of some help not only in
simplifying the notation, but also in understanding the subtle
interplay between the quantum high temperature behavior and the
classical one. Interplay which is very important in the heavy ions
scattering field. We did not address  this last issue here because this
paper is only aimed at giving the formal diagrammatics necessary in this
field.

What we will do in a forthcoming paper \cite{Cattaruzzaetal} is to
develop a formalism analog to the one of the "effective action"  and
try to give something like a  "renormalization group" approach. All this is done
in the spirit  of provinding the  formal tools which could  later be used by physicists
for more phenomenological applications.

\section*{Acknowledgments}

AFN acknowledges hospitality and financial support from ICTP and
INFN in Trieste and financial support from CNPq-Brazil under the
grant 307824/2009-8. The work of EG and EC has been supported by
grants from the MIUR (Prin 2008) and INFN (IS GE41) of Italy.
Last, but not least, we thank  G.Aarts for helpful emails exchanges and
R. Penco and M.Reuter for their suggestions,  support and generosity,

\appendix
\makeatletter \@addtoreset{equation}{section} \makeatother
\renewcommand{\theequation}{\thesection.\arabic{equation}}


\section{A Fundamental Identity.
}\label{Deltabetaborn}

In this appendix we will show the details on how to go from
(\ref{TildeZ_beta[Jphi]}) to (\ref{TildeZ_beta[Jphi]2})
If we insert  (\ref{phi0aphiapi}) into (\ref{TildeZ_beta[Jphi]}),
we get:
\begin{equation}
\begin{array}{lll}
\mathcal{\tilde{Z}}_{\beta}[J_{\phi}]
&=&\displaystyle\int {\mathscr D}\phi_i(\vec{x}) \exp\left\{
-\beta \left[\displaystyle\int d^3\vec{x}\frac{(\nabla \phi_i)^2}{2}+\frac{m^2\phi_i^2}{2}\right.\right.\vspace{2mm}\\
&&\left.\times i\displaystyle\int d^4xJ_{\phi}(x)\int d^3\vec{x}'\phi_i(x')a_{\phi}
(\vec{x}-\vec{x}',t)\right\}\vspace{2mm}\\
&&\times\displaystyle\int {\mathscr D}\pi_i(\vec{x}) \exp\left[
-\beta \displaystyle\int d^3\vec{x}\frac{(\nabla \pi_i)^2}{2}\right.\vspace{2mm}\\
&&\left.\times i\displaystyle\int d^4xJ_{\phi}(x)\int d^3\vec{x}'\pi_i(\vec{x}')a_{\pi}
(\vec{x}-\vec{x}',t)\right].
\end{array}.
\end{equation} Let us now define the following quantities:
\begin{equation}\label{Japhi}
J^a_{\phi}(\vec{x}')
=i\displaystyle\int dtd^3\vec{x}J_{\phi}(\vec{x},t)a_{\phi}(\vec{x}-\vec{x}',t),
\end{equation}
\begin{equation}\label{Japi}
J^a_{\pi}(\vec{x}')
=i\displaystyle\int dtd^3\vec{x}J_{\phi}(\vec{x},t)a_{\pi}(\vec{x}-\vec{x}',t)
\end{equation}  and let us rewrite the normalized $\mathcal{\tilde{Z}}_{\beta}[J_{\phi}]$
using (\ref{Japhi}) and (\ref{Japi}):
\begin{eqnarray}
&&\nonumber\mathcal{\tilde{Z}}_{\beta}[J_{\phi}]\vspace{2mm}\\
&&=\displaystyle\frac{1}{\mathcal{\tilde{Z}}_{\beta}[0]^{\phi}}
\displaystyle\int {\mathscr D}\phi_i(\vec{x}) \exp\left\{
-\beta \displaystyle\int d^3\vec{x}\left[\frac{(\nabla \phi_i)^2}{2}+\frac{m^2\phi_i^2}{2}\right]\right.
\nonumber\\
&&\label{TildeZ_beta[Jphi]phi}\left.+ \displaystyle\int d^3\vec{x}\phi_i(\vec{x})J_{\phi}^{a}(\vec{x})\right\}\vspace{2mm}\\
&&\label{TildeZ_beta[Jphi]pi}\times\displaystyle\frac{1}{\mathcal{\tilde{Z}}_{\beta}[0]^{\pi}}\displaystyle\int {\mathscr D}
\pi_i(\vec{x}) \exp\left\{
-\beta \displaystyle\int d^3\vec{x}\frac{(\nabla \pi_i)^2}{2}
+ \displaystyle\int d^3\vec{x}\pi_i(\vec{x})J_{\pi}^{a}(\vec{x})\right\}.
\end{eqnarray}

where
\begin{equation}
\mathcal{\tilde{Z}}_{\beta}[0]^{\phi}
=\displaystyle\int {\mathscr D}\phi_i(\vec{x}) \exp\left\{
-\beta \displaystyle\int d^3\vec{x}\left[\frac{(\nabla \phi_i)^2}{2}+\frac{m^2\phi_i^2}{2}\right]\right\}
\end{equation}
\begin{equation}
\mathcal{\tilde{Z}}_{\beta}[0]^{\pi}
=\displaystyle\int {\mathscr D}\pi_i(\vec{x}) \exp
\left\{-\beta \left[\displaystyle\int d^3\vec{x}\frac{(\nabla \pi_i)^2}{2}\right]\right\}.
\end{equation} Performing a partial integration for the kinetic piece in Eq. (\ref{TildeZ_beta[Jphi]phi})
 we get :
\begin{equation}
\displaystyle\int {\mathscr D}\phi_i(\vec{x}) \exp\left\{-\frac{1}{2}
\displaystyle\int d^3\vec{x} d^3\vec{x}'\phi_i(\vec{x})
\left[\beta\delta(\vec{x}-\vec{x}')\left(-\nabla^{2'}+m^2\right)\right]\phi_i(\vec{x}')
+ \displaystyle\int d^3\vec{x}\phi_i(\vec{x})J_{\phi}^{a}(\vec{x})\right\}.
\end{equation} As the weight is quadratic, the integration can be formally done for all
(\ref{TildeZ_beta[Jphi]phi}) and the numerator
of (\ref{TildeZ_beta[Jphi]phi}) turns out to be:
\begin{equation}\label{numTildeZ_beta[Jphi]phi}
\exp\left[\displaystyle\frac{1}{2}\displaystyle\int d^3\vec{x}
d^3\vec{x}'J_{\phi}^a(\vec{x})A^{-1}(\vec{x},\vec{x}')J_{\phi}^a(\vec{x}')\right]
\tilde{Z}_{\beta}[0]^{\phi}
\end{equation} where $A(\vec{x},\vec{x}')=\beta
\delta(\vec{x}-\vec{x}')(-\nabla^{2'}+m^2)$ and it obeys the equation:
\begin{equation}\label{AinvA=delta}
\begin{array}{lll}
&&\int d^3\vec{x}'A(\vec{x},\vec{x}')A^{-1}(\vec{x}',\vec{x}'')=\delta(\vec{x}-\vec{x}'')\vspace{2mm}\\
&&=\int d^3\vec{x}'\left[\beta \delta(\vec{x}-\vec{x}')(-\nabla^{2'}+m^2)\right]
A^{-1}(\vec{x}',\vec{x}'')=\delta(\vec{x}-\vec{x}'')
\end{array}
\end{equation} so
\begin{equation}
\left[\beta (-\nabla^2+m^2)A^{-1}\right](\vec{x},\vec{x}'')=\delta(\vec{x}-\vec{x}'').
\end{equation} Performing  the Fourier transform of $A^{-1}$ we get:
\begin{equation}
A^{-1}(\vec{x},\vec{x}'')
=\frac{1}{(2\pi)^3}\int d^3 \vec{p}
e^{-i\vec{p}.(\vec{x}''-\vec{x})}{\tilde A}^{-1}(\vec{p}).
\end{equation} It follows from (\ref{AinvA=delta}) that
\begin{equation}
{\tilde A}^{-1}(\vec{p})
=\frac{1}{\beta(\vec{p}^2+m^2)}.
\end{equation} Using this and the expressions (\ref{Japhi}), (\ref{Japi}), (\ref{a_phi}), (\ref{a_pi}) we obtain,
for the argument in the exponent of
(\ref{numTildeZ_beta[Jphi]phi},) the following expression:
\begin{equation}
\begin{array}{lll}
&&\displaystyle \int d^3\vec{x}d^3\vec{x}'
\left\{i\displaystyle\int d^3\vec{x}_1dt_1
J_{\phi}(\vec{x}_1,t_1)\displaystyle\int \displaystyle\frac{d^3\vec{p}_1}{(2\pi)^3}
e^{i\vec{p}_1.(\vec{x}_1-\vec{x})}\cos \left[E_{\vec{p}_1}(t_1-t_i)\right]\right\}\vspace{2mm}\\
&&\times\left[\displaystyle\int \displaystyle\frac{d^3\vec{p}}{(2\pi)^3}
e^{i\vec{p}.(\vec{x}'-\vec{x})}\tilde{A}^{-1}(\vec{p})\right]
\left\{i\displaystyle\int d^3x_2dt_2
J_{\phi}(\vec{x}_2,t_2)\displaystyle\int \displaystyle\frac{d^3p_1}{(2\pi)^3}
e^{i\vec{p}_2.(\vec{x}_2-\vec{x}')}\cos \left[E_{\vec{p}_2}(t_2-t_i)\right]\right\}
\end{array}
\end{equation} where $t_i$ is the initial time appearing in (\ref{tildephi0momentum}). Performing, in the
expression above, the integration over $\vec{x}$ and $\vec{x}'$ we get:
\begin{equation}
-\displaystyle\int d^4x_1d^4x_2 J_{\phi}(x_1) J_{\phi}(x_2)
\displaystyle
\int \displaystyle\frac{d^3\vec{p}}{(2\pi)^3}e^{-i\vec{p}.(\vec{x}_1-\vec{x}_2)}\tilde{A}^{-1}(\vec{p})
\cos \left[E_{\vec{p}}(t_1-t_i)\right] \cos \left[E_{\vec{p}}(t_2-t_i)\right].
\end{equation} So (\ref{TildeZ_beta[Jphi]phi}) becomes:
\begin{equation}\label{TildeZ_beta[Jphi]phi2}
\begin{array}{lll}
&&\displaystyle\frac{1}{\mathcal{\tilde{Z}}_{\beta}[0]^{\phi}}
\displaystyle\int {\mathscr D}\phi_i(\vec{x}) \exp\left\{
-\beta \displaystyle\int d^3\vec{x}\left[\frac{(\nabla \phi_i)^2}{2}+\frac{m^2\phi_i^2}{2}\right]
+ \displaystyle\int d^3\vec{x}\phi_i(\vec{x})J_{\phi}^{a}(\vec{x})\right\}\vspace{2mm}\\
&&=\exp \left\{-\frac{1}{2}\displaystyle\int d^3\vec{x}d^3\vec{x}'
J_{\phi}(\vec{x})\left\{\displaystyle\int
\displaystyle\frac{d^3\vec{p}}{(2\pi)^3}\tilde{A}^{-1}(\vec{p})
\cos \left[E_{\vec{p}}(t-t_i)\right]\right.\right. \vspace{2mm}\\
&&\left.\left.\times\cos \left[E_{\vec{p}}(t'-t_i)\right]
e^{i\vec{p}.(\vec{x}'-\vec{x})}\right\}J_{\phi}(\vec{x}')\right\}
\end{array}
\end{equation} Let us now do an analog set of manipulations for  (\ref{TildeZ_beta[Jphi]pi})
\begin{equation}\label{expJapi}
\begin{array}{lll}
&&\displaystyle\frac{1}{\mathcal{\tilde{Z}}_{\beta}[0]^{\pi}}\displaystyle\int {\mathscr D}
\pi_i(\vec{x}) \exp\left[
-\beta \displaystyle\int d^3\vec{x}\frac{(\nabla \pi_i)^2}{2}
+ \displaystyle\int d^3\vec{x}\pi_i(\vec{x})J_{\pi}^{a}(\vec{x})\right]\vspace{2mm}\\
&&=\exp \left[\displaystyle\int d^3\vec{x}d^3\vec{x}' J_{\pi}^a(\vec{x})
B^{-1}(\vec{x},\vec{x}')J_{\pi}^a(\vec{x}')\right]\vspace{2mm}\\
&&=\exp\left\{\displaystyle\frac{1}{2\beta}\displaystyle
\int d^3\vec{x}\left[J_{\pi}^a(\vec{x})\right]^2\right\}
\end{array}
\end{equation} where $B(\vec{x},\vec{x}')\equiv \beta\delta(\vec{x}-\vec{x}')$. Using the
expression (\ref{Japhi}), (\ref{Japi}), (\ref{a_phi}),
(\ref{a_pi}) the exponent in (\ref{expJapi}) becomes:
\begin{equation}\label{expJapi2}
\begin{array}{lll}
&&\displaystyle\int d^3\vec{x}\left[J_{\pi}^a(\vec{x})\right]^2\vspace{2mm}\\
&&=\displaystyle\int d^3\vec{x}
\left[i\displaystyle\int dt_1d^3\vec{x}_1J_{\phi}(\vec{x}_1,t_1)a_{\pi}(\vec{x}_1-\vec{x},t)\right]
\left[i\displaystyle\int dt_2d^3\vec{x}_2J_{\phi}(\vec{x}_2,t_2)a_{\pi}(\vec{x}_2-\vec{x},t)\right]
\vspace{2mm}\\
&&=-\displaystyle\int dx_1dx_2 J_{\phi}(x_1)J_{\phi}(x_2)\left\{
\displaystyle\int
\displaystyle\frac{d^3\vec{p}}{(2\pi)^3}
\frac{\sin \left[E_{\vec{p}}(t_1-t_i)\right]}{E_{\vec{p}}^2}
\frac{\sin \left[E_{\vec{p}}(t_2-t_i)\right]}{E_{\vec{p}}^2}e^{-i\vec{p}.
(\vec{x}_2-\vec{x}_1)}
\right\}\end{array}
\end{equation} Inserting (\ref{expJapi2}) and (\ref{TildeZ_beta[Jphi]phi2}) in (\ref{TildeZ_beta[Jphi]phi}) and
(\ref{TildeZ_beta[Jphi]pi}) we obtain:
\begin{equation}
\tilde{Z}_{\beta}[J_{\phi}]
=\exp\left\{-\frac{1}{2}
\displaystyle\int d^4x d^4x'
J_{\phi}(x)\left\{\displaystyle\frac{d^3\vec{p}}{(2\pi)^3}\displaystyle
\frac{\cos \left[E_{\vec{p}}(t-t')\right]}
{\beta \left(\vec{p}^2+m^2\right)}
e^{i\vec{p}.(\vec{x}-\vec{x}')}\right\}J_{\phi}(x')\right\}.
\end{equation} Via the identity
\begin{equation}
\displaystyle\int dp^0
\frac{1}{|p^0|}e^{-ip^0(t-t')}\delta(p^2-m^2)
=\frac{\cos\left[E_{\vec{p}}\left(t-t'\right)\right]}{E_{\vec{p}}^2}
\end{equation} we can rewrite $\tilde{Z}_{\beta}[J_{\phi}]$ as
\begin{equation}
\tilde{Z}_{\beta}[J_{\phi}]
=\exp \left[-\frac{1}{2}\displaystyle\int d^4x d^4x'
J_{\phi}(x)\Delta_{\beta}(x-x')J_{\phi}(x')\right]
\end{equation} where
\begin{equation}
\Delta_{\beta}(x)
=\displaystyle\int \frac{d^4 p}{(2\pi)^4}
\displaystyle\frac{2\pi}{\beta |p^0|}\delta(p^2-m^2)
e^{-ip.x}.
\end{equation} This is what we have used in (\ref{TildeZ_beta[Jphi]2}).

\section{Fields-Correlations Via Super-Field Projections.}
\label{Projectorsuperphi2ndorder}

In this appendix we will present  some of the detailed calculations
that we skipped in the text of the paper. Here, in particular, we
will report the calculations which support the results in
(\ref{projsuperzzzz'z'z'}) and (\ref{projsuperzz'z'z'z'z}).
Let us start from (\ref{projsuperzzzz'z'z'}):

\begin{equation}\nonumber
\displaystyle \int
d \theta_1 d \bar{\theta}_1 d \theta_2 d \bar{\theta}_2
\left(\bar{\theta}_1 \theta_1\right)\left(\bar{\theta}_2 \theta_2\right) \Bigg[\frac{1}{(2!)^2}\quad\parbox{32mm}{\fmfreuse{DiagPag53.6}}\quad\quad\quad\Bigg].
\end{equation} Using the fact that $\mathcal{G}(z,z)=0$ and the rules of Grassmannian
integrations, we get for the expression above:
\vspace{-2cm}
\begin{equation}
\begin{array}{lll}
&&\displaystyle \int
d \theta_1 d \bar{\theta}_1 d \theta_2 d \bar{\theta}_2
\left(\bar{\theta}_1 \theta_1\right)\left(\bar{\theta}_2 \theta_2\right)
\displaystyle\int d^4x d\theta d\bar{\theta}
\displaystyle\int d^4x' d\theta' d\bar{\theta}'\left[\Delta_{\beta}(x_1-x)+G_{R}^{(B)}(x_1-x)\bar{\theta}\theta
\right.\vspace{2mm}\\
&&\left.+G_{R}^{(B)}(x-x_1)\bar{\theta}_1\theta_1
+G_{R}^{(F)}(x_1-x)\theta_1 \bar{\theta}
+G_{R}^{(F)}(x-x_1)\theta\bar{\theta}_1\right]\left[\Delta_{\beta}(x-x)\right]
\left[\Delta_{\beta}(x-x')\right.\vspace{2mm}\\
&&\left.+G_{R}^{(B)}(x-x')\bar{\theta}'\theta'
+G_{R}^{(B)}(x'-x)\bar{\theta}\theta
+G_{R}^{(F)}(x-x')\theta \bar{\theta}'
+G_{R}^{(F)}(x'-x)\theta' \bar \theta\right]\,\Delta_{\beta}(x'-x')\vspace{2mm}\\
&&\left[\Delta_{\beta}(x'-x_2)
+G_{R}^{(B)}(x'-x_2)\bar{\theta}_2\theta_2
+G_{R}^{(B)}(x_2-x')\bar{\theta}'\theta'
+G_{R}^{(F)}(x'-x_2)\theta' \bar{\theta}_2
+G_{R}^{(F)}(x_2-x')\theta_2\bar{\theta}'\right]\vspace{2mm}\\
&&=\frac{1}{4}\displaystyle \int
d \theta_1 d \bar{\theta}_1 d \theta_2 d \bar{\theta}_2
\left(\bar{\theta}_1 \theta_1\right)\left(\bar{\theta}_2 \theta_2\right)
\displaystyle\int d^4x d\theta d\bar{\theta}
\displaystyle\int d^4x' d\theta' d\bar{\theta}'
\left[\Delta_{\beta}(x_1-x)
\Delta_{\beta}(x'-x_2)\right.\vspace{2mm}\\
&&\left.+\Delta_{\beta}(x_1-x)G_{R}^{(B)}(x_2-x')\bar{\theta}'\theta'
+G_{R}^{(B)}(x_1-x)\bar{\theta}\theta \Delta_{\beta}(x'-x_2)
+G_{R}^{(B)}(x_1-x)\bar{\theta} \theta\right.\vspace{2mm}\\
&&\left.G_{R}^{(B)}(x_2-x')\bar{\theta}'\theta'\right]\Delta_{\beta}(x-x)
\times\left[\Delta_{\beta}(x-x')+G_{R}^{(B)}(x-x')\bar{\theta}'\theta'
+G_{R}^{(B)}(x'-x)\bar{\theta}\theta\right.\vspace{2mm}\\
&&\left.+G_{R}^{(F)}(x-x')\theta \bar{\theta}'
+G_{R}^{(F)}(x'-x)\theta' \bar \theta\right]\Delta_{\beta}(x'-x')\,\vspace{2mm}\\
&&=\frac{1}{4}\displaystyle \int
d \theta_1 d \bar{\theta}_1 d \theta_2 d \bar{\theta}_2
\left(\bar{\theta}_1 \theta_1\right)\left(\bar{\theta}_2 \theta_2\right)
\displaystyle\int d^4x d\theta d\bar{\theta}
\displaystyle\int d^4x' d\theta' d\bar{\theta}'\vspace{2mm}\\
&&\left\{\Delta_{\beta}(x_1-x)\Delta_{\beta}(x'-x_2)\Delta_{\beta}(x-x)
\left[\Delta_{\beta}(x-x')+\mathcal{G}(z,z')\right]\Delta_{\beta}(x'-x')\right.\vspace{2mm}\\
&&+\left[\Delta_{\beta}(x_1-x)G_R^{(B)}(x_2-x')\bar{\theta}'\theta'
G_R^{(B)}(x'-x)\bar{\theta}\theta
+G_R^{(B)}(x_1-x)\bar{\theta}\theta\Delta_{\beta}(x'-x_2)G_R^{(B)}(x-x') \bar{\theta}'\theta'\right.\vspace{2mm}\\
&&\left.\left.+G_R^{(B)}(x_1-x)\bar{\theta}\theta
G_R^{(B)}(x_2-x')\bar{\theta}'\theta'\Delta_{\beta}(x-x')\right]
\Delta_{\beta}(x-x)\Delta_{\beta}(x'-x')\right\}\vspace{2mm}\\
&&=\frac{1}{4}\displaystyle \int d^4x \displaystyle \int d^4x'
\Delta_{\beta}(x_1-x) \Delta_{\beta}(x-x)
G_{R}^{(B)}(x'-x)\Delta_{\beta}(x'-x') G_{R}^{(B)}(x_2-x')\vspace{2mm}\\
&&+\frac{1}{4}\displaystyle \int d^4x \displaystyle \int d^4x'
\Delta_{\beta}(x_2-x) \Delta_{\beta}(x-x) G_{R}^{(B)}(x'-x)\Delta_{\beta}(x'-x') G_{R}^{(B)}(x_1-x')\vspace{2mm}\\
&&+\frac{1}{4}\displaystyle \int d^4x \displaystyle \int d^4x'
G_{R}^{(B)}(x_1-x)\Delta_{\beta}(x-x)\Delta_{\beta}(x-x')\Delta_{\beta}(x'-x') G_{R}^{(B)}(x_2-x')
\end{array}
\end{equation} The three diagrams presented in the last expression
are precisely
\begin{equation}\nonumber
\hspace{-0.8cm}\Bigg[\frac{1}{(2!)^2}\quad\parbox{30mm}{\vspace{-0cm}\fmfreuse{DiagPag46.2}}\hspace{2.3cm}+
\frac{1}{(2!)^2}\quad\parbox{30mm}{\vspace{-0cm}\fmfreuse{DiagPag45.3}}\hspace{2.4cm}+
\end{equation}
\begin{equation}\nonumber
\hspace{-8cm}+\frac{1}{(2!)^2}\quad\parbox{30mm}{\vspace{-0cm}\fmfreuse{DiagPag46.4}}\hspace{2.3cm}\Bigg]
\end{equation}
which is exactly the result we claimed in
(\ref{projsuperzzzz'z'z'}).

Let us now give the proof of (\ref{projsuperzz'z'z'z'z}). Again we
will make use of $\mathcal{G}(z,z)=0$ and of the standard rules of
integration over Grassmannian coordinates.
\vspace{3cm}
\begin{equation}
\hspace{-4cm}\displaystyle \int
d \theta_1 d \bar{\theta}_1 d \theta_2 d \bar{\theta}_2
\left(\bar{\theta}_1 \theta_1\right)\left(\bar{\theta}_2 \theta_2\right)
\,\Bigg[\frac{1}{(2!)^2}\quad\parbox{32mm}{\vspace{-1.2cm}\fmfreuse{DiagPag53.7}}\Bigg]\vspace{2mm}=
\end{equation}
\begin{equation}
\begin{array}{lll}
&&=\cfrac{1}{4}\displaystyle \int
d \theta_1 d \bar{\theta}_1 d \theta_2 d \bar{\theta}_2
\left(\bar{\theta}_1 \theta_1\right)\left(\bar{\theta}_2 \theta_2\right)
\displaystyle\int d^4x d\theta d\bar{\theta}
\displaystyle\int d^4x' d\theta' d\bar{\theta}'\vspace{2mm}\\
&&\times\left[\Delta_{\beta}(x_1-x)+G_{R}^{(B)}(x_1-x)\bar{\theta}\theta
+G_{R}^{(B)}(x-x_1)\bar{\theta}_1\theta_1
+G_{R}^{(F)}(x_1-x)\theta_1 \bar{\theta}
+G_{R}^{(F)}(x-x_1)\theta\bar{\theta}_1\right]\vspace{2mm}\\
&&\times\left[\Delta_{\beta}(x-x')+\mathcal{G}(z,z')\right]
\left[\Delta_{\beta}(x'-x')+\mathcal{G}(z',z')\right]
\left[\Delta_{\beta}(x'-x)+\mathcal{G}(z',z)\right]\vspace{2mm}\\
&&\times\left[\Delta_{\beta}(x-x_2)
+G_{R}^{(B)}(x-x_2)\bar{\theta}_2\theta_2
+G_{R}^{(B)}(x_2-x)\bar{\theta}\theta
+G_{R}^{(F)}(x-x_2)\theta \bar{\theta}_2
+G_{R}^{(F)}(x_2-x)\theta_2\bar{\theta}\right]\vspace{2mm}\\
&&=\cfrac{1}{4}\displaystyle \int
d \theta_1 d \bar{\theta}_1 d \theta_2 d \bar{\theta}_2
\left(\bar{\theta}_1 \theta_1\right)\left(\bar{\theta}_2 \theta_2\right)
\displaystyle\int d^4x d\theta d\bar{\theta}
\displaystyle\int d^4x' d\theta' d\bar{\theta}'\vspace{2mm}\\
&&\left[\Delta_{\beta}(x_1-x)\Delta_{\beta}(x-x_2)
+\Delta_{\beta}(x_1-x)G_{R}^{(B)}(x_2-x)\bar{\theta}\theta
+G_{R}^{(B)}(x_1-x)\bar{\theta}\theta\Delta_{\beta}(x-x_2)\right.\vspace{2mm}\\
&&\left.+G_{R}^{(B)}(x_1-x)\bar{\theta}\theta G_{R}^{(B)}(x_2-x)\bar{\theta}\theta\right]
\left[\Delta_{\beta}(x-x')+\mathcal{G}(z,z')\right]
\Delta_{\beta}(x'-x')
\left[\Delta_{\beta}(x'-x)+\mathcal{G}(z',z)\right]\vspace{2mm}\\
&&=\cfrac{1}{4}\displaystyle \int
d \theta_1 d \bar{\theta}_1 d \theta_2 d \bar{\theta}_2
\left(\bar{\theta}_1 \theta_1\right)\left(\bar{\theta}_2 \theta_2\right)
\displaystyle\int d^4x d\theta d\bar{\theta}
\displaystyle\int d^4x' d\theta' d\bar{\theta}'\vspace{2mm}\\
&&\left\{\left[\Delta_{\beta}(x_1-x)G_{R}^{(B)}(x_2-x)\bar{\theta}\theta\right]
\left[\Delta_{\beta}(x-x')G_{R}^{(B)}(x-x')\bar{\theta}'\theta'\right]
\Delta_{\beta}(x'-x')\right.\vspace{2mm}\\
&&\left.+\left[G_{R}^{(B)}(x_1-x)\bar{\theta}\theta\Delta_{\beta}(x-x_2)\right]
\left[2\Delta_{\beta}(x-x')G_{R}^{(B)}(x-x')\bar{\theta}'\theta'\right]
\Delta_{\beta}(x'-x')\right\}\vspace{2mm}\\
&&=\displaystyle\cfrac{1}{2}\displaystyle \int d^4x d^4x'
\Delta_{\beta}(x_1-x) G_{R}^{(B)}(x-x') \Delta_{\beta}(x'-x')
\Delta_{\beta}(x'-x) G_{R}^{(B)}(x_2-x) \vspace{2mm}\\
&&+\displaystyle\cfrac{1}{2}\displaystyle \int d^4x d^4x'
G_{R}^{(B)}(x_1-x) G_{R}^{(B)}(x-x') \Delta_{\beta}(x'-x')
\Delta_{\beta}(x'-x) \Delta_{\beta}(x-x_2).
\end{array}
\end{equation}
\vspace{0.8cm}
\begin{equation}\nonumber
\hspace{-4.cm}=\Bigg[\frac{1}{2!}\quad\parbox{30mm}{\vspace{-1.2cm}\fmfreuse{DiagPag46.1}}\hspace{0.5cm}+
\frac{1}{2!}\quad\parbox{30mm}{\vspace{-1.2cm}\fmfreuse{DiagPag45.2}}\,\,\,\Bigg]\,.
\end{equation}

This proves (\ref{projsuperzz'z'z'z'z}).

In both of these proofs we have been rather pedantic with the details of the calculations. We did that because we wanted to help the reader not familiar with our formalism in going through the details of the calculations.

\section{Loop Identities.}\label{Loopstuff}

In this appendix we will prove various identities which, among
other things, will help us in "proving" Eq.
(\ref{LoopInternalField}). We have put quotation marks around "proving", because ours
will not be an analytic proof. Anyhow the identities  that we will get will turn out to be rather
usefull. Let us use the following notatio: $d \mu \equiv d \theta d \bar{\theta}$, $d
\mu_1 \equiv d \theta_1 d \bar{\theta}_1$, etc.
The first identity we want  to prove is:
\begin{equation}\label{Loop0}
\displaystyle\int d\mu d \mu_1\ldots d\mu_{n}
\mathbb{G}(z,z_1)\mathbb{G}(z_1,z_2)
\ldots \mathbb{G}(z_n,z)=0.
\end{equation} We shall call it the ``first {\it loop} identity'', because the
strings of $\mathbb{G}$ makes a super-loop. Let us first observe
that
\begin{equation}\label{conv2}
\begin{array}{lll}
\displaystyle\int dz
\mathbb{G}(z_1,z)\mathbb{G}(z,z_2)
&=&\displaystyle\int d^4 x
\left[G_R^{(B)}(x_1-x)\Delta_{\beta}(x-x_2)+G_R^{(B)}(x_2-x)\Delta_{\beta}(x_1-x)\right]\vspace{2mm}\\
&&+\displaystyle\int d^4x \left[G_R^{(B)}(x_1-x)G_R^{(B)}(x-x_2)\right]\bar{\theta}_2 \theta_2\vspace{2mm}\\
&&+\displaystyle\int d^4x \left[G_R^{(B)}(x_2-x)G_R^{(B)}(x-x_1)\right]\bar{\theta}_1 \theta_1\vspace{2mm}\\
&&+\displaystyle\int d^4x \left[G_R^{(F)}(x_1-x)G_R^{(F)}(x-x_2)\right]\theta_1 \bar{\theta}_2\vspace{2mm}\\
&&+\displaystyle\int d^4x \left[G_R^{(F)}(x_2-x)G_R^{(F)}(x-x_1)\right]\theta_2 \bar{\theta}_1.
\end{array}
\end{equation} If we add to the l.h.s. of Eq. (\ref{conv2}) one more $\mathbb{G}$ and build the
expression

\begin{equation}
\int dz dz'\mathbb{G}(z_1,z)\mathbb{G}(z,z')\mathbb{G}(z',z_2),
\end{equation}
we can then do the integration over the Grassmannian variables and easily prove  that:
\begin{equation}\label{conv3}
\begin{array}{lll}
&&\displaystyle\int dz dz'
\mathbb{G}(z_1,z)\mathbb{G}(z,z')\mathbb{G}(z',z_2)\vspace{2mm}\\
&&=\displaystyle\int d^4 x
\left[G_R^{(B)}(x_1-x)\Delta_{\beta}(x-x')G_R^{(B)}(x_2-x')
+\Delta_{\beta}(x_1-x)G_R^{(B)}(x'-x)G_R^{(B)}(x'-x_2)\right.\vspace{2mm}\\
&&\left.+G_R^{(B)}(x_1-x)G_R^{(B)}(x-x')\Delta_{\beta}(x'-x_2)\right]
+\displaystyle\int d^4x \left[G_R^{(B)}(x_1-x)G_R^{(B)}(x-x')G_R^{(B)}(x'-x_2)\right]\bar{\theta}_2 \theta_2\vspace{2mm}\\
&&+\displaystyle\int d^4x \left[G_R^{(B)}(x_2-x')G_R^{(B)}(x'-x)G_R^{(B)}(x-x_1)\right]\bar{\theta}_1 \theta_1
+\displaystyle\int d^4x \left[G_R^{(F)}(x_1-x)G_R^{(F)}(x-x')\right.\vspace{2mm}\\
&&\left.\times G_R^{(F)}(x'-x_2)\right]\theta_1 \bar{\theta}_2
+\displaystyle\int d^4x \left[G_R^{(F)}(x_2-x')G_R^{(F)}(x'-x)G_R^{(F)}(x-x_1)\right]\theta_2 \bar{\theta}_1.
\end{array}
\end{equation}

In general, with a string of all $\mathbb{G}$, we can prove by
induction that the following relation holds (where $z\equiv z_0$, $z'\equiv z_{n+1}$):
\begin{equation}\label{convn}
\begin{array}{lll}
&&\displaystyle\int dz_1 dz_2\ldots dz_n
\mathbb{G}(z,z_1)\mathbb{G}(z_1,z_2)\ldots \mathbb{G}(z_n,z')\vspace{2mm}\\
&&=\left(G_R^{(B)}\star \ldots \star \Delta_{\beta\,i,i+1}\star
\ldots \star G_R^{(B)}\right)(x_0,x_{n+1})\vspace{2mm}\\
&&+\left(G_R^{(B)}\ast \ldots \ast G_R^{(B)}\right)(x_0,x_{n+1})\bar{\theta}_{n+1} \theta_{n+1}\vspace{2mm}\\
&&+\left(G_R^{(B)}\ast \ldots \ast G_R^{(B)}\right)(x_{n+1},x_0)\bar{\theta}_0 \theta_0\vspace{2mm}\\
&&+\left(G_R^{(F)}\ast \ldots \ast G_R^{(F)}\right)(x_0,x_{n+1})\theta_0 \bar{\theta}_{n+1}\vspace{2mm}\\
&&+\left(G_R^{(F)}\ast \ldots \ast G_R^{(F)}\right)(x_{n+1},x_0)\theta_{n+1} \bar{\theta}_0.
\end{array}
\end{equation} The various symbols appearing above are defined as follows:
\begin{equation}\label{usualconv}
\left(G_R^{(B)}\ast \ldots \ast G_R^{(B)}\right)(x_0,x_{n+1})
=\displaystyle\int d^4x_1 \ldots d^4x_n
G_R^{(B)}(x_0-x_1)G_R^{(B)}(x_1-x_2)\ldots G_R^{(B)}(x_n-x_{n+1}).
\end{equation} (we call this {\it ``convolution''.})
While the first expression on the r.h.s. of Eq. (\ref{convn}) is defined
as:
\begin{equation}\label{orientedconv}
\begin{array}{lll}
&&\left(G_R^{(B)}\star \ldots \star \Delta_{\beta\,i,i+1}\star
\ldots \star G_R^{(B)}\right)(x_0,x_{n+1})\vspace{2mm}\equiv\\
&&\equiv\displaystyle\int d^4x_1 \ldots d^4x_n
G_R^{(B)}(x_0-x_1)G_R^{(B)}(x_1-x_2)\ldots
G_R^{(B)}(x_{i-2}-x_{i-1})G_R^{(B)}(x_{i-1}-x_i)\vspace{2mm}\\
&&\times\Delta_{\beta}(x_i-x_{i+1})G_R^{(B)}(x_{i+2}-x_{i+1})
G_R^{(B)}(x_{i+3}-x_{i+2})\ldots
G_R^{(B)}(x_n-x_{n+1})
\end{array}
\end{equation} (we call this ``{\it orientedÊ convolution}''). Eq. (\ref{convn}) can be proved by induction
as follows. Let us take $P(n)=$ (Proposition of order $n$) to be
given by Eq. (\ref{convn}). We already know, from the beginning of
this appendix, that $P(1)$ holds true [see Eq. (\ref{conv2})].
Next, we need to show that $P(n)\Rightarrow P(n+1)$. Indeed, take
$z'\equiv z_{n+1}$ in Eq. (\ref{convn}), multiply it by
$\mathbb{G}(z_{n+1},z')$ and integrate over $z_{n+1}$. Using the
definitions in Eq. (\ref{usualconv}) and Eq. (\ref{orientedconv})
the results follows in a straightforward manner.

Let bus now go back to (\ref{convn}), take $z=z'$ and integrate over $z$,
using the definitions (\ref{usualconv}) and
(\ref{orientedconv}), we easily get:
\begin{equation}\label{LoopI}
\displaystyle\int dzdz_{1}\ldots dz_{n}
\mathbb{G}(z,z_1)\mathbb{G}(z_1,z_2)
\ldots \mathbb{G}(z_n,z)=0.
\end{equation} A second loop equality which can be proved is
\begin{equation}\label{LoopII}
\displaystyle\int d\mu d\mu_{1}\ldots d\mu_{n}
\mathbb{G}^m(z,z_1)\mathbb{G}^{m_1}(z_1,z_2)
\ldots \mathbb{G}^{m_n}(z_n,z)=0.
\end{equation} This is a generalization of (\ref{LoopI}) which is recovered
when $m=m_1=\ldots =m_n=1$.

Using the property {\bf 4)} of Eq. (\ref{Prop4}) we get that the l.h.s.
of Eq. (\ref{LoopII}) is equal to:

\begin{equation}
\begin{array}{lll}
&&\displaystyle\int d\mu d\mu_{1}\ldots d\mu_{n}
\mathbb{G}^m(z,z_1)\mathbb{G}^{m_1}(z_1,z_2)
\ldots \mathbb{G}^{m_n}(z_n,z)\vspace{2mm}\\
&&=\displaystyle\int d\mu d\mu_{1}\ldots d\mu_{n}
\left[\Delta_{\beta}^m(x-x_1)+m\Delta_{\beta}^{m-1}(x-x_1)\mathcal{G}(z,z_1)\right]\vspace{2mm}\\
&&\times\left[\Delta_{\beta}^{m_1}(x_1-x_2)+m_1\Delta_{\beta}^{m_1-1}(x_1-x_2)\mathcal{G}(z_1,z_2)\right]\vspace{2mm}\\
&&\times \ldots \times\left[\Delta_{\beta}^{m_n-1}(x_n-x)+m_n\Delta_{\beta}^{m_n-1}(x_n-x)
\mathcal{G}(z_n,z)\right]\vspace{2mm}\\
&&=mm_1\ldots m_n\Delta_{\beta}^{m-1}(x-x_1)\Delta_{\beta}^{m_1-1}(x_1-x_2)\ldots\Delta_{\beta}^{m_n-1}(x_n-x)
\vspace{2mm}\\
&&\times\displaystyle\int d\mu d\mu_{1}\ldots d\mu_{n}
\left[\mathcal{G}(z,z_1)\mathcal{G}(z_1,z_2)\ldots\mathcal{G}(z_n,z)\right]=0.
\end{array}
\end{equation}

In the steps above we have made use of (\ref{LoopI}), and of the relations:
\begin{equation}\label{LoopOneDelta}
\displaystyle\int d\mu d\mu_{1}\ldots d\mu_{n}
\mathcal{G}(z,z_1)\mathcal{G}(z_1,z_2)\ldots
\mathcal{G}(z_{i-1},z_i)\Delta_{\beta}^{m_i}(x_i-x_{i+1})\mathcal{G}(z_{i+1},z_{i+2})
\ldots \mathcal{G}(z_n,z)=0
\end{equation}and
\begin{equation}\label{LoopAllDelta}
\displaystyle\int d\mu d\mu_{1}\ldots d\mu_{n}
\Delta_{\beta}^m(z,z_1)\Delta_{\beta}^{m_1}(z_1,z_2)
\ldots \Delta_{\beta}^{m_n}(z_n,z)=0.
\end{equation} Relations (\ref{LoopOneDelta}) and (\ref{LoopAllDelta}) can be easily proved considering  the mismatch in the number of integration variables with respect to the variables
present in the argument.

We should not forget that we have to prove relation
(\ref{LoopInternalField}). The $\mathbb{F}^{(n)}(z,z')$ will be
any combination of two-point functions with any number of loops
inserted in order to reach the order n-th in perturbation theory.

Let us start from $n=1$ which is:
\vspace{0.6cm}
\begin{equation}
\frac{1}{2!}\quad\parbox{32mm}{\fmfreuse{DiagPag53.1}}\,\,\,=\mathbb{F}^{(1)}(z,z').
\end{equation} It is easy to prove (\ref{LoopInternalField}) for this $\mathbb{F}^{(1)}$, in fact:
\begin{equation}
\begin{array}{lll}
&&\displaystyle\int d\theta_1 d\theta_2 d\bar{\theta}_1 d\bar{\theta}_2\quad\parbox{32mm}{\fmfreuse{DiagPag53.1}}\vspace{2mm}\\
&&=-\quad\begin{minipage}[c]{32mm}
\vspace{0.55cm}
\begin{fmffile}{DiagPagEF6.1}
\begin{fmfgraph*}(32,10)
\fmfkeep{DiagPagEF6.1}
\fmfleft{i1}
\fmfright{o1}
\fmfdot{i1,o1}
\fmfv{label=$x_1$,label.angle=-90}{i1}
\fmfv{label=$x_2$,label.angle=-90}{o1}
\fmf{phantom}{i1,v,o1}
\fmf{phantom,tag=2}{v,o1}
\fmfdot{v}
\fmf{phantom,tag=1}{i1,v}
\fmfposition
\fmfipath{p[]}
\fmfiset{p1}{vpath1(__i1,__v)}
\fmfi{dashes}{subpath (0,length(p1)*0.60.) of p1}
\fmfi{plain}{subpath (length(p1)*0.6,length(p1)) of p1}
\fmf{plain,tension=0.9}{v,v}
\fmfiset{p2}{vpath2(__v,__o1)}
\fmfi{plain}{subpath (0,length(p2)*0.40.) of p2}
\fmfi{dashes}{subpath (length(p2)*0.4,length(p2)) of p2}
\fmfv{label=$x$,label.angle=-90}{v}
\end{fmfgraph*}
\end{fmffile}
\end{minipage}\\
&&=-\displaystyle\int dzG_R^{(B)}(x-x_1)\Delta_{\beta}(x-x)G_R^{(B)}(x-x_2)=0
\end{array}
\end{equation} This result comes from the fact that the integration contains  $d\theta d\bar{\theta}$,
but the integrant does not contain any $\theta$, $\bar{\theta}$.

Let us pass  to the  $\mathbb{F}^{(2)}(z,z')$ functions. We could choose , for example,
the following one:
\begin{equation}\label{Loopsunset}
\begin{array}{lll}
&&\displaystyle\int
d \theta_1 d\theta_2 d\bar{\theta}_1 d\bar{\theta}_2\quad\parbox{32mm}{\fmfreuse{DiagPag53.5}}\hspace{1cm}=\vspace{2mm}\\
&&=-\quad\parbox{33mm}{\vspace{0.cm}
\begin{fmffile}{DiagPagEF7.1}
\begin{fmfgraph*}(40,20)
\fmfkeep{DiagPagEF7.1}
\fmfleft{i1,i2,i3}
\fmfright{o1,o2,o3}
\fmfdot{i2,o2}
\fmffreeze
\fmf{phantom}{i1,v11,v12,v13,v14,v15,v16,o1}
\fmffreeze
\fmf{phantom}{i3,v31,v32,v33,v34,v35,v36,o3}
\fmffreeze
\fmf{phantom,tag=1}{i2,v21}
\fmf{dbl_plain}{v21,v22}
\fmf{phantom,tag=2}{v22,o2}
\fmffreeze
\fmf{dbl_plain,left=0.9}{v21,v22}
\fmf{dbl_plain,right=0.9}{v21,v22}
\fmfposition
\fmfipath{p[]}
\fmfiset{p1}{vpath1(__i2,__v21)}
 \fmfi{plain}{subpath (0,length(p1)*0.4) of p1}
 \fmfi{dashes}{subpath (length(p1)*0.4,length(p1)) of p1}
\fmfiset{p2}{vpath2(__v22,__o2)}
 \fmfi{plain}{subpath (0,length(p2)*0.4) of p2}
 \fmfi{dashes}{subpath (length(p2)*0.4,length(p2)) of p2}
\fmffreeze
\fmfdot{v21,v22}
\fmfposition
\fmffreeze
\fmfv{label=$x_1$,label.angle=-90}{i2}
\fmfv{label=$x_2$,label.angle=-90}{o2}
\fmfv{label=$z$,label.angle=-120}{v21}
\fmfv{label=$z'$,label.angle=-60,label.dist=3}{v22}
\end{fmfgraph*}
\end{fmffile}}\\
&&=-\displaystyle\int dz dz' G_R^{(B)}(x-x_1) \left[\mathbb{G}(z,z')\right]^3G_R^{(B)}(x'-x_2)\vspace{2mm}\\
&&=-\displaystyle\int d^4x d^4x'
G_R^{(B)}(x-x_1)G_R^{(B)}(x'-x_2)
\left[\displaystyle\int d\mu d\mu'\mathbb{G}^3(z,z')\right]\vspace{2mm}\\
&&=-\displaystyle d^4x d^4x'
G_R^{(B)}(x-x_1)G_R^{(B)}(x'-x_2)
\left[\displaystyle\int d\mu d\mu'\mathbb{G}^2(z,z')\mathbb{G}(z',z)\right]
\end{array}
\end{equation} In the last step we have used property {\bf 1)} of Eq. (\ref{Prop1}). The piece in
the last equality of Eq. (\ref{Loopsunset}) contained in square brackets is zero
because of the loop identity in Eq. (\ref{LoopI}). So, everything
turns out to be zero.
Another example of $\mathbb{F}^{(2)}(z,z')$ function is
\vspace{1.8cm}
\begin{equation}
\begin{array}{lll}
&&\displaystyle\int
d \theta_1 d\theta_2 d\bar{\theta}_1 d\bar{\theta}_2\quad\parbox{32mm}{\vspace{-1.2cm}\fmfreuse{DiagPag53.7}}=-\quad
\begin{minipage}[c]{30mm}
\vspace{-1.2cm}
 \begin{fmffile}{DiagPagEF8.1}
\begin{fmfgraph*}(25,15)
\fmfkeep{DiagPagEF8.1}
\fmftop{t0,t1}
\fmfbottom{b0,b1}
\fmf{phantom}{t0,v1,t1}
\fmf{phantom}{b0,v2,b1}
\fmffreeze
\fmf{phantom,tag=1}{b0,v2}
\fmf{phantom,tag=2}{v2,b1}
\fmfdot{b1}
\fmfdot{b0}
\fmfposition
\fmfipath{p[]}
\fmfiset{p1}{vpath1(__b0,__v2)}
 \fmfi{dashes}{subpath (0,length(p1)*0.6) of p1}
 \fmfi{plain}{subpath (length(p1)*0.6,length(p1)) of p1}
\fmfiset{p2}{vpath2(__v2,__b1)}
 \fmfi{plain}{subpath (0,length(p2)*0.4) of p2}
 \fmfi{dashes}{subpath (length(p2)*0.4,length(p2)) of p2}
\fmf{dbl_plain,tension=0.8}{v1,v1}
\fmf{dbl_plain,tension=0.8,right=0.6}{v1,v2}
\fmf{dbl_plain,tension=0.8,left=0.6}{v1,v2}
\fmfdot{v1,v2}
\fmfv{label=$z'$,label.angle=180}{v1}
\fmfv{label=$z$,label.angle=-90}{v2}
\fmfv{label=$x_1$,label.angle=-90}{b0}
\fmfv{label=$x_2$,label.angle=-90}{b1}
\end{fmfgraph*}
 \end{fmffile}
 \end{minipage}\vspace{0.5cm}\\
&&=-\displaystyle\int dz dz'
G_R^{(B)}(x-x_1)\mathbb{G}(z,z')\mathbb{G}(z',z')\mathbb{G}(z',z)G_R^{(B)}(x-x_2)\vspace{2mm}\\
&&=-\displaystyle d^4x d^4x'
G_R^{(B)}(x-x_1)G_R^{(B)}(x'-x_2)
\left[\displaystyle\int d\mu d\mu'\mathbb{G}(z,z')\mathbb{G}(z',z')\mathbb{G}(z',z)\right].
\end{array}
\end{equation} Once  again the piece above contained in square brackets is zero because
of the loop identity of Eq. (\ref{LoopI}).

We checked many two-point functions of the third order in
perturbation theory and we got zero because of the loop identity
(\ref{LoopI}). We do not have an analytical proof to all orders,
but we believe, from the many examples we worked out, that
(\ref{LoopInternalField}) is true.

\section{Jacobi Fields Correlations.}\label{averagecciszero}

In this appendix , to all orders in perturbation theory, we will prove thart:
\begin{equation}\label{2ptcc}
\langle c^{\phi}(x_1)c^{\phi}(x_2)\rangle_{\beta}=0
\end{equation} to all orders in perturbation theory. The reader
could think that in $\mathcal{\tilde{L}}$ the $c^{\phi}$ and $c^{\phi}$
never interact and this  could be the real reason behind (\ref{2ptcc}). But, actually,
the $c^{\phi}$ interact with $\phi$ and so, trough the $\phi$, they could interact among
themselves.

We know from (\ref{Proj<superphisuperphi>=<cc>}) that
\begin{equation}
\displaystyle\int d\theta_1 d\theta_2 d\bar{\theta}_1 d\bar{\theta}_2
 \bar{\theta}_2\theta_1\langle \Phi(x_1) \Phi(x_2)\rangle_{\beta}=
\langle c^{\phi}(x_1)c^{\phi}(x_2)\rangle_{\beta}.
\end{equation} The l.h.s at the n-th order of perturbation theory will have the general form:
\begin{equation}\label{2ptccFn}
\displaystyle\int d\theta_1 d\theta_2 d\bar{\theta}_1 d\bar{\theta}_2
\displaystyle\int dz dz'\mathbb{G}(z_1,z)\mathbb{F}^{(n)}(z,z')\mathbb{G}(z',z_2)
\end{equation} where $\mathbb{F}^{(n)}$ is a function $\mathcal{O}(g^n)$ that
we do not write down explicitly. For  the first and second order its explicit
form is
\begin{equation}
\mathbb{F}^{(1)}=\frac{1}{2}\mathbb{G}(z,z)\delta(z-z')
\end{equation}

\begin{equation}
\mathbb{F}^{(2)}
=\left\{\begin{array}{ll}
&\frac{1}{6}\mathbb{G}^3(z,z')\vspace{2mm}\\
&\frac{1}{4}\mathbb{G}(z,z)\mathbb{G}(z,z')\mathbb{G}(z',z')\vspace{2mm}\\
&\frac{1}{4}\displaystyle\int dz''\mathbb{G}(z,z'')\mathbb{G}(z'',z'')\mathbb{G}(z'',z)
\delta(z-z')
\end{array}\right.
\end{equation}
The explicit form of (\ref{2ptccFn}) turns out to be:
\begin{equation}\label{filo}
\begin{array}{lll}
(\ref{2ptccFn})&=&
\displaystyle\int d\theta_1 d\theta_2 d\bar{\theta}_1 d\bar{\theta}_2 \bar{\theta}_2\theta_1
\displaystyle\int d^4 x d\theta d\bar{\theta}
\displaystyle\int d^4 x' d\theta' d\bar{\theta}'\vspace{2mm}\\
&&\times\left[\Delta_{\beta}(x_1-x)+G_{R}^{(B)}(x_1-x)\bar{\theta}\theta
+G_{R}^{(B)}(x-x_1)\bar{\theta}_1\theta_1\right.\vspace{2mm}\\
&&\left.+G_{R}^{(F)}(x_1-x)\theta_1 \bar{\theta}
+G_{R}^{(F)}(x-x_1)\theta\bar{\theta}_1\right]
\mathbb{F}^{(n)}(z,z')\vspace{2mm}\\
&&\times\left[\Delta_{\beta}(x'-x_2)+G_{R}^{(B)}(x'-x_2)\bar{\theta}_2\theta_2
+G_{R}^{(B)}(x_2-x')\bar{\theta}'\theta'\right.\vspace{2mm}\\
&&\left.+G_{R}^{(F)}(x'-x_2)\theta' \bar{\theta}_2
+G_{R}^{(F)}(x_2-x')\theta_2\bar{\theta}'\right]\vspace{2mm}\\
&=&\displaystyle\int dzdz'\displaystyle\int d\theta_1 d\theta_2 d\bar{\theta}_1 d\bar{\theta}_2\bar{\theta}_2
\theta_1\left[G_R^{(F)}(x_1-x)\theta_1 \bar{\theta}\right]
\mathbb{F}^{(n)}(z,z')\left[G_R^{(F)}(x_2-x')\theta_2 \bar{\theta}'\right]\vspace{2mm}\\
&=&\displaystyle\int dzdz' \theta \bar{\theta}'G_R^{(F)}(x_1-x)
\mathbb{F}^{(n)}(z,z')G_R^{(F)}(x_2-x')
\end{array}
\end{equation}

The last integral above is zero. The reason is the following: $\mathbb{F}^{(n)}$ comes
from the integration of products of various $G_R$. As each $G_R$ comes with an even number
of $\hat{\theta}$ (where $\hat\theta$ indicates either $\theta$ or $\bar{\theta}$) attached , also
$\mathbb{F}^{(n)}$ will have an even number of $\hat{\theta}$ because the integrations take
away always an even number of $\hat{\theta}$. As $\mathbb{F}^{(n)}$ depends only on $z$ and
$z'$ and has an even number of $\hat{\theta}$, the only possible contributions must have the following strings
of $\hat{\theta}$:

\begin{itemize}
\item[1)] $\theta \bar{\theta} \theta' \bar{\theta}'$ which gives zero because $\theta'$
is attached  also to the final $G_R^{(B)}$  in the second equality in (\ref{filo}).

\item[2)] $\theta \bar{\theta}$ which gives zero because of  the initial $G_R^{(B)}$ (the
term $\bar{\theta}_1 \theta$ in $G_R^{(B)}$ gives zero because of the projector
$\bar{\theta}_1\theta_2$ and the same for $\theta_2$).

\item[3)] $\theta' \bar{\theta}'$ same as in 2) but because of the final $G_R^{(B)}$.

\item[4)] $\theta \bar{\theta}'$ same as in 3).

\item[5)] $\bar{\theta} \theta'$ same as in 2)
\end{itemize}

\section{Third Order Results For The Two-point Correlation.}\label{thirdorder}

In this appendix we will provide the third order perturbative
calculations for the $\langle \phi(x_1) \phi(x_2)\rangle_{\beta}$.
correlation using the super-field technique. We will list
below the super-diagrams that are needed together with the corresponding symmetry
factors. These we calculate using the algebraic rules given in
\cite{PalmerCarr}. The set of super-diagrams needed are
\vspace{-0.1cm}
\begin{equation}\nonumber
\Bigg[\frac{1}{3!}\quad
\begin{minipage}[c]{60mm}
\vspace{0.5cm}
\begin{fmffile}{DiagPagE2.1}
\begin{fmfgraph*}(60,20)
\fmfkeep{DiagPagE2.1}
\fmfleft{i1}
\fmfright{o1}
\fmf{dbl_plain}{i1,v1,v2,v3,o1}
\fmffreeze
\fmf{dbl_plain,tension=1/3,right=0.8}{v1,v2}
\fmf{dbl_plain,tension=1/3,left=0.8}{v1,v2}
\fmf{dbl_plain,tension=0.75}{v3,v3}
\fmfdot{i1,v1,v2,v3,o1}
\fmfv{label=$z_1$,label.angle=-90}{i1}
\fmfv{label=$z_2$,label.angle=-90}{o1}
\fmfv{label=$z$,label.angle=-120,label.dist=10}{v1}
\fmfv{label=$z'$,label.angle=-60}{v2}
\fmfv{label=$z''$,label.angle=-90}{v3}
\end{fmfgraph*}
 \end{fmffile}
\end{minipage}\quad+
\frac{1}{(2!)^3}\quad
\begin{minipage}[c]{60mm}
\vspace{0.5cm}
\begin{fmffile}{DiagPagE2.2}
\begin{fmfgraph*}(60,20)
\fmfkeep{DiagPagE2.2}
\fmfleft{i1}
\fmfright{o1}
\fmf{dbl_plain}{i1,v1,v2,v3,o1}
\fmffreeze
\fmf{dbl_plain,tension=0.75}{v1,v1}
\fmf{dbl_plain,tension=0.75}{v2,v2}
\fmf{dbl_plain,tension=0.75}{v3,v3}
\fmfdot{i1,v1,v2,v3,o1}
\fmfv{label=$z_1$,label.angle=-90,}{i1}
\fmfv{label=$z_2$,label.angle=-90}{o1}
\fmfv{label=$z$,label.angle=-90,label.dist=9.5}{v1}
\fmfv{label=$z'$,label.angle=-90}{v2}
\fmfv{label=$z''$,label.angle=-90}{v3}
\end{fmfgraph*}
 \end{fmffile}
\end{minipage}\,\,\,+
\end{equation}
\vspace{2.2cm}
\begin{equation}\label{SuperDiag3order}
+\frac{1}{(2!)^3}\quad\begin{minipage}[c]{60mm}
\vspace{-2.4cm}
\begin{fmffile}{DiagPagE2.3}
\begin{fmfgraph*}(40,30)
\fmfkeep{DiagPagE2.3}
\fmfbottom{b0,b1}
\fmfleft{i1}
\fmfright{o1}
\fmftop{t0,t1}
\fmf{dbl_plain}{b0,v1,b1}
\fmf{phantom}{i1,v2,o1}
\fmf{phantom}{t0,v3,t1}
\fmffreeze
\fmf{dbl_plain,tension=1/3,right=0.6}{v1,v2}
\fmf{dbl_plain,tension=1/3,left=0.6}{v1,v2}
\fmf{dbl_plain,tension=1/3,right=0.6,tag=1}{v2,v3}
\fmf{dbl_plain,tension=1/3,left=0.6}{v2,v3}
\fmf{dbl_plain,tension=0.85}{v3,v3}
\fmfv{label=$z_1$,label.angle=-90,}{b0}
\fmfv{label=$z_2$,label.angle=-90}{b1}
\fmfv{label=$z$,label.angle=-90}{v1}
\fmfv{label=$z'$,label.angle=-180}{v2}
\fmfv{label=$z''$,label.angle=-180}{v3}
\fmfdot{b0,v1,b1}
\fmfdot{v2}
\fmfdot{v3}
\end{fmfgraph*}
 \end{fmffile}
\end{minipage}\hspace{-1.5cm}+\frac{1}{(2!)^2}\quad
\begin{minipage}[c]{60mm}
\vspace{-1.6cm}
\begin{fmffile}{DiagPagE1.1}
\begin{fmfgraph*}(40,20)
\fmfkeep{DiagPagE1.1}
\fmftop{t0,t1}
\fmfbottom{b0,b1}
\fmf{phantom}{t0,v1,v0,t1}
\fmffreeze
\fmf{dbl_plain}{b0,v2,v3,b1}
\fmffreeze
\fmf{dbl_plain,tension=0.6}{v1,v1}
\fmf{dbl_plain,tension=0.6}{v3,v3}
\fmf{dbl_plain,tension=1/3,right=0.6,tag=1}{v1,v2}
\fmf{dbl_plain,tension=1/3,left=0.6}{v1,v2}
\fmfv{label=$z_1$,label.angle=-90,label.dist=9.5}{b0}
\fmfv{label=$z_2$,label.angle=-90,label.dist=9.5}{b1}
\fmfv{label=$z$,label.angle=-90,label.dist=9.5}{v2}
\fmfv{label=$z'$,label.angle=-180}{v1}
\fmfv{label=$z''$,label.angle=-90}{v3}
\fmfdot{v1,v2,b0,b1,v3}
\end{fmfgraph*}
 \end{fmffile}
\end{minipage}\hspace{-1.5cm}+
\vspace{1cm}
\end{equation}
\vspace{1.7cm}
\begin{equation}\nonumber
+\frac{1}{(2!)^2}\quad\begin{minipage}[c]{60mm}
\vspace{-1.6cm}
\begin{fmffile}{DiagPagE1.2}
\begin{fmfgraph*}(40,20)
\fmfkeep{DiagPagE1.2}
\fmfbottom{b1,b2}
\fmftop{t1,t2}
\fmf{phantom}{t1,vt1,t2}
\fmf{dbl_plain}{b1,vb1,vb2,b2}
\fmffreeze
\fmfdot{b1,b2,vt1,vb1,vb2}
\fmffreeze
\fmf{dbl_plain,left=0.7}{vb1,vt1}
\fmf{dbl_plain}{vb1,vt1}
\fmf{dbl_plain,right=0.7}{vb2,vt1}
\fmf{dbl_plain}{vb2,vt1}
\fmfv{label=$z_1$,label.angle=-90,label.dist=9.5}{b1}
\fmfv{label=$z_2$,label.angle=-90,label.dist=9.5}{b2}
\fmfv{label=$z$,label.angle=-90,label.dist=9.5}{vb1}
\fmfv{label=$z'$,label.angle=90}{vt1}
\fmfv{label=$z''$,label.angle=-90}{vb2}
\end{fmfgraph*}
\end{fmffile}
\end{minipage}
\hspace{-1.5cm}+\frac{1}{(2!)^2}\quad\begin{minipage}[c]{60mm}
\vspace{-1.2cm}
\begin{fmffile}{DiagPagE1.3}
\begin{fmfgraph*}(50,15)
\fmfkeep{DiagPagE1.3}
 \fmftop{t0,t1}
\fmfbottom{b0,b1}
\fmf{phantom}{t0,tv1,t1}
\fmf{dbl_plain}{b0,v1,v2,v3,b1}
\fmfdot{tv1}
\fmfdot{b0,v1,v3,b1}
\fmffreeze
\fmf{dbl_plain}{v1,tv1}
\fmf{dbl_plain}{v3,tv1}
\fmf{dbl_plain,tension=0.7}{tv1,tv1}
\fmf{dbl_plain,right=0.7}{v1,v3}
\fmfv{label=$z_1$,label.angle=-90,label.dist=9.5}{b0}
\fmfv{label=$z_2$,label.angle=-90,label.dist=9.5}{b1}
\fmfv{label=$z$,label.angle=-120}{v1}
\fmfv{label=$z'$,label.angle=-180}{tv1}
\fmfv{label=$z''$,label.angle=-30}{v3}
\end{fmfgraph*}
\end{fmffile}
\end{minipage}\hspace{-0.8cm}+
\end{equation}
\vspace{2.7cm}
\begin{equation}\nonumber\hspace{-1cm}+\frac{1}{3!}\,\frac{1}{2!}\quad
\begin{minipage}[c]{60mm}
\vspace{-1.6cm}
\begin{fmffile}{DiagPagE1.4}
\begin{fmfgraph*}(40,20)
\fmfkeep{DiagPagE1.4}
\fmftop{t0,t1}
\fmfbottom{b0,b1}
\fmf{dbl_plain}{b0,bv1,b1}
\fmf{phantom}{t0,tv1,tv2,tv3,t1}
\fmfdot{tv1,tv3}
\fmfdot{b0,bv1,b1}
\fmffreeze
\fmf{dbl_plain}{tv1,tv3}
\fmf{dbl_plain,right=0.5}{tv1,bv1}
\fmf{dbl_plain,right=0.8}{tv1,tv3}
\fmf{dbl_plain,left=0.8}{tv1,tv3}
\fmf{dbl_plain,left=0.5}{tv3,bv1}
\fmffreeze
\fmfv{label=$z_1$,label.angle=-90}{b0}
\fmfv{label=$z_2$,label.angle=-90}{b1}
\fmfv{label=$z$,label.angle=-90}{bv1}
\fmfv{label=$z'$,label.angle=-180}{tv1}
\fmfv{label=$z''$,label.angle=0}{tv3}
 \end{fmfgraph*}
 \end{fmffile}
\end{minipage}
-\hspace{-2cm}
+\frac{1}{(2!)^3}\quad
\begin{minipage}[c]{60mm}
\vspace{-1.6cm}
\begin{fmffile}{DiagPagE1.5}
\begin{fmfgraph*}(40,20)
\fmfkeep{DiagPagE1.5}
\fmfbottom{b1,b2}
\fmftop{t1,t2}
\fmf{phantom}{t1,vt1,vt2,t2}
\fmffreeze
\fmf{dbl_plain}{vt1,vt2}
\fmf{dbl_plain}{b1,vb1,b2}
\fmfdot{b1,b2,vt1,vt2,vb1}
\fmffreeze
\fmf{dbl_plain}{vb1,vt1}
\fmf{dbl_plain}{vb1,vt2}
\fmf{dbl_plain,tension=0.7}{vt1,vt1}
\fmf{dbl_plain,tension=0.7}{vt2,vt2}
\fmfv{label=$z_1$,label.angle=-90}{b1}
\fmfv{label=$z_2$,label.angle=-90}{b2}
\fmfv{label=$z$,label.angle=-90}{vb1}
\fmfv{label=$z'$,label.angle=-180}{vt1}
\fmfv{label=$z''$,label.angle=0}{vt2}
 \end{fmfgraph*}
 \end{fmffile}
\end{minipage}\hspace{-1.7cm}\Bigg]
\end{equation}
Also, by considering the first and fourth diagrams in eq. (\ref{SuperDiag3order}), we note that the diagrams
obtained by the interchange ($z_1\leftrightarrow z_2$) gives a contribution. So all together we end up with ten
superdiagrams at the third order (
In order to get the $\langle \phi \phi \rangle$ components of
these graphs we have to integrate them with the projector
$\bar{\theta}_1 \theta_1 \bar{\theta}_2 \theta_2$ like we did in
(\ref{<superphisuperphi>(1)})

The calculations are long but straightforward and the result is
\vspace{0.2cm}
\begin{equation}\nonumber
\hspace{-2.cm}\int d\bar{\theta}_1 d\theta_1 d\bar{\theta}_2 d\theta_2
\bar{\theta}_1 \theta_1 \bar{\theta}_2 \theta_2\quad\Bigg[\frac{1}{3!}\,\,\,
\parbox{60mm}{\fmfreuse{DiagPagE2.1}}\,\,\,\Bigg]=
\end{equation}
\begin{equation}\nonumber
\hspace{1.7cm}=\Bigg[\frac{1}{2!}\quad
\begin{minipage}[c]{60mm}
\vspace{0.5cm}
\begin{fmffile}{DiagPagE2.1Proj1}
\begin{fmfgraph*}(60,20)
\fmfkeep{DiagPagE2.1Proj1}
\fmfleft{i1}
\fmfright{o1}
\fmf{phantom}{i1,v1,v2,v3,o1}
\fmffreeze
\fmf{plain}{i1,v1}
\fmf{phantom,tag=1}{v1,v2}
\fmf{phantom,tag=2}{v2,v3}
\fmf{phantom,tag=3}{v3,o1}
\fmfposition
\fmfipath{p[]}
\fmfiset{p1}{vpath1(__v1,__v2)}
\fmfi{dashes}{subpath (0,length(p1)*0.6) of p1}
\fmfi{plain}{subpath (length(p1)*0.6,length(p1)) of p1}
\fmfiset{p2}{vpath2(__v2,__v3)}
\fmfi{dashes}{subpath (0,length(p2)*0.6) of p2}
\fmfi{plain}{subpath (length(p2)*0.6,length(p2)) of p2}
\fmfiset{p3}{vpath3(__v3,__o1)}
\fmfi{dashes}{subpath (0,length(p3)*0.6) of p3}
\fmfi{plain}{subpath (length(p3)*0.6,length(p3)) of p3}
\fmf{plain,tension=1/3,right=0.8}{v1,v2}
\fmf{plain,tension=1/3,left=0.8}{v1,v2}
\fmf{plain,tension=0.75}{v3,v3}
\fmfdot{i1,v1,v2,v3,o1}
\fmfv{label=$x_1$,label.angle=-90}{i1}
\fmfv{label=$x_2$,label.angle=-90}{o1}
\fmfv{label=$x$,label.angle=-120,label.dist=12}{v1}
\fmfv{label=$x'$,label.angle=-60,label.dist=8}{v2}
\fmfv{label=$x''$,label.angle=-90}{v3}
\end{fmfgraph*}
 \end{fmffile}
\end{minipage}\quad+
\end{equation}
\begin{equation}\nonumber
\hspace{1.7cm}+\frac{1}{2!}\quad
\begin{minipage}[c]{60mm}
\vspace{0.5cm}
\begin{fmffile}{DiagPagE2.1CorrectionProj2}
\begin{fmfgraph*}(60,20)
\fmfleft{i1}
\fmfright{o1}
\fmf{phantom}{i1,v1,v2,v3,o1}
\fmffreeze
\fmf{phantom,tag=1}{i1,v1}
\fmf{phantom,tag=2}{v1,v2}
\fmf{phantom,tag=3}{v2,v3}
\fmf{phantom,tag=4}{v3,o1}
\fmfposition
\fmfipath{p[]}
\fmfiset{p1}{vpath1(__i1,__v1)}
\fmfi{plain}{subpath (0,length(p1)*0.4) of p1}
\fmfi{dashes}{subpath (length(p1)*0.4,length(p1)) of p1}
\fmfiset{p2}{vpath2(__v1,__v2)}
\fmfi{plain}{subpath (0,length(p2)*0.4) of p2}
\fmfi{dashes}{subpath (length(p2)*0.4,length(p2)) of p2}
\fmfiset{p3}{vpath3(__v2,__v3)}
\fmfi{plain}{subpath (0,length(p3)*0.4) of p3}
\fmfi{plain}{subpath (length(p3)*0.4,length(p3)) of p3}

\fmfiset{p4}{vpath4(__v3,__o1)}
\fmfi{dashes}{subpath (0,length(p4)*0.6) of p4}
\fmfi{plain}{subpath (length(p4)*0.6,length(p4)) of p4}

\fmf{plain,tension=1/3,right=0.8}{v1,v2}
\fmf{plain,tension=1/3,left=0.8}{v1,v2}
\fmf{plain,tension=0.75}{v3,v3}
\fmfdot{i1,v1,v2,v3,o1}
\fmfv{label=$x_1$,label.angle=-90}{i1}
\fmfv{label=$x_2$,label.angle=-90}{o1}
\fmfv{label=$x$,label.angle=-120,label.dist=12}{v1}
\fmfv{label=$x'$,label.angle=-60,label.dist=8}{v2}
\fmfv{label=$x''$,label.angle=-90}{v3}
\end{fmfgraph*}
 \end{fmffile}
\end{minipage}\quad+
\end{equation}
\begin{equation}\nonumber
\hspace{1.7cm}+\frac{1}{2!}\quad
\begin{minipage}[c]{60mm}
\vspace{0.5cm}
\begin{fmffile}{DiagPagE2.1Proj2}
\begin{fmfgraph*}(60,20)
\fmfkeep{DiagPagE2.1Proj2}
\fmfleft{i1}
\fmfright{o1}
\fmf{phantom}{i1,v1,v2,v3,o1}
\fmffreeze
\fmf{plain}{v3,o1}
\fmf{phantom,tag=1}{i1,v1}
\fmf{phantom,tag=2}{v1,v2}
\fmf{phantom,tag=3}{v2,v3}
\fmfposition
\fmfipath{p[]}
\fmfiset{p1}{vpath1(__i1,__v1)}
\fmfi{plain}{subpath (0,length(p1)*0.4) of p1}
\fmfi{dashes}{subpath (length(p1)*0.4,length(p1)) of p1}
\fmfiset{p2}{vpath2(__v1,__v2)}
\fmfi{plain}{subpath (0,length(p2)*0.4) of p2}
\fmfi{dashes}{subpath (length(p2)*0.4,length(p2)) of p2}
\fmfiset{p3}{vpath3(__v2,__v3)}
\fmfi{plain}{subpath (0,length(p3)*0.4) of p3}
\fmfi{dashes}{subpath (length(p3)*0.4,length(p3)) of p3}
\fmf{plain,tension=1/3,right=0.8}{v1,v2}
\fmf{plain,tension=1/3,left=0.8}{v1,v2}
\fmf{plain,tension=0.75}{v3,v3}
\fmfdot{i1,v1,v2,v3,o1}
\fmfv{label=$x_1$,label.angle=-90}{i1}
\fmfv{label=$x_2$,label.angle=-90}{o1}
\fmfv{label=$x$,label.angle=-120,label.dist=12}{v1}
\fmfv{label=$x'$,label.angle=-60,label.dist=8}{v2}
\fmfv{label=$x''$,label.angle=-90}{v3}
\end{fmfgraph*}
 \end{fmffile}
\end{minipage}\quad+
\end{equation}
\begin{equation}\nonumber
\hspace{1.7cm}+\frac{1}{3!}\quad
\begin{minipage}[c]{60mm}
\vspace{0.5cm}
\begin{fmffile}{DiagPagE2.1Proj3}
\begin{fmfgraph*}(60,20)
\fmfkeep{DiagPagE2.1Proj3}
\fmfleft{i1}
\fmfright{o1}
\fmf{phantom}{i1,v1,v2,v3,o1}
\fmffreeze
\fmf{phantom,tag=1}{i1,v1}
\fmf{plain}{v1,v2}
\fmf{phantom,tag=2}{v2,v3}
\fmf{phantom,tag=3}{v3,o1}
\fmfposition
\fmfipath{p[]}
\fmfiset{p1}{vpath1(__i1,__v1)}
\fmfi{plain}{subpath (0,length(p1)*0.4) of p1}
\fmfi{dashes}{subpath (length(p1)*0.4,length(p1)) of p1}
\fmfiset{p2}{vpath2(__v2,__v3)}
\fmfi{dashes}{subpath (0,length(p2)*0.6) of p2}
\fmfi{plain}{subpath (length(p2)*0.6,length(p2)) of p2}
\fmfiset{p3}{vpath3(__v3,__o1)}
\fmfi{dashes}{subpath (0,length(p3)*0.6) of p3}
\fmfi{plain}{subpath (length(p3)*0.6,length(p3)) of p3}
\fmf{plain,tension=1/3,right=0.8}{v1,v2}
\fmf{plain,tension=1/3,left=0.8}{v1,v2}
\fmf{plain,tension=0.75}{v3,v3}
\fmfdot{i1,v1,v2,v3,o1}
\fmfv{label=$x_1$,label.angle=-90}{i1}
\fmfv{label=$x_2$,label.angle=-90}{o1}
\fmfv{label=$x$,label.angle=-120,label.dist=12}{v1}
\fmfv{label=$x'$,label.angle=-60,label.dist=8}{v2}
\fmfv{label=$x''$,label.angle=-90}{v3}
\end{fmfgraph*}
 \end{fmffile}
\end{minipage}\,\,\,\Bigg]
\end{equation}
\begin{center}
\line(1,0){150}\\
\vspace{-0.3cm}
\line(1,0){140}
\end{center}
\vspace{1cm}
\begin{equation}\nonumber
\hspace{-2.7cm}\int d\bar{\theta}_1 d\theta_1 d\bar{\theta}_2 d\theta_2
\bar{\theta}_1 \theta_1 \bar{\theta}_2 \theta_2\quad\Bigg[\frac{1}{(2!)^3}\quad
\parbox{60mm}{\fmfreuse{DiagPagE2.2}}\quad\Bigg]=
\end{equation}
\vspace{-0.5cm}
\begin{equation}\nonumber
\hspace{1cm}=\Bigg[\frac{1}{(2!)^3}\quad
\begin{minipage}[c]{60mm}
\vspace{0.5cm}
\begin{fmffile}{DiagPagE2.2Proj1}
\begin{fmfgraph*}(60,20)
\fmfkeep{DiagPagE2.2Proj1}
\fmfleft{i1}
\fmfright{o1}
\fmf{phantom}{i1,v1,v2,v3,o1}
\fmffreeze
\fmf{plain,tension=0.75}{v1,v1}
\fmf{plain,tension=0.75}{v2,v2}
\fmf{plain,tension=0.75}{v3,v3}
\fmfdot{i1,v1,v2,v3,o1}
\fmfv{label=$x_1$,label.angle=-90,}{i1}
\fmfv{label=$x_2$,label.angle=-90}{o1}
\fmfv{label=$x$,label.angle=-90,label.dist=9.5}{v1}
\fmfv{label=$x'$,label.angle=-90}{v2}
\fmfv{label=$x''$,label.angle=-90}{v3}
\fmf{plain}{i1,v1}
\fmf{phantom,tag=1}{v1,v2}
\fmf{phantom,tag=2}{v2,v3}
\fmf{phantom,tag=3}{v3,o1}
\fmfposition
\fmfipath{p[]}

\fmfiset{p1}{vpath1(__v1,__v2)}
\fmfi{dashes}{subpath (0,length(p1)*0.6) of p1}
\fmfi{plain}{subpath (length(p1)*0.6,length(p1)) of p1}

\fmfiset{p2}{vpath2(__v2,__v3)}
\fmfi{dashes}{subpath (0,length(p2)*0.6) of p2}
\fmfi{plain}{subpath (length(p2)*0.6,length(p2)) of p2}

\fmfiset{p3}{vpath3(__v3,__o1)}
\fmfi{dashes}{subpath (0,length(p3)*0.6) of p3}
\fmfi{plain}{subpath (length(p3)*0.6,length(p3)) of p3}
\end{fmfgraph*}
\end{fmffile}
\end{minipage}\quad+
\end{equation}
\vspace{-1.cm}
\begin{equation}\nonumber
\hspace{0.8cm}+\frac{1}{(2!)^3}\quad
\begin{minipage}[c]{60mm}
\vspace{0.5cm}
\begin{fmffile}{DiagPagE2.2Proj2}
\begin{fmfgraph*}(60,20)
\fmfkeep{DiagPagE2.2Proj2}
\fmfleft{i1}
\fmfright{o1}
\fmf{phantom}{i1,v1,v2,v3,o1}
\fmffreeze
\fmf{plain,tension=0.75}{v1,v1}
\fmf{plain,tension=0.75}{v2,v2}
\fmf{plain,tension=0.75}{v3,v3}
\fmfdot{i1,v1,v2,v3,o1}
\fmfv{label=$x_1$,label.angle=-90,}{i1}
\fmfv{label=$x_2$,label.angle=-90}{o1}
\fmfv{label=$x$,label.angle=-90,label.dist=9.5}{v1}
\fmfv{label=$x'$,label.angle=-90}{v2}
\fmfv{label=$x''$,label.angle=-90}{v3}
\fmf{phantom,tag=1}{i1,v1}
\fmf{phantom,tag=2}{v1,v2}
\fmf{phantom,tag=3}{v2,v3}
\fmf{plain}{v3,o1}
\fmfposition
\fmfipath{p[]}

\fmfiset{p1}{vpath1(__i1,__v1)}
\fmfi{plain}{subpath (0,length(p1)*0.4) of p1}
\fmfi{dashes}{subpath (length(p1)*0.4,length(p1)) of p1}

\fmfiset{p2}{vpath2(__v1,__v2)}
\fmfi{plain}{subpath (0,length(p2)*0.4) of p2}
\fmfi{dashes}{subpath (length(p2)*0.4,length(p2)) of p2}

\fmfiset{p3}{vpath3(__v2,__v3)}
\fmfi{plain}{subpath (0,length(p3)*0.4) of p3}
\fmfi{dashes}{subpath (length(p3)*0.4,length(p3)) of p3}
\end{fmfgraph*}
 \end{fmffile}
\end{minipage}\quad+
\end{equation}
\vspace{-0.5cm}
\begin{equation}\nonumber
\hspace{0.8cm}+\frac{1}{(2!)^3}\quad
\begin{minipage}[c]{60mm}
\vspace{0.5cm}
\begin{fmffile}{DiagPagE2.2Proj3}
\begin{fmfgraph*}(60,20)
\fmfkeep{DiagPagE2.2Proj3}
\fmfleft{i1}
\fmfright{o1}
\fmf{phantom}{i1,v1,v2,v3,o1}
\fmffreeze
\fmf{plain,tension=0.75}{v1,v1}
\fmf{plain,tension=0.75}{v2,v2}
\fmf{plain,tension=0.75}{v3,v3}
\fmfdot{i1,v1,v2,v3,o1}
\fmfv{label=$x_1$,label.angle=-90,}{i1}
\fmfv{label=$x_2$,label.angle=-90}{o1}
\fmfv{label=$x$,label.angle=-90,label.dist=9.5}{v1}
\fmfv{label=$x'$,label.angle=-90}{v2}
\fmfv{label=$x''$,label.angle=-90}{v3}
\fmf{phantom,tag=1}{i1,v1}
\fmf{plain}{v1,v2}
\fmf{phantom,tag=2}{v2,v3}
\fmf{phantom,tag=3}{v3,o1}
\fmfposition
\fmfipath{p[]}

\fmfiset{p1}{vpath1(__i1,__v1)}
\fmfi{plain}{subpath (0,length(p1)*0.4) of p1}
\fmfi{dashes}{subpath (length(p1)*0.4,length(p1)) of p1}

\fmfiset{p2}{vpath2(__v2,__v3)}
\fmfi{dashes}{subpath (0,length(p2)*0.6) of p2}
\fmfi{plain}{subpath (length(p2)*0.6,length(p2)) of p2}

\fmfiset{p3}{vpath3(__v3,__o1)}
\fmfi{dashes}{subpath (0,length(p3)*0.6) of p3}
\fmfi{plain}{subpath (length(p3)*0.6,length(p3)) of p3}
\end{fmfgraph*}
 \end{fmffile}
\end{minipage}\quad+
\end{equation}
\vspace{-0.5cm}
\begin{equation}\nonumber
\hspace{0.7cm}+\frac{1}{(2!)^3}\quad
\begin{minipage}[c]{60mm}
\vspace{0.5cm}
\begin{fmffile}{DiagPagE2.2Proj4}
\begin{fmfgraph*}(60,20)
\fmfkeep{DiagPagE2.2Proj4}
\fmfleft{i1}
\fmfright{o1}
\fmf{phantom}{i1,v1,v2,v3,o1}
\fmffreeze
\fmf{plain,tension=0.75}{v1,v1}
\fmf{plain,tension=0.75}{v2,v2}
\fmf{plain,tension=0.75}{v3,v3}
\fmfdot{i1,v1,v2,v3,o1}
\fmfv{label=$x_1$,label.angle=-90,}{i1}
\fmfv{label=$x_2$,label.angle=-90}{o1}
\fmfv{label=$x$,label.angle=-90,label.dist=9.5}{v1}
\fmfv{label=$x'$,label.angle=-90}{v2}
\fmfv{label=$x''$,label.angle=-90}{v3}
\fmf{phantom,tag=1}{i1,v1}
\fmf{phantom,tag=2}{v1,v2}
\fmf{plain}{v2,v3}
\fmf{phantom,tag=3}{v3,o1}
\fmfposition
\fmfipath{p[]}

\fmfiset{p1}{vpath1(__i1,__v1)}
\fmfi{plain}{subpath (0,length(p1)*0.4) of p1}
\fmfi{dashes}{subpath (length(p1)*0.4,length(p1)) of p1}

\fmfiset{p2}{vpath2(__v1,__v2)}
\fmfi{plain}{subpath (0,length(p2)*0.4) of p2}
\fmfi{dashes}{subpath (length(p2)*0.4,length(p2)) of p2}

\fmfiset{p3}{vpath3(__v3,__o1)}
\fmfi{dashes}{subpath (0,length(p3)*0.6) of p3}
\fmfi{plain}{subpath (length(p3)*0.6,length(p3)) of p3}
\end{fmfgraph*}
 \end{fmffile}
\end{minipage}\quad\Bigg]
\end{equation}
\begin{center}
\vspace{-0.3cm}
\line(1,0){150}\\
\vspace{-0.3cm}
\line(1,0){140}
\end{center}
\hspace{4cm}
\vspace{3cm}
\begin{equation}\nonumber
\hspace{-3cm}\int d\bar{\theta}_1 d\theta_1 d\bar{\theta}_2 d\theta_2
\bar{\theta}_1 \theta_1 \bar{\theta}_2 \theta_2\,\Bigg[\,\,\frac{1}{(2!)^3}\quad
\parbox{60mm}{\vspace{-2.4cm}\fmfreuse{DiagPagE2.3}}\hspace{-1.7cm}\Bigg]=
\end{equation}
\vspace{3cm}
\begin{equation}\nonumber
\hspace{-2.5cm}=\Bigg[\frac{1}{2!}\quad\begin{minipage}[c]{60mm}
\vspace{-2.4cm}
\begin{fmffile}{DiagPagE2.3Proj1}
\begin{fmfgraph*}(40,30)
\fmfkeep{DiagPagE2.3}
\fmfbottom{b0,b1}
\fmfleft{i1}
\fmfright{o1}
\fmftop{t0,t1}
\fmf{phantom}{b0,v1,b1}
\fmf{phantom}{i1,v2,o1}
\fmf{phantom}{t0,v3,t1}
\fmffreeze
\fmf{plain,tension=1/3,right=0.6,tag=1}{v1,v2}
\fmf{dashes,tension=1/3,left=0.6,tag=2}{v1,v2}
\fmf{plain,tension=1/3,right=0.6,tag=3}{v2,v3}
\fmf{dashes,tension=1/3,left=0.6,tag=4}{v2,v3}
\fmf{plain,tension=0.85}{v3,v3}
\fmf{plain}{b0,v1}
\fmf{phantom,tag=5}{v1,b1}
\fmfposition
\fmfipath{p[]}

\fmfiset{p2}{vpath2(__v1,__v2)}
\fmfi{plain}{subpath (0,length(p2)*0.4) of p2}

\fmfiset{p4}{vpath4(__v2,__v3)}
\fmfi{plain}{subpath (0,length(p4)*0.4) of p4}

\fmfiset{p5}{vpath5(__v1,__b1)}
\fmfi{dashes}{subpath (0,length(p5)*0.6) of p5}
\fmfi{plain}{subpath (length(p5)*0.6,length(p5)) of p5}

\fmfv{label=$x_1$,label.angle=-90,}{b0}
\fmfv{label=$x_2$,label.angle=-90}{b1}
\fmfv{label=$x$,label.angle=-90}{v1}
\fmfv{label=$x'$,label.angle=-180}{v2}
\fmfv{label=$x''$,label.angle=-180}{v3}
\fmfdot{b0,v1,b1}
\fmfdot{v2}
\fmfdot{v3}
\end{fmfgraph*}
 \end{fmffile}
\end{minipage}
\hspace{-1.8cm}+\frac{1}{2!}\quad\begin{minipage}[c]{60mm}
\vspace{-2.4cm}
\begin{fmffile}{DiagPagE2.3Proj2}
\begin{fmfgraph*}(40,30)
\fmfkeep{DiagPagE2.3Proj2}
\fmfbottom{b0,b1}
\fmfleft{i1}
\fmfright{o1}
\fmftop{t0,t1}
\fmf{phantom}{b0,v1,b1}
\fmf{phantom}{i1,v2,o1}
\fmf{phantom}{t0,v3,t1}
\fmffreeze
\fmf{plain,tension=1/3,right=0.6,tag=1}{v1,v2}
\fmf{dashes,tension=1/3,left=0.6,tag=2}{v1,v2}
\fmf{plain,tension=1/3,right=0.6,tag=3}{v2,v3}
\fmf{dashes,tension=1/3,left=0.6,tag=4}{v2,v3}
\fmf{plain,tension=0.85}{v3,v3}
\fmf{plain}{v1,b1}
\fmf{phantom,tag=5}{b0,v1}
\fmfposition
\fmfipath{p[]}

\fmfiset{p2}{vpath2(__v1,__v2)}
\fmfi{plain}{subpath (0,length(p2)*0.4) of p2}

\fmfiset{p4}{vpath4(__v2,__v3)}
\fmfi{plain}{subpath (0,length(p4)*0.4) of p4}

\fmfiset{p5}{vpath5(__b0,__v1)}
\fmfi{plain}{subpath (0,length(p5)*0.4) of p5}
\fmfi{dashes}{subpath (length(p5)*0.4,length(p5)) of p5}

\fmfv{label=$x_1$,label.angle=-90,}{b0}
\fmfv{label=$x_2$,label.angle=-90}{b1}
\fmfv{label=$x$,label.angle=-90}{v1}
\fmfv{label=$x'$,label.angle=-180}{v2}
\fmfv{label=$x''$,label.angle=-180}{v3}
\fmfdot{b0,v1,b1}
\fmfdot{v2}
\fmfdot{v3}
\end{fmfgraph*}
 \end{fmffile}
\end{minipage}\hspace{-1.5cm}\Bigg]
\end{equation}
\vspace{0.7cm}
\begin{center}
\line(1,0){150}\\
\vspace{-0.3cm}
\line(1,0){140}
\end{center}
\vspace{3.5cm}
\begin{equation*}
\vspace{1.5cm}
\quad\quad
\end{equation*}
\begin{equation}\nonumber
\hspace{-0cm}\int d\bar{\theta}_1 d\theta_1 d\bar{\theta}_2 d\theta_2
\bar{\theta}_1 \theta_1 \bar{\theta}_2 \theta_2\,\,\Bigg[\frac{1}{(2!)^2}\quad
\parbox{60mm}{\vspace{-1.6cm}
\fmfreuse{DiagPagE1.1}}\hspace{-1.6cm}\Bigg]=
\end{equation}
\vspace{3cm}
\parbox{60mm}{
\vspace{2.5cm}
\begin{equation}\nonumber
\hspace{2cm}=\Bigg[\frac{1}{2!}\quad
\begin{minipage}[c]{60mm}
\vspace{-1.6cm}
\begin{fmffile}{DiagPagE1.1Proj1}
\begin{fmfgraph*}(40,20)
\fmfkeep{DiagPagE1.1Proj1}
\fmftop{t0,t1}
\fmfbottom{b0,b1}
\fmf{phantom}{t0,v1,v0,t1}
\fmffreeze
\fmf{phantom}{b0,v2,v3,b1}
\fmffreeze
\fmf{plain,tension=0.6}{v1,v1}
\fmf{plain,tension=0.6}{v3,v3}
\fmf{plain}{b0,v2}
\fmf{dashes,tension=1/3,right=0.6,tag=1}{v1,v2}
\fmf{plain,tension=1/3,left=0.6}{v1,v2}
\fmf{phantom,tag=2}{v2,v3}
\fmf{phantom,tag=3}{v3,b1}

\fmfposition
\fmfipath{p[]}
\fmfiset{p1}{vpath1(__v1,__v2)}
\fmfi{plain}{subpath (length(p1)*0.6,length(p1)) of p1}

\fmfiset{p2}{vpath2(__v2,__v3)}
\fmfi{dashes}{subpath (0,length(p2)*0.6) of p2}
\fmfi{plain}{subpath (length(p2)*0.6,length(p2)) of p2}

\fmfiset{p3}{vpath3(__v3,__b1)}
\fmfi{dashes}{subpath (0,length(p3)*0.6) of p3}
\fmfi{plain}{subpath (length(p3)*0.6,length(p3)) of p3}

\fmfv{label=$x_1$,label.angle=-90,label.dist=9.5}{b0}
\fmfv{label=$x_2$,label.angle=-90,label.dist=9.5}{b1}
\fmfv{label=$x$,label.angle=-90,label.dist=9.5}{v2}
\fmfv{label=$x'$,label.angle=-180}{v1}
\fmfv{label=$x''$,label.angle=-90}{v3}
\fmfdot{v1,v2,b0,b1,v3}
\end{fmfgraph*}
 \end{fmffile}
\end{minipage}
\hspace{-1.2cm}+\frac{1}{2!}\quad
\begin{minipage}[c]{60mm}
\vspace{-1.6cm}
\begin{fmffile}{DiagPagE1.1Proj2}
\begin{fmfgraph*}(40,20)
\fmfkeep{DiagPagE1.1Proj2}
\fmftop{t0,t1}
\fmfbottom{b0,b1}
\fmf{phantom}{t0,v1,v0,t1}
\fmffreeze
\fmf{phantom}{b0,v2,v3,b1}
\fmffreeze
\fmf{plain,tension=0.6}{v1,v1}
\fmf{plain,tension=0.6}{v3,v3}
\fmf{plain}{v3,b1}
\fmf{dashes,tension=1/3,right=0.6,tag=1}{v1,v2}
\fmf{plain,tension=1/3,left=0.6}{v1,v2}
\fmf{phantom,tag=2}{b0,v2}
\fmf{phantom,tag=3}{v2,v3}

\fmfposition
\fmfipath{p[]}
\fmfiset{p1}{vpath1(__v1,__v2)}
\fmfi{plain}{subpath (length(p1)*0.6,length(p1)) of p1}

\fmfiset{p2}{vpath2(__b0,__v2)}
\fmfi{plain}{subpath (0,length(p2)*0.4) of p2}
\fmfi{dashes}{subpath (length(p2)*0.4,length(p2)) of p2}

\fmfiset{p3}{vpath3(__v2,__v3)}
\fmfi{plain}{subpath (0,length(p3)*0.4) of p3}
\fmfi{dashes}{subpath (length(p3)*0.4,length(p3)) of p3}

\fmfv{label=$x_1$,label.angle=-90,label.dist=9.5}{b0}
\fmfv{label=$x_2$,label.angle=-90,label.dist=9.5}{b1}
\fmfv{label=$x$,label.angle=-90,label.dist=9.5}{v2}
\fmfv{label=$x'$,label.angle=-180}{v1}
\fmfv{label=$x''$,label.angle=-90}{v3}
\fmfdot{v1,v2,b0,b1,v3}
\end{fmfgraph*}
 \end{fmffile}
\end{minipage}\hspace{-1.7cm}+
\end{equation}
\vspace{3cm}
\begin{equation}\nonumber
\hspace{2cm}+\,\frac{1}{2!}\quad
\begin{minipage}[c]{60mm}
\vspace{-1.6cm}
\begin{fmffile}{DiagPagE1.1Proj3}
\begin{fmfgraph*}(40,20)
\fmfkeep{DiagPagE1.1Proj3}
\fmftop{t0,t1}
\fmfbottom{b0,b1}
\fmf{phantom}{t0,v1,v0,t1}
\fmffreeze
\fmf{phantom}{b0,v2,v3,b1}
\fmffreeze
\fmf{plain,tension=0.6}{v1,v1}
\fmf{plain,tension=0.6}{v3,v3}
\fmf{plain}{v2,v3}
\fmf{dashes,tension=1/3,right=0.6,tag=1}{v1,v2}
\fmf{plain,tension=1/3,left=0.6}{v1,v2}
\fmf{phantom,tag=2}{b0,v2}
\fmf{phantom,tag=3}{v3,b1}
\fmfposition
\fmfipath{p[]}
\fmfiset{p1}{vpath1(__v1,__v2)}
\fmfi{plain}{subpath (length(p1)*0.6,length(p1)) of p1}
\fmfiset{p2}{vpath2(__b0,__v2)}
\fmfi{plain}{subpath (0,length(p2)*0.4) of p2}
\fmfi{dashes}{subpath (length(p2)*0.4,length(p2)) of p2}

\fmfiset{p3}{vpath3(__v3,__b1)}
\fmfi{dashes}{subpath (0,length(p3)*0.6) of p3}
\fmfi{plain}{subpath (length(p3)*0.6,length(p3)) of p3}

\fmfv{label=$x_1$,label.angle=-90,label.dist=9.5}{b0}
\fmfv{label=$x_2$,label.angle=-90,label.dist=9.5}{b1}
\fmfv{label=$x$,label.angle=-90,label.dist=9.5}{v2}
\fmfv{label=$x'$,label.angle=-180}{v1}
\fmfv{label=$x''$,label.angle=-90}{v3}
\fmfdot{v1,v2,b0,b1,v3}
\end{fmfgraph*}
 \end{fmffile}
\end{minipage}\hspace{-1.5cm}\Bigg]
\end{equation}
}
\vspace{-2.5cm}
\begin{center}
\line(1,0){150}\\
\vspace{-0.3cm}
\line(1,0){140}
\end{center}
\vspace{1.6cm}
\begin{equation}\nonumber
\hspace{0cm}\int d\bar{\theta}_1 d\theta_1 d\bar{\theta}_2 d\theta_2
\bar{\theta}_1 \theta_1 \bar{\theta}_2 \theta_2\,\Bigg[\,\,\frac{1}{(2!)^2}\quad
\parbox{60mm}{\vspace{-1.6cm}\fmfreuse{DiagPagE1.2}}
\hspace{-1.8cm}\,\,\Bigg]=
\end{equation}
\vspace{1cm}
\begin{equation}\nonumber
=\Bigg[\frac{1}{2!}\quad\begin{minipage}[c]{60mm}
\vspace{-1.6cm}
\begin{fmffile}{DiagPagE1.2Proj1}
\begin{fmfgraph*}(40,20)
\fmfbottom{b1,b2}
\fmftop{t1,t2}
\fmf{phantom}{t1,vt1,t2}
\fmf{phantom}{b1,vb1,vb2,b2}
\fmffreeze
\fmfdot{b1,b2,vt1,vb1,vb2}
\fmffreeze
\fmf{phantom,left=0.7,tag=1}{vb1,vt1}
\fmf{phantom,tag=2}{vb1,vt1}
\fmf{dashes,right=0.7,tag=3}{vb2,vt1}
\fmf{phantom,tag=4}{vb2,vt1}
\fmf{phantom,tag=5}{b1,vb1}
\fmf{phantom,tag=6}{vb1,vb2}
\fmf{phantom,tag=7}{vb2,b2}

\fmfposition
\fmfipath{p[]}

\fmfiset{p1}{vpath1(__vb1,__vt1)}
\fmfi{plain}{subpath (0,length(p1)*0.6) of p1}
\fmfi{plain}{subpath (length(p1)*0.6,length(p1)) of p1}

\fmfiset{p2}{vpath2(__vb1,__vt1)}
\fmfi{plain}{subpath (0,length(p2)*0.6) of p2}
\fmfi{plain}{subpath (length(p2)*0.6,length(p2)) of p2}

\fmfiset{p3}{vpath3(__vb2,__vt1)}
\fmfi{plain}{subpath (0,length(p3)*0.4) of p3}
\fmfi{plain}{subpath (length(p3)*0.4,length(p3)) of p3}

\fmfiset{p4}{vpath4(__vb2,__vt1)}
\fmfi{plain}{subpath (0,length(p4)*0.4) of p4}
\fmfi{dashes}{subpath (length(p4)*0.4,length(p4)) of p4}

\fmfiset{p5}{vpath5(__b1,__vb1)}
\fmfi{plain}{subpath (0,length(p5)*0.4) of p5}
\fmfi{plain}{subpath (length(p5)*0.4,length(p5)) of p5}

\fmfiset{p6}{vpath6(__vb1,__vb2)}
\fmfi{dashes}{subpath (0,length(p6)*0.6) of p6}
\fmfi{plain}{subpath (length(p6)*0.6,length(p6)) of p6}

\fmfiset{p7}{vpath7(__vb2,__b2)}
\fmfi{dashes}{subpath (0,length(p7)*0.6) of p7}
\fmfi{plain}{subpath (length(p7)*0.6,length(p7)) of p7}

\fmfv{label=$x_1$,label.angle=-90,label.dist=9.5}{b1}
\fmfv{label=$x_2$,label.angle=-90,label.dist=9.5}{b2}
\fmfv{label=$x$,label.angle=-90,label.dist=9.5}{vb1}
\fmfv{label=$x'$,label.angle=90}{vt1}
\fmfv{label=$x''$,label.angle=-90}{vb2}
\end{fmfgraph*}
\end{fmffile}
\end{minipage}
\hspace{-1.6cm}+\frac{1}{2!}\quad\begin{minipage}[c]{60mm}
\vspace{-1.6cm}
\begin{fmffile}{DiagPagE1.2Proj8}
\begin{fmfgraph*}(40,20)
\fmfbottom{b1,b2}
\fmftop{t1,t2}
\fmf{phantom}{t1,vt1,t2}
\fmf{phantom}{b1,vb1,vb2,b2}
\fmffreeze
\fmfdot{b1,b2,vt1,vb1,vb2}
\fmffreeze
\fmf{phantom,left=0.7,tag=1}{vb1,vt1}
\fmf{phantom,tag=2}{vb1,vt1}
\fmf{phantom,right=0.7,tag=3}{vb2,vt1}
\fmf{phantom,tag=4}{vb2,vt1}
\fmf{phantom,tag=5}{b1,vb1}
\fmf{phantom,tag=6}{vb1,vb2}
\fmf{phantom,tag=7}{vb2,b2}

\fmfposition
\fmfipath{p[]}

\fmfiset{p1}{vpath1(__vb1,__vt1)}
\fmfi{plain}{subpath (0,length(p1)*0.6) of p1}
\fmfi{plain}{subpath (length(p1)*0.6,length(p1)) of p1}

\fmfiset{p2}{vpath2(__vb1,__vt1)}
\fmfi{plain}{subpath (0,length(p2)*0.4) of p2}
\fmfi{dashes}{subpath (length(p2)*0.4,length(p2)) of p2}

\fmfiset{p3}{vpath3(__vb2,__vt1)}
\fmfi{plain}{subpath (0,length(p3)*0.6) of p3}
\fmfi{plain}{subpath (length(p3)*0.6,length(p3)) of p3}

\fmfiset{p4}{vpath4(__vb2,__vt1)}
\fmfi{plain}{subpath (0,length(p4)*0.4) of p4}
\fmfi{plain}{subpath (length(p4)*0.4,length(p4)) of p4}

\fmfiset{p5}{vpath5(__b1,__vb1)}
\fmfi{plain}{subpath (0,length(p5)*0.4) of p5}
\fmfi{dashes}{subpath (length(p5)*0.4,length(p5)) of p5}

\fmfiset{p6}{vpath6(__vb1,__vb2)}
\fmfi{plain}{subpath (0,length(p6)*0.4) of p6}
\fmfi{plain}{subpath (length(p6)*0.4,length(p6)) of p6}

\fmfiset{p7}{vpath7(__vb2,__b2)}
\fmfi{dashes}{subpath (0,length(p7)*0.6) of p7}
\fmfi{plain}{subpath (length(p7)*0.6,length(p7)) of p7}

\fmfv{label=$x_1$,label.angle=-90,label.dist=9.5}{b1}
\fmfv{label=$x_2$,label.angle=-90,label.dist=9.5}{b2}
\fmfv{label=$x$,label.angle=-90,label.dist=9.5}{vb1}
\fmfv{label=$x'$,label.angle=90}{vt1}
\fmfv{label=$x''$,label.angle=-90}{vb2}
\end{fmfgraph*}
\end{fmffile}
\end{minipage}\hspace{-1.7cm}+
\end{equation}
\vspace{1.4cm}
\begin{equation}\nonumber
+\frac{1}{2!}\quad\begin{minipage}[c]{60mm}
\vspace{-1.6cm}
\begin{fmffile}{DiagPagE1.2Proj2}
\begin{fmfgraph*}(40,20)
\fmfbottom{b1,b2}
\fmftop{t1,t2}
\fmf{phantom}{t1,vt1,t2}
\fmf{phantom}{b1,vb1,vb2,b2}
\fmffreeze
\fmfdot{b1,b2,vt1,vb1,vb2}
\fmffreeze
\fmf{phantom,left=0.7,tag=1}{vb1,vt1}
\fmf{phantom,tag=2}{vb1,vt1}
\fmf{phantom,right=0.7,tag=3}{vb2,vt1}
\fmf{phantom,tag=4}{vb2,vt1}
\fmf{phantom,tag=5}{b1,vb1}
\fmf{phantom,tag=6}{vb1,vb2}
\fmf{phantom,tag=7}{vb2,b2}

\fmfposition
\fmfipath{p[]}

\fmfiset{p1}{vpath1(__vb1,__vt1)}
\fmfi{plain}{subpath (0,length(p1)*0.6) of p1}
\fmfi{plain}{subpath (length(p1)*0.6,length(p1)) of p1}

\fmfiset{p2}{vpath2(__vb1,__vt1)}
\fmfi{plain}{subpath (0,length(p2)*0.4) of p2}
\fmfi{dashes}{subpath (length(p2)*0.4,length(p2)) of p2}

\fmfiset{p3}{vpath3(__vb2,__vt1)}
\fmfi{plain}{subpath (0,length(p3)*0.4) of p3}
\fmfi{plain}{subpath (length(p3)*0.4,length(p3)) of p3}

\fmfiset{p4}{vpath4(__vb2,__vt1)}
\fmfi{plain}{subpath (0,length(p4)*0.4) of p4}
\fmfi{plain}{subpath (length(p4)*0.4,length(p4)) of p4}

\fmfiset{p5}{vpath5(__b1,__vb1)}
\fmfi{plain}{subpath (0,length(p5)*0.4) of p5}
\fmfi{plain}{subpath (length(p5)*0.4,length(p5)) of p5}

\fmfiset{p6}{vpath6(__vb1,__vb2)}
\fmfi{dashes}{subpath (0,length(p6)*0.6) of p6}
\fmfi{plain}{subpath (length(p6)*0.6,length(p6)) of p6}

\fmfiset{p7}{vpath7(__vb2,__b2)}
\fmfi{dashes}{subpath (0,length(p7)*0.6) of p7}
\fmfi{plain}{subpath (length(p7)*0.6,length(p7)) of p7}

\fmfv{label=$x_1$,label.angle=-90,label.dist=9.5}{b1}
\fmfv{label=$x_2$,label.angle=-90,label.dist=9.5}{b2}
\fmfv{label=$x$,label.angle=-90,label.dist=9.5}{vb1}
\fmfv{label=$x'$,label.angle=90}{vt1}
\fmfv{label=$x''$,label.angle=-90}{vb2}
\end{fmfgraph*}
\end{fmffile}
\end{minipage}
\hspace{-1.5cm}+\quad\begin{minipage}[c]{60mm}
\vspace{-1.6cm}
\begin{fmffile}{DiagPagE1.2Proj3}
\begin{fmfgraph*}(40,20)
\fmfbottom{b1,b2}
\fmftop{t1,t2}
\fmf{phantom}{t1,vt1,t2}
\fmf{phantom}{b1,vb1,vb2,b2}
\fmffreeze
\fmfdot{b1,b2,vt1,vb1,vb2}
\fmffreeze
\fmf{phantom,left=0.7,tag=1}{vb1,vt1}
\fmf{phantom,tag=2}{vb1,vt1}
\fmf{phantom,right=0.7,tag=3}{vb2,vt1}
\fmf{phantom,tag=4}{vb2,vt1}
\fmf{phantom,tag=5}{b1,vb1}
\fmf{phantom,tag=6}{vb1,vb2}
\fmf{phantom,tag=7}{vb2,b2}

\fmfposition
\fmfipath{p[]}

\fmfiset{p1}{vpath1(__vb1,__vt1)}
\fmfi{plain}{subpath (0,length(p1)*0.6) of p1}
\fmfi{plain}{subpath (length(p1)*0.6,length(p1)) of p1}

\fmfiset{p2}{vpath2(__vb1,__vt1)}
\fmfi{dashes}{subpath (0,length(p2)*0.6) of p2}
\fmfi{plain}{subpath (length(p2)*0.6,length(p2)) of p2}

\fmfiset{p3}{vpath3(__vb2,__vt1)}
\fmfi{plain}{subpath (0,length(p3)*0.4) of p3}
\fmfi{plain}{subpath (length(p3)*0.4,length(p3)) of p3}

\fmfiset{p4}{vpath4(__vb2,__vt1)}
\fmfi{plain}{subpath (0,length(p4)*0.4) of p4}
\fmfi{dashes}{subpath (length(p4)*0.4,length(p4)) of p4}

\fmfiset{p5}{vpath5(__b1,__vb1)}
\fmfi{plain}{subpath (0,length(p5)*0.4) of p5}
\fmfi{plain}{subpath (length(p5)*0.4,length(p5)) of p5}

\fmfiset{p6}{vpath6(__vb1,__vb2)}
\fmfi{plain}{subpath (0,length(p6)*0.6) of p6}
\fmfi{plain}{subpath (length(p6)*0.6,length(p6)) of p6}

\fmfiset{p7}{vpath7(__vb2,__b2)}
\fmfi{dashes}{subpath (0,length(p7)*0.6) of p7}
\fmfi{plain}{subpath (length(p7)*0.6,length(p7)) of p7}

\fmfv{label=$x_1$,label.angle=-90,label.dist=9.5}{b1}
\fmfv{label=$x_2$,label.angle=-90,label.dist=9.5}{b2}
\fmfv{label=$x$,label.angle=-90,label.dist=9.5}{vb1}
\fmfv{label=$x'$,label.angle=90}{vt1}
\fmfv{label=$x''$,label.angle=-90}{vb2}
\end{fmfgraph*}
\end{fmffile}
\end{minipage}\hspace{-1.7cm}+
\end{equation}
\vspace{1cm}
\begin{equation}\nonumber
\hspace{0.4cm}+\frac{1}{2!}\quad\begin{minipage}[c]{60mm}
\vspace{-1.6cm}
\begin{fmffile}{DiagPagE1.2Proj4}
\begin{fmfgraph*}(40,20)
\fmfbottom{b1,b2}
\fmftop{t1,t2}
\fmf{phantom}{t1,vt1,t2}
\fmf{phantom}{b1,vb1,vb2,b2}
\fmffreeze
\fmfdot{b1,b2,vt1,vb1,vb2}
\fmffreeze
\fmf{phantom,left=0.7,tag=1}{vb1,vt1}
\fmf{phantom,tag=2}{vb1,vt1}
\fmf{phantom,right=0.7,tag=3}{vb2,vt1}
\fmf{phantom,tag=4}{vb2,vt1}
\fmf{phantom,tag=5}{b1,vb1}
\fmf{phantom,tag=6}{vb1,vb2}
\fmf{phantom,tag=7}{vb2,b2}

\fmfposition
\fmfipath{p[]}

\fmfiset{p1}{vpath1(__vb1,__vt1)}
\fmfi{plain}{subpath (0,length(p1)*0.6) of p1}
\fmfi{plain}{subpath (length(p1)*0.6,length(p1)) of p1}

\fmfiset{p2}{vpath2(__vb1,__vt1)}
\fmfi{plain}{subpath (0,length(p2)*0.6) of p2}
\fmfi{plain}{subpath (length(p2)*0.6,length(p2)) of p2}

\fmfiset{p3}{vpath3(__vb2,__vt1)}
\fmfi{plain}{subpath (0,length(p3)*0.4) of p3}
\fmfi{plain}{subpath (length(p3)*0.4,length(p3)) of p3}

\fmfiset{p4}{vpath4(__vb2,__vt1)}
\fmfi{plain}{subpath (0,length(p4)*0.4) of p4}
\fmfi{dashes}{subpath (length(p4)*0.4,length(p4)) of p4}

\fmfiset{p5}{vpath5(__b1,__vb1)}
\fmfi{plain}{subpath (0,length(p5)*0.4) of p5}
\fmfi{dashes}{subpath (length(p5)*0.4,length(p5)) of p5}

\fmfiset{p6}{vpath6(__vb1,__vb2)}
\fmfi{plain}{subpath (0,length(p6)*0.4) of p6}
\fmfi{dashes}{subpath (length(p6)*0.4,length(p6)) of p6}

\fmfiset{p7}{vpath7(__vb2,__b2)}
\fmfi{plain}{subpath (0,length(p7)*0.6) of p7}
\fmfi{plain}{subpath (length(p7)*0.6,length(p7)) of p7}

\fmfv{label=$x_1$,label.angle=-90,label.dist=9.5}{b1}
\fmfv{label=$x_2$,label.angle=-90,label.dist=9.5}{b2}
\fmfv{label=$x$,label.angle=-90,label.dist=9.5}{vb1}
\fmfv{label=$x'$,label.angle=90}{vt1}
\fmfv{label=$x''$,label.angle=-90}{vb2}
\end{fmfgraph*}
\end{fmffile}
\end{minipage}
\quad
\hspace{-2.cm}+\frac{1}{2!}\quad\begin{minipage}[c]{60mm}
\vspace{-1.6cm}
\begin{fmffile}{DiagPagE1.2Proj5}
\begin{fmfgraph*}(40,20)
\fmfbottom{b1,b2}
\fmftop{t1,t2}
\fmf{phantom}{t1,vt1,t2}
\fmf{phantom}{b1,vb1,vb2,b2}
\fmffreeze
\fmfdot{b1,b2,vt1,vb1,vb2}
\fmffreeze
\fmf{phantom,left=0.7,tag=1}{vb1,vt1}
\fmf{phantom,tag=2}{vb1,vt1}
\fmf{phantom,right=0.7,tag=3}{vb2,vt1}
\fmf{phantom,tag=4}{vb2,vt1}
\fmf{phantom,tag=5}{b1,vb1}
\fmf{phantom,tag=6}{vb1,vb2}
\fmf{phantom,tag=7}{vb2,b2}

\fmfposition
\fmfipath{p[]}

\fmfiset{p1}{vpath1(__vb1,__vt1)}
\fmfi{plain}{subpath (0,length(p1)*0.6) of p1}
\fmfi{plain}{subpath (length(p1)*0.6,length(p1)) of p1}

\fmfiset{p2}{vpath2(__vb1,__vt1)}
\fmfi{plain}{subpath (0,length(p2)*0.4) of p2}
\fmfi{dashes}{subpath (length(p2)*0.4,length(p2)) of p2}

\fmfiset{p3}{vpath3(__vb2,__vt1)}
\fmfi{plain}{subpath (0,length(p3)*0.4) of p3}
\fmfi{plain}{subpath (length(p3)*0.4,length(p3)) of p3}

\fmfiset{p4}{vpath4(__vb2,__vt1)}
\fmfi{plain}{subpath (0,length(p4)*0.4) of p4}
\fmfi{plain}{subpath (length(p4)*0.4,length(p4)) of p4}

\fmfiset{p5}{vpath5(__b1,__vb1)}
\fmfi{plain}{subpath (0,length(p5)*0.4) of p5}
\fmfi{dashes}{subpath (length(p5)*0.4,length(p5)) of p5}

\fmfiset{p6}{vpath6(__vb1,__vb2)}
\fmfi{plain}{subpath (0,length(p6)*0.4) of p6}
\fmfi{dashes}{subpath (length(p6)*0.4,length(p6)) of p6}

\fmfiset{p7}{vpath7(__vb2,__b2)}
\fmfi{plain}{subpath (0,length(p7)*0.6) of p7}
\fmfi{plain}{subpath (length(p7)*0.6,length(p7)) of p7}

\fmfv{label=$x_1$,label.angle=-90,label.dist=9.5}{b1}
\fmfv{label=$x_2$,label.angle=-90,label.dist=9.5}{b2}
\fmfv{label=$x$,label.angle=-90,label.dist=9.5}{vb1}
\fmfv{label=$x'$,label.angle=90}{vt1}
\fmfv{label=$x''$,label.angle=-90}{vb2}
\end{fmfgraph*}
\end{fmffile}
\end{minipage}\hspace{-1.7cm}+
\end{equation}
\vspace{1cm}
\begin{equation}\nonumber
\hspace{-0.cm}+\quad\begin{minipage}[c]{60mm}
\vspace{-1.6cm}
\begin{fmffile}{DiagPagE1.2Proj6}
\begin{fmfgraph*}(40,20)
\fmfbottom{b1,b2}
\fmftop{t1,t2}
\fmf{phantom}{t1,vt1,t2}
\fmf{phantom}{b1,vb1,vb2,b2}
\fmffreeze
\fmfdot{b1,b2,vt1,vb1,vb2}
\fmffreeze
\fmf{phantom,left=0.7,tag=1}{vb1,vt1}
\fmf{phantom,tag=2}{vb1,vt1}
\fmf{phantom,right=0.7,tag=3}{vb2,vt1}
\fmf{phantom,tag=4}{vb2,vt1}
\fmf{phantom,tag=5}{b1,vb1}
\fmf{phantom,tag=6}{vb1,vb2}
\fmf{phantom,tag=7}{vb2,b2}

\fmfposition
\fmfipath{p[]}

\fmfiset{p1}{vpath1(__vb1,__vt1)}
\fmfi{plain}{subpath (0,length(p1)*0.6) of p1}
\fmfi{plain}{subpath (length(p1)*0.6,length(p1)) of p1}

\fmfiset{p2}{vpath2(__vb1,__vt1)}
\fmfi{plain}{subpath (0,length(p2)*0.4) of p2}
\fmfi{dashes}{subpath (length(p2)*0.4,length(p2)) of p2}

\fmfiset{p3}{vpath3(__vb2,__vt1)}
\fmfi{plain}{subpath (0,length(p3)*0.6) of p3}
\fmfi{plain}{subpath (length(p3)*0.6,length(p3)) of p3}

\fmfiset{p4}{vpath4(__vb2,__vt1)}
\fmfi{dashes}{subpath (0,length(p4)*0.6) of p4}
\fmfi{plain}{subpath (length(p4)*0.6,length(p4)) of p4}

\fmfiset{p5}{vpath5(__b1,__vb1)}
\fmfi{plain}{subpath (0,length(p5)*0.4) of p5}
\fmfi{dashes}{subpath (length(p5)*0.4,length(p5)) of p5}

\fmfiset{p6}{vpath6(__vb1,__vb2)}
\fmfi{plain}{subpath (0,length(p6)*0.4) of p6}
\fmfi{plain}{subpath (length(p6)*0.4,length(p6)) of p6}

\fmfiset{p7}{vpath7(__vb2,__b2)}
\fmfi{plain}{subpath (0,length(p7)*0.6) of p7}
\fmfi{plain}{subpath (length(p7)*0.6,length(p7)) of p7}

\fmfv{label=$x_1$,label.angle=-90,label.dist=9.5}{b1}
\fmfv{label=$x_2$,label.angle=-90,label.dist=9.5}{b2}
\fmfv{label=$x$,label.angle=-90,label.dist=9.5}{vb1}
\fmfv{label=$x'$,label.angle=90}{vt1}
\fmfv{label=$x''$,label.angle=-90}{vb2}
\end{fmfgraph*}
\end{fmffile}
\end{minipage}
\hspace{-1.5cm}+\frac{1}{2!}\quad\begin{minipage}[c]{60mm}
\vspace{-1.6cm}
\begin{fmffile}{DiagPagE1.2Proj7}
\begin{fmfgraph*}(40,20)
\fmfbottom{b1,b2}
\fmftop{t1,t2}
\fmf{phantom}{t1,vt1,t2}
\fmf{phantom}{b1,vb1,vb2,b2}
\fmffreeze
\fmfdot{b1,b2,vt1,vb1,vb2}
\fmffreeze
\fmf{phantom,left=0.7,tag=1}{vb1,vt1}
\fmf{phantom,tag=2}{vb1,vt1}
\fmf{phantom,right=0.7,tag=3}{vb2,vt1}
\fmf{phantom,tag=4}{vb2,vt1}
\fmf{phantom,tag=5}{b1,vb1}
\fmf{phantom,tag=6}{vb1,vb2}
\fmf{phantom,tag=7}{vb2,b2}

\fmfposition
\fmfipath{p[]}

\fmfiset{p1}{vpath1(__vb1,__vt1)}
\fmfi{plain}{subpath (0,length(p1)*0.6) of p1}
\fmfi{plain}{subpath (length(p1)*0.6,length(p1)) of p1}

\fmfiset{p2}{vpath2(__vb1,__vt1)}
\fmfi{plain}{subpath (0,length(p2)*0.4) of p2}
\fmfi{plain}{subpath (length(p2)*0.4,length(p2)) of p2}

\fmfiset{p3}{vpath3(__vb2,__vt1)}
\fmfi{plain}{subpath (0,length(p3)*0.6) of p3}
\fmfi{plain}{subpath (length(p3)*0.6,length(p3)) of p3}

\fmfiset{p4}{vpath4(__vb2,__vt1)}
\fmfi{plain}{subpath (0,length(p4)*0.4) of p4}
\fmfi{dashes}{subpath (length(p4)*0.4,length(p4)) of p4}

\fmfiset{p5}{vpath5(__b1,__vb1)}
\fmfi{plain}{subpath (0,length(p5)*0.4) of p5}
\fmfi{dashes}{subpath (length(p5)*0.4,length(p5)) of p5}

\fmfiset{p6}{vpath6(__vb1,__vb2)}
\fmfi{plain}{subpath (0,length(p6)*0.4) of p6}
\fmfi{plain}{subpath (length(p6)*0.4,length(p6)) of p6}

\fmfiset{p7}{vpath7(__vb2,__b2)}
\fmfi{dashes}{subpath (0,length(p7)*0.6) of p7}
\fmfi{plain}{subpath (length(p7)*0.6,length(p7)) of p7}

\fmfv{label=$x_1$,label.angle=-90,label.dist=9.5}{b1}
\fmfv{label=$x_2$,label.angle=-90,label.dist=9.5}{b2}
\fmfv{label=$x$,label.angle=-90,label.dist=9.5}{vb1}
\fmfv{label=$x'$,label.angle=90}{vt1}
\fmfv{label=$x''$,label.angle=-90}{vb2}
\end{fmfgraph*}
\end{fmffile}
\end{minipage}\hspace{-1.7cm}\Bigg]
\end{equation}
\begin{center}
\line(1,0){150}\\
\vspace{-0.3cm}
\line(1,0){140}
\end{center}
\vspace{1cm}
\parbox{60mm}{
\vspace{2cm}
\begin{equation}\nonumber
\hspace{0.5cm}\int d\bar{\theta}_1 d\theta_1 d\bar{\theta}_2 d\theta_2
\bar{\theta}_1 \theta_1 \bar{\theta}_2 \theta_2\,\,\Bigg[\frac{1}{(2!)^2}\quad
\parbox{60mm}{\vspace{-1.2cm}
\fmfreuse{DiagPagE1.3}}\hspace{-0.8cm}\Bigg]=
\end{equation}
}
\vspace{3cm}
\begin{equation}\nonumber
\hspace{0cm}=\Bigg[\frac{1}{(2!)^2}\quad\begin{minipage}[c]{60mm}
\vspace{-1.2cm}
\begin{fmffile}{DiagPagE1.3Proj1}
\begin{fmfgraph*}(50,15)
 \fmftop{t0,t1}
\fmfbottom{b0,b1}
\fmf{phantom}{t0,tv1,t1}
\fmf{phantom}{b0,v1,v2,v3,b1}
\fmfdot{tv1}
\fmfdot{b0,v1,v3,b1}
\fmffreeze

\fmf{phantom,tag=1}{v1,tv1}
\fmf{phantom,tag=2}{v3,tv1}
\fmf{plain,tension=0.7}{tv1,tv1}
\fmf{phantom,right=0.7,tag=3}{v1,v3}
\fmf{phantom,tag=4}{b0,v1}
\fmf{phantom,tag=5}{v1,v3}
\fmf{phantom,tag=6}{v3,b1}

\fmfposition
\fmfipath{p[]}

\fmfiset{p1}{vpath1(__v1,__tv1)}
\fmfi{dashes}{subpath (0,length(p1)*0.6) of p1}
\fmfi{plain}{subpath (length(p1)*0.6,length(p1)) of p1}

\fmfiset{p2}{vpath2(__v3,__tv1)}
\fmfi{plain}{subpath (0,length(p2)*0.4) of p2}
\fmfi{dashes}{subpath (length(p2)*0.4,length(p2)) of p2}

\fmfiset{p3}{vpath3(__v1,__v3)}
\fmfi{plain}{subpath (0,length(p3)*0.6) of p3}
\fmfi{plain}{subpath (length(p3)*0.6,length(p3)) of p3}

\fmfiset{p4}{vpath4(__b0,__v1)}
\fmfi{plain}{subpath (0,length(p4)*0.6) of p4}
\fmfi{plain}{subpath (length(p4)*0.6,length(p4)) of p4}

\fmfiset{p5}{vpath5(__v1,__v3)}
\fmfi{plain}{subpath (0,length(p5)*0.6) of p5}
\fmfi{plain}{subpath (length(p5)*0.6,length(p5)) of p5}

\fmfiset{p6}{vpath6(__v3,__b1)}
\fmfi{dashes}{subpath (0,length(p6)*0.6) of p6}
\fmfi{plain}{subpath (length(p6)*0.6,length(p6)) of p6}

\fmfv{label=$x_1$,label.angle=-90,label.dist=9.5}{b0}
\fmfv{label=$x_2$,label.angle=-90,label.dist=9.5}{b1}
\fmfv{label=$x$,label.angle=-120}{v1}
\fmfv{label=$x'$,label.angle=-180}{tv1}
\fmfv{label=$x''$,label.angle=-30}{v3}
\end{fmfgraph*}
\end{fmffile}
\end{minipage}
\vspace{0.1cm}
\hspace{-0.7cm}+\frac{1}{(2!)^2}\,\quad\begin{minipage}[c]{60mm}
\vspace{-1.2cm}
\begin{fmffile}{DiagPagE1.3Proj2}
\begin{fmfgraph*}(50,15)
 \fmftop{t0,t1}
\fmfbottom{b0,b1}
\fmf{phantom}{t0,tv1,t1}
\fmf{phantom}{b0,v1,v2,v3,b1}
\fmfdot{tv1}
\fmfdot{b0,v1,v3,b1}
\fmffreeze

\fmf{phantom,tag=1}{v1,tv1}
\fmf{phantom,tag=2}{v3,tv1}
\fmf{plain,tension=0.7}{tv1,tv1}
\fmf{phantom,right=0.7,tag=3}{v1,v3}
\fmf{phantom,tag=4}{b0,v1}
\fmf{phantom,tag=5}{v1,v3}
\fmf{phantom,tag=6}{v3,b1}

\fmfposition
\fmfipath{p[]}

\fmfiset{p1}{vpath1(__v1,__tv1)}
\fmfi{plain}{subpath (0,length(p1)*0.4) of p1}
\fmfi{dashes}{subpath (length(p1)*0.4,length(p1)) of p1}

\fmfiset{p2}{vpath2(__v3,__tv1)}
\fmfi{dashes}{subpath (0,length(p2)*0.6) of p2}
\fmfi{plain}{subpath (length(p2)*0.6,length(p2)) of p2}

\fmfiset{p3}{vpath3(__v1,__v3)}
\fmfi{plain}{subpath (0,length(p3)*0.6) of p3}
\fmfi{plain}{subpath (length(p3)*0.6,length(p3)) of p3}

\fmfiset{p4}{vpath4(__b0,__v1)}
\fmfi{plain}{subpath (0,length(p4)*0.4) of p4}
\fmfi{dashes}{subpath (length(p4)*0.4,length(p4)) of p4}

\fmfiset{p5}{vpath5(__v1,__v3)}
\fmfi{plain}{subpath (0,length(p5)*0.4) of p5}
\fmfi{plain}{subpath (length(p5)*0.4,length(p5)) of p5}

\fmfiset{p6}{vpath6(__v3,__b1)}
\fmfi{plain}{subpath (0,length(p6)*0.6) of p6}
\fmfi{plain}{subpath (length(p6)*0.6,length(p6)) of p6}

\fmfv{label=$x_1$,label.angle=-90,label.dist=9.5}{b0}
\fmfv{label=$x_2$,label.angle=-90,label.dist=9.5}{b1}
\fmfv{label=$x$,label.angle=-120}{v1}
\fmfv{label=$x'$,label.angle=-180}{tv1}
\fmfv{label=$x''$,label.angle=-30}{v3}
\end{fmfgraph*}
\end{fmffile}
\end{minipage}\hspace{-0.5cm}+
\end{equation}
\vspace{3cm}
\begin{equation}\nonumber
\hspace{0.5cm}+\frac{1}{(2!)^2}\quad\begin{minipage}[c]{60mm}
\vspace{-1.2cm}
\begin{fmffile}{DiagPagE1.3Proj3}
\begin{fmfgraph*}(50,15)
 \fmftop{t0,t1}
\fmfbottom{b0,b1}
\fmf{phantom}{t0,tv1,t1}
\fmf{phantom}{b0,v1,v2,v3,b1}
\fmfdot{tv1}
\fmfdot{b0,v1,v3,b1}
\fmffreeze

\fmf{phantom,tag=1}{v1,tv1}
\fmf{phantom,tag=2}{v3,tv1}
\fmf{plain,tension=0.7}{tv1,tv1}
\fmf{phantom,right=0.7,tag=3}{v1,v3}
\fmf{phantom,tag=4}{b0,v1}
\fmf{phantom,tag=5}{v1,v3}
\fmf{phantom,tag=6}{v3,b1}

\fmfposition
\fmfipath{p[]}

\fmfiset{p1}{vpath1(__v1,__tv1)}
\fmfi{plain}{subpath (0,length(p1)*0.4) of p1}
\fmfi{plain}{subpath (length(p1)*0.4,length(p1)) of p1}

\fmfiset{p2}{vpath2(__v3,__tv1)}
\fmfi{plain}{subpath (0,length(p2)*0.4) of p2}
\fmfi{dashes}{subpath (length(p2)*0.4,length(p2)) of p2}

\fmfiset{p3}{vpath3(__v1,__v3)}
\fmfi{plain}{subpath (0,length(p3)*0.6) of p3}
\fmfi{plain}{subpath (length(p3)*0.6,length(p3)) of p3}

\fmfiset{p4}{vpath4(__b0,__v1)}
\fmfi{plain}{subpath (0,length(p4)*0.4) of p4}
\fmfi{dashes}{subpath (length(p4)*0.4,length(p4)) of p4}

\fmfiset{p5}{vpath5(__v1,__v3)}
\fmfi{plain}{subpath (0,length(p5)*0.6) of p5}
\fmfi{plain}{subpath (length(p5)*0.6,length(p5)) of p5}

\fmfiset{p6}{vpath6(__v3,__b1)}
\fmfi{dashes}{subpath (0,length(p6)*0.6) of p6}
\fmfi{plain}{subpath (length(p6)*0.6,length(p6)) of p6}

\fmfv{label=$x_1$,label.angle=-90,label.dist=9.5}{b0}
\fmfv{label=$x_2$,label.angle=-90,label.dist=9.5}{b1}
\fmfv{label=$x$,label.angle=-120}{v1}
\fmfv{label=$x'$,label.angle=-180}{tv1}
\fmfv{label=$x''$,label.angle=-30}{v3}
\end{fmfgraph*}
\end{fmffile}
\end{minipage}
\hspace{-0.8cm}+\frac{1}{(2!)^2}\quad\begin{minipage}[c]{60mm}
\vspace{-1.2cm}
\begin{fmffile}{DiagPagE1.3Proj4}
\begin{fmfgraph*}(50,15)
 \fmftop{t0,t1}
\fmfbottom{b0,b1}
\fmf{phantom}{t0,tv1,t1}
\fmf{phantom}{b0,v1,v2,v3,b1}
\fmfdot{tv1}
\fmfdot{b0,v1,v3,b1}
\fmffreeze

\fmf{phantom,tag=1}{v1,tv1}
\fmf{phantom,tag=2}{v3,tv1}
\fmf{plain,tension=0.7}{tv1,tv1}
\fmf{phantom,right=0.7,tag=3}{v1,v3}
\fmf{phantom,tag=4}{b0,v1}
\fmf{phantom,tag=5}{v1,v3}
\fmf{phantom,tag=6}{v3,b1}

\fmfposition
\fmfipath{p[]}

\fmfiset{p1}{vpath1(__v1,__tv1)}
\fmfi{plain}{subpath (0,length(p1)*0.4) of p1}
\fmfi{dashes}{subpath (length(p1)*0.4,length(p1)) of p1}

\fmfiset{p2}{vpath2(__v3,__tv1)}
\fmfi{plain}{subpath (0,length(p2)*0.6) of p2}
\fmfi{plain}{subpath (length(p2)*0.6,length(p2)) of p2}

\fmfiset{p3}{vpath3(__v1,__v3)}
\fmfi{plain}{subpath (0,length(p3)*0.6) of p3}
\fmfi{plain}{subpath (length(p3)*0.6,length(p3)) of p3}

\fmfiset{p4}{vpath4(__b0,__v1)}
\fmfi{plain}{subpath (0,length(p4)*0.4) of p4}
\fmfi{dashes}{subpath (length(p4)*0.4,length(p4)) of p4}

\fmfiset{p5}{vpath5(__v1,__v3)}
\fmfi{plain}{subpath (0,length(p5)*0.6) of p5}
\fmfi{plain}{subpath (length(p5)*0.6,length(p5)) of p5}

\fmfiset{p6}{vpath6(__v3,__b1)}
\fmfi{dashes}{subpath (0,length(p6)*0.6) of p6}
\fmfi{plain}{subpath (length(p6)*0.6,length(p6)) of p6}

\fmfv{label=$x_1$,label.angle=-90,label.dist=9.5}{b0}
\fmfv{label=$x_2$,label.angle=-90,label.dist=9.5}{b1}
\fmfv{label=$x$,label.angle=-120}{v1}
\fmfv{label=$x'$,label.angle=-180}{tv1}
\fmfv{label=$x''$,label.angle=-30}{v3}
\end{fmfgraph*}
\end{fmffile}
\end{minipage}\hspace{-0.5cm}+
\end{equation}
\vspace{3.5cm}
\begin{equation}\nonumber
+\frac{1}{2!}\quad\begin{minipage}[c]{60mm}
\vspace{-1.2cm}
\begin{fmffile}{DiagPagE1.3Proj5}
\begin{fmfgraph*}(50,15)
 \fmftop{t0,t1}
\fmfbottom{b0,b1}
\fmf{phantom}{t0,tv1,t1}
\fmf{phantom}{b0,v1,v2,v3,b1}
\fmfdot{tv1}
\fmfdot{b0,v1,v3,b1}
\fmffreeze

\fmf{phantom,tag=1}{v1,tv1}
\fmf{phantom,tag=2}{v3,tv1}
\fmf{plain,tension=0.7}{tv1,tv1}
\fmf{phantom,right=0.7,tag=3}{v1,v3}
\fmf{phantom,tag=4}{b0,v1}
\fmf{phantom,tag=5}{v1,v3}
\fmf{phantom,tag=6}{v3,b1}

\fmfposition
\fmfipath{p[]}

\fmfiset{p1}{vpath1(__v1,__tv1)}
\fmfi{plain}{subpath (0,length(p1)*0.4) of p1}
\fmfi{plain}{subpath (length(p1)*0.4,length(p1)) of p1}

\fmfiset{p2}{vpath2(__v3,__tv1)}
\fmfi{plain}{subpath (0,length(p2)*0.4) of p2}
\fmfi{dashes}{subpath (length(p2)*0.4,length(p2)) of p2}

\fmfiset{p3}{vpath3(__v1,__v3)}
\fmfi{plain}{subpath (0,length(p3)*0.6) of p3}
\fmfi{plain}{subpath (length(p3)*0.6,length(p3)) of p3}

\fmfiset{p4}{vpath4(__b0,__v1)}
\fmfi{plain}{subpath (0,length(p4)*0.6) of p4}
\fmfi{plain}{subpath (length(p4)*0.6,length(p4)) of p4}

\fmfiset{p5}{vpath5(__v1,__v3)}
\fmfi{dashes}{subpath (0,length(p5)*0.6) of p5}
\fmfi{plain}{subpath (length(p5)*0.6,length(p5)) of p5}

\fmfiset{p6}{vpath6(__v3,__b1)}
\fmfi{dashes}{subpath (0,length(p6)*0.6) of p6}
\fmfi{plain}{subpath (length(p6)*0.6,length(p6)) of p6}

\fmfv{label=$x_1$,label.angle=-90,label.dist=9.5}{b0}
\fmfv{label=$x_2$,label.angle=-90,label.dist=9.5}{b1}
\fmfv{label=$x$,label.angle=-120}{v1}
\fmfv{label=$x'$,label.angle=-180}{tv1}
\fmfv{label=$x''$,label.angle=-30}{v3}
\end{fmfgraph*}
\end{fmffile}
\end{minipage}
\hspace{-0.8cm}+\frac{1}{2!}\quad\begin{minipage}[c]{60mm}
\vspace{-1.2cm}
\begin{fmffile}{DiagPagE1.3Proj6}
\begin{fmfgraph*}(50,15)
 \fmftop{t0,t1}
\fmfbottom{b0,b1}
\fmf{phantom}{t0,tv1,t1}
\fmf{phantom}{b0,v1,v2,v3,b1}
\fmfdot{tv1}
\fmfdot{b0,v1,v3,b1}
\fmffreeze

\fmf{phantom,tag=1}{v1,tv1}
\fmf{phantom,tag=2}{v3,tv1}
\fmf{plain,tension=0.7}{tv1,tv1}
\fmf{phantom,right=0.7,tag=3}{v1,v3}
\fmf{phantom,tag=4}{b0,v1}
\fmf{phantom,tag=5}{v1,v3}
\fmf{phantom,tag=6}{v3,b1}

\fmfposition
\fmfipath{p[]}

\fmfiset{p1}{vpath1(__v1,__tv1)}
\fmfi{plain}{subpath (0,length(p1)*0.4) of p1}
\fmfi{dashes}{subpath (length(p1)*0.4,length(p1)) of p1}

\fmfiset{p2}{vpath2(__v3,__tv1)}
\fmfi{plain}{subpath (0,length(p2)*0.6) of p2}
\fmfi{plain}{subpath (length(p2)*0.6,length(p2)) of p2}

\fmfiset{p3}{vpath3(__v1,__v3)}
\fmfi{plain}{subpath (0,length(p3)*0.6) of p3}
\fmfi{plain}{subpath (length(p3)*0.6,length(p3)) of p3}

\fmfiset{p4}{vpath4(__b0,__v1)}
\fmfi{plain}{subpath (0,length(p4)*0.4) of p4}
\fmfi{plain}{subpath (length(p4)*0.4,length(p4)) of p4}

\fmfiset{p5}{vpath5(__v1,__v3)}
\fmfi{dashes}{subpath (0,length(p5)*0.4) of p5}
\fmfi{plain}{subpath (length(p5)*0.4,length(p5)) of p5}

\fmfiset{p6}{vpath6(__v3,__b1)}
\fmfi{dashes}{subpath (0,length(p6)*0.6) of p6}
\fmfi{plain}{subpath (length(p6)*0.6,length(p6)) of p6}

\fmfv{label=$x_1$,label.angle=-90,label.dist=9.5}{b0}
\fmfv{label=$x_2$,label.angle=-90,label.dist=9.5}{b1}
\fmfv{label=$x$,label.angle=-120}{v1}
\fmfv{label=$x'$,label.angle=-180}{tv1}
\fmfv{label=$x''$,label.angle=-30}{v3}
\end{fmfgraph*}
\end{fmffile}
\end{minipage}\hspace{-0.5cm}+
\end{equation}
\newpage
\hspace{13cm}\quad\quad
\hspace{13cm}
\hspace{13cm}
\hspace{13cm}
\hspace{13cm}
\begin{equation}\nonumber
+\frac{1}{2!}\quad\begin{minipage}[c]{60mm}
\vspace{-1.2cm}
\begin{fmffile}{DiagPagE1.3Proj7}
\begin{fmfgraph*}(50,15)
 \fmftop{t0,t1}
\fmfbottom{b0,b1}
\fmf{phantom}{t0,tv1,t1}
\fmf{phantom}{b0,v1,v2,v3,b1}
\fmfdot{tv1}
\fmfdot{b0,v1,v3,b1}
\fmffreeze

\fmf{phantom,tag=1}{v1,tv1}
\fmf{phantom,tag=2}{v3,tv1}
\fmf{plain,tension=0.7}{tv1,tv1}
\fmf{phantom,right=0.7,tag=3}{v1,v3}
\fmf{phantom,tag=4}{b0,v1}
\fmf{phantom,tag=5}{v1,v3}
\fmf{phantom,tag=6}{v3,b1}

\fmfposition
\fmfipath{p[]}

\fmfiset{p1}{vpath1(__v1,__tv1)}
\fmfi{plain}{subpath (0,length(p1)*0.4) of p1}
\fmfi{plain}{subpath (length(p1)*0.4,length(p1)) of p1}

\fmfiset{p2}{vpath2(__v3,__tv1)}
\fmfi{plain}{subpath (0,length(p2)*0.4) of p2}
\fmfi{dashes}{subpath (length(p2)*0.4,length(p2)) of p2}

\fmfiset{p3}{vpath3(__v1,__v3)}
\fmfi{plain}{subpath (0,length(p3)*0.6) of p3}
\fmfi{plain}{subpath (length(p3)*0.6,length(p3)) of p3}

\fmfiset{p4}{vpath4(__b0,__v1)}
\fmfi{plain}{subpath (0,length(p4)*0.4) of p4}
\fmfi{dashes}{subpath (length(p4)*0.4,length(p4)) of p4}

\fmfiset{p5}{vpath5(__v1,__v3)}
\fmfi{plain}{subpath (0,length(p5)*0.4) of p5}
\fmfi{dashes}{subpath (length(p5)*0.4,length(p5)) of p5}

\fmfiset{p6}{vpath6(__v3,__b1)}
\fmfi{plain}{subpath (0,length(p6)*0.6) of p6}
\fmfi{plain}{subpath (length(p6)*0.6,length(p6)) of p6}

\fmfv{label=$x_1$,label.angle=-90,label.dist=9.5}{b0}
\fmfv{label=$x_2$,label.angle=-90,label.dist=9.5}{b1}
\fmfv{label=$x$,label.angle=-120}{v1}
\fmfv{label=$x'$,label.angle=-180}{tv1}
\fmfv{label=$x''$,label.angle=-30}{v3}
\end{fmfgraph*}
\end{fmffile}
\end{minipage}
+\frac{1}{2!}\quad\begin{minipage}[c]{60mm}
\vspace{-1.2cm}
\begin{fmffile}{DiagPagE1.3Proj8}
\begin{fmfgraph*}(50,15)
 \fmftop{t0,t1}
\fmfbottom{b0,b1}
\fmf{phantom}{t0,tv1,t1}
\fmf{phantom}{b0,v1,v2,v3,b1}
\fmfdot{tv1}
\fmfdot{b0,v1,v3,b1}
\fmffreeze

\fmf{phantom,tag=1}{v1,tv1}
\fmf{phantom,tag=2}{v3,tv1}
\fmf{plain,tension=0.7}{tv1,tv1}
\fmf{phantom,right=0.7,tag=3}{v1,v3}
\fmf{phantom,tag=4}{b0,v1}
\fmf{phantom,tag=5}{v1,v3}
\fmf{phantom,tag=6}{v3,b1}

\fmfposition
\fmfipath{p[]}

\fmfiset{p1}{vpath1(__v1,__tv1)}
\fmfi{plain}{subpath (0,length(p1)*0.4) of p1}
\fmfi{dashes}{subpath (length(p1)*0.4,length(p1)) of p1}

\fmfiset{p2}{vpath2(__v3,__tv1)}
\fmfi{plain}{subpath (0,length(p2)*0.6) of p2}
\fmfi{plain}{subpath (length(p2)*0.6,length(p2)) of p2}

\fmfiset{p3}{vpath3(__v1,__v3)}
\fmfi{plain}{subpath (0,length(p3)*0.6) of p3}
\fmfi{plain}{subpath (length(p3)*0.6,length(p3)) of p3}

\fmfiset{p4}{vpath4(__b0,__v1)}
\fmfi{plain}{subpath (0,length(p4)*0.4) of p4}
\fmfi{dashes}{subpath (length(p4)*0.4,length(p4)) of p4}

\fmfiset{p5}{vpath5(__v1,__v3)}
\fmfi{plain}{subpath (0,length(p5)*0.4) of p5}
\fmfi{dashes}{subpath (length(p5)*0.4,length(p5)) of p5}

\fmfiset{p6}{vpath6(__v3,__b1)}
\fmfi{plain}{subpath (0,length(p6)*0.6) of p6}
\fmfi{plain}{subpath (length(p6)*0.6,length(p6)) of p6}

\fmfv{label=$x_1$,label.angle=-90,label.dist=9.5}{b0}
\fmfv{label=$x_2$,label.angle=-90,label.dist=9.5}{b1}
\fmfv{label=$x$,label.angle=-120}{v1}
\fmfv{label=$x'$,label.angle=-180}{tv1}
\fmfv{label=$x''$,label.angle=-30}{v3}
\end{fmfgraph*}
\end{fmffile}
\end{minipage}\hspace{-0.5cm}\Bigg]
\end{equation}
\vspace{0.2cm}
\begin{center}
\line(1,0){150}\\
\vspace{-0.3cm}
\line(1,0){140}
\end{center}
\vspace{2cm}
\begin{equation}\nonumber
\hspace{-4cm}\int d\bar{\theta}_1 d\theta_1 d\bar{\theta}_2 d\theta_2
\bar{\theta}_1 \theta_1 \bar{\theta}_2 \theta_2\,\Bigg[\,\,\frac{1}{3!}\,\frac{1}{2!}\quad
\parbox{60mm}{\vspace{-1.6cm}
\fmfreuse{DiagPagE1.4}}\hspace{-1.5cm}\Bigg]=
\end{equation}
\vspace{2cm}
\begin{equation}\nonumber
\hspace{1.3cm}\Bigg[\frac{1}{2!}\quad\begin{minipage}[c]{60mm}
\vspace{-1.6cm}
\begin{fmffile}{DiagPagE1.4Proj1}
\begin{fmfgraph*}(40,20)
\fmfkeep{DiagPagE1.4Proj1}
\fmftop{t0,t1}
\fmfbottom{b0,b1}
\fmf{phantom}{b0,bv1,b1}
\fmf{phantom}{t0,tv1,tv2,tv3,t1}
\fmfdot{tv1,tv3}
\fmfdot{b0,bv1,b1}
\fmffreeze
\fmf{phantom}{tv1,tv3}
\fmf{plain,right=0.5}{tv1,bv1}
\fmf{plain,right=0.8}{tv1,tv3}
\fmf{plain,left=0.8}{tv1,tv3}
\fmf{dashes,left=0.5,tag=1}{tv3,bv1}
\fmf{phantom,tag=2}{tv1,tv3}
\fmf{phantom,tag=3}{b0,bv1}
\fmf{phantom,tag=4}{bv1,b1}

\fmffreeze

\fmfposition
\fmfipath{p[]}

\fmfiset{p1}{vpath1(__tv3,__bv1)}
\fmfi{plain}{subpath (length(p1)*0.6,length(p1)) of p1}

\fmfiset{p2}{vpath2(__tv1,__tv3)}
\fmfi{dashes}{subpath (0,length(p2)*0.6) of p2}
\fmfi{plain}{subpath (length(p2)*0.6,length(p2)) of p2}

\fmfiset{p3}{vpath3(__b0,__bv1)}
\fmfi{plain}{subpath (0,length(p3)*0.4) of p3}
\fmfi{dashes}{subpath (length(p3)*0.4,length(p3)) of p3}

\fmfiset{p4}{vpath4(__bv1,__b1)}
\fmfi{plain}{subpath (0,length(p4)*0.6) of p4}
\fmfi{plain}{subpath (length(p4)*0.6,length(p4)) of p4}

\fmfv{label=$x_1$,label.angle=-90}{b0}
\fmfv{label=$x_2$,label.angle=-90}{b1}
\fmfv{label=$x$,label.angle=-90}{bv1}
\fmfv{label=$x'$,label.angle=-180}{tv1}
\fmfv{label=$x''$,label.angle=0}{tv3}
 \end{fmfgraph*}
 \end{fmffile}
\end{minipage}
\hspace{-1.8cm}+\frac{1}{2!}\quad\begin{minipage}[c]{60mm}
\vspace{-1.6cm}
\begin{fmffile}{DiagPagE1.4Proj2}
\begin{fmfgraph*}(40,20)
\fmfkeep{DiagPagE1.4Proj2}
\fmftop{t0,t1}
\fmfbottom{b0,b1}
\fmf{phantom}{b0,bv1,b1}
\fmf{phantom}{t0,tv1,tv2,tv3,t1}
\fmfdot{tv1,tv3}
\fmfdot{b0,bv1,b1}
\fmffreeze
\fmf{phantom}{tv1,tv3}
\fmf{plain,right=0.5}{tv1,bv1}
\fmf{plain,right=0.8}{tv1,tv3}
\fmf{plain,left=0.8}{tv1,tv3}
\fmf{dashes,left=0.5,tag=1}{tv3,bv1}
\fmf{phantom,tag=2}{tv1,tv3}
\fmf{phantom,tag=3}{b0,bv1}
\fmf{phantom,tag=4}{bv1,b1}

\fmffreeze

\fmfposition
\fmfipath{p[]}

\fmfiset{p1}{vpath1(__tv3,__bv1)}
\fmfi{plain}{subpath (length(p1)*0.6,length(p1)) of p1}

\fmfiset{p2}{vpath2(__tv1,__tv3)}
\fmfi{dashes}{subpath (0,length(p2)*0.6) of p2}
\fmfi{plain}{subpath (length(p2)*0.6,length(p2)) of p2}

\fmfiset{p3}{vpath3(__b0,__bv1)}
\fmfi{plain}{subpath (0,length(p3)*0.6) of p3}
\fmfi{plain}{subpath (length(p3)*0.6,length(p3)) of p3}

\fmfiset{p4}{vpath4(__bv1,__b1)}
\fmfi{dashes}{subpath (0,length(p4)*0.6) of p4}
\fmfi{plain}{subpath (length(p4)*0.6,length(p4)) of p4}

\fmfv{label=$x_1$,label.angle=-90}{b0}
\fmfv{label=$x_2$,label.angle=-90}{b1}
\fmfv{label=$x$,label.angle=-90}{bv1}
\fmfv{label=$x'$,label.angle=-180}{tv1}
\fmfv{label=$x''$,label.angle=0}{tv3}
 \end{fmfgraph*}
 \end{fmffile}
\end{minipage}\hspace{-1.5cm}+
\end{equation}
\vspace{1.8cm}
\begin{equation}\nonumber
\hspace{1.3cm}\frac{1}{3!}\,\frac{1}{2!}\quad\begin{minipage}[c]{60mm}
\vspace{-1.6cm}
\begin{fmffile}{DiagPagE1.4Proj3}
\begin{fmfgraph*}(40,20)
\fmfkeep{DiagPagE1.4Proj3}
\fmftop{t0,t1}
\fmfbottom{b0,b1}
\fmf{phantom}{b0,bv1,b1}
\fmf{phantom}{t0,tv1,tv2,tv3,t1}
\fmfdot{tv1,tv3}
\fmfdot{b0,bv1,b1}
\fmffreeze
\fmf{phantom}{tv1,tv3}
\fmf{dashes,right=0.5,tag=5}{tv1,bv1}
\fmf{plain,right=0.8}{tv1,tv3}
\fmf{plain,left=0.8}{tv1,tv3}
\fmf{dashes,left=0.5,tag=1}{tv3,bv1}
\fmf{phantom,tag=2}{tv1,tv3}
\fmf{phantom,tag=3}{b0,bv1}
\fmf{phantom,tag=4}{bv1,b1}

\fmffreeze

\fmfposition
\fmfipath{p[]}

\fmfiset{p1}{vpath1(__tv3,__bv1)}
\fmfi{plain}{subpath (length(p1)*0.5,length(p1)) of p1}

\fmfiset{p5}{vpath5(__tv1,__bv1)}
\fmfi{plain}{subpath (length(p5)*0.5,length(p5)) of p5}

\fmfiset{p2}{vpath2(__tv1,__tv3)}
\fmfi{plain}{subpath (0,length(p2)*0.6) of p2}
\fmfi{plain}{subpath (length(p2)*0.6,length(p2)) of p2}

\fmfiset{p3}{vpath3(__b0,__bv1)}
\fmfi{plain}{subpath (0,length(p3)*0.4) of p3}
\fmfi{dashes}{subpath (length(p3)*0.4,length(p3)) of p3}

\fmfiset{p4}{vpath4(__bv1,__b1)}
\fmfi{plain}{subpath (0,length(p4)*0.6) of p4}
\fmfi{plain}{subpath (length(p4)*0.6,length(p4)) of p4}

\fmfv{label=$x_1$,label.angle=-90}{b0}
\fmfv{label=$x_2$,label.angle=-90}{b1}
\fmfv{label=$x$,label.angle=-90}{bv1}
\fmfv{label=$x'$,label.angle=-180}{tv1}
\fmfv{label=$x''$,label.angle=0}{tv3}
 \end{fmfgraph*}
 \end{fmffile}
\end{minipage}
\hspace{-1.8cm}+\frac{1}{3!}\,\frac{1}{2!}\quad\begin{minipage}[c]{60mm}
\vspace{-1.6cm}
\begin{fmffile}{DiagPagE1.4Proj4}
\begin{fmfgraph*}(40,20)
\fmfkeep{DiagPagE1.4Proj4}
\fmftop{t0,t1}
\fmfbottom{b0,b1}
\fmf{phantom}{b0,bv1,b1}
\fmf{phantom}{t0,tv1,tv2,tv3,t1}
\fmfdot{tv1,tv3}
\fmfdot{b0,bv1,b1}
\fmffreeze
\fmf{phantom}{tv1,tv3}
\fmf{dashes,right=0.5,tag=5}{tv1,bv1}
\fmf{plain,right=0.8}{tv1,tv3}
\fmf{plain,left=0.8}{tv1,tv3}
\fmf{dashes,left=0.5,tag=1}{tv3,bv1}
\fmf{phantom,tag=2}{tv1,tv3}
\fmf{phantom,tag=3}{b0,bv1}
\fmf{phantom,tag=4}{bv1,b1}

\fmffreeze

\fmfposition
\fmfipath{p[]}

\fmfiset{p1}{vpath1(__tv3,__bv1)}
\fmfi{plain}{subpath (length(p1)*0.5,length(p1)) of p1}

\fmfiset{p5}{vpath5(__tv1,__bv1)}
\fmfi{plain}{subpath (length(p5)*0.5,length(p5)) of p5}

\fmfiset{p2}{vpath2(__tv1,__tv3)}
\fmfi{plain}{subpath (0,length(p2)*0.6) of p2}
\fmfi{plain}{subpath (length(p2)*0.6,length(p2)) of p2}

\fmfiset{p3}{vpath3(__b0,__bv1)}
\fmfi{plain}{subpath (0,length(p3)*0.6) of p3}
\fmfi{plain}{subpath (length(p3)*0.6,length(p3)) of p3}

\fmfiset{p4}{vpath4(__bv1,__b1)}
\fmfi{dashes}{subpath (0,length(p4)*0.6) of p4}
\fmfi{plain}{subpath (length(p4)*0.6,length(p4)) of p4}

\fmfv{label=$x_1$,label.angle=-90}{b0}
\fmfv{label=$x_2$,label.angle=-90}{b1}
\fmfv{label=$x$,label.angle=-90}{bv1}
\fmfv{label=$x'$,label.angle=-180}{tv1}
\fmfv{label=$x''$,label.angle=0}{tv3}
 \end{fmfgraph*}
 \end{fmffile}
\end{minipage}\hspace{-1.5cm}\Bigg]
\end{equation}
\vspace{0.5cm}
\begin{center}
\line(1,0){150}\\
\vspace{-0.3cm}
\line(1,0){140}
\end{center}

\vspace{2cm}
\begin{equation}\nonumber
\hspace{-4.5cm}\int d\bar{\theta}_1 d\theta_1 d\bar{\theta}_2 d\theta_2
\bar{\theta}_1 \theta_1 \bar{\theta}_2 \theta_2\,\Bigg[\,\,\frac{1}{(2!)^3}\quad
\parbox{60mm}{\vspace{-1.6cm}
\fmfreuse{DiagPagE1.5}}\hspace{-1.8cm}\Bigg]=
\end{equation}
\newpage
\vspace{5.5cm}
\begin{equation*}
\quad
\end{equation*}
\begin{equation}\nonumber
\hspace{-1.cm}=\Bigg[\frac{1}{(2!)^2}\quad
\begin{minipage}[c]{60mm}
\vspace{-1.6cm}
\begin{fmffile}{DiagPagE1.5Proj1}
\begin{fmfgraph*}(40,20)
\fmfkeep{DiagPagE1.5Proj1}
\fmfbottom{b1,b2}
\fmftop{t1,t2}
\fmf{phantom}{t1,vt1,vt2,t2}
\fmffreeze
\fmf{phantom,tag=1}{vt1,vt2}
\fmf{phantom}{b1,vb1,b2}
\fmfdot{b1,b2,vt1,vt2,vb1}
\fmffreeze
\fmf{phantom,tag=2}{vb1,vt1}
\fmf{phantom,tag=3}{vb1,vt2}
\fmf{phantom,tag=4}{b1,vb1}
\fmf{phantom,tag=5}{vb1,b2}
\fmf{plain,tension=0.7}{vt1,vt1}
\fmf{plain,tension=0.7}{vt2,vt2}

\fmfposition
\fmfipath{p[]}

\fmfiset{p1}{vpath1(__vt1,__vt2)}
\fmfi{dashes}{subpath (0,length(p1)*0.6) of p1}
\fmfi{plain}{subpath (length(p1)*0.6,length(p1)) of p1}

\fmfiset{p2}{vpath2(__vb1,__vt1)}
\fmfi{plain}{subpath (0,length(p2)*0.6) of p2}
\fmfi{plain}{subpath (length(p2)*0.6,length(p2)) of p2}

\fmfiset{p3}{vpath3(__vb1,__vt2)}
\fmfi{plain}{subpath (0,length(p3)*0.4) of p3}
\fmfi{dashes}{subpath (length(p3)*0.4,length(p3)) of p3}

\fmfiset{p4}{vpath4(__b1,__vb1)}
\fmfi{plain}{subpath (0,length(p4)*0.4) of p4}
\fmfi{plain}{subpath (length(p4)*0.4,length(p4)) of p4}

\fmfiset{p5}{vpath5(__vb1,__b2)}
\fmfi{dashes}{subpath (0,length(p5)*0.6) of p5}
\fmfi{plain}{subpath (length(p5)*0.6,length(p5)) of p5}

\fmfv{label=$x_1$,label.angle=-90}{b1}
\fmfv{label=$x_2$,label.angle=-90}{b2}
\fmfv{label=$x$,label.angle=-90}{vb1}
\fmfv{label=$x'$,label.angle=-180}{vt1}
\fmfv{label=$x''$,label.angle=0}{vt2}
 \end{fmfgraph*}
 \end{fmffile}
\end{minipage}
\hspace{-1.6cm}+\frac{1}{(2!)^2}\quad
\begin{minipage}[c]{60mm}
\vspace{-1.6cm}
\begin{fmffile}{DiagPagE1.5Proj2}
\begin{fmfgraph*}(40,20)
\fmfkeep{DiagPagE1.5Proj2}
\fmfbottom{b1,b2}
\fmftop{t1,t2}
\fmf{phantom}{t1,vt1,vt2,t2}
\fmffreeze
\fmf{phantom,tag=1}{vt1,vt2}
\fmf{phantom}{b1,vb1,b2}
\fmfdot{b1,b2,vt1,vt2,vb1}
\fmffreeze
\fmf{phantom,tag=2}{vb1,vt1}
\fmf{phantom,tag=3}{vb1,vt2}
\fmf{phantom,tag=4}{b1,vb1}
\fmf{phantom,tag=5}{vb1,b2}
\fmf{plain,tension=0.7}{vt1,vt1}
\fmf{plain,tension=0.7}{vt2,vt2}

\fmfposition
\fmfipath{p[]}

\fmfiset{p1}{vpath1(__vt1,__vt2)}
\fmfi{dashes}{subpath (0,length(p1)*0.6) of p1}
\fmfi{plain}{subpath (length(p1)*0.6,length(p1)) of p1}

\fmfiset{p2}{vpath2(__vb1,__vt1)}
\fmfi{plain}{subpath (0,length(p2)*0.6) of p2}
\fmfi{plain}{subpath (length(p2)*0.6,length(p2)) of p2}

\fmfiset{p3}{vpath3(__vb1,__vt2)}
\fmfi{plain}{subpath (0,length(p3)*0.4) of p3}
\fmfi{dashes}{subpath (length(p3)*0.4,length(p3)) of p3}

\fmfiset{p4}{vpath4(__b1,__vb1)}
\fmfi{plain}{subpath (0,length(p4)*0.4) of p4}
\fmfi{dashes}{subpath (length(p4)*0.4,length(p4)) of p4}

\fmfiset{p5}{vpath5(__vb1,__b2)}
\fmfi{plain}{subpath (0,length(p5)*0.4) of p5}
\fmfi{plain}{subpath (length(p5)*0.4,length(p5)) of p5}

\fmfv{label=$x_1$,label.angle=-90}{b1}
\fmfv{label=$x_2$,label.angle=-90}{b2}
\fmfv{label=$x$,label.angle=-90}{vb1}
\fmfv{label=$x'$,label.angle=-180}{vt1}
\fmfv{label=$x''$,label.angle=0}{vt2}
 \end{fmfgraph*}
 \end{fmffile}
\end{minipage}\hspace{-1.5cm}+
\end{equation}
\parbox{60mm}{
\vspace{2cm}
\begin{equation}\nonumber
\hspace{1cm}+\frac{1}{(2!)^3}\quad
\begin{minipage}[c]{60mm}
\vspace{-1.6cm}
\begin{fmffile}{DiagPagE1.5Proj3}
\begin{fmfgraph*}(40,20)
\fmfkeep{DiagPagE1.5Proj3}
\fmfbottom{b1,b2}
\fmftop{t1,t2}
\fmf{phantom}{t1,vt1,vt2,t2}
\fmffreeze
\fmf{phantom,tag=1}{vt1,vt2}
\fmf{phantom}{b1,vb1,b2}
\fmfdot{b1,b2,vt1,vt2,vb1}
\fmffreeze
\fmf{phantom,tag=2}{vb1,vt1}
\fmf{phantom,tag=3}{vb1,vt2}
\fmf{phantom,tag=4}{b1,vb1}
\fmf{phantom,tag=5}{vb1,b2}
\fmf{plain,tension=0.7}{vt1,vt1}
\fmf{plain,tension=0.7}{vt2,vt2}

\fmfposition
\fmfipath{p[]}

\fmfiset{p1}{vpath1(__vt1,__vt2)}
\fmfi{plain}{subpath (0,length(p1)*0.6) of p1}
\fmfi{plain}{subpath (length(p1)*0.6,length(p1)) of p1}

\fmfiset{p2}{vpath2(__vb1,__vt1)}
\fmfi{plain}{subpath (0,length(p2)*0.4) of p2}
\fmfi{dashes}{subpath (length(p2)*0.4,length(p2)) of p2}

\fmfiset{p3}{vpath3(__vb1,__vt2)}
\fmfi{plain}{subpath (0,length(p3)*0.4) of p3}
\fmfi{dashes}{subpath (length(p3)*0.4,length(p3)) of p3}

\fmfiset{p4}{vpath4(__b1,__vb1)}
\fmfi{plain}{subpath (0,length(p4)*0.4) of p4}
\fmfi{plain}{subpath (length(p4)*0.4,length(p4)) of p4}

\fmfiset{p5}{vpath5(__vb1,__b2)}
\fmfi{dashes}{subpath (0,length(p5)*0.6) of p5}
\fmfi{plain}{subpath (length(p5)*0.6,length(p5)) of p5}

\fmfv{label=$x_1$,label.angle=-90}{b1}
\fmfv{label=$x_2$,label.angle=-90}{b2}
\fmfv{label=$x$,label.angle=-90}{vb1}
\fmfv{label=$x'$,label.angle=-180}{vt1}
\fmfv{label=$x''$,label.angle=0}{vt2}
 \end{fmfgraph*}
 \end{fmffile}
\end{minipage}
\hspace{-1.5cm}+\frac{1}{(2!)^3}\quad
\begin{minipage}[c]{60mm}
\vspace{-1.6cm}
\begin{fmffile}{DiagPagE1.5Proj4}
\begin{fmfgraph*}(40,20)
\fmfkeep{DiagPagE1.5Proj4}
\fmfbottom{b1,b2}
\fmftop{t1,t2}
\fmf{phantom}{t1,vt1,vt2,t2}
\fmffreeze
\fmf{phantom,tag=1}{vt1,vt2}
\fmf{phantom}{b1,vb1,b2}
\fmfdot{b1,b2,vt1,vt2,vb1}
\fmffreeze
\fmf{phantom,tag=2}{vb1,vt1}
\fmf{phantom,tag=3}{vb1,vt2}
\fmf{phantom,tag=4}{b1,vb1}
\fmf{phantom,tag=5}{vb1,b2}
\fmf{plain,tension=0.7}{vt1,vt1}
\fmf{plain,tension=0.7}{vt2,vt2}

\fmfposition
\fmfipath{p[]}

\fmfiset{p1}{vpath1(__vt1,__vt2)}
\fmfi{plain}{subpath (0,length(p1)*0.6) of p1}
\fmfi{plain}{subpath (length(p1)*0.6,length(p1)) of p1}

\fmfiset{p2}{vpath2(__vb1,__vt1)}
\fmfi{plain}{subpath (0,length(p2)*0.4) of p2}
\fmfi{dashes}{subpath (length(p2)*0.4,length(p2)) of p2}

\fmfiset{p3}{vpath3(__vb1,__vt2)}
\fmfi{plain}{subpath (0,length(p3)*0.4) of p3}
\fmfi{dashes}{subpath (length(p3)*0.4,length(p3)) of p3}

\fmfiset{p4}{vpath4(__b1,__vb1)}
\fmfi{plain}{subpath (0,length(p4)*0.4) of p4}
\fmfi{dashes}{subpath (length(p4)*0.4,length(p4)) of p4}

\fmfiset{p5}{vpath5(__vb1,__b2)}
\fmfi{plain}{subpath (0,length(p5)*0.4) of p5}
\fmfi{plain}{subpath (length(p5)*0.4,length(p5)) of p5}

\fmfv{label=$x_1$,label.angle=-90}{b1}
\fmfv{label=$x_2$,label.angle=-90}{b2}
\fmfv{label=$x$,label.angle=-90}{vb1}
\fmfv{label=$x'$,label.angle=-180}{vt1}
\fmfv{label=$x''$,label.angle=0}{vt2}
 \end{fmfgraph*}
 \end{fmffile}
\end{minipage}\hspace{-1.5cm}\Bigg]
\end{equation}
}

Of course using the set of super-diagrams (\ref{SuperDiag3order})
and different projectors we could get also the $\langle \phi
\lambda\rangle$ and $\langle c\bar{c}\rangle$ correlations up to the third order.

\section{Fluctuation-Dissipation Theorem Via Ward Identities.}\label{BRS}

In this appendix we will present a non-perturbative derivation of
the fluctuation-dissipation theorem and of other relations that we
proved perturbatively in the paper. All these relations are
basically Ward identities associated to the various symmetries
present in the CPI, i.e., BRS, $\overline{\textrm {BRS}}$, SUSY,
etc. Most of the calculations presented in this appendix  were done
long ago by Martin Reuter whom we wish to warmly thank for his
generosity.

In this appendix we will not work in the proper KvN formalism  with an Hilbert space,etc,etc,
but in the Liouville one where a probability density
$\tilde{\rho}(\varphi)$ is used.

Let us suppose we give the initial probability density $\tilde{\rho}(-\infty)$ at time
$t=-\infty$ and that we want to evaluate
the average of a quantity $\mathcal{O}$. The path-integral expression for  this
average is:
\begin{equation}\label{<mathcalO>}
\langle \mathcal{O}\rangle_{\tilde{\rho}(-\infty)} =
\displaystyle\int {\mathscr D}\varphi {\mathscr D}\lambda
{\mathscr D}c {\mathscr D}\bar{c}
\mathcal{O} \tilde{\rho}\left(\varphi(-\infty)
,c(-\infty)\right)\textrm{exp} \left(i\int_{-\infty}^{\infty} dt\mathcal{\tilde{L}}\right)
\end{equation}

Let us ask for which symmetry transformation (whose infinitesimal transformation we indicate with  $\delta$) and for
which $\tilde{\rho}$ we have
\begin{equation}
\langle \delta \mathcal{O}\rangle_{\tilde{\rho}(-\infty)}=0.
\end{equation} Let us assume that the integration measure ${\mathscr D}\varphi {\mathscr D}\lambda
{\mathscr D}c {\mathscr D}\bar{c}$ is invariant under the
transformation $\delta$ and that the action $\mathcal{\tilde S}=
\int \mathcal{\tilde L}dt$ changes only by a surface term
\begin{equation}
\delta \mathcal{\tilde S}= \epsilon \sigma(+\infty)-\epsilon \sigma(-\infty),
\end{equation} where $\epsilon$ is defined by $\varphi_{\epsilon}
=\varphi+\delta \epsilon$.

Let us change the variables of integration in (\ref{<mathcalO>}) going over to those obtained via a symmertry transformation. We obtain:
\begin{equation}
\begin{array}{lll}
\langle  \mathcal{O}\rangle_{\tilde{\rho}(-\infty)}
&=&\displaystyle\int {\mathscr D}\varphi \ldots
\left(\mathcal{O}+\delta \mathcal{O}\right)\left[\tilde{\rho}\left(-\infty\right)
+\delta \tilde{\rho}\left(-\infty\right)\right]
\textrm{exp}\left(i\mathcal{\tilde S}+i\epsilon \sigma(+\infty)-i\epsilon \sigma(-\infty)\right)\\
&=&\displaystyle\int {\mathscr D}\varphi \ldots
\left\{\mathcal{O}\tilde{\rho}+\mathcal{O}\delta \tilde{\rho}+\tilde{\rho}\delta \mathcal{O}
+\mathcal{O}\tilde{\rho}\left[i\epsilon \sigma(+\infty)-i\epsilon \sigma(-\infty)\right]\right\}
e^{i\mathcal{\tilde S}}.
\end{array}
\end{equation} The first term inside the integral fives  $\langle \mathcal{O}\rangle_{\tilde{\rho}(-\infty)}$
which is equal to the LHS because we did only a change in integration variables and so these two terms cancell. In this manner we get for the third term in the integral the following expression:
\begin{equation}\label{<deltamathcalO>}
\begin{array}{lll}
\langle \delta \mathcal{O}\rangle_{\tilde{\rho}(-\infty)}
&=&-\displaystyle\int {\mathscr D}\varphi \ldots
\mathcal{O}\left\{\delta\tilde{\rho}(-\infty)
+\tilde{\rho}(-\infty)i\epsilon \left[\sigma(+\infty)-\sigma(-\infty)\right]\right\}
e^{i\mathcal{\tilde S}}.
\end{array}
\end{equation} For $\delta$ let us use the transformation generated by the super-symmetric charges
\cite{GoReuTha}
\begin{equation}
Q_H=c^a\left(i\lambda_a-\beta \partial_a H\right)
\end{equation}
\begin{equation}
\bar{Q}_H=\bar{c}^a\left(i\lambda_a+\beta \partial_bH\right)
\end{equation} where $\beta$ is a real parameter which, when is set
to zero, gives the BRS, $\overline{\textrm {BRS}}$ \cite{GoReuTha}
charges . Let us now proceed to find the
surface terms associated to this symmetry transformations:
\begin{equation}
\delta_{Q_H}\mathcal{\tilde L}=-i\beta \frac{d}{dt}\epsilon N
\end{equation}
\begin{equation}
\delta_{\bar{Q}_H}\mathcal{\tilde L}=-i\beta \frac{d}{dt}\bar{\epsilon} \bar{Q}
\end{equation} where $N\equiv c^a \partial_a H$ and $\bar{Q}\equiv \bar{c}_a\omega^{ab}i\lambda_b$.
Let us now calculate the first term inside the  brackets on the r.h.s of
Eq. (\ref{<deltamathcalO>}) for the transformation generated by
$Q_H$
\begin{equation}
\delta \tilde{\rho}(-\infty)
=\left(\delta \varphi^a+\delta c^a\frac{\partial}{\partial c^a}\tilde{\rho}\right)(t=-\infty)
=\epsilon c^a(-\infty)\tilde{\rho}\left(\varphi(-\infty),c(-\infty)\right).
\end{equation} The surface term inside the brackets of Eq. (\ref{<deltamathcalO>}) gives
\begin{equation}
i\epsilon \sigma(+\infty)-i\epsilon \sigma(-\infty)
=\beta \epsilon \left(c^a\partial_aH\right)(+\infty)
-\beta \epsilon \left(c^a\partial_aH\right)(-\infty).
\end{equation} Inserting all this in (\ref{<deltamathcalO>}) we get:
\begin{equation}
\begin{array}{lll}
\langle \delta_{Q_H}\mathcal{O}\rangle_{\tilde{\rho}}
&=&-\displaystyle\int {\mathscr D}\varphi
\ldots \mathcal{O}\epsilon
\left[c^a\left(-\infty\right)\frac{\partial}{\partial \varphi^a(-\infty)}
-\beta c^a\left(-\infty\right)\partial_{a}H\left(\varphi(-\infty)\right)\right] \vspace{2mm}\\
&&\times\tilde{\rho}\left(\varphi(-\infty),c(-\infty)\right)e^{i\mathcal{\tilde{S}}}
-\displaystyle\int {\mathscr D} \varphi \mathcal{O} \rho \epsilon
c^{a}(+\infty)\partial_{a}H(+\infty)\tilde{\rho}(-\infty)e^{i\mathcal{\tilde{S}}}\vspace{2mm}\\
&=& -\displaystyle\int {\mathscr D} \varphi \mathcal{O}\epsilon \hat{Q}_{H}\tilde{\rho}
e^{i\mathcal{\tilde{S}}}-\beta\displaystyle\int
{\mathscr D} \varphi \ldots \epsilon c^a\partial_aH \mathcal{O}\tilde{\rho}
e^{i\mathcal{\tilde{S}}}\end{array}
\end{equation}
For $\beta=0$ the second  term vanish and
the first is zero only if
\begin{equation}\label{Qrho}
\hat{Q}\tilde{\rho}=0
\end{equation} where $\hat{Q}=\hat{Q}_{H}|_{\beta=0}=0$. The solution of (\ref{Qrho}) is
\begin{equation}
\tilde{\rho}=\rho(\varphi)\delta(c).
\end{equation}

Let us do the analog for the transformation generated by $\bar{Q}_H$. Calculating
the various pieces we get
\begin{equation}
\langle \delta_{\bar{Q}_H}\mathcal{O}\rangle_{\tilde{\rho}}
\equiv-\displaystyle\int {\mathscr D}\varphi \ldots
\mathcal{O}\left(\epsilon \bar{Q}_H\tilde{\rho}\right)(-\infty)
e^{i\mathcal{\tilde S}}.
\end{equation} This implies that $\langle \delta_{\bar{Q}_H}\mathcal{O}\rangle_{\tilde{\rho}}=0$
if
\begin{equation}
\bar{Q}_H\tilde{\rho}=0.
\end{equation} The solution of this equation was calculated in the second
of Ref. \cite{GoReuTha} and is the canonical distribution
\begin{equation}\label{canaver}
\tilde{\rho}(\varphi,c)=e^{-\beta H}\delta(c).
\end{equation} So the averages taken with this $\tilde{\rho}$ are exactly
the thermal averages we have calculated all-along in the core of the paper.

Let us now choose as operator $\mathcal{O}$ the following one
\begin{equation}
\mathcal{O}\equiv \varphi^a(t)c^b(0)\;\;\;\;,t>0.
\end{equation} Doing the $Q_H$ transformation of $\mathcal{O}$ we get:
\begin{equation}\label{francesco}
\delta_{Q_{H}}\mathcal{O}=
-\bar{\epsilon}\omega^{ad}\bar{c}_d(t)c^b(0)
+\varphi^a(t)\epsilon \omega^{bd}
\left[i\lambda_d(0)+\beta\partial_dH(0)\right].
\end{equation} It was proved in \cite{GozziFDT}\cite{Reuter}\cite{Liap} that
\begin{equation}\label{averbarcc}
\langle \bar{c}_d(t)c^b(0)\rangle=0
\end{equation} if $t>0$, where the average is taken with the canonical
average (\ref{canaver}).
So we get from (\ref{francesco})
\begin{equation}
\langle \delta_{Q_{H}}\mathcal{O}\rangle
=0=\epsilon \omega^{bd}
\langle \varphi^a(t)\left[i\lambda_d(0)+\beta\partial_dH(0)\right]\rangle
\end{equation} which implies:
\begin{equation}\label{averphilambda}
\langle \varphi^a(t)i\lambda_d(0)\rangle=-\beta \langle \varphi^a(t)\partial_dH(0)\rangle.
\end{equation} From the equation of motion we have
\begin{equation}
\partial_dH(0)=\omega_{de}\dot{\varphi}^{e}(0).
\end{equation} So (\ref{averphilambda}) will become
\begin{equation}\label{averphilambdaII}
\langle \varphi^a(t)i\lambda_d(0)\rangle=-\beta \omega_{de}\langle \varphi^a(t)\dot{\varphi}^{e}(0)\rangle.
\end{equation} Now in general for time-independent Hamiltonian we have, for two
observables $A(t)$ and $B(t)$, that
\begin{equation}
\langle A(t_1)B(t_2)\rangle=F(t_1-t_2)
\end{equation} so
\begin{equation}
\frac{\partial}{\partial t_1}\langle A(t_1)B(t_2)\rangle
=-\frac{\partial}{\partial t_2}\langle A(t_1)B(t_2)\rangle
\end{equation} or
\begin{equation}
\langle \dot{A}(t_1)B(t_2)\rangle
=-\langle A(t_1)\dot{B}(t_2)\rangle.
\end{equation} Applying this last relation to (\ref{averphilambdaII}) we get:
\begin{equation}\label{FDTGoReu}
\langle \varphi^a(t)i\lambda_d(0)\rangle=-\beta \omega_{de}\langle \varphi^a(t)\dot{\varphi}^{e}(0)\rangle.
\end{equation} and this is the {\it fluctuation-dissipation} theorem.

With respect to the easier proof presented in \cite{DekerHakke} our proof
has the virtue that it can be seen as a Ward-identity of Susy and that
the canonical distribution is not put by hand, but it is derived \cite{GoReuTha}
as a super-symmetric invariant state.

Let us now prove few more identities. Let us
choose $\tilde{\rho} (\varphi,c)=\delta^{2n}(c(-\infty))$. This
distribution is BRS invariant \cite{GoReuTha} so:
\begin{equation}
\langle \delta_{{\rm BRS}} \mathcal{O}\rangle
=\displaystyle\int {\mathscr D}\varphi \ldots
\left(\delta_{{\rm BRS}} \mathcal{O}\right)
\delta^{2n}c(-\infty)
e^{i\mathcal{\tilde S}}=0.\end{equation}
where the BRS transformation \cite{GoReuTha} is
$$\left\{
\begin{array}{l}
\delta \varphi^a=\epsilon c^a\\
\delta c^a=0\\
\delta \bar{c}_a=i\epsilon \lambda_a\\
\delta \lambda_a=0.\\
\end{array}\right.$$ Let us take $\hat{\mathcal{O}}=\varphi^a(t)\bar{c}_b(0)$, then
\begin{equation}
\delta   \hat{\mathcal{O}}=
-i\epsilon \left[ic^a(t)\bar{c}_b(0)-\varphi^a(t)\lambda_b(0)\right].
\end{equation} So $\langle \delta   \hat{\mathcal{O}} \rangle=0$ implies
\begin{equation}
\langle \varphi^a(t)\lambda_b(0)\rangle=i\langle c^a(t)\bar{c}_b(0)\rangle.
\end{equation} This relation tell us that  the dash-full propagator is equal
to the Fermion one and to all order in perturbation theory. This confirms the
statement that we made all-along in our paper that this identity is related to the Susy invariance
(better say to the BRS invariance which is the susy with $\beta=0$).

Let us now choose as operator
\begin{equation}
\hat{\mathcal{O}}=\lambda_a(t)\bar{c}_b(0),
\end{equation} then
\begin{equation}
\delta \hat{\mathcal{O}}=i\epsilon \lambda_a(t)\lambda_a(0),
\end{equation} and from $\langle \delta \hat{\mathcal{O}}\rangle=0$ we obtain:
\begin{equation}
\langle \lambda_a(t) \lambda_a(0)\rangle=0.
\end{equation} This is "only" similar to the one we proved in the paper
perturbatively to all orders, because there we proved it with the
weight $\rho=e^{-\beta H}\delta^n(c)$, while here we proved it
only for $\beta=0$, but we think the proof given here should go
through also for $\beta \neq 0$.

Let us now choose as operator
\begin{equation}
\hat{\mathcal{O}}=c^a(t)\varphi^b(0),
\end{equation} then
\begin{equation}
\delta \hat{\mathcal{O}}=-\epsilon c^a(t)c^b(0),
\end{equation} so we get
\begin{equation}
\langle c^a(t)c^b(0)\rangle=0.
\end{equation} This relation for $a=b=1$ it was proved  to all
orders of perturbation theory in the paper and for the canonical
distribution.Here we have given a non-perturbative proof.

We like to conclude this appendix by saying that both the simplification in the diagrammatics
that we got via the introduction of super-diagrams and the various identites we found that will further simplify
the diagrammatics, have their roots in the various symmetries of the CPI.

\section{Perturbative Check of The Fluctuation-Dissipation Theorem.}\label{ProofFDT}

In this appendix we will prove the relations (\ref{FDT(0)}) and
(\ref{FDT(1)}).

Let us remember that
\begin{equation}
\Delta_{\beta}(x_1-x_2)
=\displaystyle\int\frac{d^4 p}{(2\pi)^4}\frac{2\pi}{\beta|p^0|}
\delta(p^2-m^2)e^{-ip.x}
\end{equation} where $p.x=p^0t-\vec{p}.\vec{x}$
and $p^2=(p^0)^2-\vec{p}^2$.

Let us take the derivative with respect to $t_1$, like on the
l.h.s of Eq. (\ref{FDT(0)})
\begin{equation}\label{partialtimedeltabeta}
\frac{\partial}{\partial t_1}\Delta_{\beta}(x_1-x_2)
=-i\displaystyle\int\displaystyle\frac{d^4 p}{(2\pi)^4}\frac{2\pi}{\beta}
\displaystyle\frac{p^0}{|p^0|}
\delta(p^2-m^2)e^{-ip.x}.
\end{equation} Let us remember that:
\begin{equation}
\displaystyle\frac{p^0}{|p^0|}=\theta(p^0)-\theta(-p^0)\equiv \Theta (p^0),
\end{equation} where we have defined the new symbol $\Theta (p^0)$. Let us
also notice that:
\begin{equation}
\delta\left((p^0)^2-\vec{p}^2-m^2\right)=\displaystyle\frac{1}{2E_{\vec{p}}}
\delta\left(p^0-E_{\vec{p}}\right)-\delta\left(p^0+E_{\vec{p}}\right),
\end{equation}
and so we get:
\begin{equation}
\Theta (p^0)\delta\left((p^0)^2-\vec{p}^2-m^2\right)=\displaystyle\frac{1}{2E_{\vec{p}}}
\delta\left(p^0-E_{\vec{p}}\right)-\delta\left(p^0+E_{\vec{p}}\right).
\end{equation} We can now rewrite (\ref{partialtimedeltabeta}) as
\begin{equation}
\begin{array}{lll}\label{partialtimedeltabeta2}
\frac{\partial}{\partial t_1}\Delta_{\beta}(x_1-x_2)
&=&-i\displaystyle\frac{1}{\beta}\displaystyle\int\displaystyle\frac{d^3 {\vec p}}{(2\pi)^3}
\displaystyle\frac{p^0}{2E_{\vec{p}}}e^{-iE_{\vec{p}}(t_2-t_1)}e^{i\vec{p}(\vec{x}_1-\vec{x}_2)}\vspace{2mm}\\
&&-i\displaystyle\frac{1}{\beta}\displaystyle\int\displaystyle\frac{d^3 {\vec p}}{(2\pi)^3}
\displaystyle\frac{p^0}{2E_{\vec{p}}}e^{iE_{\vec{p}}(t_1-t_2)}e^{-i\vec{p}.(\vec{x}_1-\vec{x}_2)}
\end{array}
\end{equation} Let us now turn to the r.h.s. of Eq. (\ref{FDT(0)}) and let us
remember  that $G_R$ solves the equation:
\begin{equation}
\left(\square+m^2\right)G_R=-\delta(x)
\end{equation} with $G_R(x)=0$ for $t<0$. Let us then choose the contour
$C_R$ of integration along the real axis in the complex p plane in such a manner that
the two poles are moved
below the real axis.
\includegraphics[scale=.90]{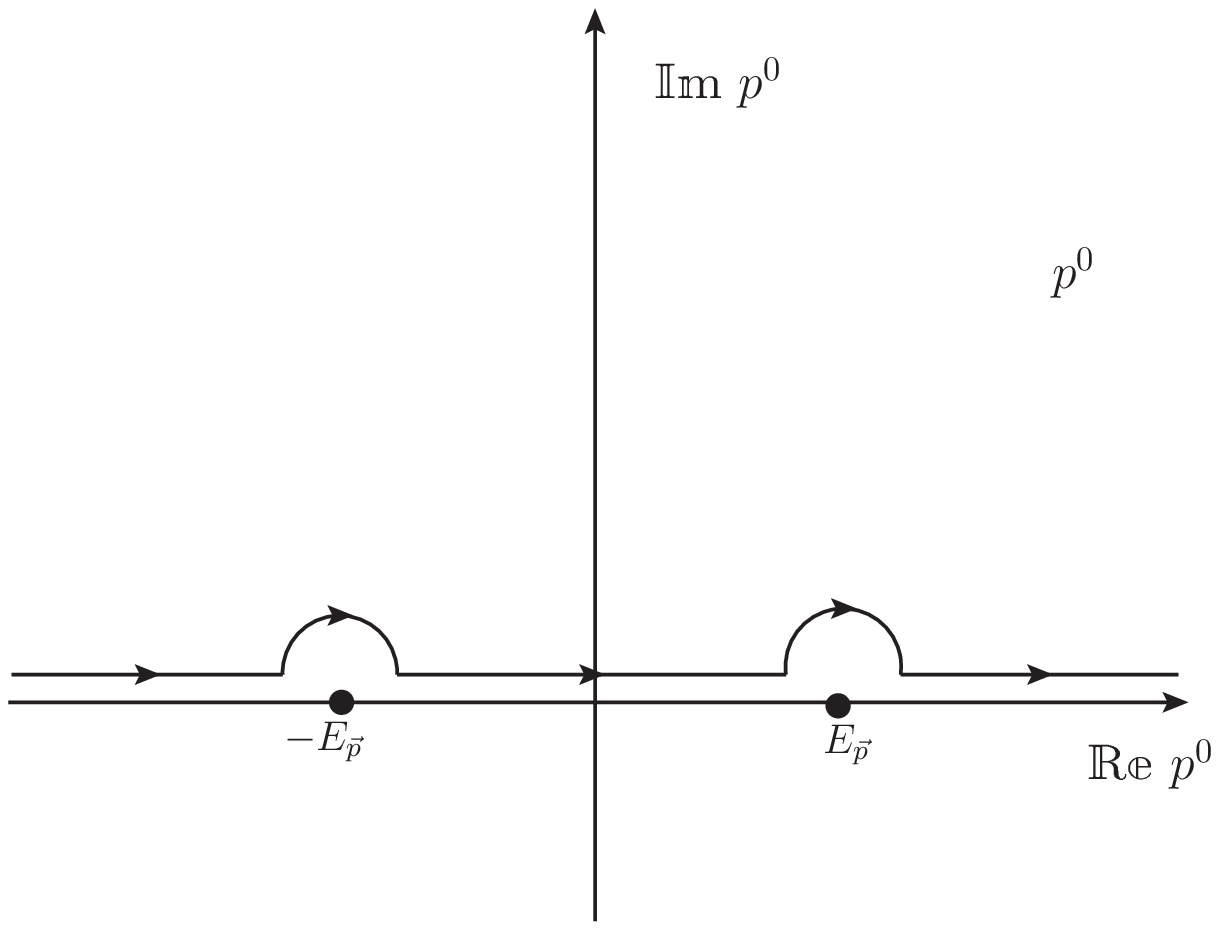}\\
For $t>0$ we close the contour in the lower
half plane:
\begin{equation}
G_R(x)=\displaystyle\int_{-\infty}^{\infty}
\frac{d^4p}{(2\pi)^4}\frac{e^{-ipx}}{\left(p^0+i\epsilon\right)^2-E_{\vec{p}}^2}
=\displaystyle\int_{C_R}
\frac{d^4p}{(2\pi)^4}\frac{e^{-ipx}}{\left(p^0+i\epsilon\right)^2-E_{\vec{p}}^2}.
\end{equation} So by the residue theorem we get:
\begin{equation}
G_R(x)=i\displaystyle\int_{-\infty}^{\infty}
\frac{d^3 \vec{p}}{(2\pi)^3}\frac{1}{2E_{\vec{p}}}\frac{e^{-ipx}}{\left(p^0+i\epsilon\right)^2-E_{\vec{p}}^2}
-i\displaystyle\int_{-\infty}^{\infty}
\frac{d^3 \vec{p}}{(2\pi)^3}\frac{1}{2E_{\vec{p}}}\frac{e^{-ipx}}{\left(p^0+i\epsilon\right)^2-E_{\vec{p}}^2}
\end{equation} and comparing with Eq. (\ref{partialtimedeltabeta2}) we obtain:
\begin{equation}
G_R(x)=G_R(x_1-x_2)=\beta \frac{\partial}{\partial t_1}\Delta_{\beta}(x_1-x_2).
\end{equation} For $t<0$ we get that $G_R(x)$ is zero, so:
\begin{equation}\label{G_R(x)}
G_R(x)=\beta \theta(t) \frac{\partial}{\partial t}\Delta_{\beta}(x).
\end{equation} In this relation  let us now turn  $x\rightarrow -x$ and
we get:
\begin{equation}
G_R(-x)=\beta \theta(-t) \frac{\partial}{\partial (-t)}\Delta_{\beta}(-x)
=\beta \theta(t) \frac{\partial}{\partial (-t)}\Delta_{\beta}(x)
=-\beta \theta(t) \frac{\partial}{\partial t}\Delta_{\beta}(x)
\end{equation} So for $t<0$ we have:
\begin{equation}\label{partialtimedeltabeta(x)}
\frac{\partial}{\partial t}\Delta_{\beta}(x)=
-\frac{1}{\beta}G_R(-x)\end{equation} Combining (\ref{G_R(x)}) with (\ref{partialtimedeltabeta(x)}) we can write:
\begin{equation}\label{partialdeltabeta=G_R(x)-G_R(-x)}
\frac{\partial}{\partial t}\Delta_{\beta}(x)=
\frac{1}{\beta}G_R(x)-\frac{1}{\beta}G_R(-x)
\end{equation} which is exactly the (\ref{FDT(0)}) or the FDT theorem at the zero
order that is what we wanted to prove.

We will now  move on to prove Eq. (\ref{FDT(1)}). Let us first note
that from (\ref{partialdeltabeta=G_R(x)-G_R(-x)}) we obtain:
\begin{equation}\label{G_R(x_1-x)}
G_R(x_1-x)=\beta \theta(x_1-x)\frac{\partial}{\partial t_1}\Delta_{\beta}(x_1-x).
\end{equation}

The first term on the l.h.s. of (\ref{FDT(1)}) has the following analytical expression:
\vspace{0.7cm}
\begin{equation}\label{1sttermFDT(1)}
\begin{array}{lll}
\displaystyle\frac{1}{2}\frac{\partial}{\partial t_1}\,\,\,\,\parbox{32mm}{\fmfreuse{DiagPag45.1}}\,\,
&=&\displaystyle\frac{1}{2}\frac{\partial}{\partial t_1}
\displaystyle \int
d^4x\Delta_{\beta}(x_1-x)
\Delta_{\beta}(x-x)
G_{R}^{(B)}(x_2-x)\vspace{2mm}\\
&=&\displaystyle\frac{1}{2}
\displaystyle \int
d^4x\displaystyle\frac{1}{\beta}\left[G_{R}^{(B)}(x_1-x)-G_{R}^{(B)}(x-x_1)\right]
\Delta_{\beta}(x-x)
G_{R}^{(B)}(x_2-x)
\end{array}
\end{equation} where we have made use of the relation (\ref{partialdeltabeta=G_R(x)-G_R(-x)}).

The second term on the l.h.s. of (\ref{FDT(1)}) has the following analytical expression:
\vspace{0.7cm}
\begin{equation}\label{2ndtermFDT(1)}
\displaystyle\frac{1}{2}\frac{\partial}{\partial t_1}\,\,\,\,\parbox{32mm}{\fmfreuse{DiagPag44.3YWithout}}\,\,
=\displaystyle\frac{1}{2}
\displaystyle \int
d^4x\displaystyle\frac{\partial}{\partial t_1}G_{R}^{(B)}(x_1-x)
\Delta_{\beta}(x-x)
\Delta_{\beta}(x-x_2).
\end{equation}

Let us now take the time derivative of (\ref{G_R(x_1-x)}):
\begin{equation}\label{partial t_1G_R(x_1-x)}
\displaystyle\frac{\partial}{\partial t_1}G_{R}^{(B)}(x_1-x)
=\beta \delta(t_1-t)
\displaystyle\frac{\partial}{\partial t_1}\Delta_{\beta}(x_1-x)
+\beta \theta(t_1-t)
\displaystyle\frac{\partial^2}{\partial t_1^2}\Delta_{\beta}(x_1-x).
\end{equation} Moreover  it is easy to prove that
\begin{equation}\label{delta partial delta_beta}
\delta(t_1-t)
\displaystyle\frac{\partial}{\partial t_1}\Delta_{\beta}(x_1-x)=0,
\end{equation} and therefore, (\ref{partial t_1G_R(x_1-x)}) becomes
\begin{equation}
\displaystyle\frac{\partial}{\partial t_1}G_{R}^{(B)}(x_1-x)=
\beta \theta(t_1-t)
\displaystyle\frac{\partial^2}{\partial t_1^2}\Delta_{\beta}(x_1-x).
\end{equation} As a consequence the (\ref{2ndtermFDT(1)}) is turned into:
\vspace{0.7cm}
\begin{equation}\label{2ndtermFDT(1)2}
\displaystyle\frac{1}{2}\frac{\partial}{\partial t_1}\,\,\,\,\parbox{32mm}{\fmfreuse{DiagPag44.3YWithout}}\,\,
=\displaystyle\frac{1}{2}\beta
\displaystyle \int
d^4x\displaystyle \theta(t_1-t)\frac{\partial^2}{\partial t_1^2}\Delta_{\beta}(x_1-x)
\Delta_{\beta}(x-x)
\Delta_{\beta}(x-x_2).
\end{equation} Performing now an integration by parts and disregarding surface terms
Eq. (\ref{2ndtermFDT(1)2}) can be rewritten as
\vspace{0.7cm}
\begin{equation}
\begin{array}{lll}
\displaystyle\frac{1}{2}\frac{\partial}{\partial t_1}\,\,\,\,\parbox{32mm}{\fmfreuse{DiagPag44.3YWithout}}\,\,
&=&-\displaystyle\frac{1}{2}\beta
\displaystyle \int
d^4x\displaystyle \delta(t_1-t)\Delta_{\beta\;t_1}(x_1-x)
\Delta_{\beta}(x-x)
\Delta_{\beta}(x_2-x)\vspace{2mm}\\
&&-\displaystyle\frac{1}{2}\beta
\displaystyle \int
d^4x\displaystyle \theta(t_1-t)\Delta_{\beta\;t_1}(x_1-x)
\Delta_{\beta}(x-x)
\Delta_{\beta\;t_2}(x_2-x)
\end{array}
\end{equation} where $\Delta_{\beta\;t_i}(x_i-x)\equiv \partial \Delta_{\beta}(x_i-x)/\partial t_i$.
Using (\ref{delta partial delta_beta}) we get:
\vspace{0.5cm}
\begin{equation}
\displaystyle\frac{1}{2}\frac{\partial}{\partial t_1}\,\,\,\,\parbox{32mm}{\fmfreuse{DiagPag44.3YWithout}}\,\,
=-\displaystyle\frac{1}{2}\beta
\displaystyle \int
d^4x\displaystyle \theta(t_1-t)\Delta_{\beta\;t_1}(x_1-x)
\Delta_{\beta}(x-x)
\Delta_{\beta\;t_2}(x_2-x).
\end{equation} If on the r.h.s. of the equation above we use
(\ref{partialdeltabeta=G_R(x)-G_R(-x)}), we obtain:
\vspace{0.5cm}
\begin{equation}\label{2ndtermFDT(1)3}
\displaystyle\frac{1}{2}\frac{\partial}{\partial t_1}\,\,\,\,\parbox{32mm}{\fmfreuse{DiagPag44.3YWithout}}\,\,
=-\displaystyle\frac{1}{2}\frac{1}{\beta}
\displaystyle \int
d^4x\displaystyle G_{R}^{(B)}(x_1-x)\Delta_{\beta}(x-x)
\left[G_{R}^{(B)}(x_2-x)-G_{R}^{(B)}(x-x_2)\right].
\end{equation} Combining (\ref{1sttermFDT(1)}) with (\ref{2ndtermFDT(1)3}) we get:
\vspace{0.5cm}
\begin{equation}
\begin{array}{lll}
\displaystyle\frac{1}{2}\frac{\partial}{\partial t_1}\,\,\,\,\parbox{32mm}{\fmfreuse{DiagPag45.1}}\,\,
&+&\displaystyle\frac{1}{2}\frac{\partial}{\partial t_1}\,\,\,\,\parbox{32mm}{\fmfreuse{DiagPag44.3YWithout}}\,\,=\vspace{0.2cm}\\
&=&\displaystyle\frac{1}{2}\frac{1}{\beta}
\displaystyle \int
d^4xG_{R}^{(B)}(x_1-x)\Delta_{\beta}(x-x)G_{R}^{(B)}(x-x_2)\vspace{4mm}\\
&&-\displaystyle\frac{1}{2}\frac{1}{\beta}
\displaystyle \int
d^4xG_{R}^{(B)}(x-x_1)\Delta_{\beta}(x-x)G_{R}^{(B)}(x_2-x)\vspace{7mm}\\
&=&\Bigg[\displaystyle\frac{i}{2\,\beta}\,\,\,\,\parbox{32mm}{\fmfreuse{DiagPag61.1}}
\,\,- \displaystyle\frac{i}{2\,\beta}\,\,\,\,\parbox{32mm}{\fmfreuse{DiagPag75.4}}\quad\Bigg].
\end{array}
\end{equation} This is exactly the relation (\ref{FDT(1)}) which is the FDT of the first
order in perturbation theory.

Long but tedious calculations allows us to derive the FDT also at the
second order in perturbation theory which is diagrammatically
represented by the the three formulas (\ref{FDT(2)1}), (\ref{FDT(2)2})
and (\ref{FDT(2)3}).

For the first order contribution to the FDT an alternative and
simpler derivation than the one presented here can be given either
using the momentum representation or the super-field formalism. We
will present them below:

{\bf Momentum space derivation}: Eq. (\ref{FDT(0)}), which is the FDT at
zero order, is:
\begin{equation}
\frac{\partial}{\partial t_1}\Delta_{\beta}(x_1-x_2)=
\frac{i}{\beta}\left[-iG_R(x_1-x_2)+iG_R(x_2-x_1)\right].
\end{equation} Going into momentum space we get:
\begin{equation}
\begin{array}{lll}
&&\frac{\partial}{\partial t_1}\displaystyle\int\displaystyle\frac{d^4p}{(2\pi)^4}
e^{-ip.(x_1-x_2)}\Delta_{\beta}(p)\vspace{2mm}\\
&&=
\displaystyle\frac{i}{\beta}\left\{\displaystyle\int\displaystyle\frac{d^4p}{(2\pi)^4}
e^{-ip.(x_1-x_2)}\left[-iG_R(p)\right]+\displaystyle\int\displaystyle\frac{d^4p}{(2\pi)^4}
e^{-ip.(x_1-x_2)}\left[iG_R(-p)\right]\right\},
\end{array}
\end{equation} or equivalently:
\begin{equation}
\begin{array}{lll}
&&\displaystyle\int\displaystyle\frac{d^4p}{(2\pi)^4}
\left[-ip^0e^{-ip.(x_1-x_2)}\right]\Delta_{\beta}(p)\vspace{2mm}\\
&&=
\displaystyle\frac{i}{\beta}\left\{\displaystyle\int\displaystyle\frac{d^4p}{(2\pi)^4}
e^{-ip.(x_1-x_2)}\left[-iG_R(p)\right]+\displaystyle\int\displaystyle\frac{d^4p}{(2\pi)^4}
e^{-ip.(x_1-x_2)}\left[iG_A(p)\right]\right\},
\end{array}
\end{equation} where $G_A$ is the advanced propagator. Formally we shall write the relation
above as:
\begin{equation}\label{FDT(0)MomentumSpace}
-ip^0\Delta_{\beta}(p)=
\displaystyle\frac{i}{\beta}\left[-iG_R(p)+iG_A(p)\right],
\end{equation} which diagrammatically is:
\begin{equation}\label{FDT(A)MomentumSpace}
-i\,p_0\,\,\,\,\begin{minipage}[c]{16mm}
\vspace{0.55cm}\begin{fmffile}{DiagPagG9.1}
\begin{fmfgraph*}(20,1)
\fmfkeep{DiagPagG9.1}
\fmfleft{i1}
\fmfright{o1}
\fmfdot{i1}
\fmfdot{o1}
\fmf{plain,label=$p$, label.side=left}{i1,o1}
\fmfv{label=$x_1$,label.angle=-90}{i1}
\fmfv{label=$x_2$,label.angle=-90}{o1}
\end{fmfgraph*}
\end{fmffile}
\end{minipage}\,\,\quad=\,\,\frac{i}{\beta}\quad\begin{minipage}[c]{16mm}
\vspace{0.6cm}\begin{fmffile}{DiagPagG9.2}
\begin{fmfgraph*}(16,1)
\fmfkeep{DiagPagG9.2}
\fmfleft{i1}
\fmfright{o1}
\fmfdot{i1}
\fmfdot{o1}
\fmf{phantom,tag=1,label=$p$,label.side=left}{i1,o1}
\fmfposition
\fmfipath{p[]}
\fmfiset{p1}{vpath1(__i1,__o1)}
\fmfi{plain}{subpath (0,length(p1)*0.40) of p1}
\fmfi{dashes}{subpath (length(p1)*0.40,length(p1)) of p1}
\fmfv{label=$x_1$,label.angle=-90}{i1}
\fmfv{label=$x_2$,label.angle=-90}{o1}
\end{fmfgraph*}
\end{fmffile}
\end{minipage}\,\,\,-\frac{i}{\beta}\quad
\begin{minipage}[c]{16mm}
\vspace{0.6cm}\begin{fmffile}{DiagPagG9.3}
\begin{fmfgraph*}(16,1)
\fmfkeep{DiagPagG9.3}
\fmfleft{i1}
\fmfright{o1}
\fmfdot{i1}
\fmfdot{o1}
\fmf{phantom,tag=1,label=$p$,label.side=left}{i1,o1}
\fmfposition
\fmfipath{p[]}
\fmfiset{p1}{vpath1(__i1,__o1)}
\fmfi{dashes}{subpath (0,length(p1)*0.5) of p1}
\fmfi{plain}{subpath (length(p1)*0.5,length(p1)) of p1}
\fmfv{label=$x_1$,label.angle=-90}{i1}
\fmfv{label=$x_2$,label.angle=-90}{o1}
\end{fmfgraph*}
\end{fmffile}
\end{minipage}.
\end{equation}
 The FDT at first order (\ref{FDT(1)}) is given by:
\vspace{0.5cm}
\begin{equation}\label{FDT(1)CoordinateSpace}
\frac{1}{2!}\frac{\partial}{\partial t_1}\left[\quad\parbox{32mm}{\fmfreuse{DiagPag45.1}}\quad+\quad\parbox{32mm}{\fmfreuse{DiagPag44.3YWithout}}\quad\right]=
\end{equation}
\vspace{0.6cm}
\begin{equation}\nonumber
\hspace{-0.3cm}=\frac{1}{2!}\,\frac{i}{\beta}\Bigg[\quad\parbox{32mm}{\fmfreuse{DiagPag61.1}}\quad-\quad\parbox{32mm}{\fmfreuse{DiagPag75.4}}\quad\Bigg]
\end{equation}
The l.h.s. of (\ref{FDT(1)CoordinateSpace}) in momentum space is:
\vspace{0.55cm}
\begin{equation}\label{LHSFDT(1)MomentumSpace}
\displaystyle\frac{1}{2!}\left(-ip^0\right)\Bigg[\,\,\,\begin{minipage}[c]{32mm}
\vspace{0.55cm}
\begin{fmffile}{DiagPagG10.1}
\begin{fmfgraph*}(32,10)
\fmfkeep{DiagPagG10.1}
\fmfleft{i1}
\fmfright{o1}
\fmfdot{i1,o1}
\fmf{phantom}{i1,v,o1}
\fmf{phantom,tag=2,label=$p$,label.side=left}{v,o1}
\fmfdot{v}
\fmf{phantom,tag=1,label=$p$,label.side=left}{i1,v}
\fmfposition
\fmfipath{p[]}
\fmfiset{p1}{vpath1(__i1,__v)}
\fmfi{plain}{subpath (0,length(p1)*0.40.) of p1}
\fmfi{plain}{subpath (length(p1)*0.4,length(p1)) of p1}
\fmf{plain,tension=0.9,label=$q$}{v,v}
\fmfiset{p2}{vpath2(__v,__o1)}
\fmfi{dashes}{subpath (0,length(p2)*0.60.) of p2}
\fmfi{plain}{subpath (length(p2)*0.6,length(p2)) of p2}
\end{fmfgraph*}
\end{fmffile}
\end{minipage}\,\,\,+\,\,\,
\begin{minipage}[c]{32mm}
\vspace{0.55cm}
\begin{fmffile}{DiagPagG10.2}
\begin{fmfgraph*}(32,10)
\fmfkeep{DiagPagG10.2}
\fmfleft{i1}
\fmfright{o1}
\fmfdot{i1,o1}
\fmf{phantom}{i1,v,o1}
\fmf{plain,label=$p$,label.side=left}{v,o1}
\fmfdot{v}
\fmf{phantom,tag=1,label=$p$,label.side=left}{i1,v}
\fmfposition
\fmfipath{p[]}
\fmfiset{p1}{vpath1(__i1,__v)}
\fmfi{plain}{subpath (0,length(p1)*0.40.) of p1}
\fmfi{dashes}{subpath (length(p1)*0.4,length(p1)) of p1}
\fmf{plain,tension=0.9,label=$q$}{v,v}
\end{fmfgraph*}
\end{fmffile}
\end{minipage}\,\,\,\,\,\,
\Bigg].
\end{equation}
Let us now make use of (\ref{FDT(0)MomentumSpace}) in the
following manner: let us replace the first leg of the first
diagram in (\ref{LHSFDT(1)MomentumSpace}) using the relation
(\ref{FDT(0)MomentumSpace}) and the same for the last leg of the
second diagram in (\ref{LHSFDT(1)MomentumSpace}). Therefore, we
get:
\vspace{1cm}
\begin{equation}
\begin{array}{lll}
(\ref{LHSFDT(1)MomentumSpace})
&=&\displaystyle\frac{1}{2!}\Bigg[\frac{i}{\beta}\quad
\begin{minipage}[c]{32mm}
\vspace{0.55cm}
\begin{fmffile}{DiagPagG10.3}
\begin{fmfgraph*}(32,10)
\fmfkeep{DiagPagG10.3}
\fmfleft{i1}
\fmfright{o1}
\fmfdot{i1,o1}
\fmf{phantom}{i1,v,o1}
\fmf{phantom,tag=2,label=$p$,label.side=left}{v,o1}
\fmfdot{v}
\fmf{phantom,tag=1,label=$p$,label.side=left}{i1,v}
\fmfposition
\fmfipath{p[]}
\fmfiset{p1}{vpath1(__i1,__v)}
\fmfi{plain}{subpath (0,length(p1)*0.40.) of p1}
\fmfi{dashes}{subpath (length(p1)*0.4,length(p1)) of p1}
\fmf{plain,tension=0.9,label=$q$}{v,v}
\fmfiset{p2}{vpath2(__v,__o1)}
\fmfi{dashes}{subpath (0,length(p2)*0.60.) of p2}
\fmfi{plain}{subpath (length(p2)*0.6,length(p2)) of p2}
\end{fmfgraph*}
\end{fmffile}
\end{minipage}\,\,\,
-\displaystyle\frac{i}{\beta}\quad
\begin{minipage}[c]{32mm}
\vspace{0.55cm}
\begin{fmffile}{DiagPagG10.4}
\begin{fmfgraph*}(32,10)
\fmfkeep{DiagPagG10.4}
\fmfleft{i1}
\fmfright{o1}
\fmfdot{i1,o1}
\fmf{phantom}{i1,v,o1}
\fmf{phantom,tag=2,label=$p$,label.side=left}{v,o1}
\fmfdot{v}
\fmf{phantom,tag=1,label=$p$,label.side=left}{i1,v}
\fmfposition
\fmfipath{p[]}
\fmfiset{p1}{vpath1(__i1,__v)}
\fmfi{dashes}{subpath (0,length(p1)*0.60) of p1}
\fmfi{plain}{subpath (length(p1)*0.6,length(p1)) of p1}
\fmf{plain,tension=0.9,label=$q$}{v,v}
\fmfiset{p2}{vpath2(__v,__o1)}
\fmfi{dashes}{subpath (0,length(p2)*0.60.) of p2}
\fmfi{plain}{subpath (length(p2)*0.6,length(p2)) of p2}
\end{fmfgraph*}
\end{fmffile}
\end{minipage}\\
&+&
\hspace{0.6cm}\displaystyle\frac{i}{\beta}\quad
\begin{minipage}[c]{32mm}
\vspace{0.55cm}
\begin{fmffile}{DiagPagG10.5}
\begin{fmfgraph*}(32,10)
\fmfkeep{DiagPagG10.5}
\fmfleft{i1}
\fmfright{o1}
\fmfdot{i1,o1}
\fmf{phantom}{i1,v,o1}
\fmf{phantom,tag=2,label=$p$,label.side=left}{v,o1}
\fmfdot{v}
\fmf{phantom,tag=1,label=$p$,label.side=left}{i1,v}
\fmfposition
\fmfipath{p[]}
\fmfiset{p1}{vpath1(__i1,__v)}
\fmfi{plain}{subpath (0,length(p1)*0.40) of p1}
\fmfi{dashes}{subpath (length(p1)*0.4,length(p1)) of p1}
\fmf{plain,tension=0.9,label=$q$}{v,v}
\fmfiset{p2}{vpath2(__v,__o1)}
\fmfi{plain}{subpath (0,length(p2)*0.40.) of p2}
\fmfi{dashes}{subpath (length(p2)*0.4,length(p2)) of p2}
\end{fmfgraph*}
\end{fmffile}
\end{minipage}\,\,\,
-\displaystyle\frac{i}{\beta}\quad
\begin{minipage}[c]{32mm}
\vspace{0.55cm}
\begin{fmffile}{DiagPagG10.6}
\begin{fmfgraph*}(32,10)
\fmfkeep{DiagPagG10.6}
\fmfleft{i1}
\fmfright{o1}
\fmfdot{i1,o1}
\fmf{phantom}{i1,v,o1}
\fmf{phantom,tag=2,label=$p$,label.side=left}{v,o1}
\fmfdot{v}
\fmf{phantom,tag=1,label=$p$,label.side=left}{i1,v}
\fmfposition
\fmfipath{p[]}
\fmfiset{p1}{vpath1(__i1,__v)}
\fmfi{plain}{subpath (0,length(p1)*0.40) of p1}
\fmfi{dashes}{subpath (length(p1)*0.4,length(p1)) of p1}
\fmf{plain,tension=0.9,label=$q$}{v,v}
\fmfiset{p2}{vpath2(__v,__o1)}
\fmfi{dashes}{subpath (0,length(p2)*0.60.) of p2}
\fmfi{plain}{subpath (length(p2)*0.6,length(p2)) of p2}
\end{fmfgraph*}
\end{fmffile}
\end{minipage}\quad\Bigg]\vspace{0.5cm}\\
&=&\hspace{0.5cm}\cfrac{i}{\beta}\,\,\,\Bigg[\,\,\,\parbox{32mm}{\fmfreuse{DiagPagG10.5}}\,\,\,-\,\,\,\parbox{32mm}{\fmfreuse{DiagPagG10.4}}\,\,\,\Bigg]
\end{array}
\end{equation} and this is exactly the momentum space
representation of (\ref{FDT(1)CoordinateSpace}).

One can notice how much simpler is this derivation than the one
in configuration space that we presented earlier.

{\bf Super-Diagram Derivation}:

Let us remember that:
\vspace{0.5cm}
\begin{equation}\label{<phi phi>(1)2}
\langle \phi(x_1) \phi(x_2)\rangle^{(1)}
=\displaystyle\frac{1}{2}
\displaystyle\int
d\nu_1d\nu_2\,\,\,\parbox{32mm}{\fmfreuse{DiagPag53.1}}
\end{equation} where we have used the notation $d \nu_i \equiv d\theta_i d\bar{\theta}_i
\bar{\theta}_i \theta_i$ ($i=1,2$) and the upper index $(1)$ on
the l.h.s. stays for first order in perturbation theory.

It is then easy to prove that:
\begin{equation}\label{<int dnu superprop>}
\begin{array}{lll}
\Bigg\langle \displaystyle\int d\nu_1\,\,\,
\begin{minipage}[c]{16mm}
\vspace{0.55cm}\begin{fmffile}{DiagPagG12.1}
\begin{fmfgraph*}(16,1)
\fmfkeep{DiagPagG12.1}
\fmfleft{i1}
\fmfright{o1}
\fmfdot{i1}
\fmfdot{o1}
\fmf{dbl_plain}{i1,o1}
\fmfv{label=$z_1$,label.angle=-90}{i1}
\fmfv{label=$z_2$,label.angle=-90}{o1}
\end{fmfgraph*}
\end{fmffile}
\end{minipage}
\hspace{0cm}\quad\Bigg\rangle_{\beta}
&=&\,\,\,\,\parbox{32mm}{\fmfreuse{DiagPag44.1}}\hspace{-1.3cm}+\,\,\,\parbox{32mm}{\fmfreuse{DiagPag25.3}}\hspace{-1.2cm}\bar{\theta}_2 \theta_2\vspace{2mm}\\
&=&\displaystyle\int d\nu_1
\left[\frac{\partial }{\partial \theta_2}
\frac{\partial }{\partial \bar{\theta}_2}\left(\bar{\theta}_2 \theta_2\,\,\parbox{16mm}{\fmfreuse{DiagPagG12.1}}\,\,\,\,\right)
+\bar{\theta}_2 \theta_2\frac{\partial }{\partial \theta_2}
\frac{\partial }{\partial \bar{\theta}_2}\left(\,\,\parbox{16mm}{\fmfreuse{DiagPagG12.1}}\,\,\,\right)\right].
\end{array}
\end{equation} The first equality above derives from the fact
that in $x_1$ only the field $\phi$ survives the integration over
$d \nu_1$. At the point $z_2$ in principle there could be any of the fields
$\phi$, $\lambda$, $c$, $\bar{c}$, but only the correlations
$\langle \phi \phi\rangle$ and $\langle \phi \lambda\rangle$
survive because the $\langle \phi c\rangle$, $\langle \phi
\bar{c}\rangle$ are zero as we proved before in the paper. The
second equality in (\ref{<int dnu superprop>}) is easy to prove:
the first term is exactly equal to the first term of the first
equality. In fact, in $z_2$ only $\phi$ survives which is the only
term which does not carry extra $\theta_2$ and in $z_1$ only
$\phi$ survives because of the integration $d \nu_1$. An analogous
reasoning applies to the second term.

Let us now go back to (\ref{<phi phi>(1)2}) and do the time
derivative of both, the l.h.s. and r.h.s.
\begin{equation}\label{partial <phi phi>}
\displaystyle\frac{\partial}{\partial t_1}
\langle \phi(x_1) \phi(x_2)\rangle^{(1)}
=\displaystyle\frac{1}{2}
\displaystyle \int d\nu_1 d\nu_2\quad
\begin{minipage}[c]{32mm}
\vspace{0.55cm}
\begin{fmffile}{DiagPagG13.1}
\begin{fmfgraph*}(32,10)
\fmfset{arrow_len}{2.5mm}
\fmfkeep{DiagPagG13.1}
\fmfleft{i1}
\fmfright{o1}
\fmfdot{i1,o1}
\fmfv{label=$z_1$,label.angle=-90}{i1}
\fmfv{label=$z_2$,label.angle=-90}{o1}
\fmf{phantom}{i1,v,o1}
\fmf{dbl_plain}{v,o1}
\fmf{dbl_plain_arrow}{i1,v}
\fmfdot{v}
\fmfv{label=$z$,label.angle=-90}{v}
\fmfposition
\fmfipath{p[]}
\fmf{dbl_plain,tension=0.9}{v,v}
\end{fmfgraph*}
\end{fmffile}
\end{minipage}
\end{equation} where with the arrow
 \,\,
\begin{minipage}[c]{16mm}
\vspace{0.55cm}\begin{fmffile}{DiagPagG13.2}
\begin{fmfgraph*}(16,1)
\fmfkeep{DiagPagG13.2}
\fmfset{arrow_len}{2.5mm}
\fmfleft{i1}
\fmfright{o1}
\fmfdot{i1}
\fmfdot{o1}
\fmf{dbl_plain_arrow}{i1,o1}
\end{fmfgraph*}
\end{fmffile}
\end{minipage}\,\,
we indicate
the time derivative. We can formally write the r.h.s. of
(\ref{partial <phi phi>}) as
\begin{equation}
\displaystyle\frac{1}{2}
\displaystyle \int d\nu_1 d\nu_2\,\,\,\parbox{32mm}{\fmfreuse{DiagPagG13.1}}\,\,\,
=\displaystyle\frac{1}{2}
\displaystyle \int d\nu_1 d\nu_2
\left(\,\,
\begin{minipage}[c]{16mm}
\vspace{0.55cm}\begin{fmffile}{DiagPagG13.3}
\begin{fmfgraph*}(16,1)
\fmfkeep{DiagPagG13.3}
\fmfset{arrow_len}{2.5mm}
\fmfleft{i1}
\fmfright{o1}
\fmfdot{i1}
\fmfdot{o1}
\fmfv{label=$z_1$,label.angle=-90}{i1}
\fmfv{label=$z$,label.angle=-90}{o1}
\fmf{dbl_plain_arrow}{i1,o1}
\end{fmfgraph*}
\end{fmffile}
\end{minipage}\,\,
\right)\hspace{-1.2cm}
\begin{minipage}[c]{32mm}
\vspace{0.55cm}
\begin{fmffile}{DiagPagG13.4}
\begin{fmfgraph*}(32,10)
\fmfset{arrow_len}{2.5mm}
\fmfkeep{DiagPagG13.4}
\fmfleft{i1}
\fmfright{o1}
\fmfdot{o1}
\fmfv{label=$z_2$,label.angle=-90}{o1}
\fmf{phantom}{i1,v,o1}
\fmf{dbl_plain}{v,o1}
\fmf{phantom}{i1,v}
\fmfdot{v}
\fmfv{label=$z$,label.angle=-90}{v}
\fmfposition
\fmfipath{p[]}
\fmf{dbl_plain,tension=0.9}{v,v}
\end{fmfgraph*}
\end{fmffile}
\end{minipage}.
\end{equation}
Using the relation (\ref{<int dnu superprop>}) on the first
term of the r.h.s. above, we get:
\begin{equation}\label{partial <phi phi>(1)}
\displaystyle\frac{\partial}{\partial t_1}
\langle \phi(x_1) \phi(x_2)\rangle^{(1)}=\displaystyle\frac{1}{2}
\displaystyle \int d\nu_1 d\nu_2
\left[\frac{\partial }{\partial \theta}
\frac{\partial }{\partial \bar{\theta}}\left(\bar{\theta} \theta\,\,\,\parbox{16mm}{\fmfreuse{DiagPagG13.3}}\,\,\,\right)\right.
\end{equation}
\vspace{0.5cm}
\begin{equation}\nonumber
\hspace{1.cm}\left.+\hspace{0cm}\bar{\theta} \theta\frac{\partial}{\partial \theta}
\frac{\partial }{\partial \bar{\theta}}\left(\,\,\,\parbox{16mm}{\fmfreuse{DiagPagG13.3}}\,\,\,\right)\right]\hspace{-1.2cm}\parbox{16mm}{\fmfreuse{DiagPagG13.4}}.
\end{equation} Let us analyze the two terms in the square brackets
on the r.h.s. of (\ref{partial <phi phi>(1)}).

The first term is equal to
\begin{equation}\label{RHSpartial<phiphi>(1)1}
\begin{array}{lll}
&&\displaystyle\frac{1}{2}
\displaystyle \int d\nu_1 d\nu_2
\left[\frac{\partial }{\partial \theta}
\frac{\partial }{\partial \bar{\theta}}\left(\bar{\theta} \theta\,\,\,\parbox{16mm}{\fmfreuse{DiagPagG13.3}}\,\,\,\right)\right]\hspace{-1.2cm}\parbox{16mm}{\fmfreuse{DiagPagG13.4}}
\vspace{6mm}\\
&&=\displaystyle\frac{1}{2}
\displaystyle \int d\nu_1 d\nu_2\left[\bar{\theta} \theta
\left(\,\,\parbox{16mm}{\fmfreuse{DiagPagG13.3}}\,\,\right)\right]\left[\frac{\partial }{\partial \theta}
\frac{\partial }{\partial \bar{\theta}}\Bigg(\,\,\,
\hspace{-1.2cm}\parbox{16mm}{\fmfreuse{DiagPagG13.4}}\hspace{1.8cm}\Bigg)\right]
\vspace{2mm}\\
&&= \displaystyle\frac{1}{2}\,\,\,\begin{minipage}[c]{32mm}
\vspace{0.55cm}
\begin{fmffile}{DiagPagG14.1}
\begin{fmfgraph*}(32,10)
\fmfkeep{DiagPagG14.1}
\fmfset{arrow_len}{2.5mm}
\fmfleft{i1}
\fmfright{o1}
\fmfdot{i1,o1}
\fmfv{label=$x_1$,label.angle=-90}{i1}
\fmfv{label=$x_2$,label.angle=-90}{o1}
\fmf{phantom}{i1,v,o1}
\fmf{phantom,tag=2}{v,o1}
\fmfdot{v}
\fmf{plain_arrow,tag=1}{i1,v}
\fmfposition
\fmfipath{p[]}
\fmf{plain,tension=0.9}{v,v}
\fmfiset{p2}{vpath2(__v,__o1)}
\fmfi{dashes}{subpath (0,length(p2)*0.60.) of p2}
\fmfi{plain}{subpath (length(p2)*0.6,length(p2)) of p2}
\end{fmfgraph*}
\end{fmffile}
\end{minipage}
\end{array}
\end{equation}
where we have used integration by parts over Grassmann variables
to obtain the first equality and we have performed the integration to
obtain the second one.

Similarly, the second term in the square brackets on the r.h.s. of
(\ref{partial <phi phi>(1)}) gives
\vspace{0.5cm}
\begin{equation}\label{RHSpartial<phiphi>(1)2}
\begin{array}{lll}
&&\displaystyle\frac{1}{2}
\displaystyle \int d\nu_1 d\nu_2
\left[\bar{\theta} \theta\frac{\partial }{\partial \theta}
\displaystyle\frac{\partial }{\partial \bar{\theta}}\left(\,\,\,\parbox{16mm}{\fmfreuse{DiagPagG13.3}}\,\,\,\right)\right]
\hspace{-1.2cm}\parbox{16mm}{\fmfreuse{DiagPagG13.4}}
\vspace{6mm}\\
&&=\displaystyle\frac{1}{2}
\displaystyle \int d\nu_1 d\nu_2
\left(\,\,\,\parbox{16mm}{\fmfreuse{DiagPagG13.3}}\,\,\,\right)\left\{\frac{\partial }{\partial \theta}
\displaystyle\frac{\partial }{\partial \bar{\theta}}
\left[\bar{\theta} \theta\left(\,\,\hspace{-1.2cm}\parbox{16mm}{\fmfreuse{DiagPagG13.4}}\hspace{2cm}\right)\right]\right\}\vspace{2mm}\\
&&=\displaystyle\frac{1}{2}\quad
\begin{minipage}[c]{32mm}
\vspace{0.55cm}
\begin{fmffile}{DiagPagG15.1}
\begin{fmfgraph*}(32,10)
\fmfkeep{DiagPagG15.1}
\fmfset{arrow_len}{2.5mm}
\fmfleft{i1}
\fmfright{o1}
\fmfdot{i1,o1}
\fmfv{label=$x_1$,label.angle=-90}{i1}
\fmfv{label=$x_2$,label.angle=-90}{o1}
\fmf{phantom}{i1,v,o1}
\fmf{plain_arrow}{v,o1}
\fmfdot{v}
\fmf{phantom,tag=1}{i1,v}
\fmfposition
\fmfipath{p[]}
\fmfiset{p1}{vpath1(__i1,__v)}
\fmfi{plain}{subpath (0,length(p1)*0.40.) of p1}
\fmfi{dashes}{subpath (length(p1)*0.4,length(p1)) of p1}
\fmf{plain,tension=0.9}{v,v}
\end{fmfgraph*}
\end{fmffile}
\end{minipage}.
\end{array}
\end{equation} Replacing now in (\ref{RHSpartial<phiphi>(1)1})
and (\ref{RHSpartial<phiphi>(1)2}) the continuous line with the
arrow (which stays for the time derivative) with the difference of the
dash-full line, like it is indicated in the FDT at zero order
(\ref{FDT(0)}) , we get that:
\begin{equation}
(\ref{RHSpartial<phiphi>(1)1})=\Bigg[
\displaystyle\frac{i}{2\,\beta}\quad
\begin{minipage}[c]{32mm}
\vspace{0.55cm}
\begin{fmffile}{DiagPagG15.2}
\begin{fmfgraph*}(32,10)
\fmfkeep{DiagPagG15.2}
\fmfleft{i1}
\fmfright{o1}
\fmfdot{i1,o1}
\fmf{phantom}{i1,v,o1}
\fmf{phantom,tag=2}{v,o1}
\fmfdot{v}
\fmf{phantom,tag=1}{i1,v}
\fmfposition
\fmfipath{p[]}
\fmfiset{p1}{vpath1(__i1,__v)}
\fmfi{plain}{subpath (0,length(p1)*0.40.) of p1}
\fmfi{dashes}{subpath (length(p1)*0.4,length(p1)) of p1}
\fmf{plain,tension=0.9}{v,v}
\fmfiset{p2}{vpath2(__v,__o1)}
\fmfi{dashes}{subpath (0,length(p2)*0.60.) of p2}
\fmfi{plain}{subpath (length(p2)*0.6,length(p2)) of p2}
\fmfv{label=$x_1$,label.angle=-90}{i1}
\fmfv{label=$x_2$,label.angle=-90}{o1}
\end{fmfgraph*}
\end{fmffile}
\end{minipage}\,\,\,
-\displaystyle\frac{i}{2\,\beta}\quad
\begin{minipage}[c]{32mm}
\vspace{0.55cm}
\begin{fmffile}{DiagPagG15.3}
\begin{fmfgraph*}(32,10)
\fmfkeep{DiagPagG15.3}
\fmfleft{i1}
\fmfright{o1}
\fmfdot{i1,o1}
\fmfv{label=$x_1$,label.angle=-90}{i1}
\fmfv{label=$x_2$,label.angle=-90}{o1}
\fmf{phantom}{i1,v,o1}
\fmf{phantom,tag=2}{v,o1}
\fmfdot{v}
\fmf{phantom,tag=1}{i1,v}
\fmfposition
\fmfipath{p[]}
\fmfiset{p1}{vpath1(__i1,__v)}
\fmfi{dashes}{subpath (0,length(p1)*0.60.) of p1}
\fmfi{plain}{subpath (length(p1)*0.6,length(p1)) of p1}
\fmfiset{p2}{vpath2(__v,__o1)}
\fmfi{dashes}{subpath (0,length(p2)*0.60.) of p2}
\fmfi{plain}{subpath (length(p2)*0.6,length(p2)) of p2}
\fmf{plain,tension=0.9}{v,v}
\end{fmfgraph*}
\end{fmffile}
\end{minipage}\,\,\,\Bigg]
\end{equation}
\begin{equation}
(\ref{RHSpartial<phiphi>(1)2})=\Bigg[
\displaystyle\frac{i}{2\,\beta}\quad
\begin{minipage}[c]{32mm}
\vspace{0.55cm}
\begin{fmffile}{DiagPagG15.4}
\begin{fmfgraph*}(32,10)
\fmfkeep{DiagPagG15.4}
\fmfleft{i1}
\fmfright{o1}
\fmfdot{i1,o1}
\fmf{phantom}{i1,v,o1}
\fmf{phantom,tag=2}{v,o1}
\fmfdot{v}
\fmf{phantom,tag=1}{i1,v}
\fmfposition
\fmfipath{p[]}
\fmfiset{p1}{vpath1(__i1,__v)}
\fmfi{plain}{subpath (0,length(p1)*0.40) of p1}
\fmfi{dashes}{subpath (length(p1)*0.4,length(p1)) of p1}
\fmf{plain,tension=0.9}{v,v}
\fmfiset{p2}{vpath2(__v,__o1)}
\fmfi{plain}{subpath (0,length(p2)*0.40.) of p2}
\fmfi{dashes}{subpath (length(p2)*0.4,length(p2)) of p2}
\fmfv{label=$x_1$,label.angle=-90}{i1}
\fmfv{label=$x_2$,label.angle=-90}{o1}
\end{fmfgraph*}
\end{fmffile}
\end{minipage}
\,\,-\displaystyle\frac{i}{2\,\beta}\quad
\begin{minipage}[c]{32mm}
\vspace{0.55cm}
\begin{fmffile}{DiagPagG15.5}
\begin{fmfgraph*}(32,10)
\fmfkeep{DiagPagG15.5}
\fmfleft{i1}
\fmfright{o1}
\fmfdot{i1,o1}
\fmf{phantom}{i1,v,o1}
\fmf{phantom,tag=2}{v,o1}
\fmfdot{v}
\fmf{phantom,tag=1}{i1,v}
\fmfposition
\fmfipath{p[]}
\fmfiset{p1}{vpath1(__i1,__v)}
\fmfi{plain}{subpath (0,length(p1)*0.40) of p1}
\fmfi{dashes}{subpath (length(p1)*0.4,length(p1)) of p1}
\fmf{plain,tension=0.9}{v,v}
\fmfiset{p2}{vpath2(__v,__o1)}
\fmfi{dashes}{subpath (0,length(p2)*0.60.) of p2}
\fmfi{plain}{subpath (length(p2)*0.6,length(p2)) of p2}
\fmfv{label=$x_1$,label.angle=-90}{i1}
\fmfv{label=$x_2$,label.angle=-90}{o1}
\end{fmfgraph*}
\end{fmffile}
\end{minipage}\,\,\,\,\Bigg].
\end{equation} Summing up the two relations above we get (\ref{FDT(1)}) which
is the FDT to first order in perturbation theory.

Both, the momentum and the super-field techniques can be used to
go to higher order in the FDT. The calculations turn out to be
simpler than in standard space, but still rather tedious, so
we will avoid to report them here.

This last part of this appendix has been somehow a demonstration of the many things one can prove by
"manipulating" the Feynman diagrams without resorting to their analytical expressions. These manupulations are "analogous" to what was done in the original "{\it diagrammar}" paper \cite{Hooftetal} for Quantum Field Theory, so we hope the use we made of the same word "{\it diagrammar}" is forgiven.



\end{document}